\documentclass[3p,times,sort&compress]{elsarticle}

\graphicspath{{Figures/}}

\usepackage{afterpage}
\usepackage{amsfonts}
\usepackage{amsmath}
\usepackage{amsthm}
\usepackage{amssymb}
\usepackage{bm}
\usepackage{bbm}
\usepackage{color}
\usepackage{dsfont}
\usepackage{graphicx}
\usepackage{hyperref}
\usepackage{mathbbol}
\usepackage{multirow}
\usepackage{pdflscape}
\usepackage{relsize}
\usepackage{slashed}
\usepackage{subcaption}
\usepackage{url}
\usepackage{xspace}
\usepackage{bbding}

\usepackage{fancyhdr}

\def\be{\begin{equation}}
\def\ee{\end{equation}}
\def\bea{\begin{eqnarray}}
\def\eea{\end{eqnarray}}

\newcommand{\smalltitle}[1]{\textbf{\emph{#1}} ---\xspace}

\parskip=\baselineskip

\def\1{{\bf 1}}
\def\2{{\bf 2}}
\def\3{{\bf 3}}
\def\4{{\bf 4}}

\def\tilde{\widetilde}

\newcommand{\bbb}[1]{\boldsymbol{#1}}

\DeclareMathOperator*{\SumInt}{%
\mathchoice%
  {\ooalign{$\displaystyle\sum$\cr\hidewidth$\displaystyle\int$\hidewidth\cr}}
  {\ooalign{\raisebox{.14\height}{\scalebox{.7}{$\textstyle\sum$}}\cr\hidewidth$\textstyle\int$\hidewidth\cr}}
  {\ooalign{\raisebox{.2\height}{\scalebox{.6}{$\scriptstyle\sum$}}\cr$\scriptstyle\int$\cr}}
  {\ooalign{\raisebox{.2\height}{\scalebox{.6}{$\scriptstyle\sum$}}\cr$\scriptstyle\int$\cr}}
}

\let\Im\relax
\DeclareMathOperator{\Im}{Im} 
\let\Re\relax
\DeclareMathOperator{\Re}{Re} 
\DeclareMathOperator{\disc}{disc} 

\newcommand*{\vcenteredhbox}[1]{\begingroup
\setbox0=\hbox{#1}\parbox{\wd0}{\box0}\endgroup}

\newcommand{\XYZ}{\ensuremath{XYZ}\space}
\newcommand{\PP}{\ensuremath{P}\xspace}
\newcommand{\XYZP}{\ensuremath{XYZ\PP}\space}

\DeclareMathOperator{\tr}{Tr}

\newcommand{\babar}{\mbox{\slshape B\kern-0.1em{\smaller A}\kern-0.1em
    B\kern-0.1em{\smaller A\kern-0.2em R}}\xspace}
\newcommand{\belle}{Belle\xspace}
\newcommand{\lhcb}{LHCb\xspace}
\newcommand{\bes}{BES~III\xspace}
\newcommand{\cdf}{CDF\xspace}
\newcommand{\cms}{CMS\xspace}
\newcommand{\atlas}{ATLAS\xspace}
\newcommand{\Dzero}{D$\varnothing$\xspace}
\newcommand{\cleoc}{CLEO-$c$\xspace}
\newcommand{\cleo}{CLEO\xspace}
\newcommand{\alice}{ALICE\xspace}

\newcommand{\sectionname}[1]{Section #1\xspace}
\newcommand{\sectionsname}[1]{Sections #1\xspace}
\renewcommand{\tablename}[1]{Table #1\xspace}
\newcommand{\tablesname}[1]{Tables #1\xspace}

\newcommand{\evnospace}{\ensuremath{\mathrm{e\kern -0.1em V}}\xspace}
\newcommand{\mevnospace}{\ensuremath{{\mathrm{Me\kern -0.1em V}}}\xspace}
\newcommand{\ev}{\ensuremath{\mathrm{\,e\kern -0.1em V}}\xspace}
\newcommand{\kev}{\ensuremath{{\mathrm{\,ke\kern -0.1em V}}}\xspace}
\newcommand{\mev}{\ensuremath{{\mathrm{\,Me\kern -0.1em V}}}\xspace}
\newcommand{\gev}{\ensuremath{{\mathrm{\,Ge\kern -0.1em V}}}\xspace}
\newcommand{\gevcc}{\ensuremath{{\mathrm{\,Ge\kern -0.1em V}}/c^2}\xspace}
\newcommand{\tev}{\ensuremath{{\mathrm{\,Te\kern -0.1em V}}}\xspace}

\newcommand{\nb}{\ensuremath{{\mathrm{\,nb}}}\xspace}
\newcommand{\fm}{\ensuremath{{\mathrm{\,fm}}}\xspace}
\newcommand{\jpsi}{\ensuremath{{J\mskip -3mu/\mskip -2mu\psi\mskip 2mu}}\xspace}
\newcommand{\psiprime}{\ensuremath{\psi(2S)}\xspace}

\newcommand{\ie}{{\it i.e.}\xspace}
\newcommand{\etal}{\xspace{\it et al.}\xspace}
\newcommand{\eg}{{\it e.g.}\xspace}
\newcommand{\BR}{\ensuremath{\mathcal{B}}\xspace}

\newcommand{\pt}{\ensuremath{p_\perp}\xspace}
\newcommand{\CL}{C.L.\xspace}
\newcommand{\lhs}{{\it lhs}\xspace}
\newcommand{\rhs}{{\it rhs}\xspace}

\newcommand{\Bz}{\ensuremath{B^0}\xspace}
\newcommand{\figref}[1]{(\figurename{~\ref{#1}})\xspace}
\newcommand{\spect}[4][]{\ensuremath{\ifthenelse{\equal{#1}{}} {} {#1\,} {}^{#2\!} {#3}_{#4}}}

\def\DeltaE     {\ensuremath{{\rm \Delta}E}\xspace}

\def\D       {\ensuremath{D}\xspace}
\newcommand{\epem}{\ensuremath{e^+e^-}\xspace}
\def\Dbar    {\kern 0.2em\bar{\kern -0.2em D}{}\xspace}
\def\Db      {\ensuremath{\Dbar}\xspace}

\def\Dz      {\ensuremath{D^0}\xspace}
\def\Dzb     {\ensuremath{\Dbar^0}\xspace}
\def\DzDzb   {\ensuremath{\Dz {\kern -0.16em \Dzb}}\xspace}
\def\Dp      {\ensuremath{D^+}\xspace}
\def\Dm      {\ensuremath{D^-}\xspace}

\def\DpDm    {\ensuremath{\Dp {\kern -0.16em \Dm}}\xspace}
\def\Dstar   {\ensuremath{D^*}\xspace}
\def\Dstarb  {\ensuremath{\Dbar^*}\xspace}
\def\Dstarz  {\ensuremath{D^{*0}}\xspace}
\def\Dstarzb {\ensuremath{\Dbar^{*0}}\xspace}
\def\Dstarp  {\ensuremath{D^{*+}}\xspace}
\def\Dstarm  {\ensuremath{D^{*-}}\xspace}

\def\Kstar  {\ensuremath{K^{*}}\xspace}
\def\Kp  {\ensuremath{K^{+}}\xspace}
\def\Km  {\ensuremath{K^{-}}\xspace}
\def\Kz  {\ensuremath{K^{0}}\xspace}
\def\kaon  {\ensuremath{K}\xspace}
\def\Bbar    {\kern 0.2em\bar{\kern -0.2em B}{}\xspace}
\def\B       {\ensuremath{B}\xspace}
\def\Bp  {\ensuremath{B^{+}}\xspace}
\def\Bz  {\ensuremath{B^{0}}\xspace}
\def\Bzb     {\ensuremath{\Bbar^0}\xspace}
\def\Bstar   {\ensuremath{\B^*}\xspace}

\def\Dsp      {\ensuremath{D^+_s}\xspace}
\def\Dsm      {\ensuremath{D^-_s}\xspace}
\def\Dssp     {\ensuremath{D^{*+}_s}\xspace}
\def\Dssm     {\ensuremath{D^{*-}_s}\xspace}

\newcommand{\Lzb}{\ensuremath{\Lambda_b^0}\xspace}

\def\KS    {\ensuremath{K^0_{\scriptscriptstyle S}}\xspace} 
\def\Kstarz  {\ensuremath{K^{*0}}\xspace}
\def\ccbar {\ensuremath{c\bar c}\xspace}

\def\Lbar     {\kern 0.2em\overline{\kern -0.2em\Lambda\kern 0.05em}\kern-0.05em{}\xspace}

\def\bpm{\begin{pmatrix}}
\def\epm{\end{pmatrix}}

\begin{document}

\begin{frontmatter}

\numberwithin{equation}{section}

\title{Multiquark Resonances}

\author[uno]{A. Esposito}
\author[due]{A. Pilloni}
\author[tre,quattro]{A.~D. Polosa}
\address[uno]{Physics Department, Center for Theoretical Physics 
\& Institute for Strings, Cosmology, and Astroparticle Physics,\\
  Columbia University, New York, NY 10027, USA}
\address[due]{Theory  Center,  Thomas  Jefferson  National  Accelerator  Facility, 
12000  Jefferson  Avenue,  Newport  News,  VA  23606, USA}
\address[tre]{Dipartimento di Fisica and INFN, ``Sapienza'' Universit\`a di Roma, P.le A. Moro 2, I-00185 Roma, Italy}
\address[quattro]{CERN, Theory Department, Geneva 1211, Switzerland}

\begin{abstract}
Multiquark resonances are undoubtedly experimentally observed. The number of states and the amount of details on their properties have been  growing over the years. It is very recent the discovery of two pentaquarks and the confirmation of four tetraquarks,  two of which had not been observed before. We mainly review the theoretical understanding of this sector of particle physics phenomenology and present some considerations
attempting a coherent description of the so called  $X$ and $Z$ resonances. The prominent problems plaguing theoretical models, like the absence of selection rules limiting the number of  states predicted,  motivate new directions in model building. Data are reviewed going 
through all of the observed resonances with particular attention to their common features and the purpose of providing a starting point to further research. 
\end{abstract}


\end{frontmatter}

\setcounter{tocdepth}{2}
\tableofcontents

\section{Introduction}
Yet another review on tetraquarks, pentaquarks, and all that?

Indeed this paper comes after a number of encyclop{\ae}dic reviews on the theme appeared since the discovery of the $X(3872)$ ---  see~\cite{Swanson:2006st,Drenska:2010kg,Brambilla:2014jmp,Esposito:2014rxa,Lebed:2016hpi} and the latest, very comprehensive~\cite{Chen:2016qju}.

Despite the considerable amount of information provided by these articles, what presented here is a further attempt to report more specifically on those efforts made to find  simple theoretical descriptions, even though incomplete, which include and explain, in a unitary picture, most of the exotic resonances observed at the time of this writing. Along these lines, some new arguments and work in progress will also be presented.

We assume that there is a common theoretical description of the $X^0(3872)$, $Z_c^{0,\pm}(3900)$,  $Z_c^{\prime 0,\pm}(4020)$, $X^0(4140)$, $Z_b^{0,\pm}(10610)$, $Z_b^{\prime 0,\pm}(10650)$, $Z(4430)$ and the pentaquarks, and we collect  those ideas which appear to us to be functional to formulate such a comprehensive picture, even if it cannot yet be considered as complete and satisfactory. %

The non-observation of $X^\pm(3872)$, the isospin violating decay pattern of $X^0$, the absence of $X^{0,\pm}$ partners in the beauty sector, which challenge the compact tetraquark interpretation~\cite{Maiani:2004vq,Maiani:2014aja}, are taken as starting points of our analysis~\footnote{We assume here that the experimental situation is definitive on these matters, being understood that any variation on these data, would revolutionize our understanding of multiquark resonances.}. On the other hand, an obvious merit of compact tetraquark models was that of strongly motivating the research on charged resonances, which were eventually copiously observed, contrary to most expectations.

The evident proximity of meson-meson thresholds to the observed states, which appears as an impressive fine-tuning in the case of the $X^0(3872)$, is also taken as a serious indication which, in the light of available data, seems unlikely to be accidental.

Meson molecule models have the same problem with the proliferation of states, despite of the attempts made to solve it. Moreover they have definitive problems at explaining prompt production at hadron colliders, especially in the case of the $X$. \emph{Bona fide} molecules like the deuteron are observed at hadron colliders, but only at very low transverse momenta, contrary to the $X$, which is detected at extremely large transverse momenta ($\pt \gtrsim 15\gev$). 
These are very serious features, often forgotten, which should not be ignored.

Because of these and other reasons, explained with more details in the text, we might conclude that none of these two models reaches a fully satisfactory or definitive description   of what observed in experiment, even though we must recognize a certain degree of success of both of them at explaining the features of some specific resonances.  

On the other hand, we are convinced that diquark degrees of freedom are essential for the construction of a theory of multiquark resonances and that most of the work done in refining the predictions of the diquarkonium picture must be part of it.  

What has  been missing in the diquarkonium model are sort of `selection rules' explaining the paucity of states which have been observed over the years. Same for the hadron molecules. Together with the presentation of hadron molecule and diquark-antidiquark models, we collect here some ideas we have been working on, which never appeared in an extensive discussion, to show one possible way to describe $X$ and $Z$ resonances as the result of an effective interaction in meson-meson channels leading to the formation of  resonant metastable states. This effective interaction is due to the presence of the discrete spectrum of diquarkonia immersed in the continuum spectrum of would-be molecules.

Special conditions are needed to switch on this interaction, but once they are met, sharp resonances are expected to be produced. These conditions strongly restrict  the number of {\it possible} states and we have used the best experimentally assessed tetraquark resonance to test the picture.  This discussion is presented in \sectionname{\ref{mol3}} and the arguments there formulated are admittedly not yet fully satisfactory, although showing rather suggestive aspects.   

The paper is organized as follows.

We first analyze the core of ideas and methods at the basis of 
meson molecule models, starting from those textbook results on low energy scattering which 
we found useful to set the frame for the arguments to be developed. We criticize the limits of the hadron molecule  approach, and discuss just some of the various technical solutions employed. This is done in \sectionsname{\ref{shallowbs} and \ref{mol2}.}

In \sectionname{\ref{diquark}} we present the diquarkonium model and the methods developed to obtain a spectrum of states which remarkably reproduces the $X,Z,Z^\prime$ mass pattern. The limitation of this model resides in the number of states predicted which, at the time of this writing, have not been observed. If in some future the experimental picture will be revolutionized by the appearance of the large number of expected resonances, together  with $X^\pm$ states (which mysteriously evaded observation until now) {\it etc.}, the diquarkonium model alone will certainly be, in our view, the strongest and most grounded option to describe this physics. 

A more theoretical discussion of tetraquarks in the $1/N$ expansion~\cite{Maiani:2016hxw} adds  confidence on the specific role of diquarks as the right degrees of  freedom to treat the emergence of tetraquark poles in meson-meson amplitudes. This is discussed in \sectionname{\ref{tqln}}, which reproduces some results recently obtained. 

 This review might be read as a standalone research paper by skipping the first five sections.
 \sectionname{\ref{mol3}} stems from some work initiated in~\cite{Papinutto:2013uya} and especially in~\cite{Esposito:2016itg}. However incomplete it might appear, the approach presented appears to us as one of the possible routes which should be explored to overcome the phenomenological problems of $XYZ$ resonances. 

In addition to this, in \sectionname{\ref{others}} we have presented known results on some interesting approaches,  which are needed to offer a coherent picture of the work done in the field, going beyond the molecule picture of hadrons held together by nuclear forces. This is also done with the purpose of underscoring  the  efforts towards a `unified' approach to the description of the observed phenomenology. 

A report on the experimental situation can be found in the last part of the review, \sectionname{\ref{sec:experiment}}. We go through  resonance by resonance, referring to the theoretical interpretations presented in the paper and highlighting connections  and common features. The aim is always that of finding more or less hidden connections among the observed resonances.
This part is intended to be a  guide to the existing spectroscopy, especially meant for `model-building' purposes. We omit a number of details which have been presented in other reviews.

The discussion on pentaquarks is  limited to a brief report on what has already been done, especially in the diquark model. Pentaquarks were, before their discovery, a highly undesirable option for hadron molecules. If new resonances with tetraquark or exaquark quantum numbers will be discovered, and if the approach described in \sectionname{\ref{mol3}} will resist to new forthcoming data on tetraquarks, it might straightforwardly be applied to more complex hadron structures.  The rules for doing these steps are explained, as we understand them at the moment. 

Most likely to explain the nature of \XYZP resonances does not require `new physics', in the most 
common adopted meaning, but likely some conceptual leap in the use of `old' strong interaction dynamics. 
 We also do not think that $XYZ$ resonances are sort of nuclear chemistry phenomena as they also occur, as witnessed by the prompt production of the $X$, at very large transverse momentum.

We hope to offer in this paper some perspective on the field and hopefully a support to identify new research directions.

\subsection{Naming and conventions}
There is some confusion about the naming of the resonances. At the beginning, the letters $X$, $Y$ and $Z$ were used with no clear criteria.
The PDG decided to call $X(\text{mass in\mev})$ all the exotic quarkoniumlike states, with the exception of the $\chi_{c0}(3915)$ (initially $X(3915)$, then promoted to the ordinary charmonium name $\chi_{c0}(2P)$ in the PDG 2014, now assessed). However, a sort of convention is generally followed in the literature, which calls $Z(\text{mass})$ the charged quarkoniumlike states, $Y(\text{mass})$ the vector $1^{--}$ ones, $P(\text{mass})$ the pentaquarks, and $X(\text{mass})$ all the other ones. We will follow this convention, with the addition of the superscripts and subscripts used by \belle and \bes, to wit $Z_c, Z_c^\prime, Z_b, Z_b^\prime, Z_1, Z_2$, as well as $P_c$ for the hidden charm pentaquarks. We call $X(4140)$ the state seen in $\jpsi\phi$ (known in the past as $Y(4140)$), $\chi_{c0}(3915)$ the state seen in $\jpsi\omega$ (known in the past as $X(3915)$, $Y(3915)$, or $Y(3940)$), $Z(4430)$ the state seen in $\psiprime \,\pi$ (sometimes called $Z(4475)$).

The charged conjugated modes are always understood, unless specified. The convention for the phases is such that $C \left| D^{0} \right\rangle = \left| \bar D^{0} \right\rangle$, while $C \left| D^{*0} \right\rangle = -\left| \bar D^{*0} \right\rangle$. The reader has to pay attention to different conventions used in the literature.

With a little abuse of notation, we will talk of $C$-parity for charged states, meaning ``the $C$-parity eigenvalue of the neutral isospin partner''. 

We do not discuss the former $Z(3930)$ and the $X(3823)$, which are good candidates for the ordinary charmonia $\chi_{c2}(2P)$ and $\psi_2(1D)$, respectively.

\subsection{A very brief survey of the experimental picture}
A detailed analysis of the experimental status of multiquark resonances will be presented in the last section of this report, together with their theoretical interpretations and references to experimental analyses.

Here we shall briefly review the basic experimental facts about the most compelling \XYZ resonances to set the stage for the discussion to follow. 

The first clearly exotic multiquark resonance is the $Z(4430)$, first claimed by \belle in 2007, but confirmed only in 2014 by the \lhcb collaboration. It was observed in the 
\begin{equation}
\bar B^0\to K^-  (\psiprime\,\pi^+)
\end{equation}
channel, \ie $Z(4430)\to \psiprime\,\pi^+$. The $s$ quark from the weak $b\to c\bar c s$ transition makes a $K^-$ with a $\bar u$ quark from a vacuum $u\bar u$ pair. The remaining $u$ and $ \bar d$ quarks, together with the $c\bar c$ pair, constitute the valence of the $(\psiprime\,\pi^+)$ resonance. How does hadronization work in this process? Is the $(\psiprime\,\pi^+)$ resonance, otherwise dubbed $Z(4430)$, a compact four-quark hadron, like a baryon or a meson, but with a different quark skeleton? Or is it just a hadron molecule kept bound for a finite lifetime by long range residual strong interactions? 

In 2013 another resonance, the $Z_c(3900)$, was observed simultaneously by  \bes and \belle, as a decay product of the $Y(4260)$
\begin{equation}
Y(4260)\to \pi^+ (\jpsi\,\pi^-)
\end{equation}
with $Y(4260)$ being a $J^{PC}=1^{--}$ {\it neutral} state, produced in initial state radiation in $e^+e^-$ collisions. $Y(4260)$ is also a multiquark resonance candidate, albeit neutral, with a $c\bar c$ quark pair in its valence.  Therefore the $(\jpsi\,\pi^-)$ resonance, otherwise dubbed $Z_c(3900)$, has again a minimal valence quark content of four, and $J^{PC}=1^{+-}$.

The $Z_c(3900)$ appears in the three states of charge, and the same occurs for $Z_c^\prime(4020)$, another, slightly heavier $J^{PC}=1^{+-}$ resonance, also found in \bes data and also unequivocally exotic.

The most aged of these resonances is however the {\it neutral} $X(3872)$, first observed by \belle in 2003 in the decays of the $B$ meson, and then confirmed by all collider experiments. This can be also be found in the decay product  
\begin{equation}
Y(4260)\to \gamma\, X(3872)
\end{equation}
as well as promptly produced at the vertex of hadronic collisions. The $X(3872)$ has $J^{PC}=1^{++}$ quantum numbers, as confirmed with a high degree of precision. This resonance encodes some very problematic features
\begin{enumerate}
\item It does not have charged partners (so far);
\item Its mass is almost perfectly fine-tuned to the mass of the $D^0\bar D^{0*}$ meson pair;
\item It decays into $\jpsi\,\rho$ and $\jpsi\,\omega$ with almost the same branching fraction;
\item It is an extremely narrow resonance, its width being $\Gamma\lesssim 1\mev$;
\item It is almost degenerate to the $J^{PC}=1^{+-}$ $Z_c(3900)$ resonance.
\end{enumerate}
 
The absolute neutralities of the $X(3872)$ and $Y(4260)$, which seem to be experimentally well assessed at the time of this writing, do not speak loud as a signal of a four-quark structure, as it is instead the case for $Z(4430)$, $Z_c$ and $Z_c^\prime$. On the other hand we have to observe that a charmonium cannot decay violating isospin. A more complex quark structure is therefore needed. The spectacular vicinity to the $D^0\bar D^{0*}$ threshold has suggested to many that the $X(3872)$ is a $D^0\bar D^{0*}$ loosely bound molecule. It is from this latter picture that we will start our discussion on models of $XYZ$ resonances, in the next section. 
 
Summarizing, in the charm sector we have $X(3872), Z_c(3900), Z_c^\prime(4020)$ and $Z(4430)$. The latter might not be included among the lowest states --- it can be considered as a radial excitation of $Z_c(3900)$.
In addition we have the $Y(4260)$, which might be considered as an orbital excitation of $X(3872)$.

Therefore, on the basis of some elementary theoretical assumptions, let us stick to the three resonances  $X(3872)$, $Z_c(3900)$, $Z_c^\prime(4020)$, reminding that the $X(3872)$ appears only in the neutral state of charge, differently from the other two. Is this pattern repeated in the beauty sector? The answer that we can give to this question, on the basis of present data, is `not entirely'.

Indeed a pair of resonances have been found in the $b$-sector, $Z_b(10610), Z_b^\prime(10650)$, very close to where expected on the basis of simple quark mass considerations --- and again very close to the $B^{(*)}\bar B^*$ thresholds --- and in the three states of charge. But no trace of a neutral or charged $X_b$ has been observed so far. 

Strange valence quarks can be found in another resonance, first observed in 2009 by the \cdf collaboration in
\begin{equation}
B\to K\,(\jpsi\,\phi)
\end{equation}
named $X(4140)$. The quark content is $c\bar cs\bar s$ and very recently the \lhcb collaboration has confirmed its existence and discovered similar resonances at higher masses.

In late 2015, two charged pentaquarks were observed in $\Lambda_b$ baryon decays
\begin{equation}
\Lambda_b\to K^-\,(J/\psi\,p)
\end{equation}
the lowest lying with  $J^P=3/2^-$ quantum numbers and the next with $J^P=5/2^+$, with masses at 4380\mev and 4550\mev respectively. The exotic nature of pentaquarks, sometimes dubbed as $\PP_c$, is very clear, as it is the case for the $Z_c$s and $Z_b$s.

On the basis of what said, we might also observe that low lying four-quark states appear with positive parities whereas five-quark ones with negative. 

\tablename{\ref{tab:shortexp}} enumerates the properties of the states we briefly discussed here.
To our understanding, those listed are the more solid multiquark resonances on experimental grounds.  Probably the amount of data available at this time is  not yet sufficient to  fully understand the nature of \XYZP resonances, but this is the challenge undertaken by many, which we mean to review in this paper.    
\begin{table}[t]
\centering
\begin{tabular}{lrclc}\hline\hline
State & $M$ (\mevnospace) & $\Gamma$ (\mevnospace)& $(I^G) J^{PC}$ & Details \\
\hline
      $X(3872)$ & $3871.69\pm0.17$ & $<1.2$ & $1^{++}$ & \sectionname{\ref{sec:X3872}}\\
      $Z_c(3900)^+$ & $3888.4\pm1.6$ & $27.9\pm2.7$ & $(1^+)1^{+-}$ & \sectionname{\ref{sec:Zc}}\\
      $Z_c^\prime(4020)^+$ & $4023.9\pm2.4$ & $10\pm6$ & $(1^+)1^{+-}$ & \sectionname{\ref{sec:Zc}}\\
      $Y(4260)$ & $4251\pm9$ & $120\pm12$ & $(0^-)1^{--}$ & \sectionname{\ref{sec:vectors}}\\
      $Z(4430)^+$ & $4478\pm17$ & $180\pm 31$ & $(1^+)1^{+-}$ & \sectionname{\ref{sec:Z4430}}\\
      $X(4140)$ & $4146.5^{+6.4}_{-5.3}$ & $83^{+30}_{-25}$ & $(0^+)1^{++}$ & \sectionname{\ref{sec:ccss}}\\
      $Z_b(10610)^+$ & $10607.2\pm2.0$ & $18.4\pm2.4$ & $(1^+)1^{+-}$ & \sectionname{\ref{sec:Zb}}\\
      $Z_b(10650)^+$ & $10652.2\pm1.5$ & $11.5\pm2.2$ & $(1^+)1^{+-}$ & \sectionname{\ref{sec:Zb}}\\
      \hline\hline
\end{tabular}
\caption{A brief summary of the exotic states we will discuss in the following. For more details, see \sectionname{\ref{sec:experiment}}, or the subsections devoted to the single particle. Because of isospin breaking, the $I^G$ quantum numbers are not defined for the $X(3872)$. With a slight abuse of notation, we will refer to the $C$ eigenvalue for charged states, as to ``the $C$ eigenvalue of the neutral isospin partner''. For the $Z_c(3900)$, we update the PDG 2014 average with the more precise results by \bes. }
\label{tab:shortexp}
\end{table}

\section{Scattering in presence of shallow bound states\label{shallowbs}}
In this  and in the following \sectionname{\ref{mol2}}, we will collect some known results from low energy scattering theory and  present, especially in~\sectionname{\ref{mol2}},  some arguments on the limitations of the loosely bound hadron molecule approach to the interpretation  of the $X(3872)$ resonance. 

The $D^0\bar D^{0*}$ molecule interpretation of the $X(3872)$ is however  very popular, and is functional to explain the $X$ isospin violation pattern.  Thus we start from our discussion on models of multiquark resonances from this interpretation.  Some general results rederived here will be of use also in other sections, when discussing the alternative approaches to the hadron molecule.

\subsection{Shallow bound states: the phase shift from wave mechanics}
\begin{figure}[t]
 \begin{center}
   \includegraphics[width=9truecm]{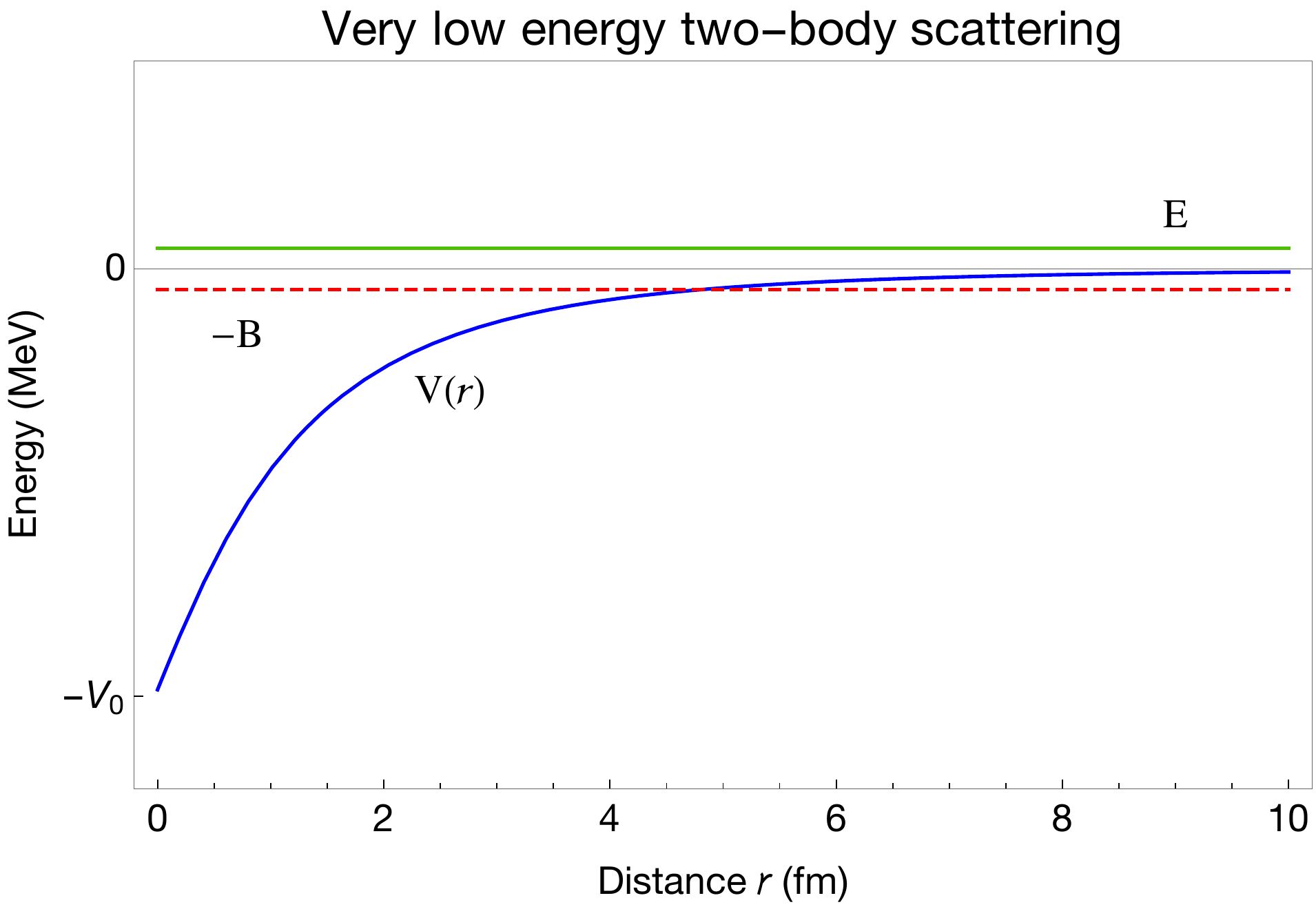}
 \end{center}
\caption{Scheme of the very low energy ($E\approx 0 $) meson-meson scattering  in a potential $V$, large-distance residual of strong interactions inside hadrons. Assume $V\gg B,E$ within its $r_0$ range, where $-B$ is a discrete level of $V$, assumed to be attractive throughout the entire $r$ range. Here it is assumed that $L=0$: no centrifugal barrier. \label{pot}}
\end{figure}

Consider two open-charm (or beauty) mesons $a=\{D^0,B^0\}$ and $b=\{\bar D^{*0},\bar B^{*0}\}$ interacting at low energies through a potential $V$ which is a weak, large distance residual of strong interactions into hadrons. Assume that $V$ allows a discrete level at $-B$ with $B\approx 0$ as in the scheme in \figurename{~\ref{pot}}. What is the effect of this bound state level on the $ab\to ab$ scattering?

The $a$ and $b$ mesons are approaching to each other from large separation $r$, where we assume they have a total energy $E\approx 0$ and $V(r\to \infty)=0$: the system with reduced mass $m=m_a m_b/(m_a+m_b)$ moves to small $r$ values, see~\figurename{~\ref{pot}}. The radial wave function $ \chi(r)=r\,R_{\ell=0}(r)$ of the system at large separation is 
\begin{equation}
\chi_{II}(r)=A\sin(kr +\delta_0)
\label{regio2}
\end{equation}
where the subscript $II$ defines a region of $r$ such that $r\gg r^*$, $r^*$ being a point where  the wave functions at short and large separations have to be matched. The region on the left of $r^*$ will be named region $I$ and $r^*$ is assumed to be quite larger than the effective range $r_0$ of  $V$ but also quite smaller than $1/k$.  The value of $k$ is $k=\sqrt{2m E}\approx 0$ and the very long-wavelength in~\eqref{regio2} allows the approximation
\begin{equation}
\left. \frac{\chi_{II}^\prime}{\chi_{II}}\right|_{r^*}\simeq \left. \frac{\chi_{II}^\prime}{\chi_{II}}\right|_{r=0}=k\cot \delta_0
\label{matchright}
\end{equation}
The determination of the phase-shift $\delta_0$ (which we will calculate in absence of orbital angular momentum, $\ell=0$) fully defines the scattering amplitude~\footnote{The scattering length $a_s$ is defined as limit $\lim_{k\to 0}k\cot \delta=-1/a_s$ so that the potential scattering cross section, in the same limit,  is given by $\sigma=4\pi a_s^2$.}
\begin{equation}
f(ab\to ab)=\frac{e^{2i\delta_0}-1}{2ik}=\frac{1}{k\cot\delta_0-i k }
\label{scattlength}
\end{equation}

The logarithmic derivative in~\eqref{matchright} has to be matched with $\chi^\prime_I/\chi_I$.  In region $I$ the Schr\"odinger equation is 
\begin{equation} \label{EqregionI}
\chi^{\prime\prime}_I-2 m V\chi_I=0\quad {\rm with}\quad  \chi(0)=0
\end{equation}
because $V\gg B, E$ all over $I$. Eq.~\eqref{EqregionI} means that the ratio $\chi^\prime_I/\chi_I$ will be independent of energy and,  in particular, it will
stay the same also at some small negative energy $\sim-B$. In that case
\begin{equation}
\left. \frac{\chi_{I}^\prime}{\chi_{I}}\right|_{r^*} \simeq 
\left. \frac{(Ce^{-\kappa r})^\prime}{Ce^{-\kappa r}}\right|_{r^*}=-\kappa=-\sqrt{2mB}
\label{matchleft}
\end{equation}
where we took $\chi_I$ to be the wave function of the stationary state with binding energy $-B$.
Therefore realizing the matching condition, {\it i.e.} equating~\eqref{matchright} and~\eqref{matchleft} we get the `universal' result
\begin{equation}
\cot \delta_0 =-\sqrt{\frac{B}{E}}
\label{ppresult}
\end{equation}

Both $B$ and $E$ are supposed to be {\it close to zero}, but in a fixed, finite, ratio. The phase shift determined does not depend on the details of the unknown potential $V(r)$, but only on its shallow discrete level at $-B$.   Any potential which allows a shallow bound state at $-B$ will have   the same scattering amplitude  at low energies. Since  strong interactions between color singlets are expected to be weak and no definite form for their potential exists, it is natural to resort to the results here described, whenever the recoil energies of the color singlets are small enough to allow the approximations needed.  

\subsection{Shallow bound states: the phase shift from scattering theory}
The same result~\eqref{ppresult}, obtained in wave mechanics as in~\cite{Landau3}, can be re-derived in scattering theory~\footnote{We follow closely the derivation in S.~Weinberg~\cite{Weinbergqm}  selecting those key  results which are most useful to our discussion.}. The advantage of the following derivation is that it allows to study directly the overlap between a superficial bound state in some potential $V$ and the continuous spectrum scattering state.

The Lippmann-Schwinger equation for the scattering in-state $\Psi^+_\alpha$ %
is 
\begin{equation}
\Psi^+_\alpha=\Phi_\alpha +\frac{1}{E_\alpha-H_0+i\epsilon}V\, \Psi^+_\alpha
\label{ls}
\end{equation}
which is the  solution of
\begin{equation}
(H_0+V)\,\Psi^+_\alpha=E_\alpha\,\Psi^+_\alpha
\end{equation}
given that the unperturbed free states $\Phi_\alpha$ are defined by
\begin{equation}
(E_\alpha-H_0)\,\Phi_\alpha=0
\end{equation}
Here $\alpha$ is a compound index labeling types and numbers of particles in the state, momenta and  spin 3-components. %
We consider $V$ in the adiabatic approximation, \ie $V \to 0$ for $t \to -\infty$. This implies that $\Psi^+_\alpha = \Phi_\alpha$ in the far past $t\to -\infty$, and that the energy eigenvalue $E_\alpha$, which is independent of time, is the same for the two states.

Equation~\eqref{ls} can be multiplied on the left by $V$
\begin{equation}
V\,\Psi^+_\alpha=V\,\Phi_\alpha +V\frac{1}{E_\alpha-H_0+i\epsilon}V\, \Psi^+_\alpha
\label{vls}
\end{equation}
and a formal solution of the latter equation is 
\begin{equation}
V\,\Psi^+_\alpha=T(E_\alpha+i\epsilon)\,\Phi_\alpha
\label{sol1}
\end{equation}
where 
\begin{equation}
T(W)=V+V\frac{1}{W-H_0}T(W)
\label{sol2}
\end{equation}
as is verified by substituting~\eqref{sol1} into~\eqref{vls} and  using~\eqref{sol2} in the resulting  \lhs of~\eqref{vls}.  The expression in~\eqref{sol2} can be solved obtaining
\begin{equation}
T(W)=\left(1- V \frac{1}{W-H_0}\right)^{-1} V=V (W-H_0) \frac{1}{W-H}= V+V\frac{1}{W-H}V
\label{solls}
\end{equation}
where in the last equality one uses $(W-H_0)=(W-H+V)$. The in-state $\Psi^+_\alpha$, which looks like a free particle state at $t\to -\infty$, at later times $t\to +\infty$ looks like~\footnote{Given that out-states $\Psi^-_\alpha$ look like $\Phi_\alpha$ at late times}
\begin{equation}
\Psi^+_\beta=\int_\alpha S_{\alpha\beta}\, \Phi_\alpha
\label{psipls}
\end{equation}
where the $S$-matrix is found to be~\footnote{Substitute~\eqref{smatrix} and~\eqref{tls} into~\eqref{psipls} finding \mbox{$\Psi_\beta^+=\Phi_\beta-2\pi i \int_\alpha\delta(E_\beta-E_\alpha)\langle\Phi_\alpha|V\, \Psi^+_\beta\rangle\Phi_\alpha$}. This is the same result that would be obtained by $i)$ integrating (in $\beta$) the Lippmann-Schwinger equation~(\ref{ls}) with  weights $f(\beta)\, e^{-i E_\beta t}$, where $f$ is a smooth function $ii)$ using completeness on $\Phi_\alpha$ in the second term on the \rhs of~(\ref{ls}) $iii)$ taking the  $t\to +\infty$ limit, {\it i.e.} computing the contour integral on the $E_\beta$ lower half plane, keeping $\int_\beta\delta(E_\beta-E_\alpha)\cdots$ to enforce the result of the residue theorem. The factor $e^{-i E_\beta t}$ is needed to perform the contour integral. }
\begin{equation}
S_{\alpha\beta}=\delta(\alpha-\beta)-2\pi i\, \delta(E_\beta-E_\alpha)\, T_{\alpha\beta}
\label{smatrix}
\end{equation}
and 
\begin{equation}
T_{\alpha\beta}\equiv\langle \Phi_\alpha | T  \,\Phi_\beta\rangle=\langle\Phi_\alpha|V \, \Psi_\beta^+
\label{tls}
\rangle
\end{equation}
according to~\eqref{sol1}. 

The second term on the \rhs of  Eq.~\eqref{solls} allows two terms by completeness: one can be obtained summing on bound states $\psi _i$, if any,  the other summing on in-states $\Psi^+_\gamma$. Thus we get the Low equation,
\begin{equation}
T_{\alpha\beta}(E_\alpha+i\epsilon)= V_{\alpha\beta}+\SumInt_{i} \frac{\langle \Phi_\beta|V\,\psi_i\rangle\langle \Phi_\alpha |V \, \psi_i\rangle^*}{E_\alpha-E_i}+\int_\gamma \frac{T_{\alpha\gamma} T_{\beta\gamma}^*}{E_\alpha-E_\gamma+i\epsilon}
\label{low}
\end{equation}
where we sum over both the discrete and continuum labels (like momentum) characterizing the bound states $\psi_i$ and we defined $V_{\alpha\beta}=\langle \Phi_\alpha|V\, \Phi_\beta\rangle$. The completeness relation over the states in the discrete and continuous spectra requires a contribution to the $T$ matrix also from bound states: the form $\Psi^+_\alpha$ will assume at later times $t\to +\infty$ (scattering) is affected by bound states in the potential $V$. We showed above that even not knowing the details of   $V(r)$, the scattering phase shift  can be determined if it is assumed that the potential admits a shallow bound state at $E_1=-B$ and scattering occurs at very low energy $E$. Following the derivation in~\cite{Weinbergqm} we will re-obtain the same result~\eqref{ppresult} for the phase shift by solving the Low equation~\eqref{low}, \ie determining $T_{\alpha\beta}$ from the equation above, under the same general assumptions on $V$ and $E$. 

At very low energy $E_\alpha$, and in presence of a single shallow bound state at $-B$, very close to the onset of the continuum, the second term on the \rhs of~\eqref{low} will be dominated by~\footnote{For the sake of comparison with~(\ref{boundfeshx}) we have the interesting case $\alpha=\beta$ in which $T_{\alpha\alpha}=\frac{|\langle\psi_i|V\,\Phi_\alpha\rangle |^2}{E_\alpha-E_i+i\epsilon}$.}  
\begin{equation}
\frac{\langle \Phi_\beta|V\, \psi_i\rangle\langle \Phi_\alpha|V\,\psi_i\rangle^*}{E_\alpha+B}
\label{bound}
\end{equation}
with $E_\alpha \approx B\approx 0$. This also dominates over transitions in the continuum spectrum $V_{\alpha\beta}$, thus we will neglect the latter.  This term is particularly relevant to  our discussion and will be recalled in \sectionname{\ref{mol3}}. 

By $\Phi_\alpha$ we mean a two-particle state with some total energy $E=E_\alpha$ relative to the rest frame and zero total momentum $\bm P$. 
In this discussion we will only consider $S$-wave states. To simplify the notation, we will not explicit spin degrees of freedom.  Therefore $\Phi_\alpha\equiv\Phi_{E,\bm P=0}$ in the center of mass of the two-particle system.  These states will  have the convenient normalization 
\begin{equation}
\langle\Phi_{E^\prime,\bm P^\prime}|\Phi_{E,\bm P}\rangle=\delta(E^\prime-E)\, \delta(\bm P^\prime-\bm P)\, 
\end{equation}
if 
\begin{equation}
\Phi_{E,\bm P}=\int d^3p\,\frac{1}{\sqrt{\mu |\bm p|}}\, \delta(E-E_1-E_2)\, \Phi_{\bm p\,;\, \bm P-\bm p}
\label{normaphi}
\end{equation}
where the latter two-particle states  $\Phi_{\bm p\,;\, \bm P-\bm p}$ have conventional continuous spectrum normalization
\begin{equation}
\langle\Phi_{\bm p^\prime_1 \,;\, \bm p^\prime_2}|\Phi_{\bm p_1\,;\, \bm p_2}\rangle=\delta(\bm p_1^\prime-\bm p_1)\, \delta(\bm p_2^\prime-\bm p_2)
\label{norma2p}
\end{equation}
Here $\mu=E_1E_2/(E_1+E_2)$ derives from the definition of relative velocity in the center of mass
\begin{equation}
v=\frac{|\bm p|}{E_1}+\frac{|\bm p|}{E_2}
\end{equation}
and in general $E_1=\sqrt{m_1^2+\bm p^2}$,  $E_2=\sqrt{m_2^2+(\bm P-\bm p)^2}$ ($\bm P=0$ in the center of mass). In the non-relativistic limit $\mu\equiv m$, the reduced mass~\footnote{
Indeed, using~\eqref{norma2p} we find that 
\begin{align}
\langle \Phi_{E^\prime,\bm P^\prime}|  \Phi_{E,\bm 0}\rangle &= \int d^3 p \,\frac{1}{\sqrt{\mu |\bm p|}} \, \delta(E^\prime -E_1-E_2)\int d^3 p^\prime \,\frac{1}{\sqrt{\mu |\bm p^\prime|}} \, \delta(E -E_1^\prime-E_2^\prime)  \langle \Phi_{\bm p\,;\, \bm P^\prime-\bm p}| \Phi_{\bm p^\prime\,;\, -\bm p^\prime}\rangle=\notag\\
&=  \int d^3 p \,\frac{1}{\mu |\bm p|} \, \delta(E^\prime -E_1-E_2)\,\delta(E -E_1-E_2)\,\delta(\bm P^\prime)=\notag\\
&\equiv \frac{\delta (\bm P^\prime)}{\mu} \int p\,dp\, \delta(E^\prime-E)\delta(E-E_1-E_2)= \frac{ \delta (\bm P^\prime)}{\mu}\,\delta(E^\prime-E)\int E_1\, dE_1 \, \delta(E-E_1-E_2)
\label{intx}
\end{align}
and the last integral gives exactly $\mu$ as can bee seen by replacing $E_2=\sqrt{E_1^2-m_1^2+m_2^2}$}.

The numerator in~\eqref{bound}, because of the $1/\!\sqrt{|\bm p|}$ factor in~\eqref{normaphi},  would be singular for very low recoils $|\bm p|\to 0$, which contradicts the analyticity requirement on the $S$-matrix. On the other hand  
\begin{equation}
\langle\Phi_{\bm p\, ;\, -\bm p}|V\, \psi_{\bm P}\rangle 
\label{prev}
\end{equation}
has to be an analytic function when $|\bm p|\to 0$ --- roughly speaking a constant  at $|\bm p|=0$. Here the bound state is defined by the total momentum $\bm P$ which in~(\ref{prev}) is set to be zero (see~(\ref{coupling})). Therefore using~\eqref{normaphi} in the determination of  
\begin{equation}
\langle\Phi_\beta|V\, \psi \rangle\equiv\langle\Phi_{E^\prime,\bm 0}|V\, \psi_{\bm P}\rangle 
\end{equation}
we find 
\begin{equation}
\langle\Phi_{E^\prime,\bm 0}|V\, \psi_{\bm P}\rangle \sim \, \delta(\bm P)\, \int \sqrt{p}\, E_1 dE_1\, \delta(E^\prime-E_1-E_2) =g\, \delta(\bm P)\, \sqrt{|\bm p(E^\prime)|}
\label{coupling}
\end{equation}
where $p=\sqrt{2\mu E}$ and $\mu$ is  the reduced mass~\footnote{We have also used the last integral in~\eqref{intx} and $\bm p^2/2m_1+\bm p^2/2m_2=E$}.  
Here  the constant $g$ measures the overlap of the bound state $\psi_{\bm P=0}$ to the continuum one $\Phi_{E^\prime,\bm 0}$. The constant $g$ can be determined as a function of $\mu$ and $B$ by substituting $V=H-H_0$ in~\eqref{coupling}. We report this solution below.

The bound state term~\eqref{bound} can therefore be rewritten as
\begin{equation}
\frac{\sqrt{p(E)p(E^\prime)}}{E+B}|g|^2\,\delta(\bm P)\, \delta(\bm P^\prime)
\end{equation}
The denominator is the energy gap between $E$ and $-B$ as in \figurename{~\ref{pot}}. The smaller this energy gap, the better is the approximation of neglecting $V_{\alpha\beta}$ in the Low equation, which otherwise would require the knowledge of $V$. Since we have in mind the residual strong interactions between color neutral hadrons, and have little clue on the explicit form of $V$, it is better to be in the approximation of small $E+B$ gap.

As for the $T$-matrix elements in the Low equation~\eqref{low} they are expressed in the center-of-mass of the two-particle system as
\begin{equation}
T_{\alpha\beta}=T_{E,\bm 0\, ;\, E^\prime, \bm P^\prime}\equiv  T(E,E^\prime)\, \delta(\bm P)
\end{equation}
so that~\eqref{low} becomes 
\begin{equation}
T(E,E^\prime)=\frac{\sqrt{p(E)p(E^\prime)}}{E+B}|g|^2 + \int_0^\infty dE^{\prime\prime} \frac{T(E,E^{\prime\prime})T^*(E^\prime,E^{\prime\prime})}{E-E^{\prime\prime}+i\epsilon}
\label{lowsolve}
\end{equation}
This version of the Low equation admits a solution 
\begin{align}
T(E,E^\prime)=&\sqrt{p(E)p(E^\prime)} \, t(E)\notag \\ 
t(E) =& \frac{|g|^2}{E+B} +\int_0^\infty dE^\prime \frac{p(E^\prime)}{E-E^\prime+i\epsilon} |t(E^{\prime})|^2
\label{te}
\end{align}
as can be seen by direct substitution in~\eqref{lowsolve}. The function $t(E)$ can be expressed in terms of $p(E)=\sqrt{2\mu E}$ as follows
\begin{equation}
t(E)=\left(\frac{E+B}{|g|^2}+(E+B)^2\int_0^\infty dE^\prime \frac{\sqrt{2\mu E^\prime}}{(E^\prime+B)^2(E^\prime-E-i\epsilon)}\right)^{-1}
\label{solinteq}
\end{equation}
The derivation of the previous solution is based on dispersion theory integrals and can be found in~\cite{Weinbergqm}. As explained there, it is not unique as it relies on some analyticity requirements. However, all the different solutions converge to Eq.~\eqref{solinteq} in the $B\to0$ limit.

The latter integral can be computed in the complex plane with a cut on the real positive axis and using the residue theorem at the poles $E^\prime=-B$ and $E^\prime = E+i\epsilon$. 
This calculation gives
\begin{equation}
t(E)=\left(\frac{E+B}{|g|^2}+\frac{\pi(B-E)}{2}\sqrt{\frac{2\mu}{B}}+i\pi\sqrt{2\mu E}\right)^{-1}
\end{equation}

As a last step towards the solution of the Low equation, let us analyze the coupling $g$ reconsidering 
Eq.~\eqref{coupling} 
\begin{equation}
g\, \delta(\bm P)\, \sqrt{|\bm p(E)|}=\langle\Phi_{E,\bm 0}|V\, \psi_{\bm P}\rangle = \langle\Phi_{E,\bm 0}|(H-H_0)\, \psi_{\bm P}\rangle=(-B-E) \langle\Phi_{E,\bm 0}| \psi_{\bm P}\rangle
\end{equation}
thus 
\begin{equation}
\langle\Phi_{E,\bm 0}| \psi_{\bm P}\rangle= - g\, \delta(\bm P)\, \frac{\sqrt{|\bm p(E)|}}{E+B}
\label{geqphipsi}
\end{equation}
On the other hand
\begin{equation}
\delta(\bm P)=\langle \psi_{\bm P}|\psi_{\bm 0}\rangle= 
\int_{E,\, \bm Q}\langle \psi_{\bm P}|\Phi_{E,\bm Q }\rangle\, \langle \Phi_{E,\bm Q}|\psi_{\bm 0}\rangle=
|g|^2 \delta(\bm P)\int_E \left( \frac{\sqrt{|\bm p(E)|}}{E+B}\right)^2
\end{equation}
or 
\begin{equation}
1=
|g|^2 \int_0^\infty dE \,\left( \frac{\sqrt{|\bm p(E)|}}{E+B}\right)^2
\end{equation}
with the solution~\footnote{More comments on this formula can be found in~\cite{Weinberg:1965zz}.} 
\begin{equation}
|g|^2=\sqrt{\frac{2 B}{\mu \,\pi^2}}
\label{bgrel0}
\end{equation}
which gives for the elastic $T$-matrix 
\begin{equation}
T(E,E)=\frac{1}{\pi}\frac{\sqrt{E}}{\sqrt{B}+i\sqrt{E}}
\label{soluzioneel}
\end{equation}
We observe here the $B\sim |g|^4$ dependency which will be reobtained later, in a similar form~(\ref{bgrel}), when connecting the binding energy of the $X(3872)$, interpreted as  a loosely bound molecule of open charm mesons, and the effective strong coupling regulating its decay in the constituent mesons.

The phase shifts for elastic scattering, in the center of mass,  are defined by the equation for the scattering amplitude $M$
\begin{equation}
M_{E,0\,;\,E,0}=e^{2i\delta(E)}-1= -2\pi i \, T(E,E)
\label{mmatrix}
\end{equation}
where 
\begin{equation}
S_{\alpha\beta}=\delta(\alpha-\beta)+ \delta(E_\alpha-E_\beta)\,\delta(\bm P_\alpha-\bm P_\beta)\, M_{\alpha\beta}
\end{equation}
to be compared with the expression in~\eqref{smatrix}.

Combining~\eqref{soluzioneel} and~\eqref{mmatrix} gives again~\eqref{ppresult} 
\begin{equation}
\cot \delta =-\sqrt{\frac{B}{E}}
\label{ppresult2} 
\end{equation}
where $B,E\to 0$ remaining in a fixed ratio.

We use this result on the phase shifts in the next section. We notice again that the formula obtained does not depend on the details of the potential, but only on the position of the bound state energy level $B$ compared to the energy $E$ of the scattering system.   This is particularly useful in the case of strong interactions for which we will not need to model the static potential between two color-neutral hadrons. We might simply assume that there is some residual strong interaction at long distances between two hadrons, like the $D^0$ and $D^{*0}$ mesons in our specific case, and that the potential is deep enough to accommodate a superficial (close to continuum) discrete level at energy $-B$.     

From Eq.~(\ref{bgrel0}) we see that the overlap between the bound state $\psi_{P=0}$ and the the continuum state $\Phi_{E^\prime,\bm 0}$ goes to zero as $B\to 0$.

\section{Loosely bound molecules\label{mol2}}
This section is  devoted to the molecular picture of the $X(3872)$, trying to cover 
most of its implications.
We will also  discuss how the relation between binding energy and total decay rate constrains the most simple hadron molecule interpretation, with particular reference to the case of the $X(3872)$, and we will remark that a deuteron-like interpretation of $X$ has to confront with data on the production of light nuclei in in $pp$ and Pb-Pb collisions at the LHC. 

\subsection{Relation between binding energy and the decay rate of $X(3872)$\label{enratex}}
We can use the result~\eqref{ppresult2} in~\eqref{scattlength}, or equivalently divide the \rhs of~\eqref{mmatrix} by $2ik$,  to obtain 
\begin{equation}
f(ab\to ab)=\frac{1}{\sqrt{2m E}} \frac{-\sqrt{E}}{\sqrt{B}+i \sqrt{E}}=-\frac{1}{\sqrt{2m}}\frac{\sqrt{B}-i\sqrt{E}}{E+B}
\label{unof}
\end{equation}
where $\mu=m$, the usual reduced mass. The scattering amplitude displays a pole at $E=-B$, and the cross section $4\pi |f|^2$ has a resonant character
\begin{equation}
\sigma=\frac{2\pi}{m}\frac{1}{E+B}
\label{xsecten}
\end{equation}
at $E\sim B\approx 0$. Within an  interval of time $\tau\sim 1/(E+B)$  the free particle can be `locked' in the finite motion region $I$ in  \figurename{~\ref{pot}} and behave temporarily as a bound state $c$. 
The value of $\sigma$ close to the resonance, is larger than $4\pi r_0^2$ (the `natural' size for the cross section), where $r_0$ is the range of the potential, because $k r_0\ll 1$ meaning $r_0^2\ll 1/E\sim \sigma$ at resonance.  
We limit our considerations to the $S$-wave cross section because we have in mind the problem of the $X(3872)$ as a resonance in the $DD^*$ scattering at low energies $k\to 0$ where $\delta_\ell\sim k^{2\ell+1}$.  Also, since $X$ has $1^{++}$  quantum numbers whereas $D$, $D^*$ are $0^-$ and $1^-$ respectively  a $P$-wave scattering is forbidden by parity, while the $D$-wave option would give a very small phase shift.

Here comes indeed the main point: in quantum mechanics the formation of a resonance requires the interplay between the attractive potential and a repulsive centrifugal barrier, especially if the potential is weak, as in the case of strong interactions between color singlets. This does not happen for an $S$-wave scattering, and hence it does not happen for the $X(3872)$. 

In presence of a centrifugal barrier, a `quasi-discrete' (metastable state) positive energy level at $\epsilon>0$ would be possible, allowing a  narrow resonance: the energy region in which $f_\ell\sim 1/k$, \ie large compared with $r_0$, has a relative width $\Delta E/\epsilon\sim k^{\2\ell-1}$ for $\ell\neq 0$~\footnote{At low energies $f_\ell\approx \delta_\ell/k\sim k^{2\ell}$. Thus, if we write $f_\ell=1/(g_\ell-i k)$,  the $g_\ell$ expansion {\it starts}  at $k^{-2\ell}$ ($k^{-2\ell}\gg k$), larger terms like $k^{-2\ell-1}\cdots$ are neglected. Writing $g_\ell$ as $g_\ell\sim B(E-\epsilon)+\cdots$,
with $B\sim k^{-2\ell}$, we are retaining the first two terms, $k^{-2\ell}$ and the subdominant $k^{-2\ell+1}$.  Interpret $\epsilon>0$ as the quasi-discrete energy level quoted above. In order to have $f_\ell\sim 1/k$, \ie large compared with $r_0$, we need $E\sim\epsilon$. %
This region has a width $\Delta E\sim k^{2\ell+1}$, and the relative width is $\Delta E/\epsilon=\DeltaE /k^2\sim k^{2\ell-1}$. 
We finally observe that the low energy behavior $\tan\delta_\ell\approx\delta_\ell\sim k^{2\ell+1}$ does not hold for $B=0$ in the case of~\eqref{ppresult} where $\tan\delta_0\to \infty$.}. 

Stated differently, even though the cross section enhancement~\eqref{xsecten}  might be possible (although $E\sim B$ is a rare event in hadron collisions with hard cuts, see \ref{appmol}), it cannot anyway generate {\it narrow} resonances  which would require higher partial waves. In the case of the $X(3872)$ at least $\ell=2$ would be required: but such partial waves do not give significant  contributions to the phase shifts when $k\to 0$. A loosely bound $DD^*$ hadron molecule in a high orbital angular momentum state would tend to be even larger than what it is in $S$-wave, damping the $J/\psi \, (\rho/\omega) $ decay rates.   

In quantum field theory, the process is $ab\to c\to ab$ where we associate the propagator function to the temporary bound state $c$. In this case, in place of the potential scattering description reported above, the interaction between $a$ and $b$ is determined by the coupling of the two components to $c$. The propagator pole dominates the scattering amplitude $M$ in~\footnote{ Dividing $M_{\alpha\beta}$  by $(2\pi)^3$ and including the wave function normalizations $1/\!\sqrt{2 {\cal E}}$ in Eq. (3.6.9) in Ref.~\cite{Weinbergqft}. Here $ {\cal E}$ is the total energy which in the non-relativistic approximation is $ {\cal E}=E+m_a+m_b$.  } 
\begin{equation}
f(\beta \to \alpha)=-\frac{1}{8\pi {\cal E}}M_{\alpha\beta}
\label{fqft}
\end{equation}
which in the standard relativistic formalism is defined by
\begin{equation}
S_{\alpha\beta}=\delta(\alpha-\beta)-i(2\pi)^4\delta^4(p_\alpha-p_\beta)\, M_{\alpha\beta}
\end{equation}
Eq.~\eqref{fqft} is for elastic two-to-two body scattering, computed with a relativistic formalism (and including the phase space calculation of the final particles). 
Therefore we can write
\begin{equation}
f(ab\to c\to ab)=\frac{1}{8\pi (m_a+m_b+E)} G^2\frac{1}{(p_a+p_b)^2-m_c^2}
\end{equation}
where $E$ has the same meaning given in the previous section: total energy relative to the rest mass. Since $E\ll m_{a,b}$ we will neglect it in the first factor on the \lhs. The constant $G$ here defines the strong coupling of the free components $ab$ to the bound state $c$, $G\equiv G_{abc}$. 

Recall now that 
\begin{align}
m_c&=m_a+m_b-B\\
(p_a+p_b)&\simeq m_a+m_b+E
\end{align}
so that 
\begin{equation}
(p_a+p_b)^2-m_c^2\simeq2(m_a+m_b)(E+B)
\end{equation}
where the very small terms  $B^2$ and $E^2$ have been neglected.
Therefore we have 
\begin{equation}
f(ab\to c\to ab)\simeq \frac{G^2}{16\pi (m_a+m_b)^2}\frac{1}{E+B}
\label{duef}
\end{equation}

Expressions~\eqref{unof} and~\eqref{duef} are two different descriptions of the same process whose cross section is obtained by $d\sigma/d\Omega=|f|^2$. The first one is  derived in low energy scattering theory. The second involves the propagation of a virtual state as in quantum field theory. The non-physical pole appears in both with the same power. Indeed, {\it at} the $E=-B$ pole we find a relation between   binding energy and strong coupling $G$ which can be written as~\footnote{A derivation of this formula, which is a straightforward consequence of the considerations in~\cite{Landau3} leading to~\eqref{unof},  can be found in~\cite{lastessa}.}
\begin{equation}
B\simeq \frac{G^4}{512\, \pi^2}\frac{m^5}{(m_am_b)^4}
\label{bgrel}
\end{equation}
The advantage of this expression is that the strong coupling constant $G$ is the same as that entering in the computation of the decay rate $c\to ab$, which in general is an accessible experimental information. Observe that the scaling relation between $B$ and $G$ 
\begin{equation}
B\sim G^4
\end{equation}
was already appreciated in the discussion of the Low equation (see~\eqref{bgrel0}), but in that case $g$ is normalized with the nonrelativistic convention: $G = 4\pi (m_a + m_b) g$. 
Notice that in  the calculation of the rate $\Gamma(c\to ab)=G^2\,\Phi/2 (m_a + m_b)$, $G$ has  dimensions of energy, the two body phase space $\Phi$ being dimensionless.  

We might suppose that the experimentally observed $X(3872)$ resonance is the effect of a shallow bound state in the potential $V$, describing the strong interaction within the $\bar D^0 D^{*0}$ meson pair when low energy scattering $\bar D^0 D^{*0}\to \bar D^0 D^{*0}$ occurs. As commented above, the $\bar D^0 D^{*0}$ system might temporarily behave as a bound state, with very loose binding energy $B\approx 0$. Indeed we know that $B=m_D+m_{D^*}-m_X\approx 0$. We found that a relation between $B$ and the strong coupling describing the decay of $X$ into its components $\bar D^0 D^{*0}$ holds.

In the case at hand $m_a=m_{D}$, $m_b=m_{D^*}$ and we define the strong coupling $G$ through~\footnote{Since $X$ has positive charge conjugation, the final state is $|f\rangle=(|D^0\bar D^{0*}\rangle -|\bar D^0 D^{0*}\rangle)/\!\sqrt{2}$. When extracting $g$ defined in~\eqref{coupl}  from data a factor of $\sqrt{2}$ has to be included: $G\to \sqrt{2}G$.} 
\begin{equation}
\langle D^0\bar{D}^{0*}(\epsilon,q)|X(\lambda,P)\rangle=G\,\lambda\cdot \epsilon^*
\label{coupl}
\end{equation}
\begin{figure}[htb!]
 \begin{center}
   \includegraphics[width=7truecm]{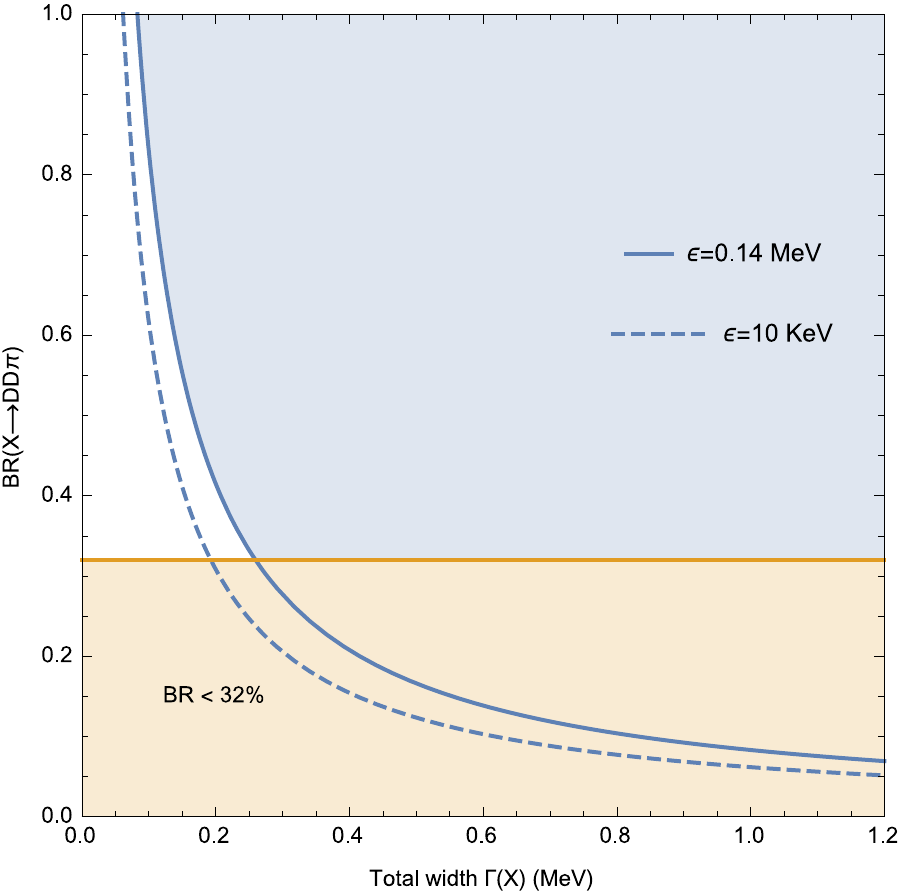}

\caption{Given the experimentally  excluded region (shaded), a loosely bound molecule, has total width $\Gamma_X$, branching ratio  $\BR(X\to \bar DD\pi)$ and binding energy $B=m_D+m_{D^*}-m_X$ as in the plot. The approximate values are used: $m_X=3871.66\mev$, $m_D=1864.84\mev$ and $m_{D^*}=2006.96\mev$ for $B=0.14\mev$ and a slightly modified value for $m_X=3871.79\mev$ in order to get $B=10$\kev. A measurement of the total width $\Gamma(X)$  ending in the shaded area would falsify the low energy resonance scattering mechanism described. \label{excluded}} \end{center}
\end{figure}
Taking into account the spin of  $D^*$ and $X$, one should rather use
\begin{equation}
G^2\to G^2\frac{1}{3}\left(2 + \frac{(m_X^2+m_{D^*}^2-m_{D}^2)^2}{4m_X^2m_{D^*}^2}\right)
\label{corr}
\end{equation}
which, however, turns out to be numerically $\simeq G^2$. The actual value of $G$  is extracted from data on the branching ratio $\BR(X\to \bar DD\pi)$, which  is measured experimentally to be larger than $32\%$.  However the total width is poorly known: $\Gamma(X)\lesssim 1.2\mev$. Using these two  extreme values and the $X\to \bar DD\pi$ decay rate  
\begin{align}
\Gamma(X\to \bar DD\pi)&=\int_{(m_D+m_\pi)^2}^{(m_X-m_D)^2}ds\,\frac{1}{3}\frac{1}{8\pi m_X^2} 3\left(G\sqrt{2}\right)^2 p^*(m_X^2,m_D^2,s)\,\times\notag\\
&\times \frac{1}{\pi} \frac{s/m_{D^*}\, \Gamma_{D^*}\,\BR(D^*\to D\pi) }{(s-m_{D^{*}}^2)^2+(s/m_{D^*}\, \Gamma_{D^*})^2} \frac{m_{D^*}}{\sqrt{s}} \frac{p^*(s,m_D^2,m_\pi^2)}{p^*(m_{D^*}^2,m_D^2,m_\pi^2)}
\label{gammaddpi}
\end{align}
where the decay momentum is $p^*(x,y,z)=\sqrt{\lambda(x,y,z)}/2\sqrt{x}$, $\lambda$ being the K\"all\'en triangular function, it is found that $G\approx 4\gev$. Considering for example a branching fraction of $\BR(X\to \bar DD\pi)\simeq 0.32$, we obtain $B=B_{\rm exp}$  on  assuming a total width of the $X$ as large as $\approx 300$\kev: lower values of $\Gamma_X$ would also be possible for higher branching ratios $\BR(X\to \bar DD\pi)$, whereas higher  $\Gamma_X$ values are excluded. These conclusions are displayed by the shaded areas in  \figurename{~\ref{excluded}}, which are limited by the hyperbolae
\begin{equation}
 \BR(X\to \bar DD\pi)\cdot \Gamma(X)\sim G^2\sim\sqrt{B}
 \end{equation}
obtained fixing the value of $B$~\cite{Polosa:2015tra}. The not-excluded region is quite small and 
a high resolution on the width of the $X(3872)$ would be necessary --- one should be able to go below $\Gamma(X)\lesssim 300\kev$. An accurate determination of $ \BR(X\to \bar DD\pi)$ and $\Gamma(X)$ would allow to test~(\ref{bgrel}) and the molecular interpretation behind it.

\subsection{Scattering length of the \texorpdfstring{$X(3872)$}{X(3872)} as a loosely bound molecule \label{ascattering}}
The extremely small binding energy of the $X(3872)$ makes it an ideal candidate for a loosely bound molecule, and a perfect example where the rigorous low energy universality could be applied. This was firstly done in~\cite{Voloshin:2003nt} and slightly later in~\cite{Braaten:2003he} to study different implications of the shallow bound state theory to the study of the $X$.

A priori, the quantum mechanical state for the $X(3872)$ can be a superposition of all the hadronic states with the right quantum numbers, $J^{PC}=1^{++}$~\cite{Braaten:2003he}
\begin{align} \label{Xket}
|X\rangle=\sqrt{Z_\text{mol}}\int \frac{d^3p}{{(2\pi)}^3}\tilde\psi(p)\frac{|D^0(\bbb{p})\bar D^{*0}(-\bbb{p})\rangle+|\bar D^0(\bbb{p})D^{*0}(-\bbb{p})\rangle}{\sqrt{2}}+\sum_H\sqrt{Z_H}|H\rangle
\end{align}
In the previous expression $\tilde\psi(p)$ is the momentum space wave function of the $D$ mesons --- see Eq.~\eqref{univpsi} below --- while $|H\rangle$ are other hadronic states, discrete or continuous. These can be other molecular components like $|D^+(\bbb p)D^{*-}(-\bbb p)\rangle$, charmonium states or even compact four-quark objects. The presence of the latter ones will play an essential role in the mechanism explained in \sectionname{\ref{mol3}}. The key question is which of these different components are relevant to the total wave function or, in other words, how the probabilities $Z_i$ look like.

The probabilities for the states scale as $Z_H\sim1/\delta_H^2a_\text{s}$, where $\delta_H$ is the mass splitting of the $X$ --- or of the loosely bound molecule in general --- with respect to the state $H$.\footnote{ The fact that $\delta_H\simeq$ 0 and 8\mev for $D^0\bar D^{*0}$ and $D^+D^{*-}$ respectively implies a suppression of the second one in Eq.~\eqref{Xket}. This explains the observed isospin violation of the $X$ from a molecular point of view.} This behavior has been computed in~\cite{Braaten:2003he} using an explicit effective field theory calculation but can also be seen qualitatively from the more general Eq.~\eqref{geqphipsi}. It tells us that
\begin{align}
\sqrt{Z_H}\sim \frac{g}{\delta_H} \,\Rightarrow\, Z_H\sim\frac{{|g|}^2}{\delta_H^2}\sim\frac{\sqrt{B}}{\delta_H^2}\sim\frac{1}{\delta_H^2a_\text{s}}
\end{align}
where we used $E+B\approx\delta_H$ and that fact that, as can be seen from \eqref{ppresult} and \eqref{ppresult2}, in the $E\sim B\sim 0$ limit the scattering length of the $ab\to ab$ process becomes
\begin{align}
a_\text{s}=\frac{1}{\sqrt{2mB}}
\end{align}
which is divergent in the strict $B\to 0$ limit. Applying this result to the case of the $X$ with $m=m_Dm_{D^*}/(m_D+m_{D^*})$ and $B=m_D+m_{D^*}-m_X$ one finds $a_\text{s}\simeq11.97$ fm, an extremely large number. It has been shown in~\cite{Esposito:2013ada} that this scattering length can hardly be reconciled with the one extrapolated from experimental data on the full width of $X(3872)$, since the latter one appears to be at least a factor of 3-4 times smaller that the one expected from shallow states theory.

Moreover, the wave function of the loosely bound molecule for separations $r\gg\ell$, $\ell$ being the typical size of the components, is universally given by~\cite{Landau3,Braaten:2004rn} (see also Eq.~\eqref{matchleft})
\begin{align} \label{univpsi}
\chi(r)\propto e^{-r/a_\text{s}}\, \Rightarrow\, R_{\ell=0}(r)=\frac{\chi(r)}{r}\propto \frac{e^{-r/a_\text{s}}}{r}
\end{align}
Therefore, the typical size of the loosely bound molecule is $a_\text{s}$ itself. The $X(3872)$ would therefore be an unnaturally extended objected, much larger than the typical hadrons (and $\sim3$ times larger than the deuteron) and of the typical range of interaction between heavy mesons, $1/m_\pi$. In other words, while the binding energy for the $X$ would be expected to be of order $m_\pi^2/2m\simeq10$\mev, while the actual value $B\simeq0.14$\mev is much smaller.
From a certain point of view this feature can explain some of the peculiar properties of this resonance but, on the other hand, it requires a good degree of fine tuning that, if not properly explained, might make the molecular interpretation questionable.

In~\cite{Braaten:2003he} two possible explanations for this phenomenon are given. The first possibility is to have a fine tuning that only interests the $D^0\bar D^{*0}$ component of \eqref{Xket}, without influencing the others. This could happen, for example, in the meson-meson potential so that it allows for a bound state very close to threshold. If the other detunings $\delta_H$ are larger than the natural energy scale, the very large scattering length would suppress all the other states $|H\rangle$. In this case, the $X(3872)$ would be a purely $D^0\bar D^{*0}$ molecular state. Although this is an appealing solution, we will see in the next section that a real, negative energy meson molecule faces severe difficulties when the \emph{prompt} production at hadron colliders is taken into account. 

The second mechanism proposed in~\cite{Braaten:2003he} to explain the large scattering length instead relies on a Feshbach resonance mechanism~\cite{leggett,petick} between the $D$ mesons and one charmonium level, specifically the $\chi_{c1}(2P)$. This situation is, however, probably too fortuitous to be taken seriously. First of all, the $\chi_{c1}(2P)$ has not been observed yet and secondly, potential models~\cite{Buchmuller:1980su,Godfrey:1985xj} predict $\delta_{\chi_{c1}(2P)}\simeq90$\mev. To bring it down to values compatible with the application of the Feshbach formalism would require quite large theoretical errors.

If one ignores for the moment the issues related to the prompt production, and assumes that the $X(3872)$ is indeed a purely molecular state, it is also rather immediate to explain the narrow width of this particle. The total width, $\Gamma_X$, will get contributions from both the decays of its constituents, $D^*\to D\pi$ and $D^*\to D\gamma$ (see \eg Eq.~\eqref{gammaddpi}), and the decays into other hadronic states, like for example lower charmonia, $X\to\psiprime\gamma$ or $X\to \jpsi\,\pi\pi$. However, since the states $\left|H\right\rangle$ in~\eqref{Xket} are suppressed by $1/a_\text{s}$, the second possibility can be neglected and one remains with $\Gamma_X\approx\Gamma_{D^*}\simeq65\kev$~\cite{Braaten:2007dw}.

The scattering amplitude~\eqref{unof} can be rewritten in terms of the scattering length as
\begin{align}
f(ab\to ab)\equiv  f(E)=-\frac{1}{\gamma+i\sqrt{2mE}}
\end{align}
where we have neglected the $i\epsilon$ regulator for convenience and defined $\gamma=1/a_\text{s}$. As already pointed out, if $\gamma>0$, $f(E)$ has a pole for $E<0$, in which case we are in presence of a \emph{real} bound state. On the other hand, if $\gamma<0$ then the pole happens still for negative values of $E$, but on the unphysical Riemann sheet, and in this case we refer to a \emph{virtual (anti-bound) state}.

This is all strictly true for a stable shallow bound state. In order to take into account the finite width of the $X(3872)$, the authors of~\cite{Braaten:2007dw} allowed both the scattering length and the reduced mass that is implicit in $E$ to have an imaginary part:
\begin{align}
f(E)=-\frac{1}{\Re\gamma+i\,\Im\gamma+i\sqrt{2m(E+\Gamma(E)/2)}}
\end{align}
with $\Gamma(E=0)=\Gamma_{D^*}$. While the latter one takes care of the direct decay of the $D^*$ constituent, $\Im\gamma$ accounts for inelastic scattering processes. It can be argued that if the total energy $E$ is close to the $D^0\bar D^{*0}$ threshold then the line shape $d\Gamma(B^+\to K^+D^0\bar D^{*0})/dE$ has energy dependence that only comes from the ${|f(E)|}^2$ factor.\footnote{For energies close to threshold the decay matrix elements get a $f(E)$ resonant factor~\cite{Braaten:2007dw}.} Using the 2003 data reported in~\cite{Choi:2003ue} by the \belle collaboration, it was shown that the $\Re\gamma>0$ situation was favored with respect to the opposite one~\cite{Braaten:2007dw}, hence suggesting for the $X$ to be a real, negative energy bound state --- again, if the other difficulties of the molecular model are ignored.

\subsection{\texorpdfstring{$X(3872)$}{X(3872)} as a deuteron-like state\label{xdeut}}

\subsubsection{Prompt production disfavors a pure molecule interpretation}

At very low energy, proton-neutron scattering $pn\to pn$, with $S=1$ as in deuteron, has a cross section 
\begin{equation}
\sigma_{pn}=\frac{2\pi}{m \, B}\simeq 2.41~\mathrm{b}
\end{equation}
using the binding energy $B=2.22\mev$ in deuteron. This agrees  fairly well with the experimental value $\sim 3.6$~b. This does not mean that a bound state deuteron gets formed, which might only happen if the $pn$ system loses energy and falls in the bound state. It is rather the manifestation of the enhancement in the scattering cross section due to the presence of a shallow bound state.

As suggested in the previous sections, the $X(3872)$ might be a deuteron-like object: a real bound state whose lifetime is given by the shorter lived component, the $D^*$. Are deuterons produced at LHC at large transverse momenta? How do their production rates compare with that of of $X(3872)$? If the nature of the two particles is indeed almost the same, then the behavior of the $X$ should not be too far away from that of the deuteron.
 
In a deuteron-like description of  $X(3872)$ based on a (three-dimensional) square well potential of depth $V_0\simeq 6.7\mev$ and range $R\simeq 3$~fm, a bound state energy $B=0.14\mev$ is obtained. 
The expectation value of the kinetic energy in the bound state $\psi$ is  found to be
\begin{equation}
\langle T\rangle_\psi\simeq 1.29~{\rm\mev} \,\;\;\; (k_{\rm rel}\simeq 50~{\rm\mev})
\label{expctsval}
\end{equation}
\begin{figure}[t]
 \begin{center}
   \includegraphics[width=7truecm]{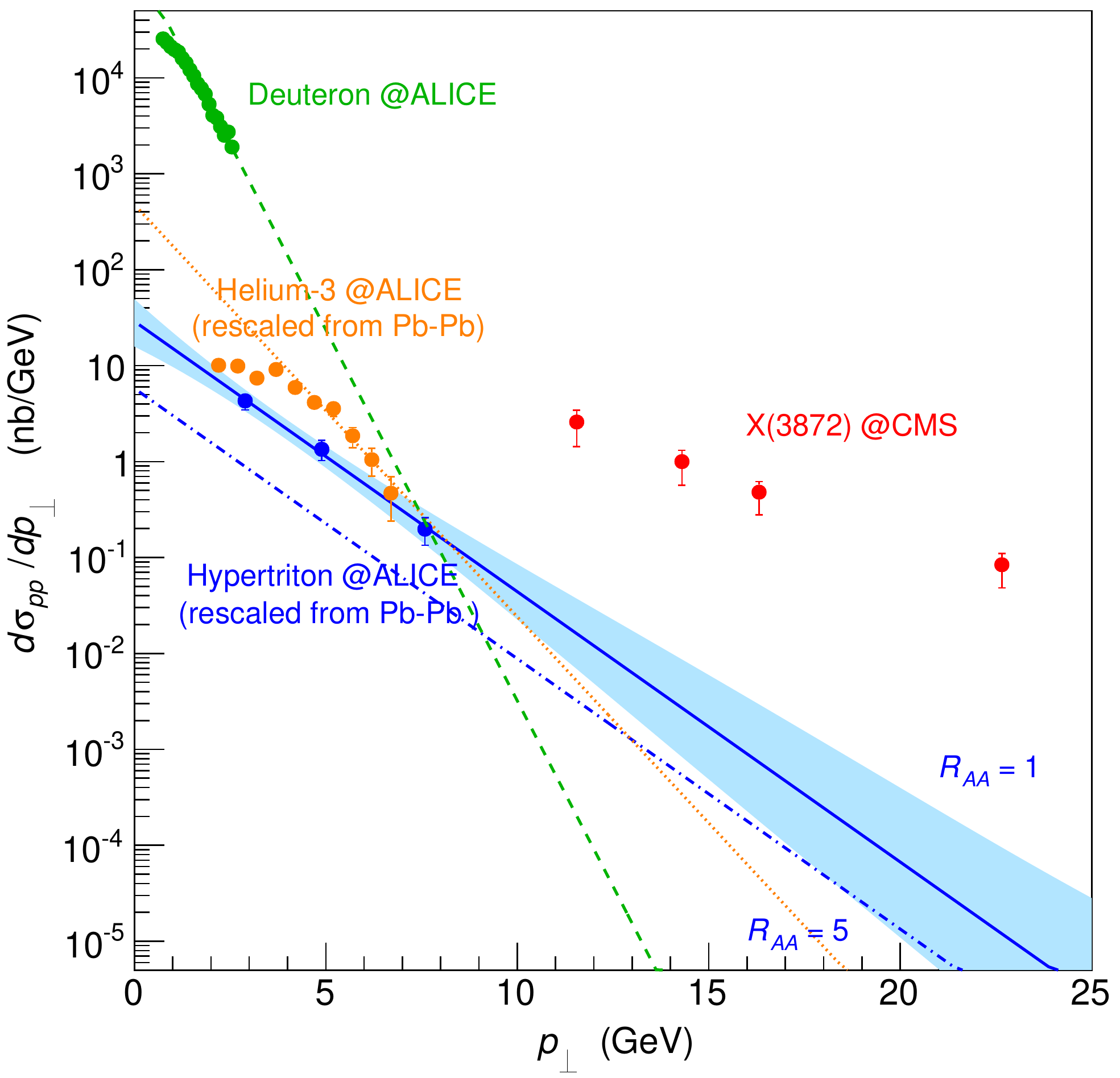} 
   \hspace{.2cm} \includegraphics[width=7truecm]{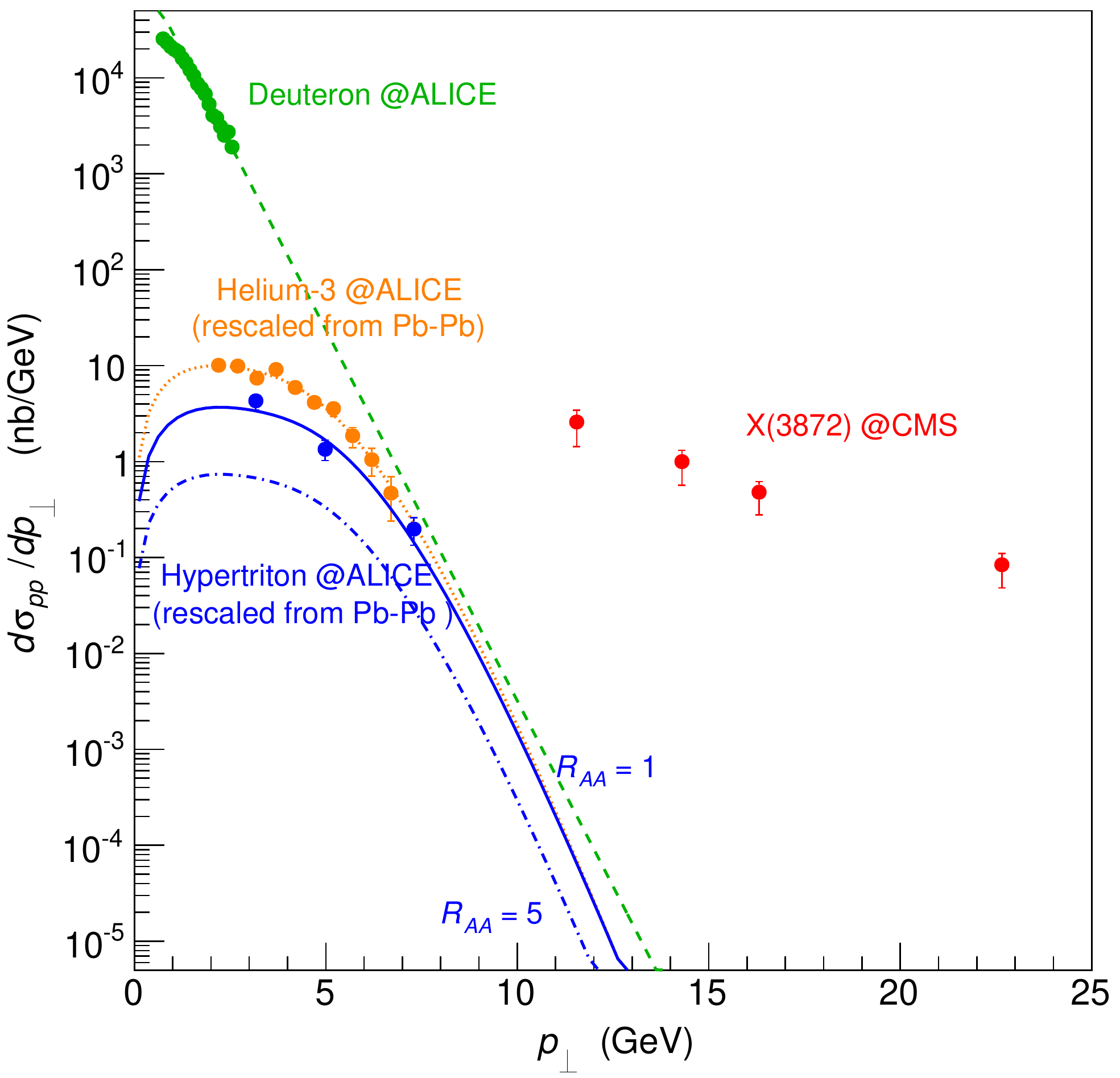}
 \end{center} 
\caption{ Comparison between the prompt production cross section in $pp$ collisions of $X(3872)$ (red), 
deuteron (green), helium (orange), and hypertriton (blue), from~\cite{Esposito:2015fsa}. The $X$ data from \cms are rescaled 
by the branching ratio $\mathcal{B}(X\to \jpsi\,\pi\pi)$. Deuteron data in $pp$ collisions are taken from 
ALICE.
The helium and hypertriton data measured by \alice in Pb-Pb collisions have been 
rescaled to $pp$ using a Glauber model.  
The dashed green line is the exponential fit to the deuteron data points in the $\pt\in[1.7,3.0]\gev$ region, 
whereas the dotted orange one is the fit to the helium data points.  The solid and dot-dashed blue lines represent 
the fits to hypertriton data with $R_{AA}=1$ (no medium effects) and a hypothetical constant value of $R_{AA}=5$. 
The hypertriton data points are horizontally shifted at the bin centers of gravity --- being defined as the point 
at which the value of the fitted function equals the mean value of the function in the bin. \label{alice}
Left panel: the hypertriton data are fitted with an exponential curve, and the light blue band is the 
$68\%$~C.L. for the extrapolated $R_{AA}=1$ curve. Helium data in the $\pt \in[4.45,6.95]\gev$ region are 
also fitted with an exponential curve. 
Right panel: the hypertriton and helium data are fitted with blast-wave functions, 
whose parameters are locked to the helium ones.
}
\end{figure}%
with an average radius of $\langle r\rangle\simeq10\fm$.\footnote{Using the central value of the latest measurement of the binding energy~\cite{Tomaradze:2015cza}, $B=(3\pm192)\kev$, one would find $k_\text{rel}\simeq20 \mev$ and a humongous size of $\langle r\rangle\simeq58\fm$!} Note that the relative momentum has a rather higher value with respect to those discussed before. However we have to observe that, in this model, the $D$ and $\bar D^*$ mesons have indeed finite negative  total energy, \ie they are stably bound together.
To make this happen, the initially free $\bar D D^*$ pair produced in $pp$ hadronization must interact with at least a third hadron, to change abruptly its relative kinetic energy and fall  in the discrete (even though superficial) level of the attractive potential~\cite{Esposito:2013ada,Guerrieri:2014gfa} --- see the discussion in \ref{appmol}. The expected $X$ width would therefore be $\Gamma_X\equiv\Gamma_{D^*}\approx 65$\kev, even at a binding energy as large as $B=0.14\mev$ (compare to \figurename{~\ref{excluded}}), for it would be  a stable bound state whose lifetime coincides with the lifetime of the shortest lived component. 

We conclude that the formation of the  $X$ resonance might occur either via a low energy resonant scattering phenomenon or via multi-body final state interactions producing a deuteron-like state whose metastability is provided by the shorter lived component in the system, namely the $D^*$ particle --- energetic arguments do not suggest that the formation of the $\bar D D^*$ molecule stabilizes the $D^*$ within the bound state. 

The deuteron-like case, might be more realistic when considering $X$ prompt production in high energy $pp$  collisions at the LHC,  with high transverse momentum cuts on hadrons. Indeed in this case it is very unlikely that the relative energy in the center of mass of the $\bar D D^*$ system is $E\sim B\sim 0$ as required by the shallow bound state formalism, especially when high transverse momentum $\pt$ cuts are considered. This latter point has thoroughly been discussed using Monte Carlo simulations~\cite{Bignamini:2009sk,Bignamini:2009fn,Esposito:2013ada,Guerrieri:2014gfa}: the main observation is that the relative momentum in the center of mass of the pair  $k_{\rm rel}$ has an average value $\langle k_{\rm rel}\rangle\approx 1.3 \pt$ which grows with $\pt$ and the $X(3872)$ has been measured with high cross section at \cms with $\pt\gtrsim 12\gev$ --- see \figurename{~\ref{alice}}~\cite{Esposito:2015fsa}. For more details on this point see \ref{appmol}.

On the other hand, if  we wish to interpret the $X(3872)$ as a deuteron-like particle formed in hadronization, then we have to confront with data on production of deuteron, or other loosely bound light nuclei, in $pp$ collisions at high transverse momenta. In particular it is useful to compare the large production cross sections of $X(3872)$ observed by \cms at high $\pt$ with the extremely small deuteron production expected in the same region extrapolating recent  data from the ALICE collaboration, shown in \figurename{~\ref{alice}}. Data points at higher $\pt$ will be collected in the future to help this comparison. 
At any rate, the extrapolation of the data on deuteron, Helium-3 and hypertriton --- all \emph{bona fide} loosely bound molecules --- clearly shows how the $X(3872)$ is too copiously produced at the vertex of high energy collisions to be easily interpreted as a meson molecule with tiny binding energy. Although more data will clearly provide an essential insight, it is our opinion that the comparison with the experiment speaks against a molecular interpretation of this resonance.

Recently, it has been proposed that, at high $p_T$, the production of exotic states depends on the short-range nature, namely on the number of quarks which compose the state, regardless of their molecular or compact nature, so that prompt production could give no information about the nature of these states.  This would agree with the predictions of counting rules~\cite{Brodsky:2015wza,Guo:2016fqg,Voloshin:2016phx}. However these rules are derived for the {\it exclusive} reactions only, where all the (relatively small, as discussed in~\cite{Voloshin:2016phx}) center of mass energy is shared among few final hadrons, and do not apply to very high energy inclusive production of exotic states. In any case, in \figurename{~\ref{alice}} one can appreciate that the slope of the deuteron is much steeper than the one of hypertriton, which is at odds with the na\"ive counting (6 quarks versus 9 quarks). The short-range nature of the production would also be inconsistent with the use of final state interactions to get the enhancement of two orders of magnitude needed to reach the experimental value of the cross section~\cite{Artoisenet:2009wk,Artoisenet:2010uu,Guo:2013ufa,Guo:2014sca}  --- see also~\ref{appmol}.

\subsubsection{Production through $c\bar c$}
An alternative idea that can justify the $X(3872)$ prompt production cross section has been proposed in~\cite{Butenschoen:2013pxa,Meng:2013gga}. As already explained in Sec.~\ref{ascattering}, the $X(3872)$ can {\it a priori} be represented by a superposition of all the hadronic states with the right $J^{PC}=1^{++}$ quantum numbers.

The authors of~\cite{Butenschoen:2013pxa,Meng:2013gga} express the $X(3872)$ state as
\begin{align} \label{Xmixed}
|X\rangle=\sqrt{Z_{c\bar c}}\,|\,\chi_{c1}(2P)\rangle+\sqrt{Z_\text{mol}}\,|\,DD^*\rangle
\end{align}
where states with a different number of `valence' quarks are superimposed. 

Here the state $|\,DD^*\rangle$ is intended as summed over momenta --- see Eq.~\eqref{Xket}. The observables like the prompt production, the $b$-production and the quark annihilating decays can be explained through the charmonium component, while the hadronic decays into $DD\,\pi$, $DD\,\gamma$, $\jpsi\rho$ and $\jpsi\omega$ through the molecular one.

The results of the two papers are  in agreement. Here we will follow the notation used in~\cite{Meng:2013gga}. The authors use the Non-Relativistic QCD (NRQCD) approach~\cite{Bodwin:1994jh} at Next-to-Leading Order (NLO), which has proven to be quite successful in reproducing the yields of \jpsi~\cite{Ma:2010yw} and $\chi_{cJ}(1P)$~\cite{Ma:2010vd} at Tevatron and LHC. The production cross section of the $X(3872)$ in the $\jpsi\pi^+\pi^-$ mode can then be written as
\begin{align}
d\sigma(pp\to X(\jpsi\pi^+\pi^-))=d\sigma(pp\to\chi_{c1}(2P))\,Z_{c\bar c}\,\mathcal{B}(X\to \jpsi\pi^+\pi^-)
\end{align}
while the cross section for the production of the charmonium can be written using the NRQCD factorization as
\begin{align} \label{dsigma}
d\sigma(pp\to\chi_{1c}(2P))=\sum_{i,j,n}\int dx_1\,dx_2 G_{i/p}(x_1)G_{j/p}(x_2)\,d\tilde\sigma(ij\to (c\bar c)_n)\langle\mathcal{O}_n^{\chi_{c1}(2P)}\rangle
\end{align}
In the previous expression $i$ and $j$ enumerate the partons, and run over gluons, light quarks ($u,\,d,\,s$) and antiquarks. Moreover, $G_{i/p}(x)$ is the Parton Distribution Function (PDF) of the proton and $d\tilde\sigma$ is the partonic cross section. The collective index $n$ instead contains color, spin and angular momentum of the intermediate $c\bar c$ pair, and $\langle\mathcal{O}_n^{\chi_{c1}(2P)}\rangle$ is a long distance matrix element embedding the conversion from $c\bar c$ to an actual charmonium. At NLO we only have $n=\,^3P_1^{[\mathbf{1}_c]},\,^3S_1^{[\mathbf{8}_c]}$.

The authors of~\cite{Meng:2013gga} then use the available data from CMS~\cite{Chatrchyan:2013cld} to fit the free parameters of Eq.~\eqref{dsigma}, and successfully reproduce \cdf data~\cite{Acosta:2003zx}\footnote{The prediction obtained from the fit underestimate the \lhcb data~\cite{Aaij:2011sn} by a factor of 2, which might be due to a large theoretical uncertainty.}. They show that both the prompt production and the production from $B$ decays can be explained with a large weight for the $\chi_{c1}(2P)$ component estimated to be
\begin{align}
Z_{c\bar c}=(28-44)\%.
\end{align}

Nevertheless  there is no reason not to include in the state~\eqref{Xmixed} a compact diquarkonium state (see Sec.~\ref{diquark}) or any compact tetraquark state. It is more natural to superimpose states having  an equal number of `valence' quarks rather than states with an increasing number quarks  as in~(\ref{Xmixed}). A compact tetraquark content would as well account for {\it both} the production and decays of the $X(3872)$, resulting in a simpler description which is able to explain the same observables. We remind here that the model for baryon production (incorporated in shower Monte Carlo libraries and generally used to study hadron collisions) is based on the diquark-quark color neutralization. In that case no meson-meson component would be needed.

\subsection{Consequences of the Heavy Quark Spin Symmetry \label{HQSS}}
The results obtained in the previous section for shallow bound states apply to every type of molecule. Also Heavy Quark Spin Symmetry (HQSS) can be used to make universal predictions.
It arises in systems with heavy $Q$ and light quarks in the $m_Q\to\infty$ limit, with $Q=\{c,b\}$. In fact, the hyperfine interaction between the spin of the heavy quark and the chromomagnetic field generated by the light degrees of freedom appears at $\mathcal{O}(1/m_Q)$ in the Heavy Quark Effective Theory (HQET) Lagrangian. Consequently, in the $m_Q\to \infty$ limit, the properties of heavy mesons and quarkonia (\eg their masses) are independent of the heavy spin. Light quarks are insensitive to the spin and flavor of the heavy quark, which behaves as a spectator.\footnote{This is exactly analogous to ordinary atomic physics where, at the lowest level, the heavy nucleus just provides a static source for the Coulomb field.} In particular, since pseudoscalar and vector heavy mesons and quarkonia are degenerate at lowest order, they can be collected in spin multiplets, \eg $(D,D^*)$, $(B,B^*)$, $(\eta_c,\jpsi)$ and $(\eta_b,\Upsilon)$. This symmetry holds for heavy meson molecules as well, and can be employed to study their spectrum and decay modes, independently of the details of the potential. Corrections to this approximations are of order $\mathcal{O}(\Lambda_{QCD}/m_Q)$, \ie $\sim25\%$ for the charm and $\sim10\%$ for the bottom, if  constituent quark masses are considered.

To lowest order, the spin of the heavy quark and that of the light one are completely decoupled. %
When this happens, the Hilbert spaces of the heavy quark spinors and of the remaining degrees of freedom are completely separated. Consequently, for pseudoscalar $(P)$ and vector $(V)$ heavy-light mesons, the total wave function can be written as the direct product of the heavy quark spinor times the wave function of the remaining degrees of freedom
\begin{align} \label{Psi}
\Psi_P=\bar \psi_{q}\, \sigma^2\, \chi_Q \quad \text{ and } \quad \bbb\Psi_V= \bar\psi_q \, \sigma^2 \bbb\sigma  \, \chi_Q
\end{align}
where $\chi_Q$ is the Pauli two-dimensional spinor of the heavy quark, $\bar \psi_{q}$ is the wave function of the bound light antiquark and the spinless heavy quark and $\bbb \sigma$ are the Pauli matrices, suitable to spin-1 states. Note that we are using a non-relativistic formalism, and hence the antiparticles are independent of the particles, \ie $\bar\psi_q\neq \psi_q^\dagger$.\footnote{$\psi_q$ only contains annihilation operators for the particle.} This implies that no creation/annihilation of pairs is allowed. In general $\bar{\psi}_{q}(\bbb r)$ will be a complicated object containing information about everything but the spin of the heavy quark, \eg spinor of the light antiquark, spatial distribution of the $Q\bar q$ pair, etc. Note also that with our notation, the wave functions for the respective antimesons are~\cite{Bondar:2011ev}
\begin{align} \label{Psibar}
\bar \Psi_{P}=\bar\chi_{Q}\,\sigma^2 \psi_q \quad \text{ and } \quad \bar{\bbb\Psi}_{V}= \bar \chi_{Q}\,\sigma^2\bbb \sigma\,\psi_q
\end{align}

Since the spin-flip of the heavy degree of freedom is suppressed by powers of $1/m_Q$, the spin of the $Q \bar Q$ pair can be used as a ``label" for the state. If one assumes  the $X$, $Z_Q$ and $Z_Q^\prime$ to be molecular states, then using the SU(2) Fierz identities as explained in \sectionname{\ref{spins}}, it is possible to rewrite the total spin wave function of the molecule in the $[Q\bar Q]$ and $[q\bar q^\prime]$ basis rather than in the $[q\bar Q]$ and $[Q\bar q^\prime]$ one. If in Eqs.~\eqref{Psi} and \eqref{Psibar} we only focus on the spin part of the full wave function, it is possible to find~\cite{Voloshin:2004mh,Bondar:2011ev}
\begin{subequations}
\begin{align}
X&\sim \bar{\bbb\Psi}_{V}\Psi_P -\bar \Psi_{P} \bbb{\Psi}_V \sim |{1}_{Q\bar Q},{1}_{q \bar q}\rangle \label{spinX} \\
Z_Q &\sim \bar{\bbb\Psi}_{V}\Psi_P +\bar \Psi_{P} \bbb\Psi_V\sim |{0}_{Q\bar Q}, {1}_{q\bar q^\prime}\rangle+ |{1}_{Q\bar Q}, {0}_{q\bar q^\prime}\rangle \label{spinZ} \\
Z_Q^\prime &\sim \bar{\bbb\Psi}_{V}\bm\times \bbb{\Psi}_V \sim |{0}_{Q\bar Q}, {1}_{q\bar q^\prime}\rangle- |{1}_{Q\bar Q}, {0}_{q\bar q^\prime}\rangle \label{spinZp}
\end{align}
\end{subequations}
where by $S_{\!q_1q_2}$ we represent the spin of the $q_1q_2$ pair, $P=(D,B)$ and $V=(D^*,B^*)$. 

The previous spin decompositions can be used to extract selection rules for the decay of the would-be meson molecules into quarkonia. For example, one readily deduces that the hadronic transition $X\to\eta_c\pi\pi$ should be suppressed because the spin of the $\eta_c$ is $S_{\!c\bar c}=0$, while the decay into spin-1 charmonia should be favored. In~\cite{Voloshin:2004mh}, Eq.~\eqref{spinX} is used to estimate the ratio between the widths $\Gamma(X\to\pi^0\chi_{c1})$ and $\Gamma(X\to\pi^+\pi^-\jpsi)$,
\begin{align}
\frac{\Gamma(X\to\pi^0\chi_{c1})}{\Gamma(X\to\pi^+\pi^-\jpsi)}\approx 0.35\left(\frac{0.5\gev}{\mu}\right)^2
\end{align}
where $\mu$ is a dimensional factor that takes into account the $P$-wave emission of the single $\pi^0$. The $X$ has, in fact, $J^{P}=1^{+}$ while the $\pi^0$ and $\chi_{c1}$ are $J^{P}=0^{-}$ and $1^{+}$ respectively. Therefore, in order to conserve both angular momentum and parity the decay must be in $P$-wave. From the above equation it is clear that for reasonable values of $\mu$ the decay into $\pi^0\chi_{c1}$ should have an observable rate.

In~\cite{Bondar:2011ev} HQSS is also employed to make several predictions about the $Z_b$ and $Z_b^{\prime}$ states. These considerations clearly apply to the charm sector as well, albeit with larger uncertainties. First of all, Eqs.~\eqref{spinZ} and \eqref{spinZp} show that the magnitude of the couplings to $J=0$ and $J=1$ bottomonia are the same in both the $Z_b$ and the $Z_b^\prime$. Moreover, in the $m_b\to \infty$ limit, the $0_{b\bar b}$ and $1_{b\bar b}$ bottomonia are degenerate, as well as the $Z_b$ and $Z_b^{\prime}$. It then follows that the two $Z_b^{(\prime)}$ states are expected to decay into degenerate states with lower masses and their widths should be roughly the same, $\Gamma(Z_b)\approx\Gamma(Z_b^\prime)$. Moreover, since the combinations \eqref{spinZ} and \eqref{spinZp} are orthogonal, the decay $Z_b^\prime\to\bar B B^*$ is forbidden. Both these predictions are in good agreement with the data. In~\cite{Bondar:2011ev} considerations about the $\Upsilon(5S)\to\Upsilon(nS)\,\pi^+\pi^-$ process as compared to the $\Upsilon(5S)\to h_b(kP)\,\pi^+\pi^-$ were also made. In particular, even though the coupling to $J=0$ and $J=1$ bottomonia have the same magnitude, they have opposite signs. This is manifested in the interference pattern for the two decay previously mentioned.

In~\cite{Guo:2009id} the HQSS was employed for the study of the $Y(4660)$. Because of its decay into $\psiprime\,\pi^+\pi^-$, it is interpreted in the molecular picture as a $\psiprime f_0(980)$ bound state.\footnote{The same authors consider the $f_0(980)$ to be a $K\bar K$ molecule~\cite{Baru:2003qq}. The $Y(4660)$ would be a sort of `matryoshka' molecule, where the interplay of the different scales involved is not clear.} The main consequence is the expected existence of the other state of the heavy spin multiplet, namely the $Y_\eta=\eta_c^\prime\, f_0(980)$. In particular, it can be estimated~\cite{Guo:2009id} that the branching ratio for the $B^+\to\eta_c^\prime \,K^+\pi^+\pi^-$ process is of order $10^{-3}$ and hence it should be very likely to observe the $Y_\eta$ in the $\eta_c^\prime\,\pi^+\pi^-$ final state of the previous decay at \belle and \babar.

\subsection{An effective field theory approach to heavy meson molecules}
\label{sec:NREFT}
Another potential-independent tool for the study of the decays of loosely bound molecules has been developed and employed in~\cite{AlFiky:2005jd,Fleming:2007rp,Cleven:2011gp,Cleven:2013sq,Wang:2013cya,Guo:2013nza,Esposito:2014hsa,Mehen:2015efa}. It is a Non-Relativistic Effective Field Theory (NREFT) obtained from the time-honored Heavy Meson Chiral Theory~\cite{Casalbuoni:1996pg}. The work on this topic is copious and it has been applied mostly to the prediction of the branching fractions for the radiative and hadronic decays of the \XYZ resonances.

\begin{figure}[t]
\centering
\includegraphics[width=\textwidth]{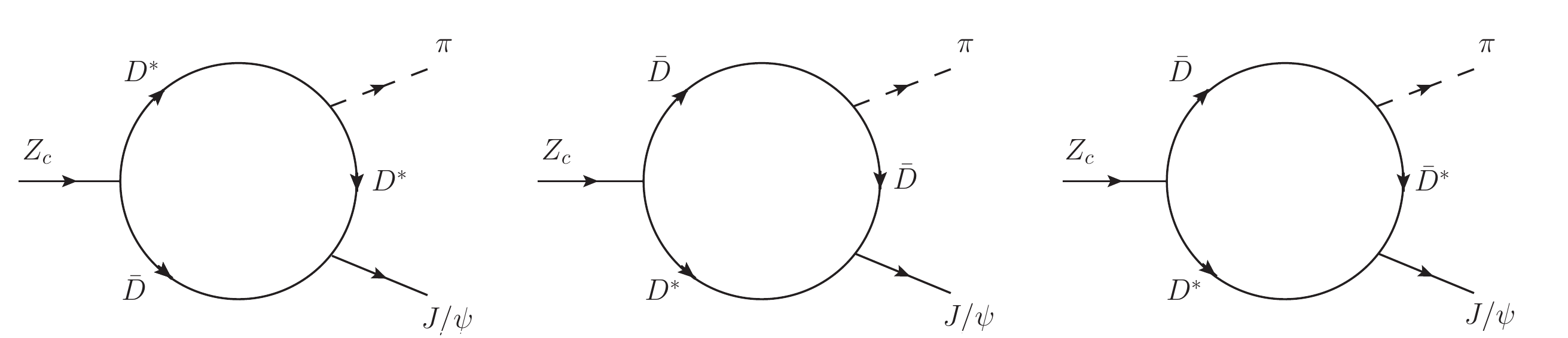}
\caption{Typical heavy meson loop used in NREFT to describe the decay of the exotic states. In this figure the $Z_c\to J\psi\,\pi$ decay happens through an intermediate $DD^*$ loop, as imposed by the molecular nature of the $Z_c(3900)$.} \label{fig:mesonloops}
\end{figure}

The hypothesis about the molecular nature of the states is implemented in the model by forcing the exotic resonances to couple to their constituents only, as a consequence of Weinberg's compositeness theorem~\cite{Weinberg:1962hj,Weinberg:1965zz}. It then follows that every decay into a final state different from the constituents themselves must happen via a heavy meson loop. In \figurename{~\ref{fig:mesonloops}} we report an example of such process for the $Z_c\to \jpsi\,\pi$ decay. 

The typical velocity of the heavy mesons involved in the previous loops is
\begin{align}
v\approx\sqrt{\frac{|M_X-2m|}{m}}
\end{align}
where $M_X$ is the mass of the exotic states and $m$ that of the open flavor mesons. Being the former ones very close to threshold, such velocities are small and hence allow for a non-relativistic treatment. This is done by letting $v_\mu\to(1,\bbb 0)$ in the HQET bi-spinors (see \sectionname{\ref{sec:1pi}} for their definitions). As a result of this procedure, the interaction of the \XYZ states with heavy pseudoscalar and vector mesons is described by the following terms
\begin{subequations} \label{Lnreft}
\begin{align}
\mathcal{L}_X&=\frac{g_X}{\sqrt{2}}X^{i\dagger}\left(\bar PV^i-P\bar V^i \right) + h.c. \\
\mathcal{L}_{Z_Q}&=\frac{g_{Z_Q}}{\sqrt{2}}Z_Q^{i\dagger}\left(\bar P V^i+P\bar V^i\right) + h.c. \\
\mathcal{L}_{Z^\prime_Q}&=ig_{Z_Q^\prime}\epsilon_{ijk}(Z_Q^\prime)^{i\dagger}\bar V^j V^k + h.c.
\end{align}
\end{subequations}
The fields $\bbb X$ and $\bbb Z_Q^{(\prime)}$ annihilate the exotic meson states, while the $P$ ($\bar P$) and $\bbb V$ ($\bar{\bbb V}$) annihilate the pseudoscalar and vector (anti)mesons respectively. In particular, the normalization of the states is such that $P|P(p)\rangle=\sqrt{m_P}|0\rangle$ and $V^i|V(p,\lambda)\rangle=\lambda^i\sqrt{m_V}|0\rangle$, where $\lambda$ is a polarization vector. The indices $i$, $j$ and $k$ are only spatial, as a consequence of the non-relativistic limit. The $g_i$'s are effective strong couplings to be fitted from the data --- see \eg~\cite{Guo:2013nza,Cleven:2013sq,Esposito:2014hsa}.

The smallness of the velocities also allows to substitute the relativistic heavy mesons propagator with its non-relativistic version
\begin{align} \label{nonrelprop}
\frac{i}{p^2-m^2+i\epsilon}\longrightarrow\frac{1}{2m}\frac{i}{p^0-\frac{\bbb p^2}{2m}-m+i\epsilon}
\end{align}
It is important to note that the effective couplings can be large and hence it is not obvious \emph{a priori} that including one-loop diagrams only is enough. However, since $v$ is here a small parameter, one can indeed apply a power counting procedure to estimate the relevance of higher order processes and consequently determine the applicability of perturbation theory.
In particular, heavy meson loops imply a non-relativistic integral over $dp^0d^3p/(4\pi)^2$ and hence count as $v^5/(4\pi)^2$ while the propagator in Eq.~\eqref{nonrelprop} contributes with a $1/v^2$, given that the states are close to threshold. Lastly, if the interaction vertices in~\eqref{Lnreft} contain derivatives they will come with either an additional power of $v$ or of the outgoing momenta $q$, depending if the derivative acts on the heavy meson running in the loop or on the external legs.

This NREFT has been employed in~\cite{Esposito:2014hsa} to show that the branching ratios for the $Z_c^{(\prime)}\to \eta_c\,\rho$ decays can provide a powerful tool of discrimination between meson molecule and compact tetraquark.
\begin{figure}[t]
\centering
\includegraphics[width=0.45\textwidth]{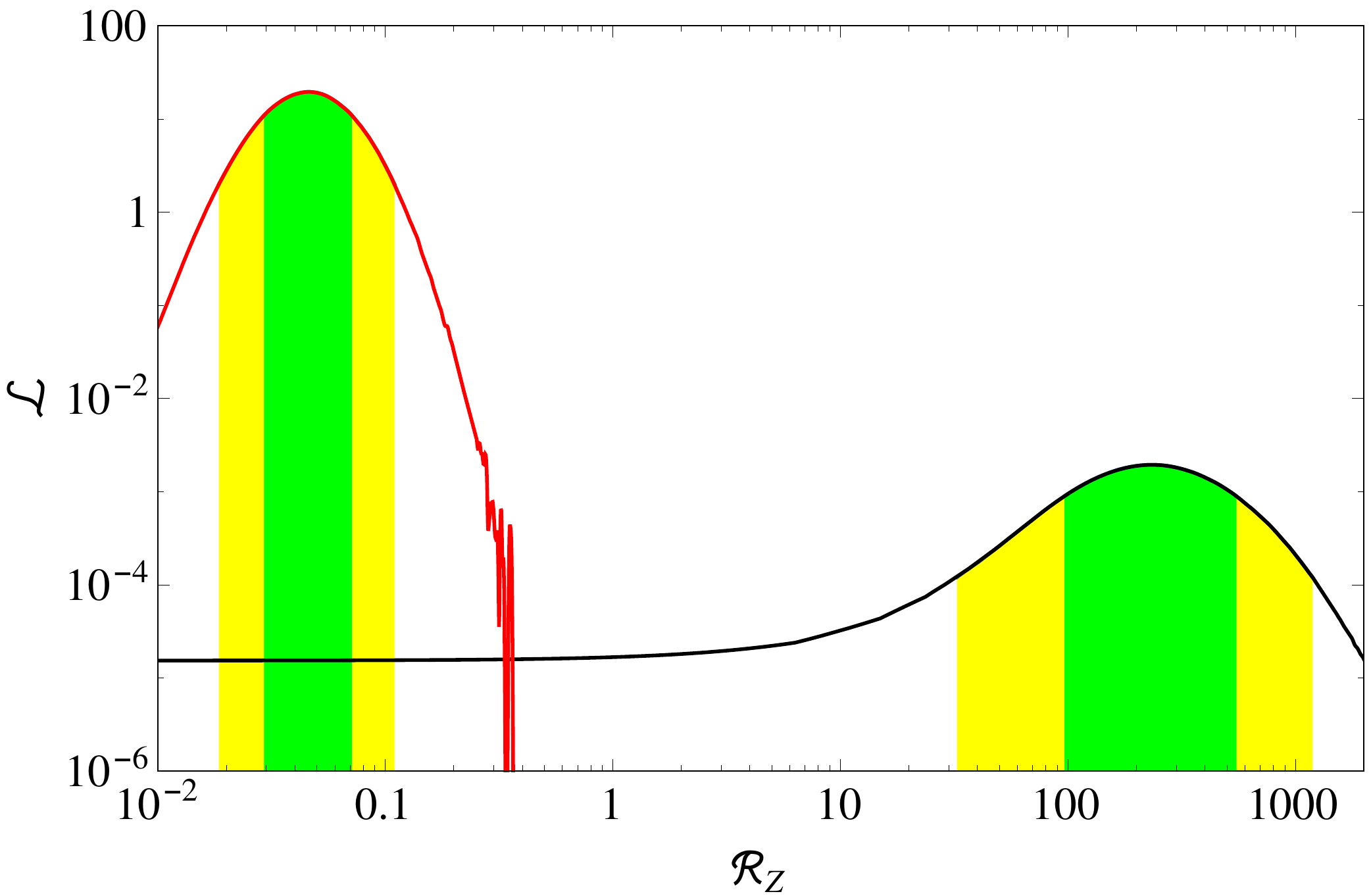}
\hspace{0.5em}
\includegraphics[width=0.45\textwidth]{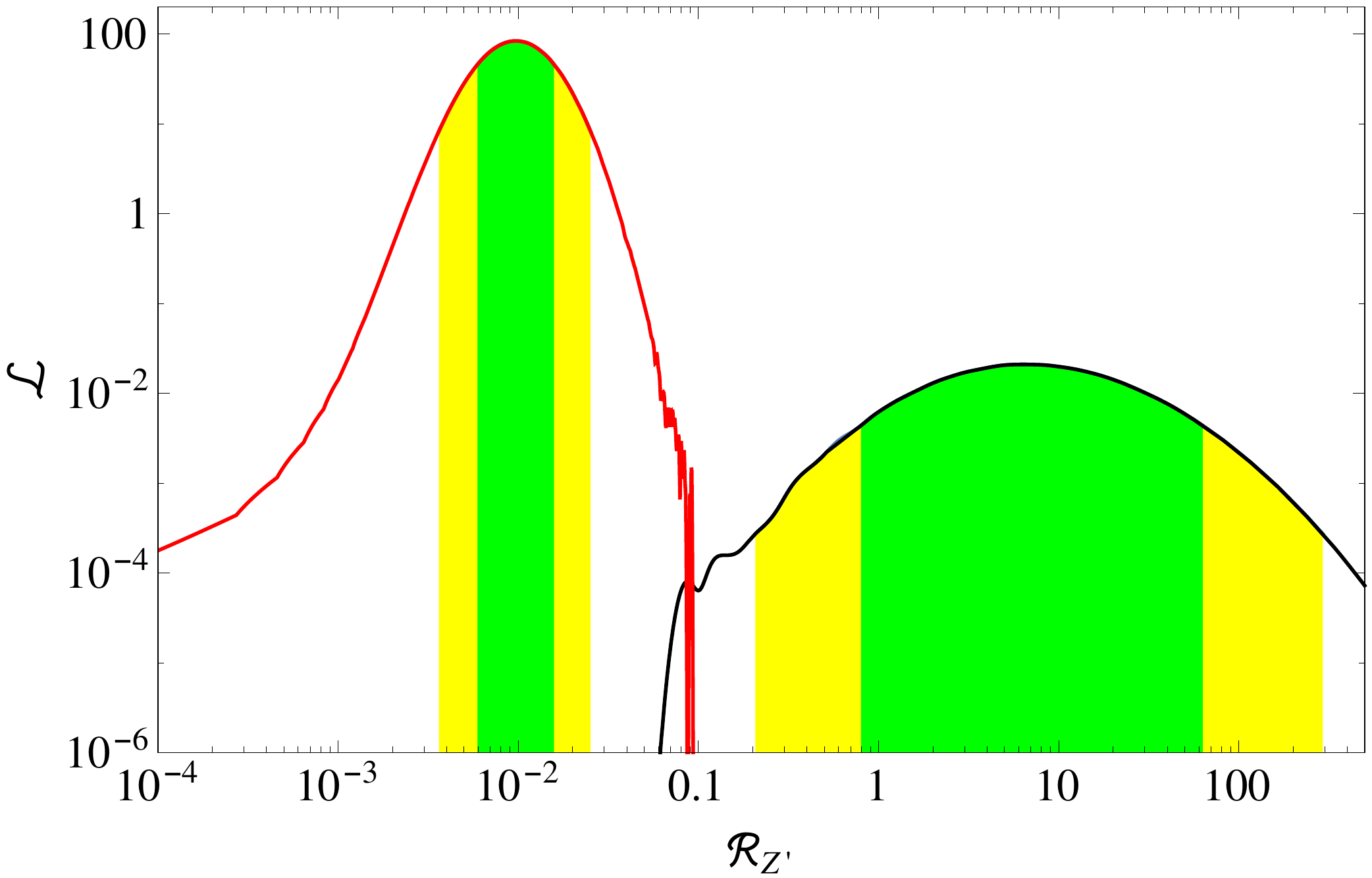}
\caption{Likelihood distributions for the ratios in Eq.~\eqref{RZ} as computed in~\cite{Esposito:2014hsa}. The red line corresponds to the results obtained using the NREFT for the molecular model, while the black one corresponds to the diquarkonium picture. The yellow (green) band corresponds to the $68\%$ ($95\%$) confidence region. The two models provide significantly different predictions.} \label{fig:etac}
\end{figure}
In \figurename{~\ref{fig:etac}} we report the likelihood distributions for the following ratios
\begin{align} \label{RZ}
\mathcal{R}_Z=\frac{\mathcal{B}(Z_c\to\eta_c\,\rho)}{\mathcal{B}(Z_c\to \jpsi\,\rho)} \quad \text{ and } \quad \mathcal{R}_{Z^\prime}=\frac{\mathcal{B}(Z^\prime_c\to\eta_c\,\rho)}{\mathcal{B}(Z^\prime_c\to h_c\,\pi)}
\end{align}
computed using the two models. As one immediately sees, the predictions differ from each other by more than $2\sigma$ and hence the $\eta_c\,\rho$ channel can be used as an experimental tool to probe the internal structure of the charged $Z_c^{(\prime)}$ states. The same authors fitted the couplings $g_{Z_c}$ and $g_{Z_c^\prime}$ from data, assuming the total width to be saturated by the $D^{(*)}D^*,\,\eta_c\,\rho,\,h_c\,\pi\,\jpsi\pi$ and $\psi(2S)\,\pi$ final states. Once the couplings are known one can compute the ratios for the decays of the two different $Z_c^{(\prime)}$ into the same final state, assuming molecular nature. One finds
\begin{align}
\frac{\mathcal{B}(Z_c\to h_c\pi)}{\mathcal{B}(Z_c^\prime\to h_c\pi)}=0.34^{+0.21}_{-0.13}\quad \text{ and } \quad \frac{\mathcal{B}(Z_c\to J/\psi\pi)}{\mathcal{B}(Z_c^\prime\to J/\psi\pi)}=0.35^{+0.49}_{-0.21}
\end{align}
The previous estimates essentially tell us that, within theoretical errors, the branching fractions for the $Z_c$ and $Z_c^\prime$ into the same final state should be of the same order of magnitude. Both should be observed in both channels.
The $h_c\pi$ channel might indeed be compatible with experimental data, where a hint of $Z_c$ is seen (albeit not statistically significant). On the other hand the $Z_c^\prime$ is completely missing from the $J/\psi\pi$ final state, which seems to be in contrast with the molecular prediction --- see \sectionname{\ref{sec:Zc}}

A slightly different approach has been used in~\cite{Braaten:2003he,Fleming:2007rp,Mehen:2015efa} to describe the $X(3872)$. In this case, the interaction Lagrangian does not contain the $\bbb X$ field explicitly but it only describes the interaction of the $D$ mesons with pions and charmonia. Instead of requiring the presence of intermediate open flavor meson loops as explained above, the authors automatically implement the molecular hypothesis by describing the $X(3872)$ with the interpolating field $\bbb X=(D\bar{\bbb D} - \bar D \bbb D)/\!\sqrt{2}$. This method has been used to describe the $X\to \bar D^0 D^0\pi^0$~\cite{Fleming:2007rp} and $X\to\chi_{cJ}\pi^0$~\cite{Mehen:2015efa} decays, as well as to provide an explicit implementation of the fine tuning required to explain the unnaturally large scattering length of the $X(3872)$~\cite{Braaten:2003he} --- see \sectionname{\ref{ascattering}}.

\subsection{Phenomenological realizations of the meson-meson potential} \label{sec:potentials}
Most of the considerations made so far about loosely bound molecules follow strictly from the application of the theory of shallow bound states, or of the HQSS. The results obtained in the two cases are universal in the $B\to0$ and $m_Q\to\infty$ limits respectively, as they do not rely on the detailed knowledge of the binding potential, but rather on the mere existence of a (loosely) bound state. The actual interaction between the two mesons is the complicated result of residual strong forces between color singlet objects. It then follows that its description must rely on some phenomenological picture and approximations, which necessarily introduce a good degree of model dependence in the final results, \eg the value of the binding energy or even the very existence of bound states.
Here we introduce the most popular models for the explicit realization of the inter-hadron potential.

\subsubsection{One-pion-exchange \label{sec:1pi}}

It is well known that the force that holds the deuteron together can be  described by a potential whose main contribution comes from the exchange of a pion between the two nucleons. With this in mind, it is rather natural to try to extend the same approach to mesons and see whether or not this so-called \emph{one-pion-exchange} potential allows for bound states. This program was first carried on in~\cite{Tornqvist:1991ks,Tornqvist:1993ng}, mainly for the study of light mesons and extended to the heavy sector by several different authors, \eg~\cite{Voloshin:1976ap,DeRujula:1976zlg,Tornqvist:2004qy,Cleven:2015era,Esposito:2014hsa}. It was already pointed out in the original references that using only this kind of potential could only provide plausibility arguments since the full treatment would require the inclusion of coupled channel effects, two-pion-exchange, etc.

The leading order contribution to this potential comes from the exchange of a single pion between the two mesons. Although the result is true also for light mesons, as computed in~\cite{Tornqvist:1993ng}, here we will derive the potential starting from the HQET formalism (see \eg~\cite{Manohar:1992nd,Casalbuoni:1996pg}). It is well known that in the $m_Q\to\infty$ limit, the interaction between the light pseudoscalar mesons and the heavy mesons can be described by the following Lagrangian
\begin{align} \label{HQET}
\mathcal{L}=-\frac{g}{f_\pi}\text{tr}\left[\bar H_aH_b\gamma_\mu\gamma_5\right]\partial^{\mu}\mathcal{M}_{ab}
\end{align}
where $f_\pi\simeq132$\mev is the pion decay constant and $g$ is some dimensionless axial coupling (typically around $0.5-0.7$~\cite{Casalbuoni:1996pg}). The expression is traced over Dirac indices. $H_a$ is a bi-spinor containing the operators $V_{a\mu}$ and $P_a$ describing respectively the vector and pseudoscalar heavy mesons as
\begin{align} \label{HQETspinor}
H_a=\frac{1+\slashed{v}}{2}\left[V_{a\mu}\gamma^\mu-P_a\gamma_5\right] \quad \text{and} \quad \bar H_a=\gamma_0H_a^\dagger\gamma_0
\end{align}
where $a$ and $b$ are flavor indices. The light pseudo-Goldstone boson fields are instead contained in
\begin{align}
\mathcal{M}=\left(\begin{array}{ccc}
\frac{1}{\sqrt{2}}\pi^0+\frac{1}{\sqrt{6}}\eta_8 & \pi^+ & K^+ \\
\pi^- & -\frac{1}{\sqrt{2}}\pi^0+\frac{1}{\sqrt{6}}\eta_8 & K^0 \\
K^- & \bar K^0 & -\sqrt{\frac{2}{3}}\eta_8
\end{array}\right)
\end{align}
The interaction in Eq.~\eqref{HQET} can then be rewritten as
\begin{align} \label{Lint}
\mathcal{L}=-\frac{2g}{f_\pi}\left(V_\mu \partial^\mu\mathcal{M}P^\dagger +h.c.\right)+\frac{2ig}{f_\pi}\epsilon_{\alpha\beta\mu\nu}V^\beta\partial^\mu\mathcal{M}V^{\dagger\alpha} v^\nu
\end{align}
where we recall that $v^\nu$ is the HQET four-velocity appearing in Eq.~\eqref{HQETspinor}.
From the previous equations one can  find the nonrelativistic momentum space potentials for the exchange of a pion between vector and pseudoscalar heavy mesons
\begin{subequations} \label{1pipot}
\begin{gather}
V_\pi(VV)=-V_\pi(V\bar V)=\frac{8g^2}{f_\pi^2}(\bbb \tau_1\cdot \bbb \tau_2)\left(\bbb \Sigma_1\cdot \bbb q\right)\left(\bbb \Sigma_2\cdot \bbb q\right)\frac{1}{\bbb q^2+m_\pi^2} \\
V_\pi(PV)=\frac{8g^2}{f_\pi^2}(\bbb \tau_1\cdot \bbb \tau_2)\,(\mathds{1}_1\cdot \bbb q)\,(\mathds{1}_2 \cdot \bbb q)\frac{1}{\bbb q^2+\mu^2}
\end{gather}
\end{subequations}
to be sandwiched between appropriate polarization vectors. Here $\bbb \tau_i$ are isospin Pauli matrices, $\bbb \Sigma_i$ and $\mathds{1}_i$ are spin-1 matrices,  and \mbox{$\mu^2=m_\pi^2-(m_V-m_P)^2$}. The definition of $\mu$ is supposed to approximately take into account recoil effects due the different masses of the constituents.\footnote{This prescription is meaningful as long as $\mu^2$ is not too far from $m_\pi^2$ as, for example, in the bottom case. Less obvious is its validity in the charm sector where $m_{D^*}-m_D\approx m_\pi$, or even for $\bar KK^*$ molecules, where $\mu^2 < 0$.} Note that $PP$ interactions in the one-pion-exchange approximation are forbidden by parity conservation. 

The two potentials are both in the form
\begin{align}
V_\pi(\bbb q)=\frac{8g^2}{f_\pi^2}(\bbb \tau_1\cdot \bbb \tau_2)\left(\bbb{\mathcal{A}}_1\cdot \bbb q\right)\left(\bbb{\mathcal{A}}_2\cdot \bbb q\right)\frac{1}{\bbb q^2+m^2}
\end{align}
with obvious definitions. The position space potential is therefore given by their Fourier transform
\begin{align}
V_\pi(\bbb x)=&\frac{8g^2}{f_\pi^2}(\bbb \tau_1\cdot \bbb \tau_2)\int \frac{d^3q}{{(2\pi)}^3}\left(\bbb{\mathcal{A}}_1\cdot \bbb q\right)\left(\bbb{\mathcal{A}}_2\cdot \bbb q\right)\frac{e^{i\bbb q\cdot \bbb x}}{\bbb q^2+m^2}=-\frac{8g^2}{f_\pi^2}(\bbb \tau_1\cdot \bbb \tau_2)\left(\mathcal{A}_{1i}\cdot \partial_i\right)\left(\mathcal{A}_{2j}\cdot \partial_j\right)\int \frac{d^3q}{{(2\pi)}^3}\frac{e^{i\bbb q\cdot \bbb x}}{\bbb q^2+m^2} \notag \\
=&-\frac{8g^2}{f_\pi^2}(\bbb \tau_1\cdot \bbb \tau_2)\left(\mathcal{A}_{1i}\cdot \partial_i\right)\left(\mathcal{A}_{2j}\cdot \partial_j\right) \frac{e^{-m r}}{4\pi r}
\end{align}
Now, recalling that $\partial^2(1/r)=-4\pi\,\delta^{(3)}(\bbb x)$, we obtain the one-pion-exchange potentials in configuration space
\begin{align}
V_\pi(\bbb x)=-\frac{8g^2}{f_\pi^2}(\bbb \tau_1\cdot \bbb \tau_2)\,\mathcal{A}_{1i}\mathcal{A}_{2j}\left\{\left[ \left(3\frac{x_ix_j}{r^2}-\delta_{ij}\right) \left(1+\frac{3}{mr}+\frac{3}{m^2r^2}\right) +\delta_{ij} \right]\frac{m^2}{3}\frac{e^{-mr}}{r}-\frac{4\pi}{3}\delta_{ij}\delta^{(3)}(\bbb x)\right\}
\end{align}
Because of the singular behavior at $r=0$ the previous potential needs to be regularized. This is typically done introducing some form factor for the $\pi V$ and $\pi P$ vertices, which also allows to neglect the $\delta$-function contribution. At any rate, this short term contribution could  be ignored since one-pion exchange is not a fully reliable model for meson-meson short range interactions.  Once the potential is specified it is just a matter of numerically solving the Schr\"{o}dinger equation and verify whether a bound state is present or not.

In the very first attempt~\cite{Tornqvist:1991ks}, the author tried to identify the observed $f_1(1420)$, $f_0(1710)$, $f_2^\prime(1525)$ and $f_0(1500)$ mesons with respectively $K\bar K^*$, $K^*\bar K^*$, $\omega\omega$ and $\rho\rho$ bound states. However, the numerical analysis later performed~\cite{Tornqvist:1993ng} showed that, although attractive, one-pion-exchange potentials alone are not enough to bind light mesons molecules. On the other hand, they are allowed in the heavy sector. In particular, charm bound states are predicted to be very near threshold while their bottom counter part should have a binding energy around $50$\mev.\footnote{In~\cite{Ericson:1993wy} it was also noticed that, even in presence of an attractive one-pion-potential, meson-meson bound states are only allowed for sufficiently large reduced mass.} The detailed analysis of the allowed bound states and their quantum numbers, in the heavy quark sector is carried out in~\cite{Tornqvist:1993ng,Tornqvist:2004qy}. The results are reported in \tablename{\ref{tab:bounds}}. One immediately notices that there are good candidates for the $X$, the $Z_c^\prime$ and $Z_b^\prime$ (even though the mass of the last one does not comply with  experimental data) while there are no bound states corresponding to the $Z_c$ and $Z_b$. The latter ones should, in fact, be $P\bar V$ molecules but with $J^{PC}=1^{+-}$. 

The absence of observed bound states from the present model does not rule out the molecular interpretation of the exotic resonances. It rather means that the long range, pion mediated interaction alone, as captured by the present model, cannot describe the observed spectrum. A contribution from short range potential is therefore needed. Such interactions are, however, complicated and their phenomenological realization can hardly be deduced from first principles. In the following section we will report one of this attempts.

\begin{table}[thb]
\centering
\begin{tabular}{ccc|ccc}
\hline\hline
Constituents & $J^{PC}$ & Expected mass (\mevnospace) & Constituents & $J^{PC}$ & Expected mass (\mevnospace) \\
\hline
$D\bar D^*$ & $0^{-+}$ & $\simeq 3870$ & $B\bar B^*$ & $0^{-+}$ & $\simeq10545$ \\
$D\bar D^*$ & $1^{++}$ & $\simeq 3870$ & $B\bar B^*$ & $1^{++}$ & $\simeq10562$ \\
\hline
$D^*\bar D^*$ & $0^{++}$ & $\simeq 4015$ & $B^*\bar B^*$ & $0^{++}$ & $\simeq 10582$ \\
$D^*\bar D^*$ & $0^{-+}$ & $\simeq 4015$ & $B^*\bar B^*$ & $0^{-+}$ & $\simeq 10590$ \\
$D^*\bar D^*$ & $1^{+-}$ & $\simeq 4015$ & $B^*\bar B^*$ & $1^{+-}$ & $\simeq 10608$ \\
$D^*\bar D^*$ & $2^{++}$ & $\simeq 4015$ & $B^*\bar B^*$ & $2^{++}$ & $\simeq 10602$ \\
\hline \hline
\end{tabular}
\caption{Bound states allowed by the one-pion-exchange potential as found in~\cite{Tornqvist:2004qy}. All states have $I=0$.} \label{tab:bounds}
\end{table}

In~\cite{Cleven:2015era} it was also observed that because of the $\bbb \tau_1\cdot\bbb \tau_2$ term in Eqs.~\eqref{1pipot}, the potential for a given molecule made of isospin-1/2 mesons, and with total isospin $I$ and third component $I_3$, comes with an overall factor
\begin{align}
C \,\langle I,I_3|\bbb \tau_1\cdot\bbb \tau_2|I,I_3\rangle=2C\left(I(I+1)-\frac{3}{2}\right)
\end{align}
where $C$ is the eigenvalue of the charge conjugation. This means that in the one-pion-exchange approximation, if the $I=0$ channel is attractive the $I=1$ is repulsive and vice versa. Hence, for a given set of quantum numbers, isosinglets and isovectors are unlikely to coexist. Indeed, in the case of the $X(3872)$ assumed as an isosinglet, we have $C=+$, $I=0$, which is attractive. For the almost degenerate $Z_c(3900)$, we have $C=-$, $I=1$, which is still attractive, although with a three times smaller potential. Indeed it was for a long time considered too weak to allow for a bound state. The argument was withdrawn as soon as the charged states were observed.
This is in contrast with the predictions obtained in the diquarkonium picture for the tetraquark model, as found in \sectionname{\ref{diquark}}. In fact, for a given $J^{PC}$, the model admits both $I=0$ and $I=1$ states. This is for example, the famous case of the charged partners of the $X(3872)$.

From the phenomenological description of the potential it is clear that meson molecules should mainly decay in their constituents. Channels with large phase space are, in fact, mediated by heavier particles, and are therefore suppressed by the large size of the bound state --- see again \sectionname{\ref{ascattering}}.

As we mentioned at the beginning of the section, the results obtained from the previous potentials are only reliable as long as they are considered as plausibility arguments. There are several instances where the one-pion-exchange model fails. For example, it has been shown~\cite{Kalashnikova:2012qf,Liu:2008fh} that the presence of a shallow bound state corresponding to the $X(3872)$ is not certain, as the results strongly depend on the value of the phenomenological cut-offs that one introduces in the regulator of the potential as well as in the vertex form factor. Furthermore~\cite{Suzuki:2005ha}, since the $X$ is very close to the $DD\pi$ threshold as well (only $\approx7\mev$), all the long range terms of the $S$-wave potential are suppressed, leaving only the contact $\delta(r)$ term in the $m_{D^*}=m_D+m_\pi$ limit.

To overcome these problems one can try to go beyond the tree level approximation including, for example, additional terms to describe the short range part of the meson-meson potential or to use a three-body formalism for the $D\bar D\pi$ system~\cite{Baru:2011rs,Baru:2013rta}. However, these approaches add in general several complications and appear to be rather ad hoc.

\subsubsection{Interaction potential at the quark level}

Because of the small pion mass, the potential described in the previous section is a long range one. An interesting model to describe the short range meson-meson interaction has been introduced in~\cite{Barnes:1991em}, and then applied to the case of the $X(3872)$ in~\cite{Swanson:2003tb} (see also~\cite{Swanson:2006st}). In this picture, the interaction between the two mesons is described as interaction between their four constituents via the following Hamiltonian
\begin{align} \label{Hswanson}
H_I=\sum_{i<j}\frac{\bbb\lambda_i}{2}\cdot\frac{\bbb\lambda_j}{2}\left(\frac{\alpha_s}{r_{ij}}-\frac{3b}{4}r_{ij}-\frac{8\pi\alpha_s}{3m_im_j}\bbb S_i\cdot \bbb S_j \,\delta^{(3)}(\bbb r_{ij}) \right)
\end{align}
where $\bbb \lambda^a$ are the Gell-Mann matrices, $\bbb r_{ij}$ is the separation between the $i$-th and $j$-th quark, $m_i$ their masses and $\bbb S_i$ their spins. The first and second terms are the phenomenologically well-known Coulomb and confining terms, while the last one is a short range spin-spin hyperfine interaction motivated by one-gluon-exchange. The point-like $\delta$-function is  regularized with a Gaussian
\begin{align}
\delta^{(3)}(\bbb r_{ij})\longrightarrow\left(\frac{\sigma^3}{\pi^{3/2}}\right)\, e^{-\sigma^2r_{ij}^2}
\end{align}
The different parameters can be estimated from the observed meson spectrum. It should be mentioned that exponential form factors produce wrong analytic properties in $r^2$ and should therefore be avoided.

The interaction in Eq.~\eqref{Hswanson} can be used to compute the scattering amplitude for four mesons in a $AB\to CD$ process~\cite{Barnes:1991em}. The scattering matrix is parametrized as
\begin{align} \label{Sswanson}
S_{AB\to CD}\equiv-2\pi i \, \delta(E_{AB}-E_{CD}) \, \delta^{(3)}(\bbb P_{AB}-\bbb P_{CD}) \, h_{AB\to CD}
\end{align}
For each Feynman diagram describing the four quarks in the $AB\to CD$ process, the amplitude can be written as
\begin{align}
h_{AB\to CD}^{\text{(diagr.)}}=S I_\text{flavor} \, I_\text{color} \, I_\text{spin} \, I_\text{space}
\end{align}
Here $S=\pm 1$ is `signature' factor that takes into account the permutation of fermion operators in the scattering matrix elements, $I_\text{flavor}$ is the overlap coefficient between the flavor states of the incoming and outgoing mesons, $I_\text{color}$ is the color factor associated with usual QCD diagrams, and $I_\text{spin}$ and $I_\text{space}$ are the overlap integrals of the meson spin and space wave functions respectively. See~\cite{Barnes:1991em} for details.

\begin{figure}[ht]
\centering
\includegraphics[width=0.5\textwidth]{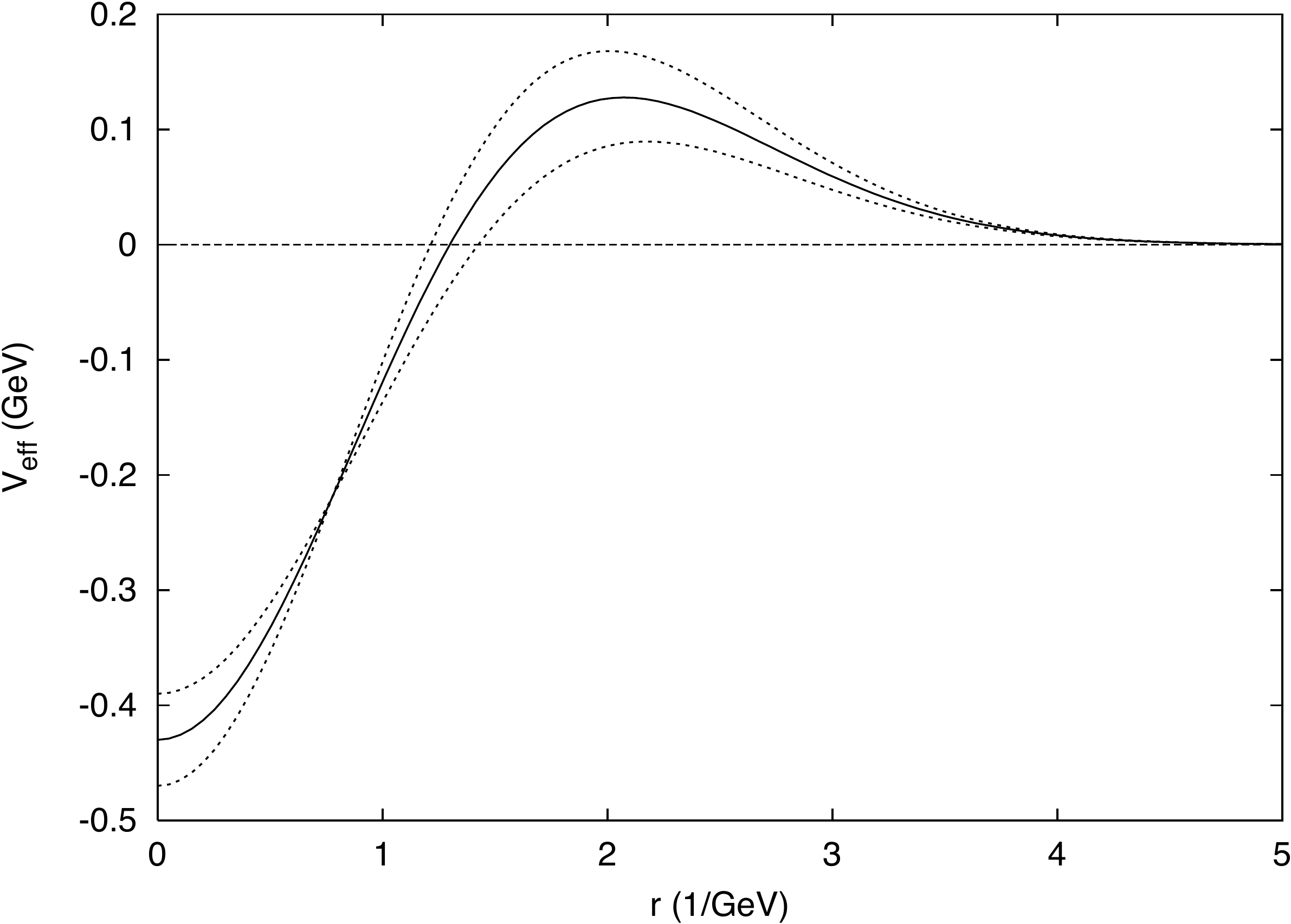}
\caption{Effective potential for the meson-meson interaction as extracted from the $D\bar D^*\to \jpsi\omega$ scattering amplitude. The dashed lines describe the theoretical error due to the approximate knowledge of the initial and final mesons wave function. See~\cite{Barnes:1991em} for more details.} \label{fig:Veff}
\end{figure}

Once the quark level amplitude $h_{AB\to CD}$ is known, one can compute the same thing using an effective meson-meson potential. This is usually parametrized as a sum of Gaussians
\begin{align}
V_\text{eff}=\sum_i \alpha_i \, e^{-r^2/2\beta_i^2}
\end{align}
Equating the results obtained with the two techniques, the $\alpha_i$ and $\beta_i$ parameters can be fixed. In \figurename{~\ref{fig:Veff}} we show an example of effective potential extracted from the $D\bar D^*\to \jpsi\omega$ process~\cite{Barnes:1991em}.

The shape of the potential shows how this short range interaction could in principle admit bound states in the $D\bar D^*$, $\jpsi\omega$ or $\jpsi\rho$ systems. However, given the quark level Hamiltonian~\eqref{Hswanson}, it is not clear how to distinguish a `molecular' bound state obtained with it from a compact tetraquark object --- see \sectionname{\ref{diquark}}. The typical distinction between the two cases is that a molecule is an extended object (several fm) while a tetraquark is an almost point-like object. A bound state of $H_I$ alone would have the typical hadronic size of 1 fm and hence would be distinguishable from a diquarkonium only because of its color structure. From a quantum field theory point of view, the latter one is a meaningless distinction since different color arrangements mix with each other under renormalization.

In any case, more detailed calculations show that the depth of $V_\text{eff}$ is not sufficient, and therefore a contribution from the one-pion-exchange potential needs to be added anyway~\cite{Swanson:2003tb}.  On the basis  of the observations made in the previous section, one can conclude that neither the long range one-pion potential, nor the short range quark potential alone are able to reproduce the observed \XYZ spectrum. If one wants to insist in a molecular description of these states, a combination of the two is needed. However, this inevitably introduces a good degree of model dependence on the physical results.

\subsection{Open problems with the molecular model}

The molecular description of the \XYZ states has many appealing advantages, as we explained during the previous section. However, it is now hopefully clear that it is also plagued by some serious limitations that make its plain application questionable.
In summary, the most relevant issues with this framework are
\begin{enumerate}
\item It is not able to convincingly solve the prompt production problem. The production cross section of a meson molecule with very small binding energy should be extremely suppressed in high energy collisions at high $\pt$. The high relative momentum between the two constituents should prevent their binding. This intuition is supported both by MC simulations (see \ref{appmol}) and experimental data on deuteron, Helium-3 and hypertriton (see \sectionname{~\ref{xdeut}}). This is in striking contrast with the prompt production of the $X(3872)$ with $\pt\gtrsim12\gev$.

\vspace{1em}

\item The theory of shallow bound states is universal in the limit %
 $E, B\to0$ (see \sectionname{~\ref{shallowbs}}). However, exception made for the $X$, the would-be binding energy of several \XYZ resonances is of some $10\mev$. This value escapes the rigorous application of the shallow bound state formalism. If this is not possible, one has to rely on phenomenological realizations of the meson-meson potential (see \sectionname{~\ref{sec:potentials}}). This introduces a large degree of model dependence on several quantities like the binding energy, but also on the very existence of bound states. This is clearly not a problem of the molecular interpretation {\it per se}, but rather of the available models, which have little predictive power.

\vspace{1em}

\item As it can be seen from \tablename{\ref{tab:bounds}}, and its obvious extensions,  the molecular model predicts a large number of states which are still largely unobserved.
\end{enumerate}

\section{Diquark building blocks \label{diquark}}
Heavy-light diquarks were introduced by Maiani~\etal in~\cite{Maiani:2004vq} to discuss the $X(3872)$ following the suggestion of Jaffe and Wilczek~\cite{Jaffe:2003sg} to use light diquarks in exotic spectroscopy, with particular reference to some experimental hints of a light pentaquark.  

\begin{figure}[t]
\centering
\includegraphics[width=.6\textwidth]{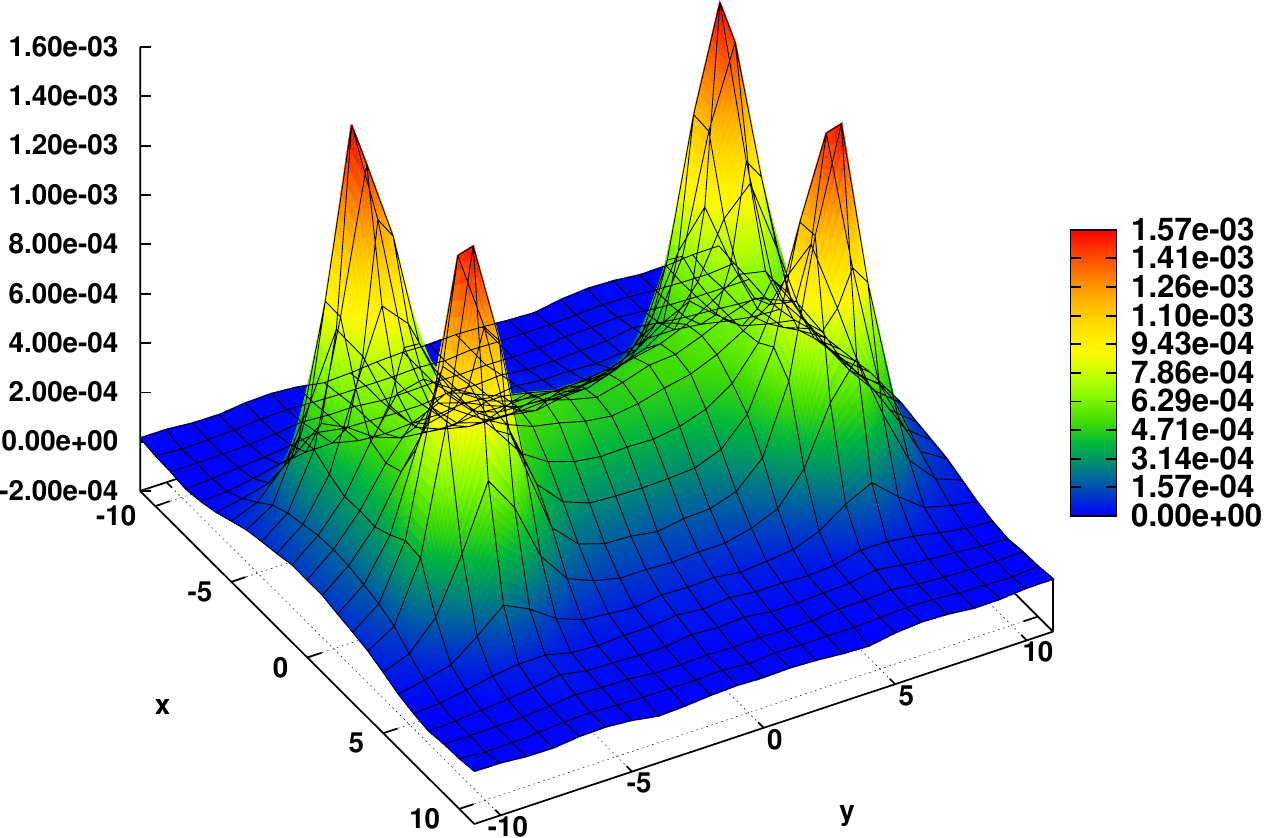}
\caption{Lagrangian density 3D plot for a four quark system, from~\cite{Cardoso:2011fq}. The meson mixing with the tetraquark is sufficiently small to produce such a clear tetraquark H flux tube. 
} \label{fig:bicudo}
\end{figure}

Evidence that in a tetraquark system the two quarks arrange their color in a diquark before interacting with the antiquarks has also been found on the lattice in the static limit~\cite{Cardoso:2011fq}. The same authors also show how the four constituents arrange themselves in a H shaped configuration, as already predicted in the literature~\cite{Rossi:1977cy,Montanet:1980te,Rossi:1977dp,Cotugno:2009ys,Rossi:2016szw} --- see Figure~\ref{fig:bicudo}.

The diquark-antidiquark model of \XYZ resonances has inspired the search of charged resonances, like the $Z(4430)$ and $Z_{c,b}$s since it straightforwardly predicts complete charge multiplets, contrarily to the molecular models applied to the description of the $X(3872)$ soon after its discovery.

We are convinced that diquarks are good degrees of freedom to understand the \XYZ phenomenology, and we will illustrate the indisputable successes of the diquark-antidiquark model, together with its obvious limitations. The attempt to incorporate diquarks in an extended picture including the role of meson-meson threshold  will be extensively discussed in~\sectionname{\ref{mol3}}. 

\subsection{General features}
As discussed in \ref{appN}, in SU(3) there is attraction between $qq$ pairs in the color antitriplet channel and this is just twice weaker than in the color singlet $q\bar q$ in the one-gluon exchange approximation. Because of this we will usually refer to diquarks in the antisymmetric color configuration 
\begin{equation}
 {d}^{A}_\Gamma=q^\alpha \Gamma q^{\prime \beta}  - q^\beta \Gamma q^{\prime \alpha}\equiv [qq^\prime]_S
\label{diqop}
\end{equation}
where $A$ stands for `antisymmetric', $\Gamma$ are matrices to characterize the diquark spin $S$ and $\alpha,\beta$ are color indices.

A kind of fermion-boson transformation can be defined
\begin{subequations}
\begin{align} 
q &\to {\bar d}^A  \label{sosq}\\
\bar q &\to d^A \label{sosqbar}
\end{align}
\end{subequations}
as a generic rule to build new structures with diquarks starting from mesons and baryons. 

In particular there is a special relation between baryons and tetraquarks. 
If we start from an antibaryon, the substitution in Eq.~\eqref{sosqbar} produces the tetraquark $d^A \bar q \bar q= d^A{\bar d}^A$. Applying~\eqref{sosqbar} once again, one obtains a {\it pentaquark}, $d^A d^A\bar q$, and finally, with a third substitution, a state with baryon number $B=2$, a {\it dibaryon} with the configuration $d^A d^A d^A$. 

It is clear that iterating this procedure a whole bunch of new multiplets of particles might be expected with a large variety of flavors and charges. The first steps in this direction were done in~\cite{Maiani:2004vq, ynoi}. The search even extended to excited states as discussed~\cite{Drenska:2008gr} and considering also strange light quarks~\cite{Drenska:2009cd}. This program was reported in~\cite{Drenska:2010kg}.
As of today the experimental situation in the field of \XYZ resonances does not seem (yet?) to be compatible with such a multitude of particles. 

The first and better known resonance, the  $X(3872)$, appears as a neutral state alone: the charged counterparts $X^+$, obviously predicted by a $d^A\bar d^A$ assignment,  have never been observed.  Moreover the $X^0$ is observed to decay with strong isospin violations, having almost the same branching ratio in the $\jpsi\rho$ and $\jpsi \omega$ decays, as if there were two almost degenerate states $X_u=[cu][\bar c\bar u]$ and $X_d=[cd][\bar c\bar d]$, causing maximal isospin violation.  These should be observed in the vicinity of  $3872\mev$~\cite{Maiani:2007vr} but, as of today, this has not been the case. 

On the other hand, the diquark approach, differently from all other models of \XYZ states discussed in the literature, was the first pointing to the existence of charged resonances in decays such as $\jpsi\,  \pi^+$~\cite{Maiani:2004vq,Maiani:2014aja, Faccini:2013lda}.  We consider a remarkable success of this model the fact that it stimulated the experimental search, and discovery, of the {\it charged} states  $Z(4430)$, $Z_c(3900)$, $Z_c (4020)$, $Z_b(10610)$ and $Z_b(10650)$.   They came as a surprise for those working with molecular models tailored to describe the special case of the $X(3872)$. 

The $X(3872)$, besides being the worst enemy against a straightforward construction of tetraquarks as diquark-antidiquark states for the reasons just reminded, has as an additional oddity which has inspired a lot of work: the  {\it double-fine-tuning} of the $X$ mass with the $\jpsi\, \rho$ and, especially with $\bar D^0D^{*0}$ threshold, which has been discussed at length in previous sections. The latter case is really remarkable, leaving a difference of few{\kev}s (compatible with zero within experimental errors).  To our knowledge there is no simple way  to explain this fact in  the diquarkonium picture. As we saw in the previous section, the tiny binding energy of the $X$ can be related to a low energy scattering enhancement in the $\bar D^0 D^{*0}$ system --- with all the difficulties we discussed.

Another straightforward prediction of the diquark-antidiquark model would be the existence of a $[bq][\bar b\bar q]$ partner in the beauty sector of the $[cq][\bar c \bar q]$ in the charm one. Also this has not been found so far.

Therefore, despite the success with the qualitative prediction of charged resonances decaying into charmonium + charged meson, the diquark model has way more states than what observed and the research of selection rules within  the diquarkonium picture has  never been really attempted.  

The large number of predicted states is a problem afflicting also models based on the idea of loosely bound molecules. The number of expected states, is basically the same, especially if a resonance is associated to each  possible meson-meson threshold. The good point about insisting on the relevance of the vicinity of the $X(3872)$ to the $\bar D^0D^{*0}$ threshold is that also the most recently discovered   
$Z(4430)$, $Z_c(3900)$, $Z_c (4020)$, $Z_b(10610)$ and $Z_b(10650)$ have meson-meson thresholds rather close to their masses. 

This could  also seem to be a casual feature due to the fact that the meson-meson thresholds from the Particle Data Book form a quasi-continuum spectrum of states.

However, looking at things in greater detail,  there are very suggestive facts. The $Z_b(10610)$ happens to be exactly at the  $\bar B^0 B^{*0}$ threshold, as is the case for the $X(3872)$. 
 $Z_c(3900)$, $Z_c (4020)$ and $Z_b(10650)$ do not feature the same impressive tuning with thresholds but are still very close to some of them, being a bit {\it heavier} --- so that no `standard' molecule interpretation can be given, the binding energy having the wrong sign!
 
 So, on one hand the diquarks seem to be the right tools to easily build a spectroscopy of exotic hadrons like tetraquarks, pentaquarks and dibaryons. On the other hand the feature of vicinity to thresholds seems to have a relevant phenomenological role. In \sectionname{\ref{diqn}}, we will examine the role of diquarks in the $1/N$ expansion of QCD showing that they are good degrees of freedom to describe tetraquarks in the $s$-channel cuts of meson-meson amplitudes. 
 
 In \sectionname{\ref{mol3}} we will present a theory which eventually might explain 
 \begin{enumerate}
 \item The absence of $X^+(3872)$;
 \item The absence of a degenerate doublet of $X^0_{1,2}$ neutral doublet, required from the na\"ive diquarkonium picture to explain the isospin violation pattern in the $X$ decay;
 \item The extremely small width of the $X(3872)$ ($\Gamma_X\lesssim1\mev$);
 \item The appearance of $Z_c$ and of $Z_b$ ;
\item The difference in total widths of these states as related to their mass values.
\end{enumerate}
 
In order to reach this  description of $XZ$ phenomenology, we will begin by reviewing the diquarkonium picture, and then show how  diquarkonium discrete levels with some $J^{PC}$ might induce an effective interaction between color neutral mesons with the same quantum numbers, leading to the formation of a resonance~\cite{Esposito:2016itg}. 

A dynamical picture for the diquarkonium has been proposed by Brodsky \etal in~\cite{Brodsky:2014xia,Brodsky:2015wza,Blitz:2015nra}, which leads to interesting predictions for the decay into charmonia, and for the production cross section at lepton colliders. Different extensions of the constituent quark models for the tetraquark spectroscopy have been explored by Valcarce \etal\cite{Barnea:2006sd,Vijande:2007fc,Vijande:2007rf,FernandezCarames:2009zz,Vijande:2013qr}, Buccella~\etal\cite{Buccella:2006fn,Abud:2009rk} and Santopinto~\etal~\cite{Santopinto:2006my,Ferretti:2013faa,Ferretti:2014xqa,Santopinto:2016pkp}. The idea of probing the number of constituents by producing tetraquark pairs has been discussed in~\cite{Brodsky:2015wza,Guo:2016fqg,Voloshin:2016phx}. A different and very interesting viewpoint is given by Rossi and Veneziano~\cite{Rossi:1977cy,Montanet:1980te,Rossi:2016szw}. The consequences of color dynamics and Rosner duality for exotic states are reviewed in~\cite{Richard:2016eis}. Lastly, the problem of light tetraquarks have been approached from a field theoretical point of view using the Dyson-Schwinger formalism in~\cite{Eichmann:2015cra}. Its application to the heavy sector as well might give an interesting contribution to the discussion.
 
In the next few sections, we will review the notion of diquark and especially the algebra needed to deal with them when discussing tetraquarks in the form of diquarkonia.  

Diquarks appear in several branches  of QCD, especially in the non-perturbative regime, see for example~\cite{Rajagopal:2000wf} and reference therein, concerning  the behavior of hadronic matter at high density studied with techniques of condensed matter theory. New phases of matter and their properties are predicted with interesting consequences also for the astrophysics of neutron stars, for example. 
Diquarks can have an interesting role also for the discussion of proton and neutron structure functions~\cite{Wilczek:2004im}, see also~\cite{Jaffe:2004ph}. 
A tetraquark condensate, built with diquarks, could be the order parameter for an additional chiral phase transition in QCD as recently discussed in~\cite{Pisarski:2016ukx}.

\subsection{Diquarks: color}
\label{sec:diquarkscolor}

In the four spinor formalism the {\it spin zero} diquark $J^P=0^+$ is 
\begin{equation} \label{defidiqu}
[cq]_i = \epsilon_{ijk}\,\bar c^{j}_{\bm c}\gamma_5 q^{k}\equiv \epsilon_{ijk}\, (c^{j})^T \,C\gamma_5\,q^k
\end{equation}
where $i$, $j$ and $k$ are color indices and $\bm c$ 
indicates the charge conjugated four-spinor~\footnote{
A light-light diquark with spin 0 has to be antisymmetric in flavor for Fermi statistics. If we consider two flavors only this would write 
\begin{equation}
[qq]_i =  \epsilon_{ijk}\,\epsilon^{ab}\, (q^{a j})^T \,C\gamma_5\,q^{b k}
\label{defidiqu2}
\end{equation}
where the flavor structure is (the transposition in~\eqref{defidiqu2} is with respect to spinor indices)
\begin{equation}
\epsilon^{ab}\, q^{a j} \,q^{b k}=q^T\, (i\sigma^2)\, q
\end{equation}
being $\epsilon^{ab}=(i\sigma^2)_{ab}$ and leaving aside non-flavors labels and matrices commuting with flavor indices. 

Under a SU(2) flavor transformation we have 
\begin{equation}
 (e^{i\bm \sigma\cdot \bm \theta} q )^T\, (i\sigma^2)\, (e^{i\bm \sigma\cdot \bm \theta} q)=(q^T e^{i\bm \sigma^T\cdot \bm \theta}  )\, (i\sigma^2)\, (e^{i\bm \sigma\cdot \bm \theta} q)
\end{equation}
Since $\sigma^2\bm \sigma=-\bm \sigma^T\sigma^2$
\begin{equation}
 e^{i\bm \sigma^T\cdot \bm \theta}  \, (i\sigma^2)\, e^{i\bm \sigma\cdot \bm \theta} \equiv i\sigma^2
 \end{equation}
we conclude that~\eqref{defidiqu2} is SU(2) invariant. With three flavors, the flavor antisymmetric diquark would be 
\begin{equation}
\zeta^a_i=[qq]_i ^a=  \epsilon_{ijk}\,\epsilon^{abc}\, (q^{b j})^T \,C\gamma_5\,q^{c k}
\label{defidiqu3}
\end{equation}
which transforms non trivially under SU(3) as  
\begin{equation}
\zeta \to V \zeta 
\end{equation}
where 
\begin{equation}
V^{ad}=\frac{1}{2}\epsilon^{abc}\, \epsilon^{db^\prime c^\prime}U^{b b^\prime}U^{c c^\prime}
\end{equation}
A tetraquark condensate can be constructed
\begin{equation}
\Phi^{ab}=(\zeta^a_{L \,i})^* \, \zeta^{bj}_R
\end{equation}
which transforms under SU(3)$_L\,\times\,$SU(3)$_R$ as
\begin{equation}
\Phi \to V_R\,\Phi \, V_L^\dag
\end{equation}
\ie like the standard quark condensate, with interesting consequences for chiral phase transitions~\cite{Pisarski:2016ukx}.
}. The charge conjugation operator is 
\begin{equation}
C=i\gamma^2\gamma^0=(\sigma^2\otimes i\tau^2)(\mathds{1}\otimes\tau^3)=(\sigma^2\otimes (-1)\tau^1)=-\bpm 0&\sigma^2\\ \sigma^2 &0\epm \notag
\end{equation}
where both $\bbb \sigma$ and $\bbb\tau$ are Pauli matrices. It follows that
\begin{equation}
C\gamma_5=(\sigma^2\otimes (-1)\tau^1)(\mathds{1}\otimes \tau^1)=-\sigma^2\otimes\mathds{1}
\end{equation}
Irrespective of the overall phase, we see that~\eqref{defidiqu} has a non-relativistic limit, which in the Pauli bispinor notation writes
\begin{equation}
[cq]_i=\epsilon_{ijk}(c^j)^T\sigma^2 q^k
\end{equation}
In the four spinor formalism the {\it spin one} diquark $J^P=1^+$ is 
\begin{equation}
[cq]_i = \epsilon_{ijk}\,\bar c^{j}_{\bm c}\bm \gamma q^{k}\equiv \epsilon_{ijk}\, (c^{j})^T \,C\bm \gamma\,q^k
\end{equation}
and
\begin{equation}
C\bm \gamma=-(\sigma^2\otimes \tau^1)(\bm\sigma\otimes i\tau^2)=(\sigma^2\bm \sigma\otimes \tau^3)
\end{equation}
Therefore a spin-1 heavy-light diquark in the non-relativistic limit is 
\begin{equation}
[cq]_i^\lambda=\epsilon_{ijk}(c^j)^T\sigma^2\sigma^\lambda q^k
\end{equation}

The formation of diquarks follows from the attractive nature of the color antitriplet $qq$ channel, when color force is mediated by the exchange of one gluon  (see also \ref{appN}). Thus diquarks are antisymmetric in color indices, and they can be antisymmetric in spin, if $S=0$, or symmetric if $S=1$. The fact that we can have both spin 0 and 1 heavy-light diquarks derives from heavy quark spin symmetry.

The color-spin Hamiltonian describing the interaction between the constituents of a tetraquark is~\cite{Jaffe:2004ph}
\begin{equation}
H=-2\sum_{i\neq j,a} \kappa_{ij} \; \bm S_i\cdot \bm S_j\;\frac{\lambda^a_i}{2}\cdot \frac{\lambda^a_j}{2}\equiv\sum_{i\neq j}H_{ij}
\label{colorspin}
\end{equation}
The indices $i,j$ run over the four quarks, while $a$ is the index of the adjoint SU(3) representation. The $\lambda^a$ are Gell-Mann matrices, the $\bbb S_i$ are spin vectors and the $\kappa_{ij}$ are unknown effective couplings. Let us discuss the color interaction for the moment. 

We introduce the (normalized) color singlet/octet states using the following notation which turns out to be rather practical for calculations
\begin{subequations}
\begin{align}
\label{singlet}
\left|\bar c c_{\bm 1},\bar q q_{\bm 1}\right\rangle& :=\frac{1}{3}\;\mathds{1}_{\bar c c}\otimes\mathds{1}_{\bar q q} \\
\left|\bar c c_{\bm 8},\bar q q_{\bm 8}\right\rangle &:= \frac{1}{4\sqrt{2}}\;\lambda^a_{\bar c c}\otimes\lambda^a_{\bar q q }
\label{octet}
\end{align}
\end{subequations}
where by $\lambda^a_{\bar c c}$, for example, we mean $\bar c_i \;(\lambda^{a})^i_{j} \;c^j$ using latin letters for color indices.

With the notation $|cq_{\bm{\bar{3}}},\bar c\bar q_{\bm 3}\rangle_{\bm 1}$ we mean (aside from overall phases) {\it an overall color singlet} state of a diquark-antidiquark pair
\begin{equation}
|cq_{\bm{\bar{3}}},\bar c\bar q_{\bm 3}\rangle_{\bm 1}\sim [cq]_i[\bar c\bar q]^i=c_j\bar c^jq_k\bar q^k- c_j\bar q^j q_k\bar c^k
\end{equation}
which, using 
\begin{equation}
\label{lfierz}
(\lambda^a)^i_{\;j}(\lambda^a)^k_{\;\,l}=2\left(\delta^i_l\delta^k_j-1/3\;\delta^i_j\delta^k_l\right)
\end{equation}
can be written as 
\begin{equation}
|cq_{\bm{\bar{3}}},\bar c\bar q_{\bm 3}\rangle_{\bm 1}\sim\frac{2}{3}\;\mathds{1}_{\bar c c}\otimes\mathds{1}_{\bar q q}-\frac{1}{2}\;\lambda^a_{\bar c c }\otimes\lambda^a_{\bar q q}=2|\bar c c_{\bm 1},\bar q q_{\bm 1}\rangle-2\sqrt{2} |\bar cc_{\bm 8},\bar q q_{\bm 8}\rangle,
\label{2octet}
\end{equation}
\ie the octet-octet component  has {\it  twice the probability}   of the singlet-singlet one. This fact may have interesting consequences. Strong interactions are known to preserve the spin of the heavy quark pair (see \sectionname{\ref{HQSS}}). 

The previous state can itself  be normalized in the following way (multiply by $1/\!\sqrt{12}$)
\begin{equation}
|cq_{\bm{\bar{3}}},\bar c\bar q_{\bm 3}\rangle_{\bm 1}
=\frac{1}{\sqrt{3}}\left(\frac{1}{3}\;\mathds{1}_{\bar c c}\otimes\mathds{1}_{\bar q q}-T^a_{\bar c c}\otimes T^a_{\bar q q}\right)
\label{atriplet}
\end{equation}
and use  the $T^a=\lambda^a/2$ matrices. 

Let us represent  states of the fundamental representation with the symbol $|_i\rangle$ whereas those of the anti-fundamental are $|^j\rangle$. Then we have
\begin{subequations}
\begin{align}
\langle_j|T^a|_i\rangle &=(T^a)^j_{\;\,i}\\
 \langle^j|T^a|^i\rangle &=-(T^a)^i_{\;j}
\end{align}
\end{subequations}
\ie one is the opposite-transposed (complex-conjugate) of the other.
From the latter equation we get for example 
\begin{equation}
T^a|^i\rangle=-|^j\rangle(T^a)^i_{\;j}
\end{equation}
Consider a generic state $|v\rangle$
\begin{equation}
|v\rangle=|^i\rangle v_i
\end{equation}
then
\begin{equation}
|T^a v\rangle=T^a|v\rangle=T^a|^i\rangle v_i=-|^j\rangle(T^a)^i_{\;j} v_i
\end{equation}
Thus we get that  (multiply the latter by $\langle_k|$ and then rename $k\to i$)
\begin{equation}
\label{anti}
T^a v_i=-(T^a)^j_{\;\,i} v_j
\end{equation}
whereas 
\begin{equation}
\label{tri}
T^a v^i=(T^a)^i_{\;\,j} v^j
\end{equation}

\begin{table}[t]
\centering
\begin{tabular}{c}
\hline\hline
Rules of color manipulation \\
\hline
$\mathcal{O}_{q_1}\,\mathcal{O}^{\,\prime}_{\bar q_2q_1}=-\left(\mathcal{O}\,\mathcal{O}^{\,\prime}\right)_{\bar q_2 q_1}$\quad \quad $\mathcal{O}_{\bar q_2}\,\mathcal{O}^{\,\prime}_{\bar q_2q_1}=\left(\mathcal{O}\,\mathcal{O}^{\,\prime}\right)_{\bar q_2 q_1}$ \\
$T^a_{\bar q_1 q_2}\otimes T^a_{\bar q_3 q_4}=\frac{1}{2}\mathds{1}_{\bar q_1 q_4}\otimes \mathds{1}_{\bar q_3 q_2}-\frac{1}{6}\mathds{1}_{\bar q_1q_2}\otimes\mathds{1}_{\bar q_3 q_4}$ \\
$\mathds{1}_{\bar q_1 q_2}\otimes\mathds{1}_{\bar q_2 q_1}=\tr(\mathds{1})$ \\
$T^a_{\bar q_1 q_2}\otimes T^b_{\bar q_2 q_1}=\tr(T^a T^b)$ \\
\hline\hline
\end{tabular}
\caption{Useful rules for the manipulation of color indices and operators. The quantities with a single quark index are operators acting on the Hilbert space of that quark, \eg $T^a_c$ acts on the charm quark, while the quantities with two quark indices are operators saturated with quark states, \eg $\mathcal{O}_{\bar q_1q_2}=\bar q_{1i}\,\mathcal{O}^i_j\,q_2^j$.} \label{tab:color}
\end{table}

It then follows that, aside from an overall minus sign, the different quark indices, \emph{e.g} $q_1q_2$ in $T^a_{q_1q_2}$, can be treated as ordinary matrix indices. In \tablename{\ref{tab:color}} we report some useful rules for the manipulation of color states and operators.
With this rules we can compute the action of of the color part of the  $H_{c q}$ Hamiltonian on a diquark state defined in~\eqref{atriplet}
\begin{equation}
H_{ c q} |cq_{\bm{\bar{3}}},\bar c\bar q_{\bm 3}\rangle_{\bm 1}\; \propto T^b_cT^b_q\frac{1}{\sqrt{3}}\left(\frac{1}{3}\mathds{1}_{\bar c c}\otimes\mathds{1}_{\bar q q}-T^a_{\bar c c}\otimes T^a_{\bar q q}\right)=\; -\frac{1}{\sqrt{3}}\left(\frac{1}{3}\; T^b_{\bar c c}\otimes T^b_{\bar q q}-\left(T^a T^b\right)_{\bar c c}\otimes \left(T^aT^b\right)_{\bar q q}\right)
\end{equation}
and thus 
\begin{align}
_{\bm 1}\langle cq_{\bm{\bar{3}}},\bar c\bar q_{\bm 3}|H_{c q} |cq_{\bm{\bar{3}}},\bar c\bar q_{\bm 3}\rangle_{\bm 1}&\;\propto\;-\frac{1}{3}\left(\frac{1}{9}\tr(T^a)\,\tr(T^a)-\frac{2}{3}\tr(T^aT^b)\tr(T^aT^b)+\tr(T^aT^bT^c)\tr(T^aT^bT^c) \right) \notag \\
&\;=\; -\frac{1}{3}\left(-2\;\frac{2}{3}-\frac{2}{3}\right)=\frac{2}{3}
\end{align}
where we have used 
\begin{equation}
\tr(T^aT^b)=\frac{1}{2}\delta^{ab}\quad \quad \mathrm{and}\quad \quad \tr(T^aT^bT^c)=\frac{1}{4}(d^{abc}+if^{abc})
\end{equation}
and
\begin{subequations}
\begin{align}
 f^{abc}f^{abd}&=3\,\delta^{ab} \\
d^{abc}d^{abd}&=\frac{5}{3}\,\delta^{ab}
\end{align}
\end{subequations}
This means that, as far as color is concerned, taking matrix elements on diquark-antidiquark color-neutral states,  amounts to redefine the chromomagnetic couplings by some representation-dependent numerical factor like   $2/3$, when the $H_{cq}$ and $H_{\bar c\bar q}$ terms are considered. Actually we will assume that the dominant couplings in the Hamiltonian are $\kappa_{cq}$ and $\kappa_{\bar c\bar q}$, \ie,  intra-diquark interactions~\cite{Maiani:2014aja}. At a first sight, this assumption may look to violate the $1/m_Q$ scaling of the couplings, which would suggest $\kappa_{q \bar q} > \kappa_{cq} > \kappa_{c \bar c}$. Instead, this is due to the very short range nature of spin-spin interactions (see also Eq.~\eqref{Hswanson} and Sec.~\ref{matricimassa}), which makes the interaction between two quarks of different diquarks vanish. This ansatz turns out to be compatible with experimental data. Charge conjugation symmetry also implies 
$\kappa=\kappa_{cq}=\kappa_{\bar c\bar q}$~\footnote{
If extra diquark couplings were considered we could determine them, \eg $\kappa_{c\bar c}$, from the masses of standard $L=0$ mesons observing that
\begin{equation}
_{\bm 1}\langle cq_{\bm{\bar{3}}},\bar c\bar q_{\bm 3}|H_{\bar c c} |cq_{\bm{\bar{3}}},\bar c\bar q_{\bm 3}\rangle_{\bm 1} = \frac{1}{4} \langle \bar c c_{\bm 1},\bar q q_{\bm 1}|H_{\bar c c} |\bar c c_{\bm 1},\bar q q_{\bm 1}\rangle
\end{equation}
}.

In view of what found in this section we understand that mass splittings among tetraquark states conceived as diquarkonia, will be determined solely from the spin-spin part of the Hamiltonian in~\eqref{colorspin}.

\subsection{Diquarks: spin \label{spins}}

Heavy-light diquarks are different from light-light ones in that they can have either spin 0 or spin 1, whereas light-light are preferably spin 0, the so called `good diquarks'~\cite{Jaffe:2004ph}. Heavy quark symmetry is such that the light degrees of freedom are blind to the flavor and spin of the heavy quark source (see again \sectionname{\ref{HQSS}}) and this spin decoupling feature makes equally possible heavy $S=0,1$. The use of diquarks in exotic hadron spectroscopy was suggested in~\cite{Jaffe:2003sg} and heavy-light diquarks, their properties, and masses were first discussed in~\cite{Maiani:2004vq}. Light-light diquarks and their possible role in light scalar meson spectroscopy are discussed in~\cite{Maiani:scalars,thscal}.

We generally use the definitions (see also~\cite{Esposito:2014rxa})
\begin{subequations} \label{spindef}
\begin{align}
\label{def1}
|1_{\bm q},0_{\bar{\bm q}}\rangle&=\frac{1}{2}\,\sigma^2\sigma^i\otimes\sigma^2  \\ 
\label{def2}
|0_{\bm q},1_{\bar{\bm q}}\rangle&=\frac{1}{2}\,\sigma^2\otimes\sigma^2\sigma^i  \\
\label{def3}
|1_{\bm q},1_{\bar{\bm q}}\rangle_{J=1}&=\frac{i}{2\sqrt{2}}\, \epsilon^{ijk}\sigma^2\sigma^j\otimes\sigma^2\sigma^k
\end{align}
\end{subequations}
With the symbol $\bm q$ we either mean a diquark in the order $cq$ or $\bar c \bar q$ or a quark-antiquark pair in the order 
$c\bar c$ or $q\bar q$. The ordering is relevant (see~\cite{Esposito:2014rxa} for normalizations~\footnote{
For example, the normalization in~\eqref{def3} is obtained by
\be
\langle 1_{\bm q},1_{\bar{\bm q}}~|1_{\bm q},1_{\bar{\bm q}}\rangle_{J=1}=-\frac{1}{4\times 2}\, \left(\tr(\sigma^{j\,T}\sigma^j)\,\tr(\sigma^{k\,T}\sigma^k)-\tr(\sigma^{j\,T}\sigma^k)\,\tr(\sigma^{k\,T}\sigma^j)\right)=-\frac{1}{4\times 2}\,(2\times 2-2\times 2\times 3)=1
\ee 
}).

In these notations a particle with the quantum numbers of $X(3872)$, would be described by
\begin{equation}
X=\frac{|1_{cq},0_{\bar c \bar q}\rangle+|0_{cq},1_{\bar c \bar q}\rangle}{\sqrt{2}}=|1_{c\bar c},1_{q\bar q}\rangle_{J=1}=\frac{|1_{c\bar q},0_{q\bar c}\rangle-|0_{c\bar q},1_{q\bar c}\rangle}{\sqrt{2}}
\label{ics}
\end{equation}
Since diquarks are defined to be positive parity states, overall we have  $J^{P}=1^+$ and   $C=+$. This diquark-antidiquark arrangement is a natural candidate to 
describe the $X(3872)$, which is a $1^{++}$ resonance decaying into $\jpsi+\rho/\omega$, compatibly  with the $|1_{c\bar c},1_{q\bar q}\rangle_{J=1}$ assignment --- especially for what concerns the heavy spin. Similarly the last term on the \rhs is compatible with the $DD^*$ decay mode of the $X(3872)$. Anyway, light quark spins in $Q\bar q$ or $\bar Q q$ configurations 
 might rearrange also to allow $DD$ or $D^*D^*$ decays but the latter is phase space forbidden and the former is simply forbidden by quantum numbers.  

The orthogonal combination to the \lhs of~\eqref{ics}  can be formed by
\begin{equation}
Z=\frac{|1_{cq},0_{\bar c \bar q}\rangle-|0_{cq},1_{\bar c \bar q}\rangle}{\sqrt{2}} = \frac{|1_{c\bar c},0_{q\bar q}\rangle-|0_{c\bar c},1_{q\bar q}\rangle}{\sqrt{2}}= |1_{c\bar q},1_{q\bar c}\rangle_{J=1}
\end{equation}
This has $J^P=1^+$ and $C=-$ for the neutral component (if an isospin triplet is to  be considered, the $G$-parity has to be $G=+$). 
The state has $C=-$ since, in the charmonium basis 
\begin{align}
C=(-1)^{L+S_{q\bar q}+S_{c\bar c}}
\end{align}

In the quark-antiquark basis there is another state with $C=-$, orthogonal to $Z$
\begin{equation}
Z^\prime= |1_{cq},1_{\bar c \bar q}\rangle_{J=1}=\frac{|1_{c\bar c},0_{q\bar q}\rangle + |0_{c\bar c},1_{q\bar q}\rangle}{\sqrt{2}}=\frac{|1_{c\bar q},0_{q\bar c}\rangle + |0_{c\bar q},1_{q\bar c}\rangle}{\sqrt{2}}
\label{zprime0}
\end{equation}
which is indeed a $1^{+-}$ state. Exchanging the coordinates, spins and charges of two fermions/bosons having each spin $s$ and total spin $S$, the total wavefunction has to be completely 
antisymmetric/symmetric under this exchange
\begin{equation}
(-1)^L(-1)^{2s+S}C=\mp 1
\end{equation}
which in the case of~\eqref{zprime0} (first term on the \rhs) is
\begin{equation}
(-1)^0(-1)^{2+1}C=+1
\end{equation}
giving $C=-$. The case of $X=|1_{c\bar c},1_{q\bar q}\rangle_{J=1}$ is different as the charge conjugation operator concerns the distinct $c\bar c$ and $q\bar q$ pairs. 

Linear combinations of $Z$ and $Z^\prime$ which diagonalize the spin-spin Hamiltonian can be  identified with $Z_c(3900)$ and $Z_c(4020)$. We will not treat here scalar or tensor states whose discussion in terms of diquarkonia can be found in~\cite{Esposito:2014rxa}. 

\subsection{Spectrum for \texorpdfstring{$L=0$}{L=0} and the \texorpdfstring{$X,Z,Z^\prime$}{X,Z,Z'} system\label{matricimassa}}
We assume that the spin-spin interactions within the diquark shells are dominant with respect to quark-antiquark interactions. This is the main assumption in~\cite{Maiani:2014aja} and, as we will see, it gives the right mass pattern for the  $X,Z_c,Z_c^\prime$ system. 

This hypothesis leads to believe that diquarks are quite separated from each other in the tetraquark and this feature could be modeled with some potential keeping the diquarks relatively apart in space --- recall that the spin-spin interaction is proportional to $\delta(r)$ as in Eq.~\eqref{Hswanson}. At the same time this potential might be responsible for the reluctance of diquarks at mixing or decaying into meson pairs. Widths of actual diquarkonia measure  this behavior. 

Then spin-spin Hamiltonian is (color factors have been reabsorbed in the definition of chromomagnetic couplings as discussed above)
\begin{equation}
H\approx 2 \kappa\, (\bm S_q\cdot \bm S_c+\bm S_{\bar q}\cdot \bm S_{\bar c})
\label{ham1}
\end{equation}
Consider for example
\begin{equation}
4\bm S_q\cdot \bm S_c |1_{cq},0_{\bar c \bar q}\rangle=\bm \sigma_{(q)}\cdot \bm \sigma_{(c)} |1_{cq},0_{\bar c \bar q}\rangle := \frac{1}{2}(\sigma^j)^T\sigma^2 \sigma^i\sigma^j\otimes\sigma^2
\end{equation}
where summation over $j$ is understood. The matrix $(\sigma^j)^T$ works on $c$ whereas $\sigma^j$ on $q$. Considering that 
\begin{equation}
\frac{1}{2}(\sigma^j)^T\sigma^2 \sigma^i\sigma^j\otimes\sigma^2=-\frac{1}{2}(\sigma^2 \sigma^j \sigma^i\sigma^j)\otimes\sigma^2= \frac{1}{2} \sigma^2\sigma^i\otimes \sigma^2=|1_{cq},0_{\bar c \bar q}\rangle
\end{equation}
where we have used $i\epsilon^{ijk}\sigma^j\sigma^k=i\epsilon^{ijk}i\epsilon^{jk\ell}\sigma^\ell=-2\sigma^i$. Thus $|1_{cq},0_{\bar c\bar q}\rangle$ is an eigenstate of $\bbb S_q\cdot \bbb S_c$ with eigenvalue $1/4$.

Considering also the antidiquark contribution one readily finds
\begin{equation}
4\bm S_{\bar q}\cdot \bm S_{\bar c} |1_{cq},0_{\bar c \bar q}\rangle=-3\,|1_{cq},0_{\bar c \bar q}\rangle
\end{equation}
thus 
\begin{equation}
4(\bm S_q\cdot \bm S_c+\bm S_{\bar q}\cdot \bm S_{\bar c}) |1_{cq},0_{\bar c \bar q}\rangle=-2\,|1_{cq},0_{\bar c \bar q}\rangle
\end{equation}
and
\begin{equation}
\label{h11}
H|1_{cq},0_{\bar c \bar q}\rangle=-\kappa\, |1_{cq},0_{\bar c \bar q}\rangle
\end{equation}
Similarly
\begin{subequations}
\begin{eqnarray}
\label{h12}
H|0_{cq},1_{\bar c \bar q}\rangle&=&-\kappa\, |0_{cq},1_{\bar c \bar q}\rangle \\
H|1_{cq},1_{\bar c \bar q}\rangle_{J=1}&=&+\kappa\, |1_{cq},1_{\bar c \bar q}\rangle_{J=1}
\end{eqnarray}
\end{subequations}

The Hamiltonian~\eqref{ham1} is diagonal in the diquark-antidiquark  basis formed by the $1^{+-}$ states  
\begin{equation}
(H)^{1^{+-}}=\begin{pmatrix} -\kappa\, & 0\\ 0 & \kappa\end{pmatrix}
\label{mat1pm}
\end{equation}
with eigenvectors $|1\rangle=1/\!\sqrt{2}(|1,0\rangle-|0,1\rangle)\equiv |Z\rangle$ and $|2\rangle=|1,1\rangle_{J=1}\equiv |Z^\prime\rangle$.  This requires $|Z\rangle$ 
to be lighter than $|Z^\prime\rangle$.
Similarly
\begin{equation}
(H)^{1^{++}}=-\kappa
\label{mat2pp}
\end{equation}
Therefore  we conclude that the $X(3872)$ and the $Z(3900)$ are degenerate in first approximation, their masses being
twice the diquark mass plus the same spin-spin interaction correction
\begin{equation}
M_X=M_Z=2m_{[cq]}-\kappa
\label{spectpos1}
\end{equation}
The $Z^\prime$ is instead {\it heavier} by a gap of $2\kappa$
\begin{equation}
M(Z^\prime)=2m_{[cq]}+\kappa
\label{spectpos2}
\end{equation}
This pattern of $X,Z,Z^\prime$ masses is observed in experimental data. In particular, the two relations in Eqs.~\eqref{spectpos1} and \eqref{spectpos2} are both satisfied within $20\mev$ for
\begin{align}
\kappa=67\mev
\end{align}

\subsection{Scalar and tensor states\label{scaltens}}
The diquark-antidiquark model also allows  $J^P=0^+,2^+$ states with $C=+$. We have two $J^P=0^+$ states and a tensor one
\begin{subequations}
\begin{align}
X_0&=|0_{cq},0_{\bar c\bar q}\rangle_{J= 0} =\frac{1}{2}|0_{c\bar c},0_{q\bar q}\rangle-\frac{\sqrt{3}}{2} |1_{c\bar c},1_{q\bar q}\rangle_{J=0}\\
X_0^\prime &=|1_{cq},1_{\bar c\bar q}\rangle_{J=0}=\frac{\sqrt{3}}{2}|0_{c\bar c},0_{q\bar q}\rangle+\frac{1}{2} |1_{c\bar c},1_{q\bar q}\rangle_{J=0}\\
X_2 &=|1_{cq},1_{\bar c\bar q}\rangle_{J=2} = |1_{c\bar c},1_{q\bar q}\rangle_{J=2}
\end{align}
\end{subequations}
which are all charge-conjugation even, $C=+$. Following the same notations introduced above we have
\begin{subequations}
\begin{align}
|0_{\bm q},0_{\bm q}\rangle_{J\equiv 0}&=\frac{1}{2}\sigma^2\otimes \sigma^2\\
\label{zero}
|1_{\bm q},1_{\bm q}\rangle_{J=0}&=\frac{1}{2\sqrt{3}}\sigma^2\sigma^i\otimes \sigma^2\sigma^i \\
|1_{\bm q},1_{\bm q}\rangle_{J=2}&=
\frac{1}{2}\left(\sigma^2\sigma^{(i}\otimes \sigma^2\sigma^{j)}-\frac{2}{3}\delta^{ij}\sigma^2\sigma^{\ell}\otimes \sigma^2\sigma^{\ell}\right)
\label{spindue}
\end{align}
\end{subequations}
where  $i,j$ indices are symmetrized in the last equation  (a factor of $1/2$ has to be included in the symmetrization) and the trace is subtracted. For the determination of the normalizations coefficients see~\cite{Esposito:2014rxa}.

We can also determine
\begin{align}
(H)^{0^{++}}&=-3\kappa\notag\\
(H)^{0^{++\prime}}&=\kappa\notag\\
(H)^{2^{++}}&=\kappa
\label{duezz}
\end{align}
There are no official candidates for these scalar and tensor states. However, there is the tantalizing possibility for the $Z_1(4050)$ state decaying into $\chi_{c1} \pi$ to be  a scalar or a tensor (see for example~\cite{Olsen:2015zcy}). In this case, it would be degenerate with the $Z_c^\prime(4020)$ within errors, and would be the natural candidate for the $0^{++\prime}$ ($2^{++}$) tetraquark.
We also remark that, if the $Z(4430)$ is considered as the radial excitation of the $Z_c(3900)$, it is natural to identify the $Z_2(4250)$ as the radial excitation of the lowest $0^{++}$ tetraquark, given that their mass difference, $m_{Z(4430)} - m_{Z_1(4250)} \simeq (230 \pm 190)\mev$ is compatible with $2\kappa \simeq 130\mev$. 

\begin{figure}[htb!]
 \begin{center}
   \includegraphics[width=9truecm]{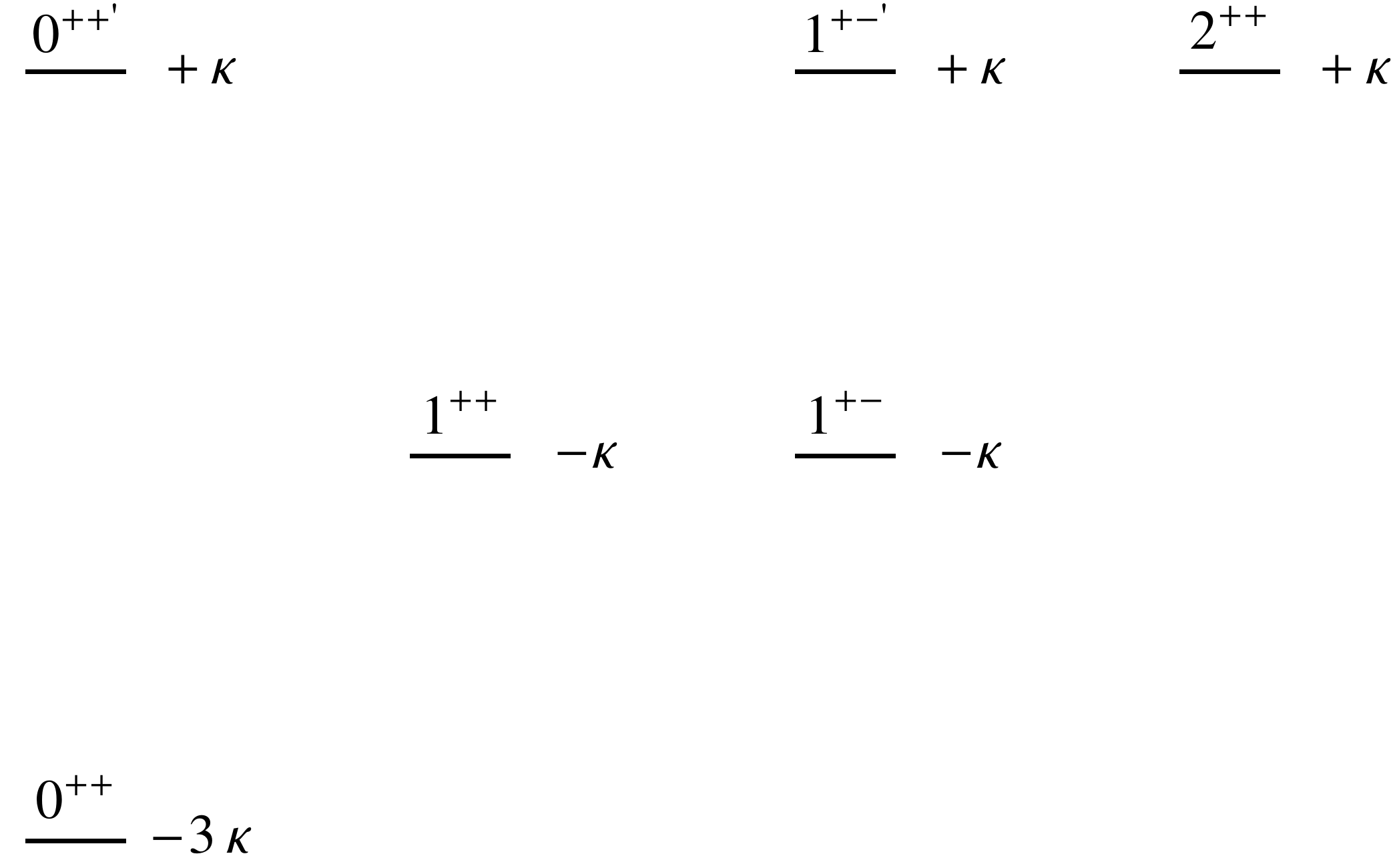}
 \end{center}
\caption{The spectrum of Hamiltonian in Eq.~\eqref{ham1}.\label{spettrot2}}   
\end{figure}

In \figurename{~\ref{spettrot2}} we show a pictorial representation the spectrum obtained from the Hamiltonian~\eqref{ham1}. In \tablename{\ref{primatab}} we instead list the predicted states and their quantum numbers.

\begin{table}[htb!]
\centering
\caption{Summary table of states. We refer here to the neutral components.\label{tavola_somm}}{
\begin{tabular}{lll}
    \hline\hline
$J^{PC}$ & $|S_{cq},S_{\bar c\bar q}\rangle$ & $|S_{c\bar c},S_{q\bar q}\rangle$\\ \hline
$0^{++}$ & $|0,0\rangle$ & $\frac{1}{2}|0,0\rangle+ \frac{\sqrt{3}}{2}|1,1\rangle_{J=0}$ \\
$0^{++}$ & $|1,1\rangle_{J=0}$ &  $\frac{\sqrt{3}}{2}|0,0\rangle-\frac{1}{2}|1,1\rangle_{J=0}$ \\
$1^{++}$ & $\frac{1}{\sqrt{2}}(|1,0\rangle+|0,1\rangle)$ & $|1,1\rangle_{J=1}$ \\
$1^{+-}$ & $\frac{1}{\sqrt{2}}(|1,0\rangle-|0,1\rangle) $ &  $\frac{1}{\sqrt{2}}(|1,0\rangle-|0,1\rangle) $ \\
$1^{+-}$ & $|1,1\rangle_{J=1}$ & $\frac{1}{\sqrt{2}}(|1,0\rangle+|0,1\rangle) $ \\
$2^{++}$ & $|1,1\rangle_{J=2}$ & $|1,1\rangle_{J=2}$ \\\hline
\hline
    \end{tabular}
\label{primatab}
}
\end{table}

\subsection{\texorpdfstring{$Y$}{Y} resonances and \texorpdfstring{$L=1$}{L=1} diquarkonia\label{yresos}}
\label{sec:tetray}
Tetraquarks with $J^{PC}=1^{--}$ can be obtained with odd values of the angular momentum. The lowest lying have $L=1$. We select states with odd charge conjugation.

In the diquark-antidiquark basis of $[cq][\bar c \bar q]$ we have
\begin{subequations}
\begin{align}
\label{y1}
Y_1&=|0_{cq},0_{\bar c \bar q}\rangle\quad\quad\quad\quad\quad\quad C=(-1)^{L=1}\\
\label{y2}
Y_2&=\frac{|1_{cq},0_{\bar c \bar q}\rangle+|0_{cq},1_{\bar c\bar q}\rangle}{\sqrt{2}}\quad  \,\,C=(-1)^{L=1}\\
Y_3&=|1_{cq},1_{\bar c\bar q}\rangle_{S=0}\quad\quad\quad  (-1)^L (-1)^{2s+S}C=+1\Rightarrow C=(-1)^1(-1)^{2\times 1+0}\\
Y_4&=|1_{cq},1_{\bar c\bar q}\rangle_{S=2}\quad\quad\quad  C=(-1)^1(-1)^{2\times 1+2}
\end{align}
\end{subequations}

The spin structure of $Y_2$ and $X$ in~\eqref{ics} is  the same and therefore the mass difference between them might entirely be attributed to the orbital excitation of $Y_2$.  The fact that $Y_2$ and $X$ have the same spin structure also suggests that radiative transitions with $\Delta L=1$ and $\Delta S_{c\bar c}=0$ might occur
\begin{equation}
Y_2\to \gamma X
\end{equation}

As an important caveat, we have to add here that the experimental assessment of the $Y$ states, at the time of this writing, is still rather unclear, even more so with the new results by \bes~\cite{Ablikim:2016qzw,BESIII:2016adj}. In particular, the very existence of a $Y(4008)$ is controversial. The identification of the structures seen in $\jpsi\,\pi\pi$ with the ones in $h_c \pi \pi$ is problematic too, and would undermine the assignment even of the well known $Y(4260)$ (see Section~\ref{sec:vectors} for the experimental details).
The assignments made below are based on the existence of these resonance, and are therefore subject to variation if these states 
turn out not to be confirmed. We preferred to stick to these hypotheses for the sake of illustration of the method of constructing the states in the diquarkonium picture.   As we will comment in \sectionname{~\ref{mol3}}, orbitally excited and radially excited resonances present more difficulties than lowest lying states.  

Anyway, if a distinct $Y(4260)$ does exist, its identification as the $Y_2$ excitation would be supported by the conspicuous radiative decay mode~\cite{Ablikim:2013dyn}
\begin{equation}
Y(4260)\to X(3872)+\gamma
\end{equation}

The $Y(4360)$ and $Y(4660)$ are interpreted as radial excitations of $Y_1=Y(4008)$ (see \tablename{\ref{secondatab}}) and $Y_2=Y(4260)$.
As a check for this assumption we may note that they indeed present mass splitting similar to the one observed between ordinary ground state and excited charmonia, \ie $M(\chi_{bJ}(2P))-M(\chi_{bJ}(1P))\simeq M(Y(4360))-M(Y(4008))$ and $M(\chi_{cJ}(2P))-M(\chi_{cJ}(1P))\simeq M(Y(4660))-M(Y(4260))$. For the identification of the $Y_3$ state as the structures seen in $e^+e^- \to h_c\, \pi \pi, \chi_{c0} \,\omega$, see~\cite{Faccini:2014pma}. 

As for the $Y(4630)$, decaying predominantly into $\Lambda_c^+ \Lambda_c^-$, we recall that there is also the possibility of its assignment to a tetraquark 
degenerate with $Y(4660)$~\cite{Cotugno:2009ys}.
\begin{table}[h!]
\centering
\caption{Possible assignments for~$Y_1,Y_2,Y_3,Y_4$. Observe that $Y_3$ 
is predicted to decay preferably in $h_c(1P)$ where $S_{c\bar c}=0$.  The state $Y(4220)$ corresponds to the narrow structure described in~\cite{Chang-Zheng:2014haa,Yuan:2014rta,BESIII:2016adj}.
Radiative decays are suggested by conservation of the heavy quark spin $\Delta S_{c\bar c}=0$. The relative probability of having spin 1 versus spin 0 in the $c\bar c$ pair as read by \tablename{\ref{tavola_somm}}.}
{    \begin{tabular}{cccc}
    \hline\hline
State &$P(S_{c\bar c}=1):P(S_{c\bar c}=0)$& Assignment & Radiative Decay\\ \hline
$Y_1$& 3:1 & $Y(4008)$ & $\gamma +X(0^{++})$ \\
$Y_2$&1:0 & $Y(4260)$ & $\gamma+X(1^{++})$\\
$Y_3$&1:3 & $Y(4220)$ & $\gamma+ X(0^{\prime ++})$ \\
$Y_4$&1:0 & $Y(4630)$ & $\gamma+X(2^{++})$\\
\hline\hline
    \end{tabular} \label{secondatab}}
\end{table}

To find the spectrum of the $Y$ states we use the same Hamiltonian form~\eqref{ham1} with the addition of a spin-orbit and a purely orbital term --- here  the chromomagnetic coupling $\kappa^\prime$ is taken to be a priori different from $\kappa$ used in~\eqref{ham1}. We already discussed that the spatial separation of the diquarks makes the other couplings vanish. We have then
\begin{equation}
H\approx 2 \kappa^\prime (\bm S_q\cdot \bm S_c+\bm S_{\bar q}\cdot \bm S_{\bar c})-2 A\, \bm S\cdot \bm L+\frac{1}{2} B\,\bm L^2
\label{ham2}
\end{equation}
in such a way that the energy increases for increasing $\bm L^2$ and $\bm S^2$, provided $\kappa^\prime, A, B$ are positive. Indeed the  masses of $Y$ states will be given by
\begin{equation}
M=M_0^\prime+\kappa^\prime (s(s+1)+\bar s(\bar s+1)-3)+A(L(L+1)+S(S+1)-2)+B\frac{L(L+1)}{2}
\end{equation}
where $s,\bar s$ are the total spins of diquark and antidiquark. The latter equation can be simplified to
\begin{equation}
M=M_0+(A+B/2)\,L(L+1)+AS(S+1)+\kappa^\prime(s(s+1)+\bar s(\bar s+1))
\label{massumm}
\end{equation}
with 
\begin{equation}
M_0=M_0^\prime-2A-3\kappa^\prime
\end{equation}
From Eq.~\eqref{massumm}, the mass of the state $Y_1$ in~\eqref{y1} is given by
\begin{equation}
M_1=M_0+2(A+B/2)
\end{equation}
for $s=\bar s=0$, therefore implying $S=0$, and $L=1$.  The $Y_2$ state in~\eqref{y2} has $s=1$ or $\bar s=1$, thus $S=1$ --- considering that $M_0$ contains $-3\kappa^\prime$ we can determine the mass gap between $Y_2$ and $Y_1$
\begin{equation}
\label{dmu}
M_2-M_1=2\kappa^\prime+2A
\end{equation}
hence implying $M_2>M_1$.
The $Y_3$ state has both spins $s=\bar s=1$ but $S=0$ so that 
\begin{equation}
\label{tmd}
M_3-M_2=2\kappa^\prime-2A
\end{equation}
which can take either sign depending on the $\kappa^\prime-A$ difference; $\kappa^\prime$ and $A$ have in principle similar sizes.
Finally $Y_4$ has both spins $s=\bar s=1$ and $S=2$ so that 
\begin{equation}
\label{qmt}
M_4-M_3=6A
\end{equation}
implying $M_4>M_3$. So the mass ordering is $M_1,M_2,M_3,M_4$ or $M_1,M_3,M_2,M_4$ from lighter to heavier. 
Using the assignments in \tablename{\ref{secondatab}}, from~\eqref{dmu} and~\eqref{tmd} we obtain~\footnote{In the original paper~\cite{Maiani:2014aja} also the possible assignment $Y_3=Y(4290)$ was considered. However, the energy scan by BES~\cite{BESIII:2016adj} pushes the mass of the state to be $100\mev$ heavier, thus making this assignment unlikely.}:
\begin{equation}
\kappa^\prime=53\mev\quad\quad \text{ and } \quad\quad A=73\mev
\end{equation}
The values found for  $\kappa^\prime$ have to be compared with the value of $\kappa=67\mev$  obtained studying the spectrum of $L=0$ states. They look reasonably consistent, also in consideration of the simplicity of the model described.
In the original formulation for the diquarkonium model for the \XYZ states, all the couplings $\kappa_{ij}$ were kept non-zero and their value was fitted from known meson and baryon masses. In that case it was found $\kappa\simeq22\mev$. We conclude that diquarks inside a tetraquark behave differently than those inside a baryon.

\subsection{Pentaquarks from diquarks}
\label{sec:pentatheor}
In the diquark composition model, pentaquarks are clusters of diquarks of the form 
\begin{equation}
\PP=d^Ad^A\bar q
\end{equation}
where $d^A$ is the antisymmetric diquark introduced in Eq.~\eqref{diqop}. There are several flavor compositions which may realize a pentaquark, and the diquark model would clearly predict a great variety of states.

Eventually there was a stunning observation by the \lhcb collaboration~\cite{Aaij:2015tga} 
 reporting the observation of two new resonances in the $\Lambda_b$ decay
\begin{equation}
\Lambda_b (bud)\to \PP^+ K^-
\label{unop}
\end{equation}
 each decaying according to
 \begin{equation}
\PP^+ \to \jpsi \, p
\label{duep}
\end{equation}
Thus the new particles carry a unit of baryonic number and feature the valence quark composition 
 \begin{equation}
\PP^+ = \bar c c u u d
\label{valencep}
\end{equation}
 whence the name \emph{pentaquarks}. The  best fit quantum numbers and masses are
 \begin{align}
 J^P&=~3/2^-\quad\quad M\simeq 4380\gev\quad\quad {\rm fract.}\simeq 8.4~\%\nonumber \ \\
 J^P&=~5/2^+\quad\quad M\simeq 4450\gev\quad\quad {\rm fract.}\simeq 4.1~\%  
 \label{5/2}
 \end{align}
The very fact that two pentaquarks have been observed with {\it opposite parities} strongly suggests that diquarks should have a role in their composition --- bound states of color neutral hadrons cannot appear in orbitally excited configurations. 

The quarks in~\eqref{valencep} have to be arranged into diquarks of different spin and flavor. The standard lore that light-light diquarks have to be spin zero, which has some support from lattice calculations~\cite{reticolo}, is dubious in the case of diquarks in baryons. Indeed spin one light-light diquarks (Jaffe's `bad diquarks'), while conspicuously absent in light meson spectroscopy, are well established in baryons as indicated by  the $\Sigma-\Lambda$ mass difference~\cite{DeRujula:1975ge} and confirmed by $\Sigma_{c,b}-\Lambda_{c,b}$ mass differences~\cite{Maiani:2004vq}. We also recall the claim of observation of an $I=2$ state (which would be made of two bad diquarks) in a phenomenological analysis of $\gamma \gamma^* \to \rho \rho$ data~\cite{Anikin:2005ur}.

At first sight, the near $70\mev$ difference between the masses in~\eqref{5/2}  does not go well with the energy associated to orbital excitation. One orbital excitation in mesons and baryons carries an energy difference which is typically of order $300$\mev, as exemplified by the  mass difference $M(\Lambda(1405))-M(\Lambda(1116))\simeq 290\mev$. Mass formulae for the orbital excitation in  \XYZ mesons are discussed in~\cite{Maiani:2014aja} and the associated energy difference is estimated to be $\Delta M ( L=0\to 1)\simeq 280\mev$. 

However, the mass difference between light-light diquarks with spin $s=1,0$~\cite{DeRujula:1975ge}, estimated from charm and beauty baryon spectra, is of order $200\mev$, \eg $M(\Sigma_c(2455))-M(\Lambda_c(2286))\simeq 170\mev$ and $M(\Sigma_b(5811))-M(\Lambda_b(5620))\simeq 190\mev$. 

If we use spin one light-light diquarks in the compositions
\begin{align}
\label{compos0} 
\PP( 3/2^-) &= \left(\bar c\, [cq]_{s=1} [q^\prime q^{\prime \prime}]_{s=1}, L=0\right)\\
\PP( 5/2^+) &=\left(\bar c\, [cq]_{s=1} [q^\prime q^{\prime \prime}]_{s=0}, L=1\right) 
\label{compos} 
\end{align}
the orbital gap is reduced to about $100\mev$, which brings it back to the range of spin-spin  and spin-orbit corrections indicated by~\eqref{5/2}. In other works the heavier diquark in the $L=0$ state shortens the mass distance from the $L=1$ state with a spin zero (\ie lighter) diquark. 

Notice that the intrinsic parity of the $L=0$ state is dictated by the presence of an antiquark: therefore, differently from standard baryons, the low lying state has negative parity. The opposite holds true for tetraquarks: the low lying ones have positive parities --- whereas the low-lying mesons have negative parities. 

\begin{figure}[t]
\centering
   \includegraphics[height=5truecm]{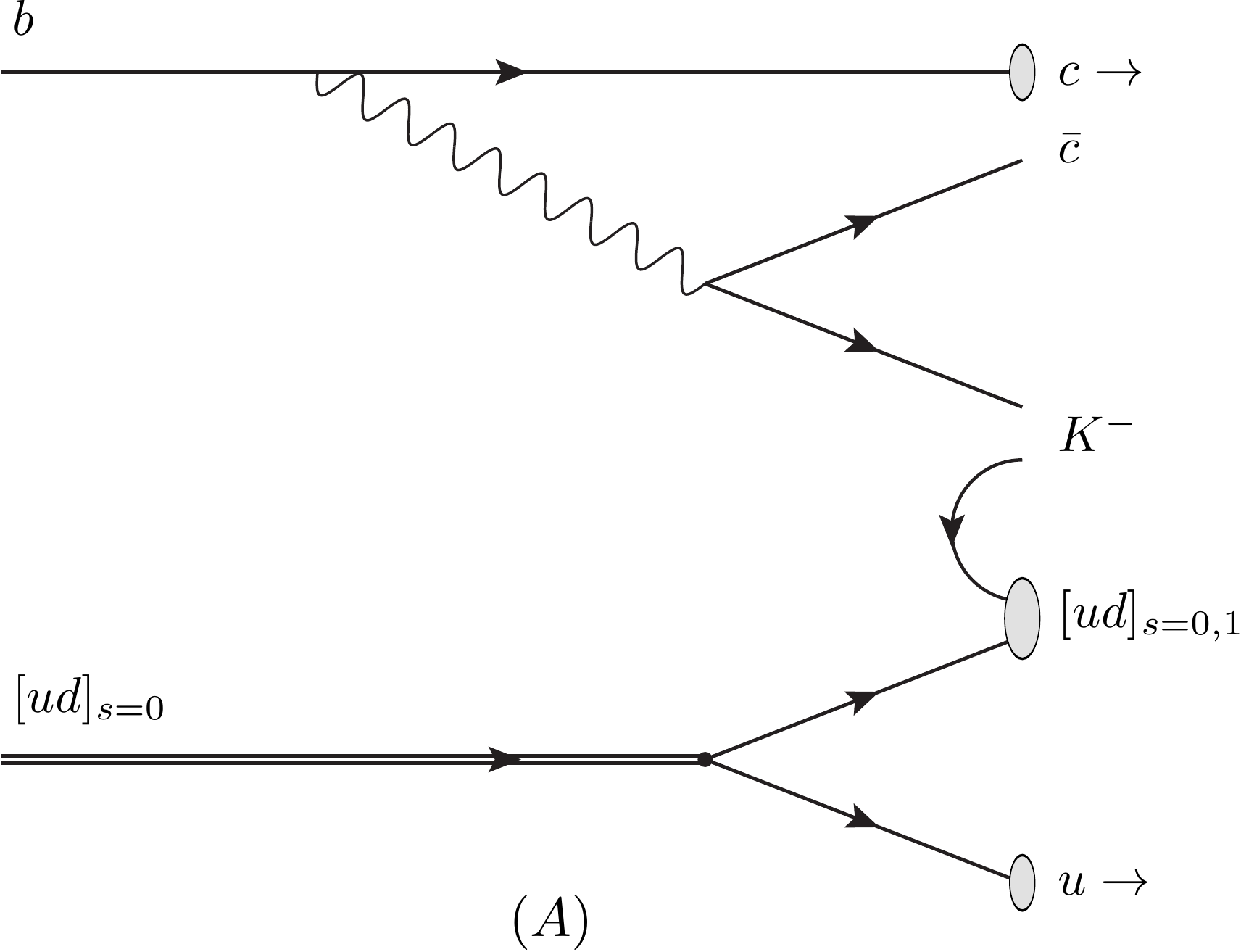} 
\hspace{1cm}
   \includegraphics[height=5truecm]{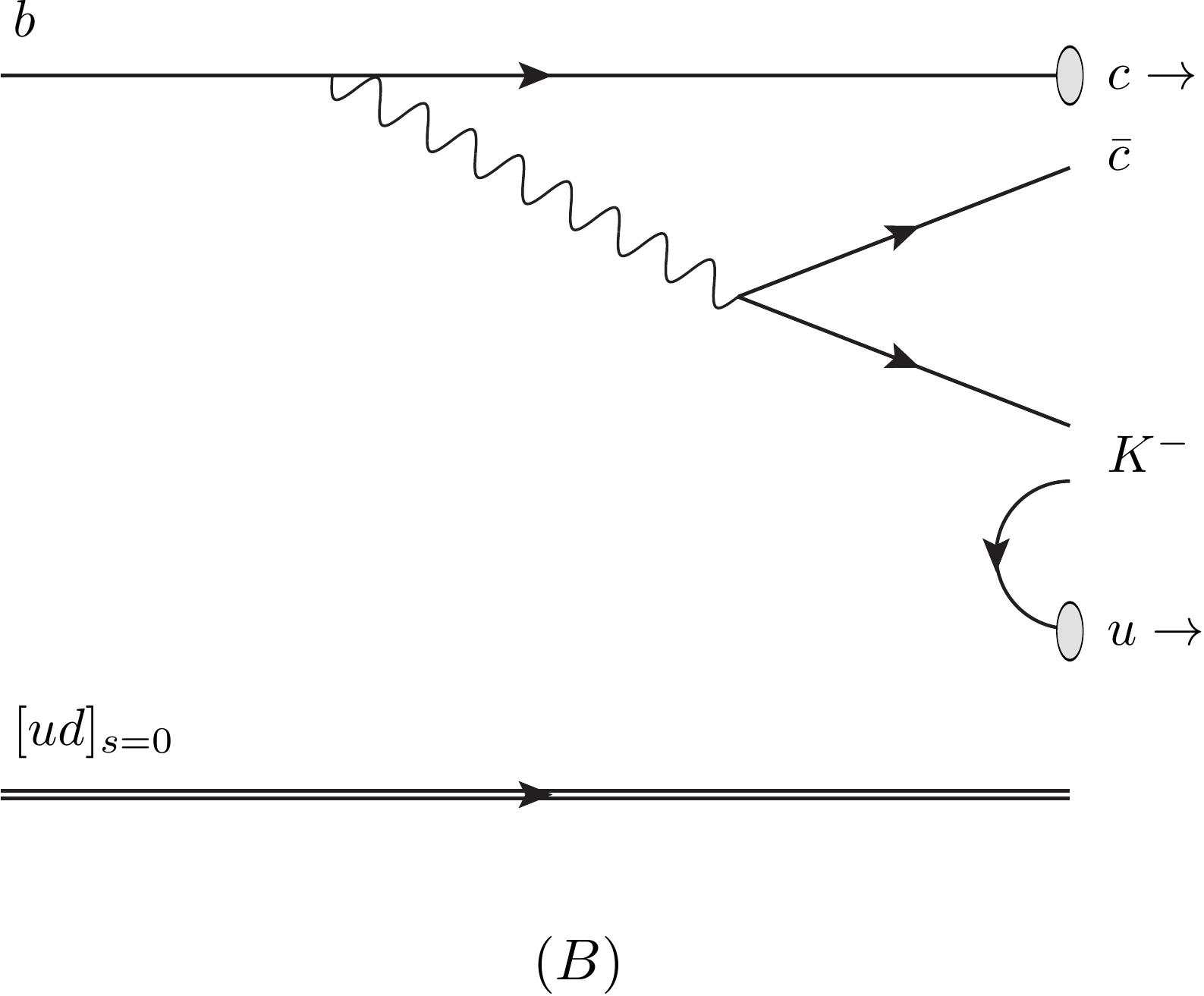}
\caption{$(A)$: The $[ud]$, spin zero diquark in the  $\Lambda_b$  is transmitted to the $\PP_u$ type pentaquark as a spectator --- it is expected that $\Lambda_b$ contains a spin zero light-light diquark of the $[ud]$ flavor.  $(B)$: The $u$ quark from the vacuum participates in the formation of the light-light diquark: spin zero and one are both permitted. Mechanism (B) may also produce a $[uu]_{s=1}$ diquark. \label{funop} }
\end{figure}

Concerning the production of a spin-1, light-light $[ud]$ diquark in  $\Lambda_b$ decay, we note that there are in fact two possible mechanisms leading to the pentaquark production --- see 
\figurename{~\ref{funop}}. In the first one (diagram $A$ in \figurename{~\ref{funop}}) the $b$-quark spin is shared between the Kaon and the $\bar c$ and $[cu]$ components. Barring angular momentum transfer due to gluon exchanges between the light diquark and light quarks from the vacuum, the final $[ud]$ diquarks has to have spin zero. In the second mechanism, however (diagram $B$), the $[ud]$ diquark is formed from the original $d$ quark and the $u$ quark from the vacuum. Angular momentum is shared among all final components and the  $[ud]$ diquark may well have spin one. 

Diquarks are antisymmetric in color indices and can be symmetric or antisymmetric in spin indices ($S=1,0$ respectively). When a diquark contains light quarks with  flavors $u,d,s$, the overall state must be antisymmetric by Fermi statistics (strong interactions do not distinguish the flavors $u,d,s$). That means 
\begin{subequations}
\begin{align}
S=0&\Rightarrow \mathrm{Flavor~Antisymmetric} \\
S=1&\Rightarrow \mathrm{Flavor~Symmetric}
\end{align}
\end{subequations}
A diquark with the composition $[uu]$, $[dd]$ or $[ss]$ must be spin 1 necessarily.
Pentaquarks realizing the valence quark structure \eqref{valencep} are therefore of two types
\begin{subequations}
\begin{align}
\label{pentau} 
\PP_u&=\epsilon^{ijk}\;\bar c_i\, [cu]_{j,\,S=0,1}\, [ud]_{k,\,S=0,1}\\
\PP_d&=\epsilon^{ijk}\;\bar c_i\, [cd]_{j,\,S=0,1}\, [uu]_{k,\,S=1}
\label{pentad}
\end{align}
\end{subequations}
where greek indices are for color, diquarks are in the color  antisymmetric, $\bar{\bm 3}$, configuration and overall antisymmetry requires flavor symmetric light-light diquark with $s= 1$.
The $\PP_d$ component can appear only in~\eqref{compos0}. 

This shows that a remarkable number of pentaquarks are immediately predicted once an SU(3) analysis is done. Depending on the flavor symmetry of the light-light quark, given that the heavy-light has flavor $\bm 3$, we have that, depending on the spin of the light-light diquark
\begin{subequations}
\begin{align}
S=0&\Rightarrow  {\bm 3}\otimes  \bar{\bm 3}= {\bm 1}\oplus {\bm 8} \\
S=1&\Rightarrow {\bm 3}\otimes {\bm 6} = {\bm 8}\oplus {\bm{10}}
\end{align}
\end{subequations}
Moreover the $J$ quantum numbers have to be considered. If we stick to the case $L=0$, depending on the spin $S$ of the light-light diquark we have
\begin{subequations}
\begin{align}
S=0&\Rightarrow J=1/2~(\mathrm{2~ways}), J=3/2~(\mathrm{1~way}) \\
S=1&\Rightarrow J=1/2~(\mathrm{3~ways}), J=3/2~(\mathrm{3~ways}),J=5/2~(\mathrm{1~way})
\end{align}
\end{subequations}
which gives a further $J$ multiplicity of 10 states. An analysis of expected states and decays in the diquark model can be found in~\cite{Maiani:2015vwa} together with an extension in~\cite{Maiani:2015iaa}. 

From these considerations we understand that selection rules in the na\"ive diquark model should be found to limit the number of pentaquark states as well. On the other hand the experimental situation is too preliminary to allow these kind of reasonings and comparisons. For this reason, the following section, which is devoted to the presentation of a different approach to the understanding of \XYZ phenomenology, does not contain any reference to pentaquarks. 

What we can learn from the recent observation of pentaquark resonances is again the indication that  diquarks must be good degrees of freedom to describe the nature of exotic hadronic resonances.  A new study on the diquark structure of pentaquarks which revises some of the conclusions drawn here is found in~\cite{alipenta}. 

\section{Tetraquarks and diquarks in the \texorpdfstring{$1/N$}{1/N} expansion\label{tqln}}
This section is especially based on a work by Maiani~\etal on tetraquarks in the $1/N$ expansion~\cite{Maiani:2016hxw}, where $N$ is the number of colors, following a stream of papers on the same subject initiated by Weinberg~\cite{Weinberg:2013cfa}. The contributions to this discussion by Knecht and Peris~\cite{Knecht:2013yqa} and Cohen and Lebed~\cite{Cohen:2014tga} were also particularly useful to us. 

The reputation of tetraquarks was somehow obscured by a theorem by S.~Coleman and E.~Witten~\cite{Coleman,Witten:1979kh} stating that: {\it tetraquarks correlators for $N \to \infty$ reduce to disconnected meson-meson propagators}.  The theorem follows from the simple fact that a four quark operator can be reduced to products of color singlet bilinears.  Connecting each bilinear with itself, one gets two disconnected one-loop diagrams, \ie a result of order $N^2$, while connected tetraquark diagrams are one-loop, thus  of order $N$.

The argument was reexamined in~\cite{Weinberg:2013cfa}, where it is argued that {\it if connected tetraquark correlators develop a pole, it will be irrelevant that its residue is subleading with respect to the disconnected parts}. After all, meson-meson scattering amplitudes are of order $1/N$, in the $N\to \infty$ limit, but we do not consider mesons to be  free particles.

The real issue, according to Weinberg, is the width of the tetraquark pole: if it increases for large $N$, the state will be undetectable for $N\to \infty$.
Weinberg finds that decay rates scale $\sim 1/N$ in the large $N$ expansion, making tetraquarks a respectable possibility.  We will show that widths are even narrower than this.  The tetraquark-meson `decoupling' we find here will be functional to the discussion in \sectionname{\ref{mol3}}.

After a concise review of the $1/N$ expansion in QCD we will show that tetraquark poles should  appear at non-planar orders of the $1/N$ expansion~\cite{Maiani:2016hxw}. In particular we are interested in the role of diquarks as tetraquark constituents. Their special role is discussed 
in~\sectionname{\ref{diqn}}. The results obtained in~\sectionname{\ref{diqn}} will be used extensively 
in~\sectionname{\ref{mol3}}.

For a comprehensive review on the papers which  followed Weinberg's remark~\cite{Weinberg:2013cfa,Knecht:2013yqa,Lebed:2013aka,Cohen:2014tga,Cohen:2014via,Cohen:2014vta} we also refer to~\cite{Esposito:2014rxa}.

\subsection{\texorpdfstring{$1/N$}{1/N} expansion in QCD: a short reminder}

\begin{figure}[htb!]
 \begin{center}
   \includegraphics[width=6truecm]{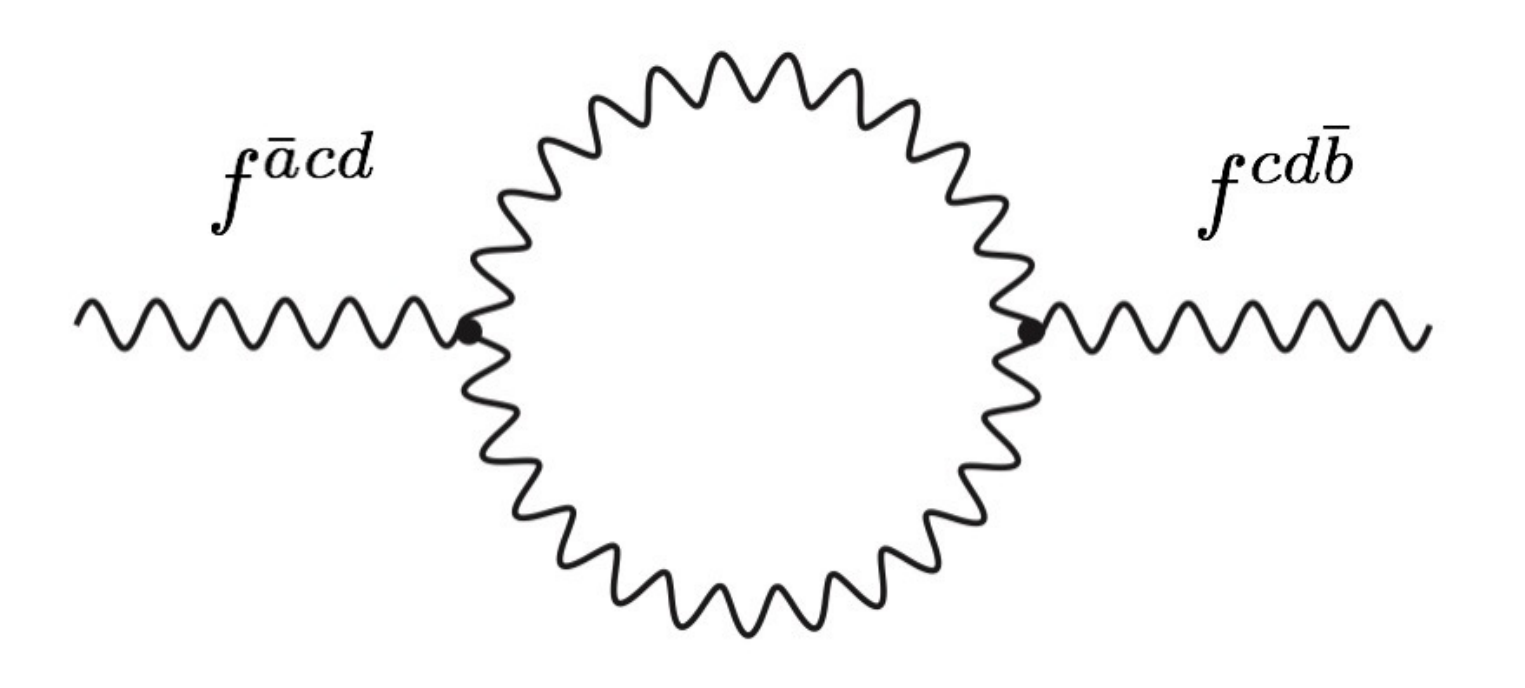}
 \end{center}
\caption{Gluon self-energy diagram with fixed colors $\bar a=\bar b$.}   
\label{gb}
\end{figure}

The behavior of QCD for $N\to \infty$ has been characterized by G.~`t~Hooft~\cite{'tHooft:1974hx}. Consider the gluon self-energy diagram in \figurename{~\ref{gb}} with external gluon colors fixed to $\bar a,\bar b $.

This diagram involves the product
\begin{equation}
\sum_{c,d}f^{\bar a cd} f^{ cd\bar b}=\tr\left(T^{\bar a}_\text{adj}T^{\bar b}_\text{adj}\right)=N\,\delta^{\bar a\bar b}
\end{equation}
in the adjoint representation of SU(N). The gluon loop therefore contains a multiplicity factor of $N$ in SU(N). To make the large $N$ limit of this diagram finite, one requires that the couplings at vertices, $g_{_{\rm QCD}}$, scale with $N$ as $g_{_{\rm QCD}}=\sqrt{\lambda/N}$ so that 
\begin{equation}
\frac{\lambda}{N}\times N=\lambda~\text{independent of $N$}
\end{equation}
where $\lambda$ is known as  `t~Hooft coupling.  
The large-$N$ limit is obtained keeping $\lambda$ fixed. 

\begin{figure}[htb!]
 \begin{center}
   \includegraphics[width=5truecm]{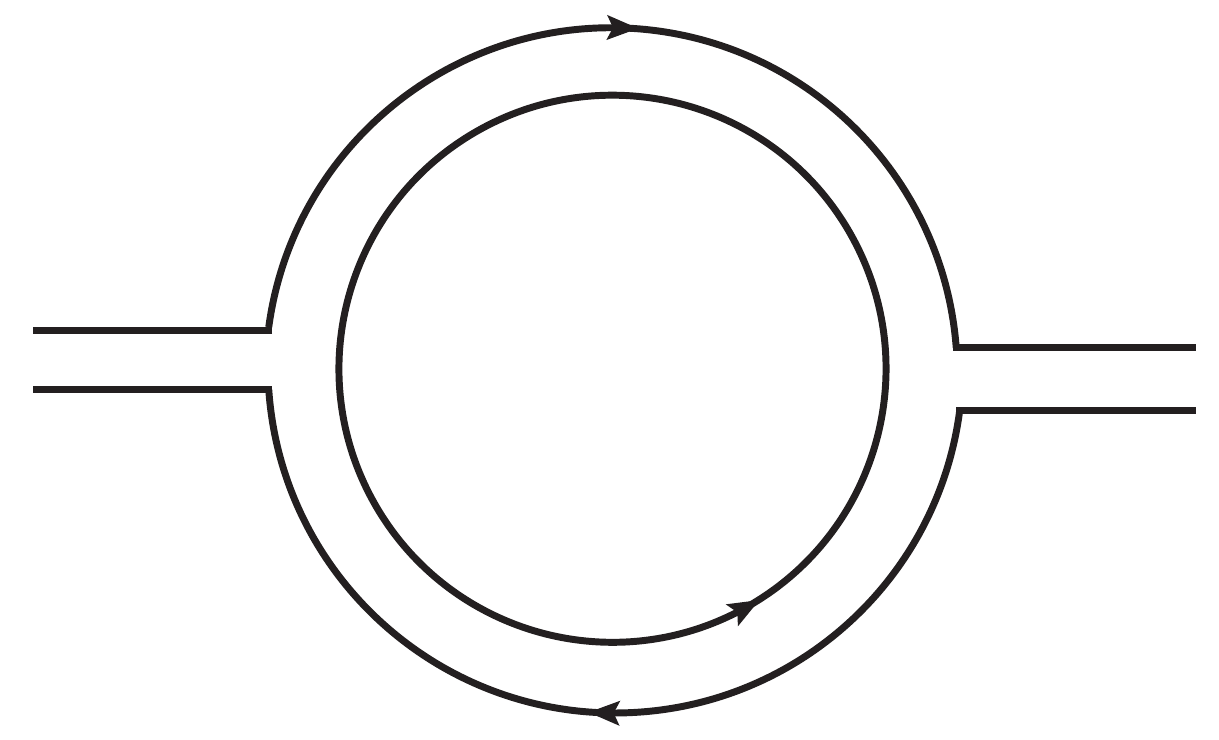}
\caption{Gluon self-energy diagram in the large $N$. With this notation the multiplicity factor  $N$ traced above in the $f$ structure constants, has a clear origin in the color loop at the center.}  %
\label{gbl}%
\end{center}%
\end{figure}

The gluon field is characterized by the color indices
\begin{equation}
(A_\mu)^i_j=A^a_\mu\, (T^a)^i_j 
\end{equation}
The number of independent components of this matrix in SU(N) is $N^2-1$. In the large $N$ limit however we can neglect the traceless condition, and treat it as a $N\times N$ matrix, with $N^2$ independent real components $A_\mu^a$, and represent the gluon line by a double color line --- carrying a pair of color indices $i,j$.
With this notation the diagram in \figurename{~\ref{gb}} can be represented as in \figurename{~\ref{gbl}}.

The origin of the multiplicity factor discussed above becomes apparent in the double-line notation: the factor of $N$ arises from the color loop in Fig.~\ref{gbl} and the factors of $1/\sqrt{N}$ at the vertices make the large $N$ limit smooth.  From the same figure we argue that the quark-gluon coupling will therefore also scale as $1/\!\sqrt{N}$ and the quadrilinear gluon coupling as $1/N$.

\begin{figure}
 \begin{center}
   \includegraphics[width=10truecm]{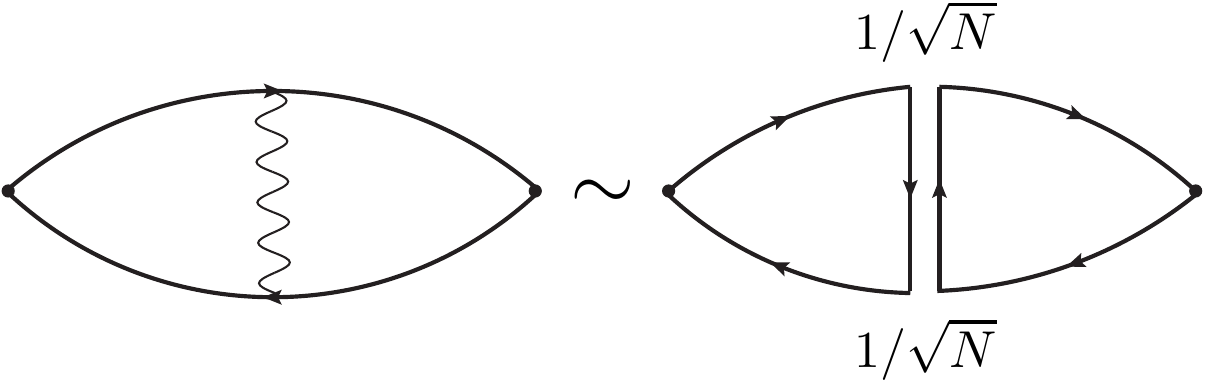}
\caption{One gluon exchange correction to the correlation function of a color singlet quark bilinear, represented by the open circle. Representing  the gluon line by two, oppositely running lines joining the quark lines on the edge, one sees that the diagram reduces, for color number counting, to a two loop diagram. Thus one recovers a result of order $N$, like the lowest order diagram, multiplied by the color reduced coupling $\lambda$. }   
\label{loopuno} \end{center}
\end{figure}

`t~Hooft shows that in the  $N\to \infty$  limit, the quark lines have to be at the edge of the diagram (valence quarks), whereas diagrams with internal quark lines are suppressed (quenched theory). Internal gluons are instead allowed, but only planar diagrams, where gluon lines do not step over each other,  survive at leading order. 
Let us consider the correlation function of a color singlet quark bilinear.
With no gluon lines, the result is obviously proportional to $N$, the number of colors  running in the loop. A gluon line traversing the loop, see \figurename{~\ref{loopuno}}, can be represented by two color lines running in opposite directions and joining the quark and antiquark lines that flow in the vertex. 
Thus we get two loops, \ie a factor of $N^2$, times the factor $1/N$, therefore a contribution of order $N$ again. The same adding trilinear or quadrilinear gluon vertices.
As long as gluons are inserted within the the surface delimited by the external quark lines and as long as they are planar, the $N$ power counting does not change. Non-planar gluons, as we will see, are instead subleading. 

The sum of all planar diagrams of this kind will again be of order $N$, times a non-perturbative function of $\lambda$, which may well develop poles for certain values of the external momentum, $q^2$. 

The sum of all planar diagrams like the one on the \lhs of \figurename{~\ref{loopuno}} is represented by
\begin{equation}
\langle 0|J(p)J^\dagger(p)|0\rangle\sim N
\end{equation}
where the operator $J^\dagger$ acts on the vacuum to create a meson state, and
 \begin{equation}
\langle 0|J(p)J^\dagger(p)|0\rangle=\sum_n \frac{\langle 0|J(p)|n\rangle\langle n|J^\dagger(p)|0\rangle}{p^2-m_n^2}
=\sum_n \frac{f_n^2}{p^2-m_n^2}
\label{spettrale}
\end{equation}
with the decay constant $f_n=\langle 0|J(p)|n\rangle$.

The behavior at large momenta of  $\langle 0|J(p)J^\dagger(p)|0\rangle$ is expected to be logarithmic~\footnote{The asymptotic behavior of the correlators is indeed driven by asymptotic freedom, and apart of a factor $p^n$ given by dimensional analysis, it is expected to scale with fractional powers of logarithms, according to the anomalous dimension of $J$, see~\cite{Bochicchio:2013tfa}.}. However the sum over meson states in Eq.~\eqref{spettrale} can scale as $\sim \ln p^2$, at large $p^2$, only if it has an infinite number of terms, as can be seen by substituting $\sum_n\to \int dm_n^2$. Thus we have an infinite number of poles, corresponding to a tower of (stable) meson states in the correlation function  $\langle 0|J(p)J^\dagger(p)|0\rangle$.  These have a given flavor content, \eg quarkonium mesons with varying spin and parity.
Meson masses are independent of $N$ and the entire $N$ dependency of the \rhs of~\eqref{spettrale} is encoded in $f_n$. In the case at hand this means that each $f_n\sim\sqrt{N}$. 

In the strict $N\to\infty$ limit, mesons are non-interacting particles and hence they are stable. This can be seen by considering a three-point function, a loop like the one in Fig.~\ref{loopuno} with three meson insertions.  The color loop of order $N$ has to be confronted with $f_n^3 g$, so that $g\sim 1/\sqrt{N}$.

\subsection{Tetraquarks in the \texorpdfstring{$1/N$}{1/N} expansion}
\subsubsection{From planar to non-planar diagrams}
\begin{figure}[htb!]
 \begin{center}
   \includegraphics[scale=.6]{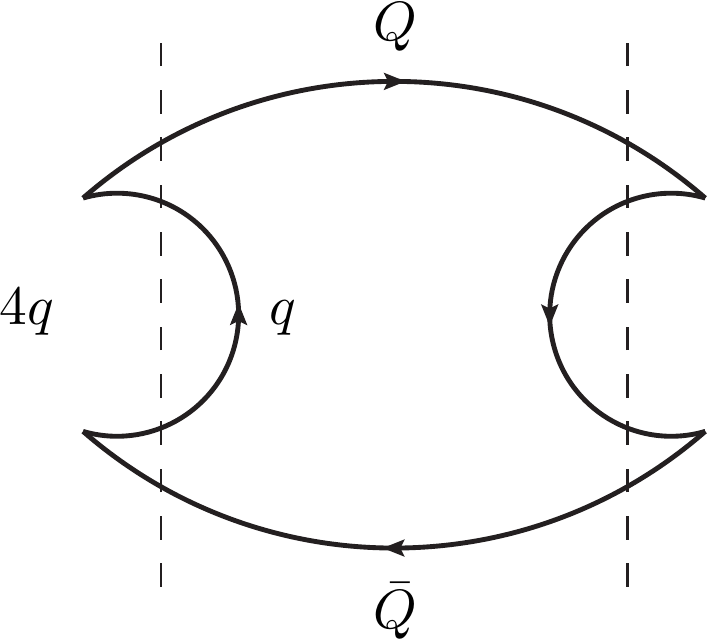}
 \caption{Free two-point correlation function of a tetraquark meson.}\label{neutro} \end{center}
\end{figure}

Following~\cite{Knecht:2013yqa}, diagrams with tetraquark operator insertions can be depicted as in \figurename{~\ref{neutro}}, where a two-point correlation function of a neutral tetraquark is represented. We do not include diagrams with heavy quark annihilations, which are expected to be suppressed (OZI rule).

The dashed lines represent the cuts where tetraquarks could be found. Apparently tetraquark intermediate states are possible, along with tetraquark-meson mixings. This is basically the point of view developed in~\cite{Knecht:2013yqa}, following Weinberg in~\cite{Weinberg:2013cfa}.

\begin{figure}[htb!]
 \begin{center}
   \includegraphics[scale=.3]{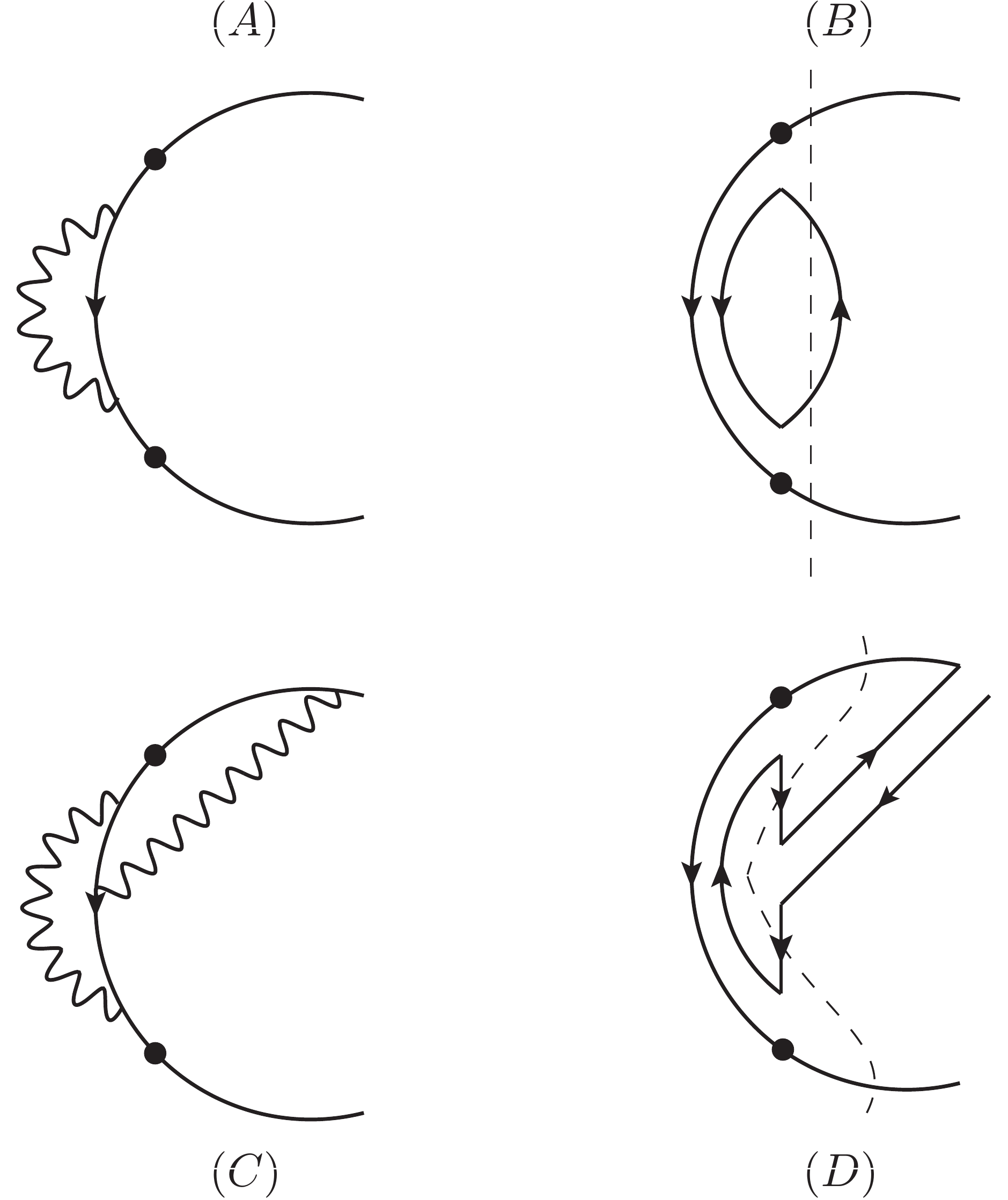}
 \caption{ $(A)$ A piece of a correlation function with two quark bilinear insertions and a planar gluon. Except for the interaction gluon, like the left hand half of the diagram in Fig.~\ref{neutro} with the light quark  line deformed to run on the perimeter of the diagram. Let the two bilinear insertions be $Q^iq^k$ and $\bar Q^\ell \bar q^m$ ($i\neq k$, $\ell\neq m$).   $(B)$ Double-line representation of $(A)$. Cuts might only  intercept color-disconnected  four-quark configurations: the cut in figure is for example $Q^i q^k\bar q^k \bar Q^r$. $(C)$ Non-planar diagram with a gluon running on the perimeter of the amplitude. $(D)$ Double-line representation of $(C)$. Cuts might intercept connected four quark configurations like $Q^i q^3\bar Q^r\bar q^3$ in figure. Although $(D)$ can topologically be deformed into a simple loop diagram --- which is what one usually do to count external color loops. Closing the right end half of the loop, the order in the $1/N$ expansion is $N\times (1/\sqrt{N})^4\sim 1/N$ instead of $N$.  Notice alternatively that diagram $(A)$ is of order $T^aT^a\times (1/\sqrt{N})^2\sim N$ whereas $(C)$ is $T^aT^bT^aT^b\times (1/\sqrt{N})^4\sim 1/N\times N^2\times1/N^2\sim 1/N$ {\it i.e.} $1/N^2$ with respect to $(A)$.
 }\label{nonplanar1} \end{center}
\end{figure}

The simplest way to include color interactions in the tetraquark cut of Fig.~\ref{neutro} is that of adding planar gluons. Consider for example one planar gluon on the light quark line as in Fig.~\ref{nonplanar1}~$(A)$. Using the double line notation, Fig.~\ref{nonplanar1}~$(B)$, one sees that the tetraquark cut contains color disconnected components. This motivated the considerations in~\cite{Maiani:2016hxw}.

The main content of~\cite{Maiani:2016hxw} is that non-planar diagrams may provide cuts with color correlated tetraquarks. 
In essence this can be seen in Fig.~\ref{nonplanar1}~$(C)$, whose double line notation version, $(D)$, shows that this is a possibility.   
Diagrammatic considerations of this kind do not provide a {\it proof} that tetraquark cuts are indeed found in the non-planar approximation, 
but at least suggest that tetraquarks are more strongly decoupled from mesons than expected.

The non planar gluon in Fig.~\ref{nonplanar1}, is like a {\it handle} applied to the diagram and the amplitude of \figurename{~\ref{nonplanar1}} goes like $1/N$, as can be  checked by direct calculation.

Even though we cannot provide a proof that diagrams with one handle have indeed a tetraquark pole,  we find that different correlation functions are consistent with the factorization of residues at the pole, as we discuss below.

\begin{figure}[t]
 \begin{center}
   \includegraphics[scale=.7]{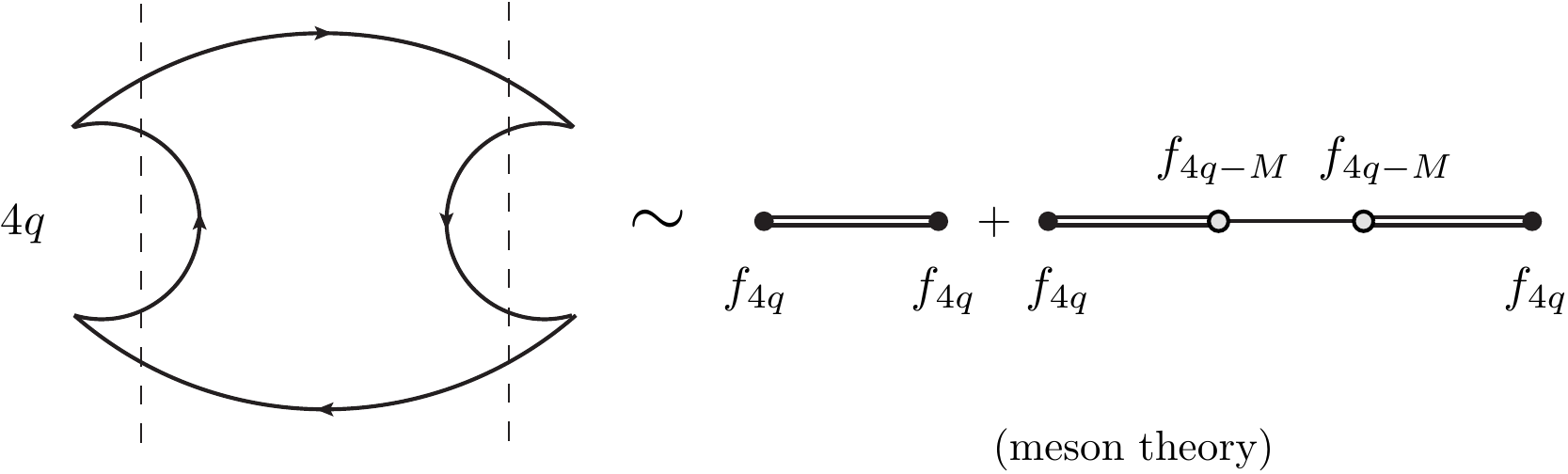}
\caption{The same diagram as in \figurename{~\ref{neutro}} with its counterpart in meson theory on the \rhs. The diagram on the \lhs is of order $N$ therefore $f_{4q}$ in the meson theory is of order $\sqrt{N}$ and the mixing $f_{4q-M}\sim N^0$. This counting relies on the fact that we might   have only the tetraquark propagator in the meson theory. If we assume instead that the loop diagram only involves $4q-M$ mixings we would simply get $f_{4q}^2f_{4q-M}^2\sim N$. In other words, diagrams on the \rhs are meant to be summed on meson/tetraquarks intermediate states. This eventually generates a contact term in the second diagram on the \rhs, which is represented by the first diagram.  This means that $f_{4q}$ in the first diagram is equivalent to $f_{4q}f_{4q-M}$ in the second. In the following we will consider mostly diagrams of the type of the second one on the \rhs.}\label{f4q} \end{center} 
\end{figure}

Let us consider again the diagram in \figurename{~\ref{neutro}}, and the corresponding effective meson theory diagrams, as done in \figurename{~\ref{f4q}}.  
\begin{figure}[ht]
 \begin{center}
   \includegraphics[scale=.7]{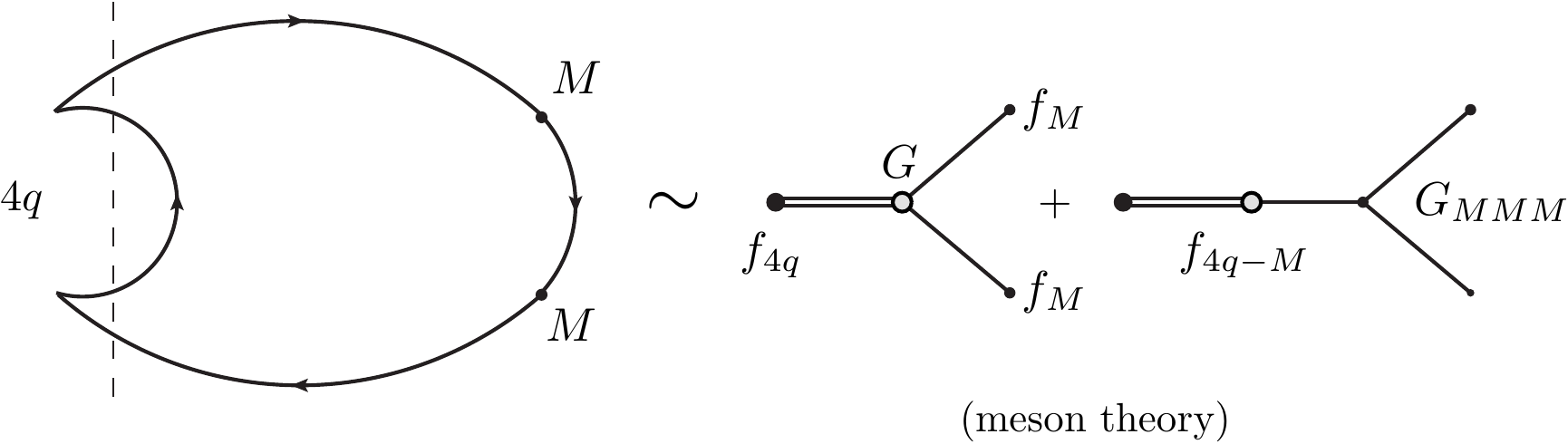}
 \end{center} \caption{The coupling $G$ is found to be of order $G\sim1/\!\sqrt{N}$ to match the quark loop diagram, which is $\sim N$. Also $G\sim N^0\times 1/\!\sqrt{N}$ using the mixing in \figurename{~\ref{f4q}}. Because of this either writing the diagram with no $4q-M$ mixing and $G$ or writhing the diagram with mixing and $f_{4q-M}/\!\sqrt{N}$ is the same thing. The $N$ order of the diagram will be $f_{4q}f_{4q-M}/\!\sqrt{N} \, (\sqrt{N})^2\sim N$
  or $f_{4q}f_{4q-M}/\!\sqrt{N}\sim 1$. Also here, as commented in \figurename{~\ref{f4q}}, the first diagram on the \rhs corresponds to the contact term generated by the infinite sum over mesons on the second diagram. This in turn corresponds to the identification $G=f_{4q-M}\, 1/\!\sqrt{N}$.}\label{gwein}
\end{figure}
Since the quark loop diagram is of order $N$ (whatever planar gluon interactions might be inserted), the $f_{4q}$ decay constant $f_{4q}=\langle0|J|4q\rangle$ has to be such that $f_{4q}^2\sim N$ or (see discussion around Eq.~\eqref{spettrale})
\begin{equation}
f_{4q}\sim \sqrt{N}
\end{equation}
 Since the \rhs in \figurename{~\ref{f4q}} also scales as $f_{4q}^2f_{4q-M}^2$,  the latter requires the tetraquark-meson mixing to be 
 \begin{equation}
 f_{4q-M}\sim N^0 
\end{equation}
We might use the result found for $f_{4q}$ in the three-point function responsible for the decay of a tetraquark into two mesons.  The $4q$-$MM$ vertex contains the coupling $G$ and $G\, f_{4q}\, f_M^2\sim N$ to match the $N$ order in the loop diagram --- see \figurename{~\ref{gwein}}. Since $f_M\sim \sqrt{N}$, we have that $G\sim 1/\!\sqrt{N}$; this can also be seen through the $4q$-$M$ mixing of the \rhs of \figurename{~\ref{gwein}}. It goes like $G\sim f_{4q-M}\, G_{MMM}\sim N^0\times 1/\!\sqrt{N}$, from the fact that the trilinear meson vertex scales as $G_{MMM}\sim1/\!\sqrt{N}$.

The conclusion that 
\begin{equation}
G\sim1/\!\sqrt{N}
\end{equation}
reached in~\cite{Weinberg:2013cfa}, implies that tetraquarks will be narrow resonances in the large $N$ limit. In other words they are states which in principle might be observed as narrow resonances,  contrarily to the fall apart decay which is intrinsic in the Coleman-Witten argument mentioned above.  In these respects if connected tetraquark correlators develop a pole, it will be irrelevant that its residue is of the subleading order $N$ (instead of $N^2$).  What is really important is that the total width will scale as $\Gamma\sim 1/N$. 

\begin{figure}[ht]
 \begin{center}
   \includegraphics[scale=.7]{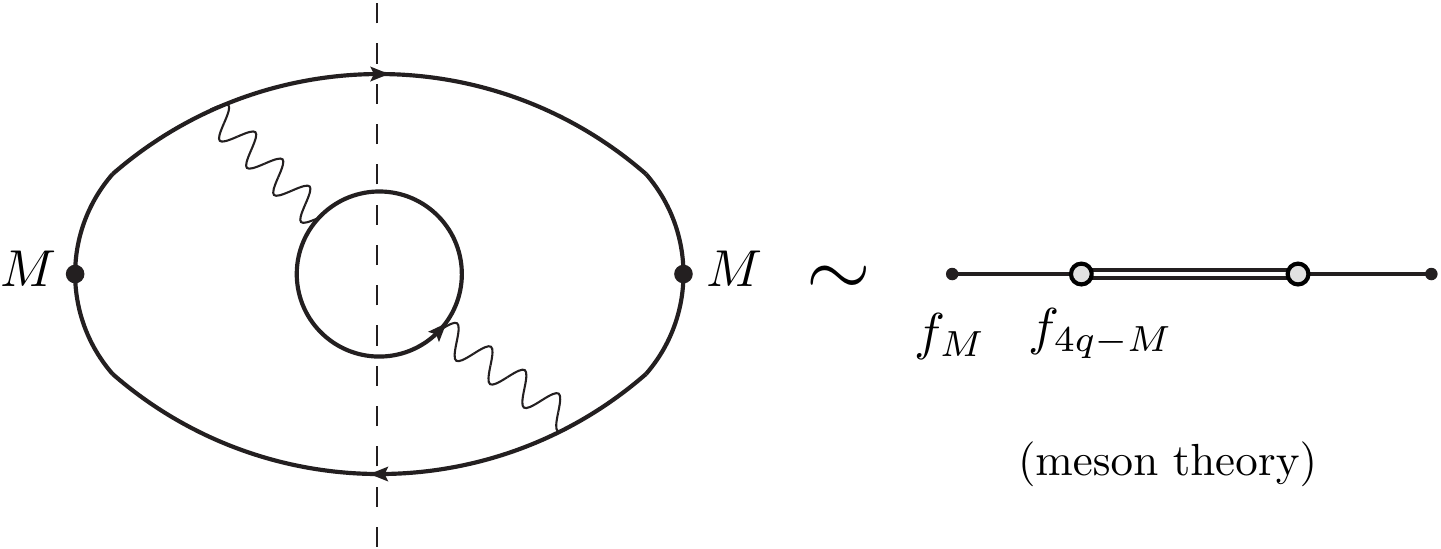}
\caption{The diagram on the \lhs is of order $N^0$ therefore $f_{4q-M}\sim 1/\!\sqrt{N}$, which is not consistent with $f_{4q-M}\sim N^0$ in \figurename{~\ref{f4q}}.}\label{fmixwei} \end{center} 
\end{figure}

If we  assumed that only  $4q$-$M$ mixings are allowed in meson theory, and that the first term on the \rhs of \figurename{~\ref{f4q}} is the contact term resulting from a (infinite) sum over the mesons on the second term,  we would simply have $f_{4q}^2f_{4q-M}^2\sim N$ which means
\begin{equation}
f_{4q}\sim \frac{\sqrt{N}}{f_{4q-M}}
\end{equation}
leading to
\begin{equation}
G\sim \frac{f_{4q-M}}{\sqrt{N}}
\end{equation}
consistent with the second term on the \rhs of \figurename{~\ref{gwein}}.
To compute the order of $G$  in the large $N$ expansion, the mixing $f_{4q-M}$ should be computed 
from another diagram.  Consider the two-point $MM$ correlator in \figurename{~\ref{fmixwei}}
where it is found that 
\begin{equation}
f_{4q-M}\sim 1/\!\sqrt{N}
\end{equation}
where it is used the standard result $f_M\sim \sqrt{N}$. This leads to $G\sim 1/N$.

From what just found, we observe an inconsistency. In fact, from \figurename{~\ref{f4q}} one deduces $f_{4q-M}^2\sim N^0$, while from \figurename{~\ref{fmixwei}} is found that $f_{4q-M}^2\sim 1/N$. 

This inconsistency is removed if  tetraquark decay and mixing constants are extracted from non-planar diagrams as shown in the next section.

\subsubsection{Non-planar diagrams \label{sec:nonplanar}} 
\begin{figure}[htb!]
 \begin{center}
   \includegraphics[scale=.7]{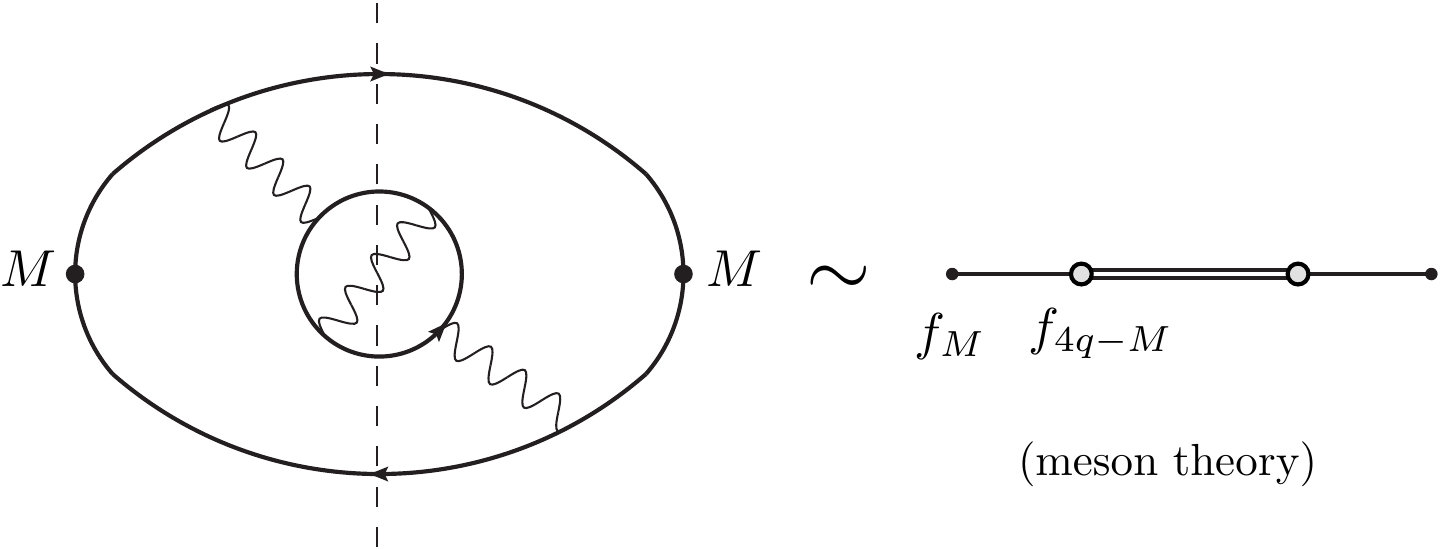}
\caption{In this non-planar diagram the quark loop order is $1/N^2$ and $f_M^2 f_{4q-M}^2\sim1/N^2$, giving $f_{4q-M}\sim 1/(N\sqrt{N})$ instead of $N^0$ as in \figurename{~\ref{fmixwei}}.  }\label{mixnonplanar} \end{center} 
\end{figure}

We suggest that tetraquark poles, if present, should appear in non-planar diagrams. The correct way to extract the decay constant involving a four-quark state is therefore to repeat the previous analysis to the first non-planar order. We consider the following diagram in \figurename{~\ref{mixnonplanar}}, which contains one handle. The $1/N$ order is determined by 
\begin{equation}
N^{2- L-2H}
\end{equation}
where $L$ is the number of quark loops and $H$ is the number of handles. In the case in \figurename{~\ref{mixnonplanar}} we have $L=2, H=1$ giving $N^{-2}$, as can be seen also using the standard perturbative counting rules introduced in \figurename{~\ref{loopuno}}. 

From the counting of powers in \figurename{~\ref{mixnonplanar}} it  results that 
\begin{equation}
f_{4q-M}\sim 1/(N\sqrt{N})
\end{equation}

\begin{figure}[htb!]
 \begin{center}
   \includegraphics[scale=.7]{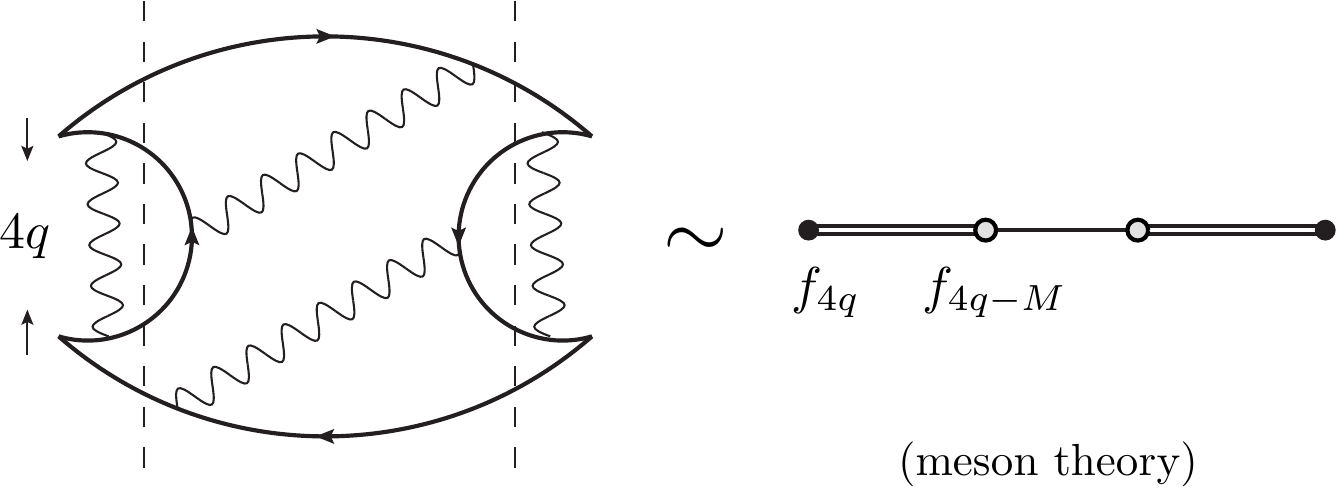}
\caption{This diagram is of order $1/N^3$ thus, in the meson theory we have $f_{4q}^2f_{4q-M}^2\sim1/N^3$. Using $f_{4q-M}=1/(N\sqrt{N})$ from \figurename{~\ref{mixnonplanar}}, we get $f_{4q}\sim N^0$ (in place of $f_{4q}\sim\sqrt{N}$ found in \figurename{~\ref{f4q}}). }\label{nonplanar3} \end{center} 
\end{figure}

We can  determine the $f_{4q}$ decay constant from the non-planar  two-point correlator of the tetraquark. In the meson  theory we have the $4q-M$ mixings --- contact terms are inessential to our discussion. The non-planar diagram in \figurename{~\ref{nonplanar3}} is of order $1/N^3=N^{2-1-2\times 2}$, which corresponds to  $f_{4q-M}^2$. Therefore we need 
\begin{equation}
f_{4q}\sim N^0 
\end{equation}
We must check if the power counting leading to $f_{4q}\sim N^0$ does not produce conflicting results in other correlation functions. 

\begin{figure}[htb!]
 \begin{center}
   \includegraphics[scale=.7]{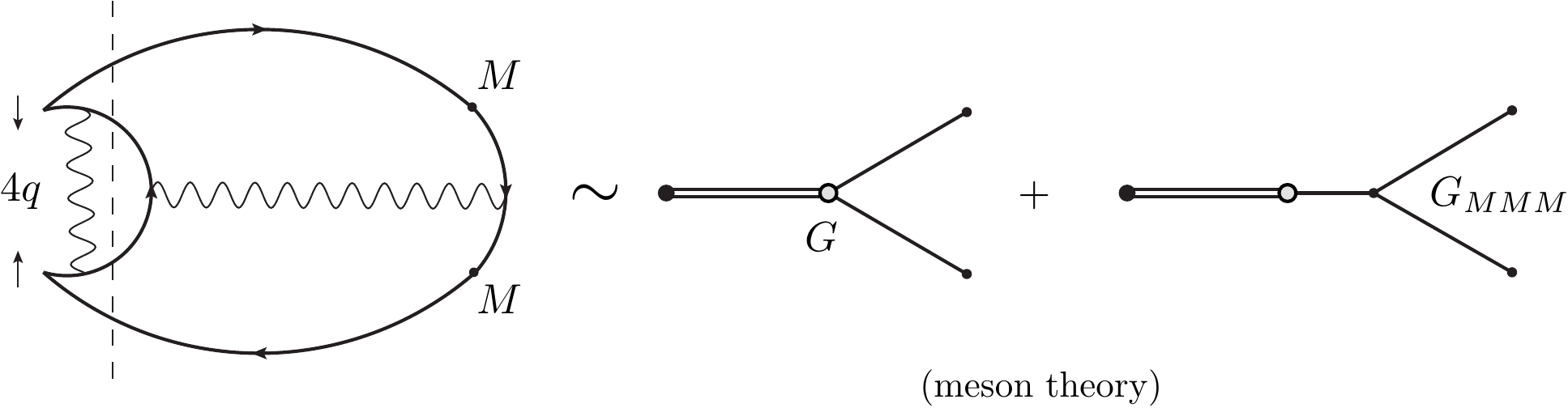}
 \end{center} \caption{The quark loop diagram  is of order $1/N$. Therefore $f_{4q}\,G\,(\sqrt{N})^2=f_{4q}\,f_{4q-M}\,1/\!\sqrt{N}\,(\sqrt{N})^2\sim 1/N$ gives $G\sim1/N^2$, taking $f_{4q}\sim N^0$. Also here, as commented in \figurename{~\ref{f4q}}, the first diagram on the \rhs corresponds to the contact term generated by the infinite sum over mesons on the second diagram. This in turn corresponds to the identification $G=f_{4q-M}\, 1/\!\sqrt{N}$. }\label{nonplanar4}
\end{figure}

For this purpose consider the trilinear correlator giving rise to the tetraquark decay into two mesons. 
This is represented in \figurename{~\ref{nonplanar4}} to the non planar order with one handle. 
We find that 
\begin{equation}
G\sim 1/N^2
\end{equation}
\begin{figure}[htb!]
 \begin{center}
   \includegraphics[scale=.7]{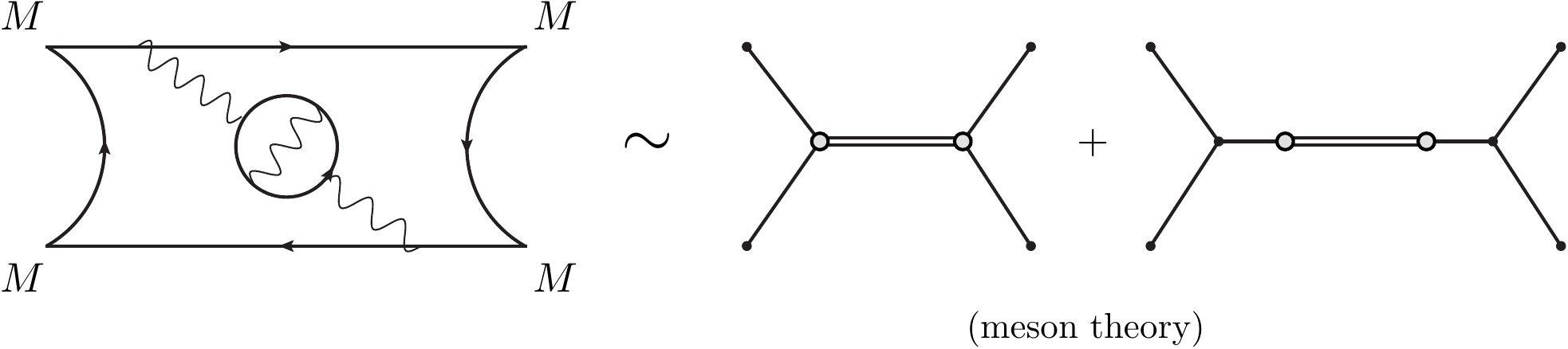}
\caption{A consistency check. Here we can use that $G\sim 1/N^2$ and $f_{4q-M}=f_{M-4q}=1/(N\sqrt{N})$ to verify that $G^2\,(\sqrt{N})^4=f_{M-4q}^2\,(1/\!\sqrt{N})^2\,(\sqrt{N})^4=1/N^2$ as in the quark-loop diagram --- we used $G\sim 1/N^2$ as found in \figurename{~\ref{nonplanar4}}. Also here, as commented in \figurename{~\ref{f4q}}, the first diagram on the \rhs corresponds to the contact term generated by the infinite sum over mesons on the second diagram. This in turn corresponds to the identification $G=f_{4q-M}\, 1/\!\sqrt{N}$.}\label{nonplanar5}
 \end{center} 
\end{figure}
using $f_{4q}$ and $f_{4q-M}$ introduced above. As a consistency check we also analyze the $MM\to MM$ four-point function finding a match between the $N$ order in the loop diagram and in the meson theory, the decay and mixing constants being fixed. This is done in \figurename{~\ref{nonplanar5}}. 

Therefore we conclude that the tetraquark states can be even narrower than what expected at the order $N$, since we found  $G\sim 1/N^2$ in place of $G\sim 1/\!\sqrt{N}$. With this approach we consistently find $f_{4q-M}^2\sim 1/N^3$. %

\subsubsection{Charged tetraquarks}
\begin{figure}[t]
 \begin{center}
   \includegraphics[scale=.6]{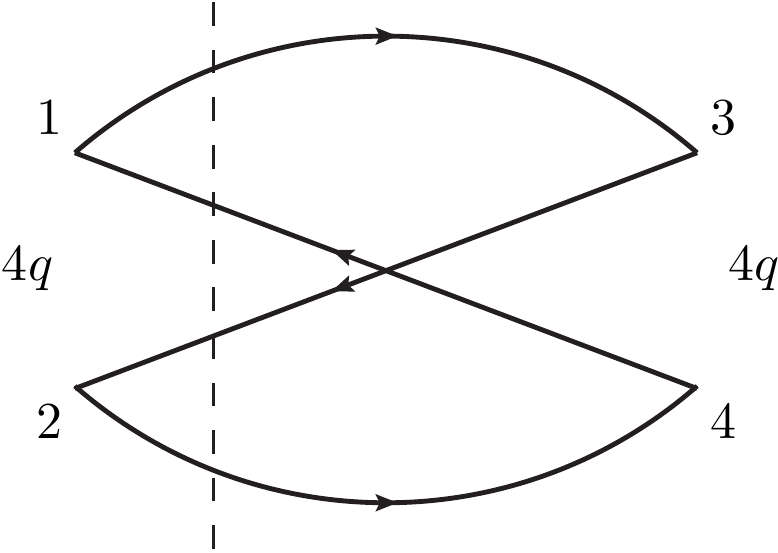}
\caption{Charged tetraquark two-point correlation function. This might be untwisted in a $MM\to MM$ meson-meson amplitude, with no tetraquarks in the $s$-channel.}\label{carico} \end{center} 
\end{figure}

\begin{figure}[t]
 \begin{center}
   \includegraphics[scale=.6]{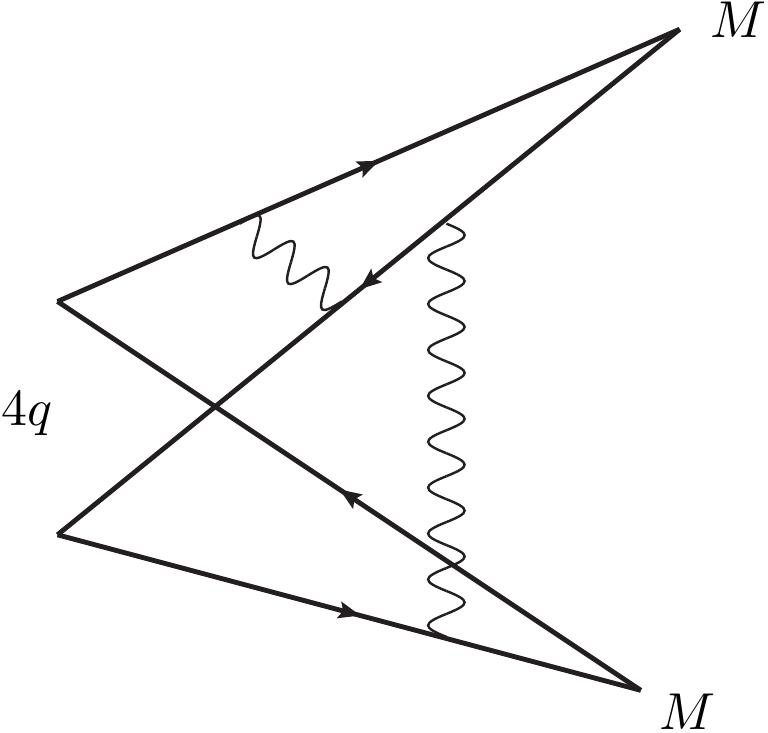}
\caption{ Decay of a charged tetraquark in two mesons in the lowest non-planar diagram. Following the same reasonings from this topology we see that $i)~f_{4q}^2\sim 1/N$ and $ii)~f_{4q}\,G\, (\sqrt{N})^2\sim 1/N$ thus $G\sim 1/(N\sqrt{N})$ (see text for more details) differently from what found for neutral tetraquarks ($G\sim 1/N^2$ in \figurename{~\ref{nonplanar4}}).}\label{chargednonplanar} \end{center} 
\end{figure}

The determination of the coupling $G$ is different if charged tetraquarks are taken into account. Consider for example the  two-point function containing the charged tetraquark $cu\bar c \bar d$, in \figurename{~\ref{carico}} --- {\it i.e.} assume that the amplitude in Fig.~\ref{carico} connects  a hidden-charm + light meson $c\bar c+ u\bar d$ pair to two open-charm mesons $c\bar d+u\bar c$.  Like in the cases examined previously,  the four quarks in the cut are found in color  disconnected configurations at the planar order, whereas color connections are forced in non-planar diagrams, see Fig.~\ref{chargednonplanar}. 

The same topology can be used to represent a two-point function of tetraquark operators, $4q\to 4q$, or a $4q\to MM$ correlator. In the first  case it is  obtained that $f_{4q}\sim1/\!\sqrt{N}$, since  the graph with one handle is of order $1/\!\sqrt{N}$. In the second case  
\begin{equation}
f_{4q}\, G\, f_M^2\sim 1/N\Rightarrow G\sim1/(N\sqrt{N})
\end{equation}
Neutral tetraquarks found in the cuts of Fig.~\ref{nonplanar3} can mix with meson-meson states, see the {\it rhs} of Fig.~\ref{nonplanar3}. Such mixings are not present when we come to topologies as the one in Fig.~\ref{chargednonplanar}. Therefore  charged tetraquarks, which do not mix with mesons, tend to be broader than  neutral ones, which do. Indeed we see that $G_{\rm charged}\sim 1/(N\!\sqrt{N}) =1/f_{4q-M}\,G_{\rm neutral}$. 

Without the color connections present at non-planar orders, these amplitudes would be saturated by meson-meson channels, as observed at the beginning of this section.  Diagrammatic arguments are not a proof of the existence of tetraquark poles, but an indication of the fact that they might occur at non-planar orders. However what found indicates  that tetraquarks are reluctant to mix with meson states.

As we will see in the next subsection, any diquarkonium tetraquark can be written in terms of its meson-meson quantum numbers (Fierz rearrangement) but this does not contain any dynamical information about the probability amplitude of a compact tetraquark to rearrange its quark content and  decay into a meson-meson state (or vice versa) or to mix with a meson state. $1/N$ expansion is used to shed some light on these aspects.

Taking the large $N$ considerations at their extreme consequences ($N\to \infty$), the fact that the $4q-M$ and $4q-MM$ mixings are found to be rather suppressed at large $N$ would make the observed prominent decay modes very unlikely. However $N=3$ and the numerators of the $1/N$ expansion are unknown. 

More interestingly for us, an enhancement mechanism of the mixing, as that proposed in \sectionname{~\ref{mol3}}, would explain why some of these very narrow  states are  being observed in experiments.

\subsection{Diquarkonia in the \texorpdfstring{$1/N$}{1/N} expansion\label{diqn}}
For $N$ strictly infinite, $q\bar q$ mesons generated by quark bilinear correlators are free particles. Interactions are generated by letting $N$ to be large but finite. Irreducible vertices with $k$ external mesons are of order $N^{1-k/2}$. 

We refer to \figurename{~\ref{carico}} for the flavor composition of a charged tetraquark. 
Non-planar quark diagrams with the same flavor structure, as in \figurename{~\ref{chargednonplanar}}, would develop tetraquark poles in the $s$-channel contributing an amplitude of order $G\sim 1/(N\sqrt{N})$.   

We consider $S$-wave meson-meson scattering in the channel with $J^P=1^+$ and the the three pairs of `in' pseudoscalar and vector mesons ($M_1\,M_2$) together with the three pairs of `out'  open charm mesons ($M_3\,M_4$) 
\begin{align}
\label{hidch}
M_1\,M_2&= \eta_c\, \rho^+,~\psi\,\pi^+, ~\psi\, \rho^+\\ %
M_3\,M_4&= {\bar D}^0\, D^{* +}, ~D^+ \, {\bar D}^{* 0},~{\bar D}^{* 0}\, D^{* +}
\label{opch}
\end{align}
As commented above, a diagram like the one in \figurename{~\ref{carico}} might be interpreted as a meson-meson $MM\to MM$ scattering diagram converting hidden charm+light mesons into open charm ones. 

To appropriately consider the quantum numbers of the $M_{1,2,3,4}$ mesons, we have to insert the spin matrices of each of them  (written as $\bar q \,\Gamma_i\, q^\prime)$ in the vertices, and obtain the spin factor~\footnote{$\gamma$-matrices associated to the quark gluon interaction $i)$ do not count since they reduce to $\gamma^0=\pm 1$, in the non-relativistic limit.}
\begin{equation}
S(12;34)={\rm Tr}\left(\Gamma_1\Gamma_3^\dagger \Gamma_2 \Gamma_4^\dagger\right)\equiv
\langle \Gamma_1\Gamma_3^\dagger \Gamma_2 \Gamma_4^\dagger \rangle
\label{cpfac}
\end{equation}
The order in the trace follows backward the arrows in the  diagram of \figurename{~\ref{carico}}. 

The quark diagram is such that it transforms the hidden charm, into open charm channels. If we take for example $M_1 M_2= \eta_c \rho^+$ and $M_3 M_4= {\bar D}^0 D^{* +}$, we obtain
\begin{equation}
S^{ab}(\eta_c \rho^+;{\bar D}^0 D^{* +}) = C\,\langle\sigma^2\cdot\sigma^2\cdot \sigma^2\sigma^a\cdot (\sigma^2\sigma^b)^\dag\rangle =  C\,\langle\sigma^a\sigma^b\rangle = 2C\,\delta^{ab}
\label{sfact}
\end{equation}
and the same  for $\eta_c \rho^+ \leftrightarrow D^+ \, {\bar D}^{* 0} $
\begin{equation}
S^{ab}(\eta_c \rho^+;D^+ \, {\bar D}^{* 0})=2C\,\delta^{ab}
\label{sfact2}
\end{equation}
For $\eta_c \rho^+ \leftrightarrow {\bar D}^{* 0}\, D^{* +} $, we obtain
\begin{equation}
S^{ab}(\eta_c \rho^+;{\bar D}^{* 0} D^{* +})= C^\prime\,\langle\sigma^2\cdot\sigma^d\sigma^2\cdot \sigma^2\sigma^a\cdot \sigma^c\sigma^2\rangle\, (-i\epsilon^{bcd})= 4C^\prime\,\delta^{ab}
\label{sfact3}
\end{equation}
According to the definitions in Eqs.~\eqref{spindef} $C=1/2$ and $C^\prime=1/\!\sqrt{2}$. We can then complete the spin factor matrix of hidden $(H)$ to open charm $(O)$ transitions $H\leftrightarrow O$
\begin{equation}
S^{ab}(H;O)=\quad \bordermatrix{~ & \bar D^0\, (D^{*+})^b & D^+\,  (\bar D^{*0})^b & (\bar D^{*0}\bm\times  D^{*+})^b \cr
                  \eta_c\,(\rho^+)^a & 1 & 1 & \sqrt{2} \cr
                  \psi^a\,\pi^+ & 1 & 1 & -\sqrt{2} \cr
                  (\psi \bm\times  \rho^+)^a & \sqrt{2} & -\sqrt{2} & 0 \cr
                  }\cdot \frac{1}{2}\delta^{ab}
\label{matrixsho}
\end{equation}
The first row is explicitly given by~\eqref{sfact}-\eqref{sfact3}.

The state in the cut must be a superposition of the `in' state  and the  `out' one to project on both. If we fix the quantum numbers of the `in' state to be $(I^G)\,J^{PC}=(1^-)\,1^{++}$, we are choosing $M_1\,M_2=\psi \bm\times  \rho^+$.  The out state can then be the superposition  $ \bar D^0\, D^{*+} - D^+\,  \bar D^{*0} $ (last row of the above matrix), which happens to have the right $J^{PC}$ quantum numbers.  For the sake of brevity, from now one we will omit the normalization factors and assume that the hadronic states are everywhere normalized to one. We then find that the state in the cut of \figurename{~\ref{carico}} should be 
\begin{equation}
\psi \bm\times  \rho^+ \pm (\bar D^0\, D^{*+} - D^+\,  \bar D^{*0})
\end{equation}
But this, apart from phases,  corresponds to the $X^+$ meson written as a compact diquarkonium (to make sure that attraction in diquarks makes sense in the $1/N$ expansion we refer to \ref{appN})
\begin{align}
X^+&= [cu]_{S=0}[\bar c\bar d]_{S=1}+[cu]_{S=1}[\bar c\bar d]_{S=0}=\notag\\
&= {(\epsilon_{\gamma\alpha\beta}c^{\alpha}\,\sigma^2\, u^{\beta})}   (\epsilon^{\gamma\alpha^\prime\beta^\prime}\bar c_{\alpha^\prime}\,\sigma^2\bm \sigma \, \bar d_{\beta^\prime}) +(\sigma^2 \leftrightarrow \sigma^2{\bm \sigma})=\notag\\
&=(c^\alpha \,\sigma^2\, u^\beta)\left[({\bar c}_\alpha\, \sigma^2{\bm \sigma}\, {\bar d}_\beta)-(\bar c_\beta\, \sigma^2{\bm \sigma}\, {\bar d}_\alpha)\right]+(\sigma^2 \leftrightarrow \sigma^2{\bm \sigma})=\nonumber \\ 
&=-i(c\,\sigma^2 {\bm \sigma}\,{\bar c})\bm\times  (u\,\sigma^2{\bm \sigma}\, {\bar d})
- \left[(c\,\sigma^2\,{\bar d})(u\,\sigma^2{\bm \sigma}\,{\bar c}) - (c\,\sigma^2{\bm \sigma}\,{\bar d})(u\,\sigma^2\, {\bar c})\right]=\notag\\
&\sim -i\, \psi \bm\times  \rho^+ + \left(D^+ {\bar D}^{* 0}-{\bar D}^0 D^{* +}\right)
\label{colsing}
\end{align}
In the first row we have used the definition of the $X^+(J^{PC}=1^{++})$ state in terms of diquarks, in the second we have assumed  attraction in the antisymmetric color representation (see \ref{appN}),  in the fourth we {\it neutralize color} composing color singlets and used the Fierz rearrangement   results from the \tablename{\ref{fierz}}
and $(\sigma^2\bm \sigma)^T=\sigma^2\bm \sigma$, $(\sigma^2)^T=-\sigma^2$. 
\begin{table}[thb!]
\centering
    \begin{tabular}{c|ccc} \hline\hline
-- &{\footnotesize$\begin{array}{c}(\sigma^2)_{ik}(\sigma^2{\bm \sigma})_{jl} \end{array}$}& {\footnotesize $\begin{array}{c}(\sigma^2{\bm \sigma})_{ik}(\sigma^2)_{jl}\end{array}$} &  {\footnotesize $\begin{array}{c} i (\sigma^2{\bm \sigma})_{ik} \bm\times  (\sigma^2{\bm \sigma})_{jl}\end{array}$ }\\ \hline
 {\footnotesize $(\sigma^2)_{ij}(\sigma^2{\bm \sigma)}_{kl} $} &{\footnotesize $+1/2$} & {\footnotesize $-1/2$} & {\footnotesize $+1/2$} \\ 
{\footnotesize $(\sigma^2{\bm \sigma})_{ij}(\sigma^2)_{kl} $} &{\footnotesize $-1/2$} & {\footnotesize $+1/2$} & {\footnotesize $+1/2$} \\
{\footnotesize $i(\sigma^2{\bm \sigma})_{ij}\bm\times  (\sigma^2{\bm \sigma})_{kl}$} &{\footnotesize $+1$} & {\footnotesize $+1$} & {\footnotesize $0$}\\
\hline\hline
\end{tabular}
\caption{Coefficients of the Fierz rearrangement of $J^P=1^+$ quadrilinears. Notice that the square of this matrix is equal to the identity. }
\label{fierz}
\end{table}
In addition to the results in \tablename{\ref{fierz}}, other useful relations which will be needed in the following, are listed below  
\begin{equation}
 (\sigma^2 \bm \sigma)_{ij} \cdot (\sigma^2 \bm \sigma)_{kl}=\frac{3}{2}(\sigma^2)_{il}(\sigma^2)_{kj}-\frac{1}{2}(\sigma^2 \bm \sigma)_{il}\cdot (\sigma^2 \bm \sigma)_{kj}
 \end{equation}
 or
\begin{equation}
 (\sigma^2 \bm \sigma)_{ij} \cdot (\sigma^2 \bm \sigma)_{kl}=-\frac{3}{2}(\sigma^2)_{ik}(\sigma^2)_{jl}-\frac{1}{2}(\sigma^2 \bm \sigma)_{ik}\cdot (\sigma^2 \bm \sigma)_{jl}
 \end{equation}
 together with
 \begin{equation}
 (\sigma^2)_{ij} (\sigma^2)_{kl}=\frac{1}{2}(\sigma^2)_{il}(\sigma^2)_{kj}+\frac{1}{2}(\sigma^2 \bm \sigma)_{il}\cdot (\sigma^2 \bm \sigma)_{kj}
 \end{equation}
 or
  \begin{equation}
 (\sigma^2)_{ij}  (\sigma^2)_{kl}=\frac{1}{2}(\sigma^2)_{ik}(\sigma^2)_{jl}-\frac{1}{2}(\sigma^2 \bm \sigma)_{ik}\cdot (\sigma^2 \bm \sigma)_{jl}
 \end{equation}
Similarly one can show that (see~\eqref{spindue}) 
 \begin{equation}
 (\sigma^2\sigma^{(a})_{ij} (\sigma^2\sigma^{b)})_{kl}-\frac{2}{3}(\sigma^2\sigma^{c})_{ij} (\sigma^2\sigma^{c})_{kl}=(\sigma^2\sigma^{(a})_{ik} (\sigma^2\sigma^{b)})_{jl}-\frac{2}{3}(\sigma^2\sigma^{c})_{ik} (\sigma^2\sigma^{c})_{jl}
 \end{equation}

A factor of two must be taken into account from each spinor exchange --- notice the $+$ sign of  the spin exchanged term. Each time spinors are exchanged a factor of $-1$ is included: a term like $(u\,\sigma^2\, {\bar c})=(\bar c\,\sigma^2\, u)\sim \bar D^0$.
In the fifth row we assume a simplified hadronization of the quark bilinears (light spin might be not conserved), and, up to an overall normalization,  restore  the $1/\!\sqrt{2}$ normalizations. We thus found that the `eigenchannel' of matrix $S(H;O)$ contains $X^+$ written as a diquarkonium state. 

Let us consider the G-parity transformation properties of the second term on the fourth row of in~\eqref{colsing}. Recall that $Gu=-\bar d, Gd=\bar u$ and $G\bar d=u,G\bar u =-d$. Then we get  that under G-parity
\begin{equation}
 \left[(c\,\sigma^2\,{\bar d})(u\,\sigma^2{\bm \sigma}\,{\bar c}) - (c\,\sigma^2{\bm \sigma}\,{\bar d})(u\,\sigma^2\, {\bar c})\right]\to  \left[(\bar c\,\sigma^2\,(-u))(\bar d\,\sigma^2{\bm \sigma}\,c) - (\bar c\,\sigma^2{\bm \sigma}\, (-u))(\bar d\,\sigma^2\, c)\right]
\end{equation}
but the latter term is equivalent to $(-1)$ times the term on the \lhs (because $(\sigma^2)^T=-\sigma^2$ and $(\sigma^2\bm \sigma)^T=\sigma^2\bm \sigma$ and a $-1$ is to be included each time spinors are exchanged). Therefore the $G$ parity is $G=-$, which means that the neutral state will have $C=+$. The same can be concluded just starting from $([cu]_0[\bar c\bar d]_1+[cu]_1[\bar c\bar d]_0)$. The $G$-parity analysis proceeds in the same way also for the two following terms. 

Similarly we can write (notice the $-$ sign in the spin exchanged term)
\begin{align}
Z^+&= (c^\alpha\, \sigma^2 \, u^\beta)\left[({\bar c}_\alpha\, \sigma^2 {\bm \sigma}\, {\bar d}_\beta)-(\bar c_\beta\, \sigma^2 {\bm \sigma} \, {\bar d}_\alpha)\right] -  (\sigma^2 \leftrightarrow \sigma^2{\bm \sigma})=\notag\\
&=-\left[(c\,\sigma^2\, {\bar c})(u\,\sigma^2{\bm \sigma}{\bar d})-(c\,\sigma^2{\bm \sigma} \,{\bar c})(u\,\sigma^2\,{\bar d})\right]  - i (c\,\sigma^2 {\bm \sigma}\,{\bar d})\bm\times  (u\,\sigma^2{\bm \sigma}\,{\bar c})\nonumber \\
&\sim \eta_c \,\rho^+ - \psi\, \pi^+ + i\,{\bar D}^{*0 } \bm\times  D^{* +}
\label{zzmesons}
\end{align}
to be compared with the difference between the first two rows in~\eqref{matrixsho},
and 
\begin{align}
Z^{\prime +}&=i(c^\alpha\, \sigma^2{\bm \sigma}\, u^\beta)\bm\times  \left[({\bar c}_\alpha\, \sigma^2{\bm \sigma}\, {\bar d}_\beta)-({\bar c}_\beta\, \sigma^2{\bm \sigma}\, {\bar d}_\alpha)\right]=\notag\\
&=-\left[(c\,\sigma^2\,{\bar c})(u\,\sigma^2\bm \sigma\,{\bar d})+(c\,\sigma^2{\bm \sigma}\, {\bar c})(u\,\sigma^2\,{\bar d})\right]+ \left[ (c\,\sigma^2 \,{\bar d})(u\,\sigma^2{\bm \sigma}\, {\bar c})+(c\,\sigma^2 {\bm \sigma}\,{\bar d})(u\,\sigma^2\, {\bar c})\right]\nonumber \\
&\sim \eta_c \,\rho^+ +\psi \,\pi^+ - \left({\bar D}^0 D^{* +} + D^+ {\bar D}^{ * 0}\right)
\label{zzpmesons}
\end{align}
from the sum of the first two rows in~\eqref{matrixsho}. Meson-meson amplitudes of  the kind described in \figurename{~\ref{carico}} contain diquarkonia in both the $J^{PC}=1^{++}$ and $J^{PC}=1^{+-}$ eigenchannels. 

As for the neutral component
\begin{align}
Z^0&\sim \eta_c \,\rho^0 - \psi\, \pi^0 + i\, {\bar D}^{*0 } \bm\times  D^{* 0} - i\,{\bar D}^{*-} \bm\times  D^{* +}
\label{zzmesons0}
\end{align}

The neutral components for the $X$ are particularly interesting.  We may have as well 
\begin{align}
X_u&=[cu]_0[\bar c\bar u]_1+[cu]_1[\bar c\bar u]_0\notag\\
X_d&=[cd]_0[\bar c\bar d]_1+[cd]_1[\bar c\bar d]_0
\end{align}
Here  $C=+$. Take the $X_d$ for example. It may be rewritten as
\begin{equation}
X_d\sim \left(D^{*-}D^{+} - D^{*+}D^{-}\right) + i \left(\psi\bm\times  \rho^0 - \psi\bm\times  \omega^0\right)
\label{icssd}
\end{equation}

whose charge conjugation $C=+$ is obtained by~\footnote{Actually, under charge conjugation $u\,\sigma^2\,\bar c\equiv u^T\,\sigma^2\,\bar c$ which is the non-relativistic limit of $\psi^T\,C\gamma_5\, \chi\to \psi^T_c\,C\gamma_5\, \chi_c=
\psi^T_c\,C\gamma_5C^{-1}C\, \chi_c=\psi^T_c\,\gamma_5^TC\, \chi_c\equiv (\psi^T_c\,\gamma_5^TC\, \chi_c)^T=\chi_c^T\, C^T\gamma_5\,\psi_c(-1)=\chi_c^T\, C\gamma_5\,\psi_c=c\,\sigma^2\bar u$. On the other hand proceeding like in Eq.~\eqref{verifychc} we would have $u\,\sigma^2\,\bar c\to \bar u\,\sigma^2\, c\equiv (\bar u^T\,\sigma^2\, c)^T =-c^T\,(\sigma^{2})^T\,\bar u=c^T\, \sigma^2\, \bar u$ corresponding to $ c\,\sigma^2\,\bar u$. When considering charge conjugation in cases like $u\,\sigma^2\bm \sigma\,\bar u$, the same techniques apply, but we also have to remember that $C\gamma_\mu C^{-1}=-\gamma_\mu^T$.}
\begin{align}
&(c\, \sigma^2\, \bar d)(d\, \sigma^2\bm \sigma\, \bar c)-(c\, \sigma^2\bm \sigma\, \bar d)(d\, \sigma^2\, \bar c)\to (\bar c\, \sigma^2\, d)(\bar d\, \sigma^2\bm \sigma\, c)-(\bar c\, \sigma^2\bm \sigma\, d)(\bar d\, \sigma^2\,  c)\notag\\
&\quad= -(d\, \sigma^2\, \bar c)(c\, \sigma^2\bm \sigma\, \bar d)+(d\, \sigma^2\bm \sigma\, \bar c)(c\, \sigma^2\,  \bar d)
\label{verifychc}
\end{align}
Given that $X_d$ contains both $I=0,1$, we include both $\omega$ and $\rho$ in Eq.~\ref{icssd}. The same for the $X_u$ state. Exchanging $d\to u$
\begin{equation}
X_u\sim \left(\bar D^{*0}D^{0} - D^{*0}\bar D^{0}\right)- i \left(\psi\bm\times  \rho^0 + \psi\bm\times  \omega^0\right)
\end{equation}

In other words, should the quark loop diagram in \figurename{~\ref{carico}}  develop a pole in one of the eigenchannels, the meson pairs coupled to the resonance would have {\it precisely  the right quantum numbers} to arise from the color Fierz rearranged diquark-antidiquark state. 

It is tempting to assume that all three channels corresponding to the antisymmetric diquark develop a pole, in which case molecules at meson-meson thresholds and  tetraquarks would coincide.

\subsection{Further remarks\label{conldiq}}
The issue of tetraquarks in the $1/N$ expansion has been commented thoroughly in a number of papers with different approaches, like putting all quarks in the antisymmetric representation, as done in~\cite{Cohen:2014via} --- the two representations coincide for $N=3$. What was found there  is that in such  large-$N$ limit, a sort of extreme version of the Corrigan-Ramond scheme~\cite{Corrigan:1979xf} (which would take only one of the quarks to transform according to the two-index antisymmetric representation and leaves the remaining ones in the fundamental representation),  one can produce tetraquarks in a completely natural way, because new color-entangled operators exist.

In~\cite{Weinberg:2013cfa} it is shown that the Coleman-Witten lore that no tetraquarks occur at large-$N$, because their residue disappears at $N\to\infty$, is flawed. In particular, to evaluate the tetraquark $n$-point correlators one has to work large but finite $N$ and only at the end send $N\to\infty$. In this case, they appear as very narrow states. Therefore tetraquarks can be made the way Weinberg suggests, but with the concerns discussed in~\cite{Lebed:2013aka,Cohen:2014tga}. All this  was of further stimulus to study if tetraquarks poles could instead be relevant at higher orders in the $1/N$ expansion, as discussed in the previous Sections. 
The fact that subleading topologies may be important, as pointed here, seems to emerge also to explain the large-$N$ behavior of the lightest scalars~\cite{Cohen:2014vta}.

Aside from the intrinsic interest of the topic, the present discussion should be considered as instrumental to the following section. The key points one should keep in mind are:
\begin{enumerate}
\item Neutral or charged tetraquarks could appear in non-planar diagrams only. For the latter ones the tetraquark contribution cannot disappear from the cut diagrams unless the order of the expansion is changed.
\item The mixing of tetraquarks with mesons and the decay constants are found to be suppressed by powers of \mbox{$f\sim 1/(N\sqrt{N})$} and \mbox{$G\sim1/N^2~(G\sim1/{(N\sqrt{N})})$} for the neutral (charged) tetraquarks. This means that production, mixing with ordinary mesons, and decays of tetraquarks are quite suppressed in the large $N$ limit, and tetraquarks might appear more like glueballs, largely decoupled from the meson sector;
\item Color meson-meson amplitudes correlating hidden-charm + meson to open charm states, especially in the large $N$ limit, display diquarkonia in their eigenchannels.
\end{enumerate}

In many respects what emerges from this discussion of tetraquarks in the $1/N$ expansion reinforces the thesis that a dynamical interplay between diquarkonia and thresholds should be at work. 

\section{An alternative picture of \texorpdfstring{$XZ$}{XZ} resonances\label{mol3}}
In this Section we propose some  arguments on an alternative explanation of the exotic hadron spectrum of $XZ$ resonances, following some early discussions appeared in~\cite{Papinutto:2013uya,Guerrieri:2014gfa,Esposito:2016itg}. Differently from other sections, this one contains a higher fraction of material not discussed before and it relies on results discussed  in \sectionsname{\ref{shallowbs}, \ref{diquark}, and \ref{tqln}}. The mechanism we are going to describe, has been considered in different setups, for ordinary hadron spectroscopy~\cite{Jaffe:2007id}, and more specifically for the \XYZ states~\cite{Braaten:2003he,Hanhart:2014ssa}.

In brief the conclusion of this section is the following: effective interactions, rearranging the quarks in meson-meson pairs into  specific compact tetraquark states 
whose mass is higher than the meson-meson threshold, are enhanced when the energy of the free meson-meson pair in the continuum spectrum matches the mass of the compact tetraquark (a discrete level in a binding potential).

Not all compact tetraquarks are experimentally visible because their mixing to mesons is extremely small --- as suggested by $1/N$ expansion arguments. In our view, only those tetraquarks that receive an enhancement from the mechanism described below will manifest themselves in the experimental data. This also explains why not all meson-meson thresholds have a resonance associated to it: simply there are no loosely bound meson-meson molecules in this picture. 
 
The attempt is that of responding to the most evident problems plaguing the compact tetraquark and molecular models with a simple scheme able to describe most of the best assessed experimental states. %

\subsection{The discrete spectrum of diquarkonia in the continuum of free meson-meson states}

Let us consider a system of four quarks produced  in a small enough  phase-space volume. The quantum state $\Psi$ describing this system can be represented as a superposition of  alternative ways of  neutralizing the color
\begin{equation}
\Psi=\alpha\, \Psi_d+\beta \,\Psi_{m}
\label{statohadr}
\end{equation}
where
\begin{equation}
\Psi_d=(\epsilon_{ijk}\,Q^j q^k)\,(\epsilon^{i m n}\,\bar Q_m\bar q_n^\prime)
\label{psid0}
\end{equation} 
is the diquark-antidiquark alternative and 
\begin{equation}
\Psi_{m}=(Q^i\bar q_i)\, (\bar Q_k q^{\prime k})~\,\,\,\text{or}\,\,\,~(Q^i\bar Q_i)\, (\bar q_k q^{\prime k})
\end{equation}
is the meson-meson alternative. There are no know superselection rules forbidding to consider such a coherent superposition. 

In the Coleman-Witten picture, the compact tetraquark is equivalent to two free mesons~\footnote{The two-point function of a tetraquark is equivalent to the product of two free meson two-point functions.}, therefore there should be no real distinction between $\Psi_m$ (two free mesons) and $\Psi_d$. But what we called a diquarkonium tetraquark in the previous section, only  appears  at subleading non-planar orders and its mixing with mesons is rather suppressed at large-$N$. This allows to keep $\Psi_m$ and $\Psi_d$ distinguished.

Let us assume that $|\alpha| \ll |\beta|$,  treating the diquarkonium state as a perturbation to the prominent free meson-meson state. 
We consider the persistent non-observation of several predicted diquarkonia to be a consequence  of this hypothesis.

The $\Psi_m$ state has energy $E_m$ in the {\it continuous} spectrum, with $E_m\geq M_1+M_2$, where $M_{1,2}$ are the masses of its {\it free} meson components. The $\Psi_d$ states, being bound states of colored objects, can have discrete values of the energy $E_d$, corresponding to the fundamental and excited diquarkonium states produced by strong interactions.  Consider values of $E_d$ such that $E_d\geq M_1+M_2$.

If the $\Psi_d$ and $\Psi_m$ states are in orthogonal spaces (see \ref{feshphen}), $m\to m$ transitions with amplitudes 
\begin{equation}
T_{m\, m}\sim \frac{|\langle \Psi_d|H_I\,\Psi_m\rangle|^2}{E_m-E_d+i\epsilon}
\label{boundfeshx}
\end{equation}
are allowed. Here we consider only that particular $E_d$, if any, which occurs close (from above) to the $M_1+M_2$ threshold. It corresponds to the mass of the would-be diquarkonium. The numerator, containing the effective interaction Hamiltonian $H_I$  in the continuous spectrum, is difficult to estimate, but expected to be small, on the basis of  large-$N$ arguments. Therefore the term $T_{m\, m}$ is not negligible only when $E_m\sim E_d$. 

Eq.~(\ref{boundfeshx}) should be confronted to Eq.~(\ref{bound}) noting that in the latter case the discrete level is found in the negative energy spectrum of some potential $V$ whereas in the case of~(\ref{boundfeshx}) the discrete level is {\it on the same side} of the continuous spectrum (therefore the $i\epsilon$ prescription is needed, as in the third term of Eq.~\eqref{low}), but {\it pertains to a different potential} with respect to that giving the onset of continuum levels\footnote{This strongly differentiates our metastable state from a true resonance. The latter is above threshold but belongs to the same potential as for the two-meson state}.

\begin{figure}[ht!]
 \begin{center}
   \includegraphics[width=9truecm]{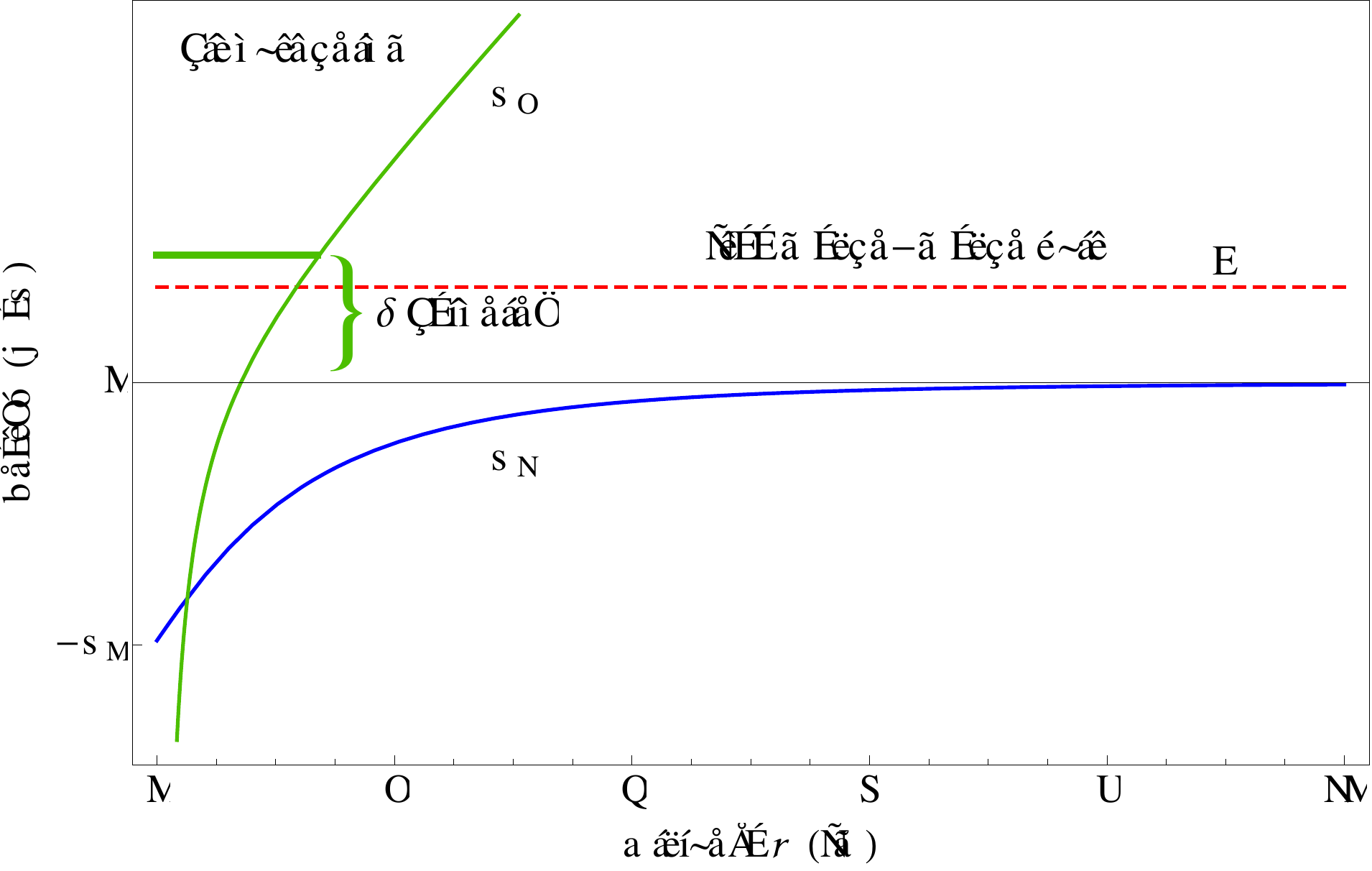}
\caption{The short range color potential has discrete levels: the lowest lying one is shown with a thicker line. Just for the sake of illustration we qualitatively sketch a quarkonium-like potential $V_2$ for the radial excitations, but do not assume any explicit functional form for it.
The detuning $\delta$ is the energy-gap measured from the onset of the continuous spectrum of the long-range potential to the mass of the diquarkonium state in $V_2$. 
\label{pot2x}} \end{center}
\end{figure}

The total rate for all reactions initiated by $\Psi_m$ in a volume $V$ ({\it i.e.} with a density of initial meson-meson pairs produced in hadronization $=\rho=1/V$) is given by unitarity
\bea
\Gamma &=& -16\pi^3 \, \rho \, \mathrm{Im} (T_{m\, m})\simeq 16\pi^3\, \rho\,\frac{\epsilon}{(E_m-M_{4q})^2+\epsilon^2}|\langle \Psi_d|H_I\,\Psi_m\rangle|^2 \notag\\
&= & 16\pi^4\,\rho\, \delta\left(\frac{p^2}{2M_1}+\frac{p^2}{2M_2}-\delta\right)|\langle \Psi_d|H_I\,\Psi_m\rangle|^2
\label{totalwf}
\eea
where the detuning parameter $\delta=M_{4q}-M_1-M_2$ is represented in Fig.~\ref{pot2x}  and it is assumed $|\bm p|\ll M_{1,2}$. Hadronization randomly generates relative momenta $\bm p$ between the components of  the meson-meson system $\Psi_m$.
The expected width $\overline{ \Gamma}$ is the average of $\Gamma$ over all $\bm p$'s within the ball $|\bm p|\lesssim \bar p$ --- where $\bar p$ is the support of the diquarkonium tetraquark wavefunction in momentum space~\footnote{In the type II model the diquarks are considered to be quite distant in space (see Section~\ref{matricimassa}). A large separation in position corresponds to a small support in momentum space.}.  %
We consider $\bar p$ to be roughly less than the average radial excitation gap  between diquarkonia levels.%

Assuming $M_1\simeq M_2\simeq M$ we have therefore~\footnote{Similarly one finds that $\bar \sigma$ does not depend on $\delta$. This has the interesting consequence that we expect the $Z$'s to be produced in prompt hadron collisions.}
\be
\overline{ \Gamma} \simeq A \sqrt{\delta}
\label{sqrtl}
\ee
provided that $\sqrt{M\delta}< \bar p$. Here 
\be
A=\frac{48\pi^4}{\bar p^3}\,\rho\, |\langle \Psi_d|H_I\,\Psi_m\rangle|^2 M^{3/2}\sim 48\pi^4 |\langle \Psi_d|H_I\,\Psi_m\rangle|^2 M^{3/2}
\ee
because the density of states  scales  like $\bar p^3$. 
Therefore the expected width scales with the square root of the detuning $\delta$ between the mass of the diquarkonium state (which can be estimated theoretically) and the nearest meson-meson threshold having the same quantum numbers. 
If $T_i^j$ is a diquarkonium tetraquark with an antidiquark of color $i$ and a diquark of color $j$, a simple ansatz for $H_I$ is
\be
H_I= B\,(T_i^j \,\epsilon_{j\ell m}\epsilon^{ikn}\,M_k^\ell M_n^m)
\label{HI}
\ee
{\it i.e.} an operator crossing the quarks from $T$ to the meson-meson $MM$ pair. The coupling $B$ has dimensions of MeV and assume that it scales as $B\sim \epsilon\,  M$, $\epsilon$ being a  small numerical coefficient. Meson fields and tetraquark fields might have spin. To avoid the momentum dependent transformation matrices associated with each particle spin, which arise under Lorentz transformations, one generally sums over spins to form the Lorentz invariant amplitude  $E_{M_1}E_{M_2}E_{T}\sum_{\rm spins} |\langle \Psi_d|H_I\,\Psi_m\rangle|^2\equiv R$.
Therefore we have $A\sim 48\pi^4 \epsilon^2 (M^2/M^3)\, R\, M^{3/2}$, {\it i.e.} the mass dependence in $A$ gets soften to $A\sim M^{1/2}$.  
This is computed in the meson-meson frame for sufficiently small recoils allowing to exchange energies with masses (see discussion in Section~3).

\begin{figure}[htb!]
 \centering
   \includegraphics[width=.45\textwidth]{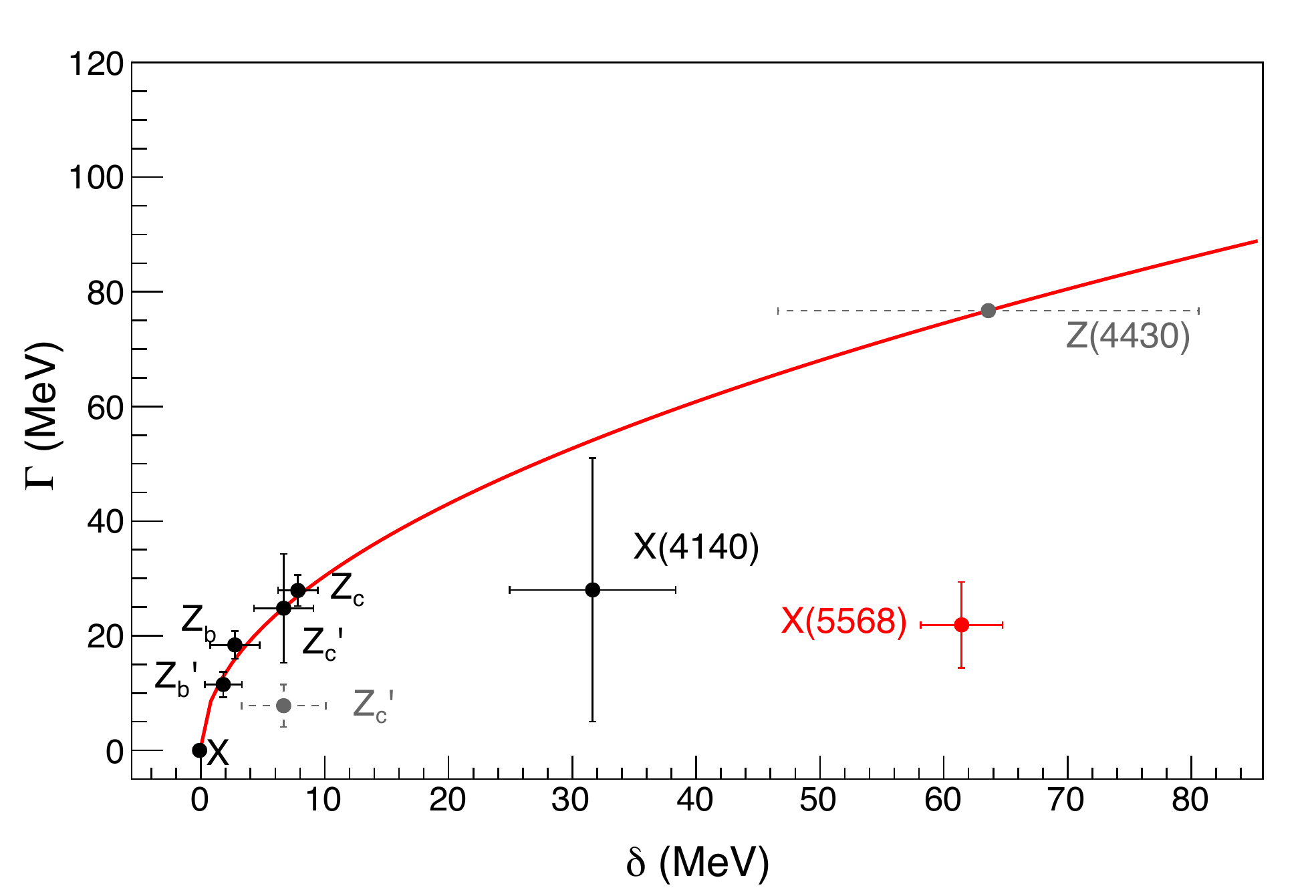} 
   \includegraphics[width=.45\textwidth]{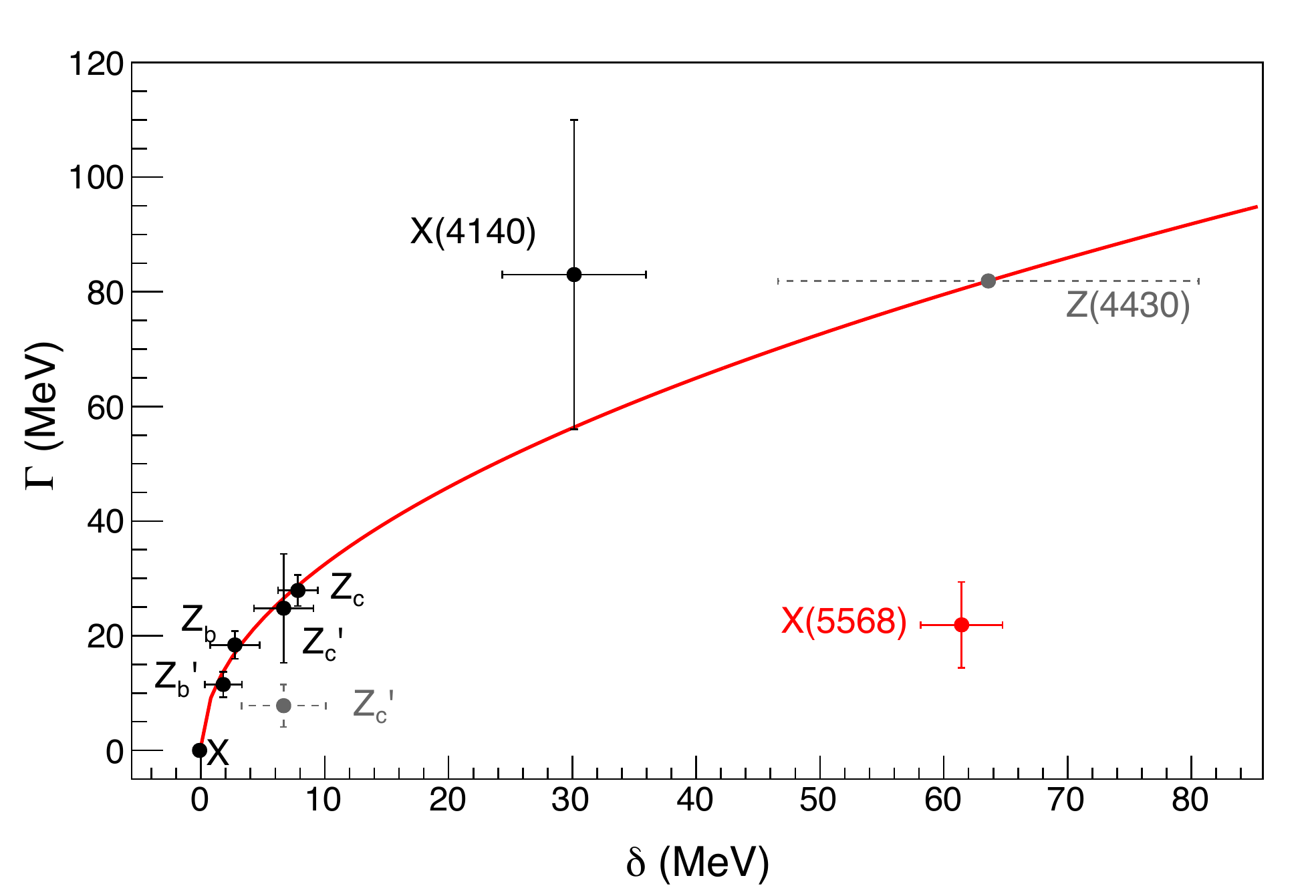}
\caption{Width of the observed exotic mesons as a function of their detuning $\delta$ to the closest, from below, two-meson thresholds. The solid curve is the fitted function $A \sqrt{\delta}$, with $A = (10.3 \pm 1.3)\mev^{1/2}$ with $\chi^2/\text{DOF} = 1.2 / 5$ (without the $X(4140)$, the quality of the fit would be $\chi^2/\text{DOF} = 0.2 / 4$). The two points associated with the $Z_c^\prime(4020)$ correspond to the two measurements of its width obtained from $\bar D^{*0}D^{*+}$ (solid black) and the $h_c\,\pi$ (dashed gray) channels and which differ at $2\sigma$ level from each other. In the fit we considered the $\Gamma(Z_c^\prime) \simeq 25\mev$ measured in the $\bar D^{*0}D^{*+}$ channel, which is consistent with the width of the neutral $Z_c^{\prime 0}$ partner.  We also show the prediction for the $Z(4430)$ width, which underestimates the total width as expected. The red point corresponds to the $X(5568)$ state whose observation has been claimed by \Dzero~\cite{D0:2016mwd}. On the left panel we use the $X(4140)$ width as measured by \cms~\cite{Chatrchyan:2013dma}. On the right we use the most recent \lhcb value~\cite{Aaij:2016nsc,Aaij:2016iza} --- see \sectionsname{\ref{specialcase4140}} and~\ref{sec:ccss}.} \label{fit_feshx}
\end{figure}

A fit to data can be done to estimate $A$ and  to verify {\it if} and to which extent it is universal among the various $X,Z$ resonances --- see Fig.~\ref{fit_feshx}.

The derivation of~(\ref{sqrtl}) is based on the assumption that
\begin{enumerate}
\item $\delta>0$: the mass of the diquarkonium must be larger than the relative meson-meson threshold --- $\delta<0$ corresponds to a zero rate
\item $\sqrt{M\delta}<\bar p$: the contribution of~(\ref{boundfeshx}) to $\overline{ \Gamma}$ comes only from small enough detunings 
\item  The diquarkonium state, when Fierz-transformed in terms of meson singlets, does not contain the meson components in the meson-meson pair~(\ref{boundfeshx})) 
\end{enumerate}

With the discovery of charged $Z_c$s and $Z_b$s resonances, the connection to open charm/beauty meson-meson  thresholds has become manifest. The $X(3872)$ mass is fine tuned with the $\bar D^0D^{*0}$ threshold.  Also  the $Z_b(10610)$, $Z_b(10650)$,  
$Z_c(3900)$ and $Z_c^\prime(4025)$ are close to the $BB^*$, $B^*B^*$, $DD^*$ and $D^*D^*$  thresholds respectively, but at energy distances of $+2.7$, $+1.8$, $+7.8$, $+6.7\mev$; this positive sign trend does not appear to  be an accidental feature. It  means that, lying {\it above} threshold values, $Z$ resonances cannot be deuteron-like states! 

As illustrated in Fig.~\ref{fit_feshx}, the $\sqrt{\delta}$ fit works rather well for $X$, $Z_b$, $Z_b^\prime$, $Z_c$, $Z_c^\prime$, and $X(4140)$. For all of them the $\delta>0$ condition is met (in the case of $Z_b$, $Z_b^\prime$, $Z_c$, and $Z_c^\prime$, the detunings are therefore $\delta= 2.7,1.8,7.8,6.7\mev$ respectively). The estimate of $A = (10.3 \pm 1.3)\mev^{1/2}$  is compatible, within $2\sigma$, with the $A\sim M^{1/2}$ behavior\footnote{Meaning that $(10.3+2.6)/(10.3-2.6)\simeq \sqrt{M(Z_b)/M(Z_c)}$.}. More will be said on condition $(3)$ in the following. 

This approach badly fails to accommodate the $X(5568)$ which was considered to falsify this approach, until its experimental disproof. 

What is called an $X,Z$ resonance could therefore result from the non-vanishing amplitude $\langle \Psi_d|H_I\,\Psi_m\rangle$ inducing an effective interaction in the continuous spectrum of meson-meson states due to a diquarkonium  $\Psi_d$ located just above threshold.  The free meson-meson state  gets temporarily locked in the discrete level within $V_2$ thanks to the interaction $H_I$ --- the diquarkonium state is produced.  If $H_I$ were zero, the only effect in the continuous spectrum of $V_1$ could come from a shallow bound state, if any, at $-B$. However anything appreciable in the cross section would need a gap $E+B\approx 0$, as reported in \sectionname{\ref{shallowbs}}, whereas, as we see from phenomenology, $E$ can go rather far from threshold in the continuous spectrum (the fact that $E+B$ can be as large as $E+B\approx 10\mev$, would contradict most of the hypotheses we have worked with in \sectionname{\ref{shallowbs}}). In absence of a level at $-B$, and assuming $H_I=0$,  we recall that $S$-wave (re)scattering of meson-meson pairs alone could not generate narrow width resonances.
 
The only fact that diquarkonium levels exist, provides an effective interaction in the meson-meson channel which allows the temporary `hybridization' of the meson-meson system into a diquarkonium. In other words, all diquarkonium tetraquarks predicted should in principle be observed, but only few of them get produced: the conditions described above should be met. Under these conditions an enhancement of the tetraquark-meson mixing (otherwise extremely small) becomes possible.

\subsection{\texorpdfstring{$\Psi_m$ and $\Psi_d$}{Psi\_m and Psi\_d}  states}
We used $\Psi_d$ for  diquarkonium states and $\Psi_{m}$ for free meson-meson states, with negligible residual strong interactions. 

Let us consider  the particular case of the $Z_c(3900)$ resonance. From $(I^G)~ J^{PC}=(1^+)~1^{+-}$ quantum numbers the $\Psi_d$ state is
\begin{equation}
\Psi_d=[cu]_{S=0}[\bar c\bar d]_{S=1}-[cu]_{S=1}[\bar c\bar d]_{S=0}
\end{equation}
and, from~\eqref{psid0}, this notation is a shorthand for
\begin{equation}
\Psi_d= \epsilon_{\alpha\beta\gamma} (c^\alpha \sigma^2 u^\beta)\,\epsilon^{\alpha^\prime\beta^\prime\gamma}({\bar c}_{\alpha^\prime} \sigma^2 {\bm \sigma}{\bar d}_{\beta^\prime})  - (\sigma^2 \leftrightarrow \sigma^2{\bm \sigma})
\end{equation}
Here $\epsilon_{\alpha\beta\gamma}(c^\alpha \sigma^2 u^\beta)$ describes a spin zero diquark whereas $\epsilon^{\alpha^\prime\beta^\prime\gamma}({\bar c}_{\alpha^\prime} \sigma^2 {\bm \sigma}{\bar d}_{\beta^\prime})$ has spin 1. Upon Fierz rearrangement we found 
\begin{equation}
\Psi_d\sim \eta_c \,\rho^+ - \psi\, \pi^+ + i\, {\bar D}^{*0 } \bm\times  D^{* +}
\label{diquarkonio0}
\end{equation}
The closer ({\it i.e.} fulfilling $\sqrt{M\delta}<\bar p$) meson-meson threshold  below the measured $Z_c(3900)$ mass is  $ \bar D^0D^{*+}$, therefore the relevant threshold is
\be
\Psi_m\sim \bar D^0D^{*+}
\label{psiemme}
\ee
This is orthogonal to $\Psi_d$: there is no $\bar D^0D^{*+}$ component in the  $\Psi_d$ Fierz-transformed state.  The $\Psi_m$ might be taken as a superposition of different meson-meson states and constructed in such a way to be orthogonal to $\Psi_d$.  However we computed the total rate for all reactions initiated by a selected meson-meson state with fixed quantum numbers and flavor, and this is why we stick to $\Psi_m$ as in~(\ref{psiemme}).  

In the $Z^\prime_c$ case we found in~\eqref{zzpmesons}
\begin{equation}
\Psi_d\sim \eta_c \,\rho^+ +\psi \,\pi^+ - \left({\bar D}^0 D^{* +} + D^+ {\bar D}^{ * 0}\right)
\end{equation}
whereas  the relevant threshold must be
\be
\Psi_m\sim {\bar D}^{0 *} D^{* +}
\ee
The case of the $Z_b$s resonances is completely analogous.

In the case of the $X(3872)$ we have  
\begin{equation}
X_d\, =[cd]_0[\bar c\bar d]_1+[cd]_1[\bar c\bar d]_0
\label{icsd1}
\end{equation}
where the subscript $d$ stands here  for $d$-quark. 
As it was shown in \sectionname{\ref{diqn}}, this latter state  can be rewritten as 
\begin{equation}
\Psi_d\sim \left(D^{*-}D^{+} - D^{*+}D^{-}\right) + i \left(\psi\bm\times  \rho^0 - \psi\bm\times  \omega^0\right)
\label{icsd2}
\end{equation}
whereas  
\be
\Psi_m\sim \bar D^0 D^{*0}
\ee
so orthogonality between $\Psi_m$ and $\Psi_d$ is at work. Any $\Psi_m\sim D^+ D^{*-}$ would be slightly heavier than the $M(\Psi_d)$ mass and is not to be considered here.

On the other hand there is also a  $X_u$ component which would Fierz to  neutral open charm states. We might observe that since the mechanism we consider for the formation of the observed resonances relies on the orthogonality between $\Psi_d$ and the open charm meson pair $D^0\bar D^{0*}$, only the $X_d$ component gets involved. This would rule out the $X_u$ component therefore inducing isospin violations in $X(3872)$ decays, with no need of any hyperfine neutral doublet close to 3872\mev. This point will be  discussed again  in \sectionname{\ref{misteri}}.  

From this discussion we see that
\begin{enumerate}
\item  We do not expect a narrow resonance in  proximity of every single meson-meson threshold but only free meson pairs unless a $\Psi_d$, with the right quantum numbers, occurs right above threshold
\item  We do not expect a resonance in correspondence of every diquarkonium state: pure diquarkonia might have small probabilities to be formed on their own given $|\alpha|\ll |\beta|$ in~(\ref{statohadr}) 
\end{enumerate}

\subsection{The \texorpdfstring{$X_b$}{Xb} in the beauty sector,  isospin violations and the non-observation of \texorpdfstring{$X^+$}{X+}\label{misteri}}
A pressing question is about whether we should expect a twin particle of the $X(3872)$ in the beauty sector. Also in this case, the determination of the discrete diquarkonium level should be precise enough to allow a prediction.   According to~\eqref{mat1pm} and~\eqref{mat2pp}, the twin of the $X$ in the beauty sector should be almost degenerate with the lighter of the two $Z_b$s and from~\eqref{spectpos1} 
\begin{equation}
M[X([bq][\bar b \bar q])]\simeq M[Z_b(10610)]= (10607\pm 2)\mev 
\end{equation}
If the relevant threshold is $\bar B^0B^{*0}$, in perfect analogy with the charm sector, we have
\begin{equation}
M(B^0)+M(B^{*0})\simeq (10604.4\pm 0.3)\mev
\label{valnelb}
\end{equation}
This alternatively tells that 
\begin{enumerate}
\item There is a very narrow partner of the $X(3872)$ in the beauty sector if indeed the diquarkonium level is higher by few{\mev}s as from the above rough estimate
\item There is not such a resonance because the diquarkonium mass would be below the value of 10604\mev in~\eqref{valnelb}
\end{enumerate}
We might prefer the second option because it is observed in the charm sector that the experimental mass difference is 
\begin{equation}
M(Z_c)-M(X)\simeq (32\pm 3)\mev
\end{equation}
whereas the diquarkonium model would predict an approximate degeneracy. 
This leads to infer that the twin diquarkonium level of the $X_b$ falls below the relevant threshold, and no such resonance will be discovered. 
On the other hand, in going from the mass value of the charm to that of the beauty, we could observe a sensibly smaller lift of the $Z_b -X_b$ degeneracy and 
consequently a narrow $X_b$ particle could be eventually found in $pp$ collisions.

As for the time of this writing, there are no experimental clues of an $X_b$ in the beauty sector located close to 10604\mev (see \sectionname{\ref{sec:Zb}}). 

Besides the absence of a relative of the $X(3872)$ in the beauty sector, another serious problem with a na\"ive diquarkonium interpretation of this resonance is the absence of a almost degenerate doublet of neutral states $X_u=[cu][\bar c\bar u]$ and $X_d=[cd][\bar c\bar d]$. 
The need of it comes from the observation of strong isospin violation in the $X$ decays~\cite{pdg}
\begin{equation}
\frac{\BR(X\to \jpsi \, \omega)}{\BR(X\to \jpsi \, \rho)}\simeq 0.8\pm 0.3
\end{equation}
If the mass matrix were of the kind
\begin{equation}
M=\begin{pmatrix} 2m_{[cq]}&0\\0&2m_{[cq]}\end{pmatrix}+\epsilon \, \begin{pmatrix}1&1\\1&1\end{pmatrix}
\end{equation}
it would have two eigenstates
\begin{align}
\frac{|1\rangle-|2\rangle}{\sqrt{2}} &\quad \text{with} \quad \lambda=2 m_{[cq]}\notag\\
\frac{|1\rangle+|2\rangle}{\sqrt{2}} &\quad \text{with} \quad \lambda=2m_{[cq]}+2\epsilon
\end{align}
corresponding to two isospin eigenstates $I=0,1$ with approximately the same mass, $\epsilon$ being a small annihilation term $u\bar u\leftrightarrow d\bar d$. Here $|1\rangle$ and $|2\rangle$ are the basis vectors
\begin{equation}
|1\rangle=\bpm 1\\0 \epm\sim X_u \quad \quad |2\rangle=\bpm 0\\1 \epm\sim X_d
\end{equation}

If $\epsilon$ is negligible with respect to $m_{[cq]}$, \ie annihilations are negligible, we can set it to zero to first approximation and this breaks isospin giving $X_u$ and an $X_d$, degenerate in mass, each maximally isospin mixed. Each of them, $X_u$ and $X_d$, can therefore decay into $\jpsi \, \omega$ and $\jpsi \, \rho$ with almost the same rate. However a hyperfine splitting between them is expected and it even seemed to be in data for some time~\cite{Maiani:2007vr} --- those results were never confirmed in successive data analyses. 

On the other hand, as discussed in Eqs.~\eqref{icsd1}, \ref{icsd2}, only the $X_d$  is subject to the hybridization phenomenon described. The state $X_u$ would instead have a projection on the $\bar D^0 D^{*0}$. Ruling out $X_u$ in favor of $X_d$ at the same time breaks isospin, as required by data, and solves the problem of the absence of a hyperfine neutral doublet at about 3872\mev.

Finally, also the problem of the non-observation of $X^+$ might be addressed within the same scheme.
The $V_2$ potential gives almost degenerate $X^+,X^0$ would-be compact tetraquark levels. On the other hand the $D^+ \bar D^{*0},\bar D^0 D^{*+}$ thresholds  happen to be slightly heavier than those. This might be the reason why we do not observe $X^+$ resonances. Most remarkably the $\Psi_d$ state for the $X^+$ is not  orthogonal to the relevant (neutral) threshold. This is summarized in \tablename{\ref{tableres}}. 

The $\jpsi \, \rho $ and $\jpsi \, \omega $ thresholds are sensibly above $M(\Psi_d)$, the mass of the diquarkonium state. This might be a reason not to include  them in this discussion. However, given the width of the $\rho$ meson especially, they could have. Yet  $\jpsi \, \rho $ and  $\jpsi \, \omega $ are not included in the total width calculation in~(\ref{totalwf}), because of~(\ref{icsd2}): the enhancement in~(\ref{totalwf}) is obtained for $\Psi_d$ and $\Psi_m$ orthogonal states only.   

Even though $X$, in this picture, is prominently produced by $DD^*$, it can {\it also} decay in $\jpsi \, \rho (\omega)$, with a phase factor penalty, as it is clear from observations on $X^0$.   

The case of  $X(3872)$, among the one examined here, is particularly extreme because of the very narrow splittings involved. %

The picture described here of a resonant hybridization with a tuned diquarkonium level in the  meson-meson channel  has been first discussed in~\cite{Esposito:2016itg}.
Some earlier  considerations on the same topic can be found in~\cite{Braaten:2004ai} and in~\cite{Papinutto:2013uya,Guerrieri:2014gfa}, although formulated differently. 

\subsection{The special case of the \texorpdfstring{$X(4140)$}{X(4140)}\label{specialcase4140}}
As discussed in \sectionname{\ref{matricimassa}}, diquarkonium spectra can be computed assuming an interaction Hamiltonian depending on the spin degrees of freedom of quarks within diquarks~\cite{Maiani:2014aja}.  From Eqs.~(\ref{spectpos1}, \ref{spectpos2})
\begin{equation}
M_{Z^\prime}-M_{Z}=2\kappa
\label{coupldq}
\end{equation}
and
\begin{equation}
M_{Z^\prime}+M_{Z}=4m_{[Qq]}
\label{massadq}
\end{equation}
so that
\begin{align}
\kappa_{cq}&\simeq 67\mev\notag\\
m_{[cq]}&\simeq1980\mev
\end{align}
We argue that  
\begin{equation}
m_{[cs]}\simeq m_{[cq]}+(m_s-m_q)\simeq 2100\mev
\label{estmcs}
\end{equation}
with  $(m_s-m_q)\simeq 120\mev$ from the SU(3) decuplet (this mass difference, when taken from $D_s$ mesons, would be lighter by $\sim 20$ MeV).  It is also worth noticing that the difference in mass between $X(4140)$ and $X(3872)$ is consistent with the $\psi-\rho$ mass difference. We use  the na\"ive scaling law to determine
\begin{equation}
\kappa_{cs}=\kappa_{cq}\, \frac{m_q}{m_s}\simeq  (45\pm 3)\mev
\label{scalingac}
\end{equation}
where the quark mass are the constituent ones as taken from Table I in~\cite{Maiani:2004vq}. 
 Therefore, from the spectrum in \figurename{~\ref{spettrot2}} we have (see 
also~\cite{Lebed:2016yvr})
\begin{equation}
M(X(4140;1^{++}))=2\times 2100 - 45= 4155\mev 
\label{massa4140}
\end{equation}
(or as light as 4410~MeV with a smaller mass difference) which, considering the approximations made, results in reasonable agreement ($4130\pm 20$~MeV) with the experimentally observed mass value. This estimate improves on what was found before the discovery of the $Z_c$s system~\cite{Drenska:2009cd}.

The $X(4140)$, also reported in \figurename{~\ref{fit_feshx}}, is a $1^{++}$ state, like the $X(3872)$, but in the diquarkonium picture it has quark content~\cite{Lebed:2016yvr,bloisjs}
\begin{equation}
X(4140)= [cs]_0[\bar c\bar s]_1+[cs]_1[\bar c\bar s]_0
\end{equation}
which is indicated by  the experimental observation of the decay mode 
\begin{equation}
X(4140)\to \jpsi \, \phi
\end{equation}
fixing $C=+$. 
 As in Eq.~\eqref{icssd} we have
\begin{equation}
\Psi_d\sim \left(D^{*-}_s D^{+}_s - D^{*+}_sD^{-}_s\right) - i \left(\psi\bm\times  \phi\right)
\label{icssdrep}
\end{equation}
On the other hand, the closest threshold from below with quark content $cs\,\bar c \bar s$ and $J^{PC}=1^{++}$ would be $J/ \psi \, \phi$, which is located just 20\mev below 4140\mev (about~40\mev higher than $D^{*-}_s D^{+}_s$ ). %
In both cases, $\Psi_m=D^{*-}_s D^{+}_s, \psi\bm\times  \phi $, would have a non-zero overlap with $\Psi_d$.
This  forbids the mechanism we presented here and it could be that this resonance needs some other explanation.  
Indeed there are other aspects making of the $X(4140)$ a special case.

Recent results by the \lhcb collaboration confirm it, but with a total width of about $(83\pm 21)\mev$, much larger 
than what  reported before.  Quantum numbers are confirmed to be $1^{++}$.  Most importantly, 
three more states are observed (with slightly less significance)
\begin{equation}
X(4274;1^{++}),\quad X(4500;0^{++}),\quad X(4700;0^{++})
\end{equation}
and similar decay modes.  A second $1^{++}$ state (heavier by $130\mev$) is {\it not} expected in the diquark-antidiquark picture as can be seen from Fig.~\ref{spettrot2}.
 
 Heavier states might be easily considered to be radial excitations\footnote{Although all these states appear to decay into $\jpsi$ rather that $\psiprime$ (in contrast to the $Z(4430)$, which decays prominently into $\psiprime\,\pi$), and the mass gap looks too small with respect to the $M(Z(4430)) - M(Z_c(3900))$ one. 
}: $X(4500;0^{++}), X(4700;0^{++})$, could be the radial excitations of the two scalar states $0^{++}$ and $0^{++\prime}$ expected by the diquarkonium model, see Eqs.~\eqref{duezz} --- the lowest of them cannot kinematically decay into $\jpsi \, \phi$~\cite{bloisjs}.%
 
But the second $1^{++}$ state remains a problem.
We might suspect that the second $1^{++}$ state, $X(4274;1^{++})$, has not been identified correctly: its quantum numbers are not $1^{++}$ but either $0^{++}$ or $2^{++}$ --- see Fig.~\ref{spettrot2}. Another possibility is that two almost unresolved peaks, one  with $0^{++}$ and the other  with $2^{++}$ quantum numbers, are observed at $\sim 4270\mev$. Angular information would be  washed out for the sum of the two states to make it appear as one with averaged spin-parity, see~\cite{bloisjs}.

Another, even stronger hypothesis, is that  (anti)quark-(anti)quark systems might, less likely, be produced also in the color symmetric configuration $\bm 6\,(\bar{\bm 6})$. We would not talk of diquarks in that case. Overall $\bm 6\times \bar {\bm 6}$ contains the color singlet and  we would have $\Psi_s$ stationary states, in some other potential $V_3$ in Fig.~\ref{pot2x}. The spectrum of  $\Psi_s$ states should be different from that of $\Psi_d$ ones, as computed in~(\ref{massa4140}) --  with an unknown mass gap.  Yet we could assume 
\begin{equation}
X(4140)=\{cs\}_0\{\bar c\bar s\}_1+\{cs\}_1\{\bar c\bar s\}_0
\label{cscs}
\end{equation}
where curly brackets denote color $\bm 6$ quark-quark pairs. The $\Psi_s$ state associated would be
\be
\Psi_s\sim \psi\bm\times \phi+D_s^{*+}\bm\times D_s^{*-}
\ee
to be compared with the meson-meson threshold 
\be
\Psi_m=D^{*-}_s D^{+}_s
\ee
This would have the interesting consequence to make the mechanism here described possible, and force to use the detuning $\delta$ to the $D^{*-}_s D^{+}_s$ threshold. This predicts $\bar \Gamma\sim 10\times \sqrt{60}\sim 77$~MeV, to be compared with the \lhcb value $\Gamma_{\rm exp}=(83\pm 21)\mev$.  On the other hand this would not solve the problem of having {\it two} observed $1^{++}$ states --- a spin misidintification is  still the favored option.   In the diquarkonium tetraquark picture~\cite{cinesi6}, this gets solved doubling the spectrum with  $\bar{\bm 3}$ diquarks and $\bm 6$ quark-quark pairs.

\subsection{Summary}
\label{sec:feshbachsummary}
In \tablename{\ref{tableres}}, for each of the known multiquark resonances, we report whether or not they fulfill the criteria 
to be produced with the mechanism described in this section.
\begin{table}[htb!] 
\centering
\begin{center}
\begin{tabular}{c|c|c|c|c|c}\hline\hline
Resonance & $\Psi_m$ & $\Psi_d$ & $M(\Psi_d)\geq M(\Psi_m)$ & $\Psi_d\perp\Psi_m$ & Exp.  \\ 
\hline \hline
\multirow{2}{*}{$X^0_u$} & \multirow{2}{*}{$\bar D^0 D^{*0}$} & $(\bar D^{*0}D^{0} - D^{*0}\bar D^{0})$& \multirow{2}{*}{yes} & \multirow{2}{*}{no} & \multirow{2}{*}{\XSolidBrush}   \\
& & $-i\, \psi\bm\times (\rho^0+\omega^0) $ & &\\ 
\hline
\multirow{2}{*}{$X^0_d$} & \multirow{2}{*}{$\bar D^0 D^{*0}$} & $ (D^{*-}D^{+} - D^{*+}D^{-})$ & \multirow{2}{*}{yes} & \multirow{2}{*}{yes}  & \multirow{2}{*}{\Checkmark} \\ 
& & $+i\psi\bm\times (\rho^0-\omega^0)$ & & \\
\hline
$X^+$ & $ D^+ {\bar D}^{* 0} \vee {\bar D}^0 D^{* +} $ & $(D^+ {\bar D}^{* 0}-{\bar D}^0 D^{* +})- i\,\psi \bm\times  \rho^+$& no & no& \XSolidBrush   \\ \hline
\multirow{2}{*}{$Z_c^0$} & \multirow{2}{*}{$\bar D^0 D^{*0}$} & $ \eta_c \,\rho^0 - \psi\pi^0 $ & \multirow{2}{*}{yes} & \multirow{2}{*}{yes} & \multirow{2}{*}{\Checkmark}   \\  
 &  & $+i \,{\bar D}^{*0 } \bm\times  D^{* 0} - i\,{\bar D}^{*- } \bm\times  D^{* +} $ &  &  &    \\  
\hline
$Z_c^+$ & $\bar D^0 D^{*+}$ & $ \eta_c \,\rho^+ - \psi\, \pi^+ + i\, {\bar D}^{*0 } \bm\times  D^{* +}$ & yes & yes & \Checkmark  \\  
\hline
\multirow{2}{*}{$Z_c^{\prime 0}$} & \multirow{2}{*}{$\bar D^{*0} D^{*0}$} & $\eta_c \,\rho^0 +\psi \pi^0+(\bar D^0D^{*0}+D^0\bar D^{*0})$ & \multirow{2}{*}{yes} & \multirow{2}{*}{yes} & \multirow{2}{*}{\Checkmark}   \\ 
& & $-(D^-D^{*+}+D^+D^{*-})$ & & \\
\hline
$Z_c^{\prime +}$ & $\bar D^{*0} D^{*+}$ & $(\eta_c \,\rho^+ +\psi \,\pi^+)- ({\bar D}^0 D^{* +} + D^+ {\bar D}^{ * 0})$ & yes & yes& \Checkmark   \\
\hline 
\multirow{2}{*}{$(X_b^0)_u$} & \multirow{2}{*}{$\bar B^0 B^{*0}$} & $(\bar B^{*0}B^{0} - B^{*0}\bar B^{0})$ & \multirow{2}{*}{no} & \multirow{2}{*}{no} & \multirow{2}{*}{\XSolidBrush} \\ 
& & $-i\, \Upsilon\bm\times (\rho^0+\omega^0)$ & & \\
\hline
\multirow{2}{*}{$(X_b^0)_d$} & \multirow{2}{*}{$\bar B^0 B^{*0}$} & $(B^{*-}B^{+} - B^{*+}B^{-})$ & \multirow{2}{*}{no} & \multirow{2}{*}{yes} & \multirow{2}{*}{\XSolidBrush} \\ 
& & $+i\, \Upsilon\bm\times (\rho^0-\omega^0)$ & & \\
\hline
$X_b^+$ & $B^+ {\bar B}^{* 0} \vee {\bar B}^0 B^{* +} $ & $(B^+ {\bar B}^{* 0}-{\bar B}^0 B^{* +})- i\,\Upsilon \bm\times  \rho^+$ & no & no & \XSolidBrush \\ 
\hline
$Z_b^{(\prime)0},\, Z_b^{(\prime)+}$ & like $Z_c$s & like $Z_c$s & yes & yes &\Checkmark \\  
\hline\hline
 \end{tabular}
 \end{center}
\caption{We resume here the mechanism described. In the last column to the right the final evaluation: is the resonance expected in data (checkmark) or not (cross)? Some relevant features: $i)$ there is only one neutral $X$ which decays violating isospin --- this is not the case for neutral $Z_c$s or $Z_b$s $ii)$ there are no neutral $X$s in the beauty sector --- one of the two states is subject to the approximate evaluation of the diquarkonium mass though $iii)$ there is no charged $X$ in the charm or in the beauty sector. \\
}\label{tableres}
\end{table}

The starting point of our discussion is that the \emph{all the observed $XZ$ states appear to be above threshold}. We are not considering this as an accidental pattern. Being the detunings fairly small, different parametrizations of the lineshapes might move the states below threshold~\cite{Albaladejo:2015lob,Guo:2016bjq,Kang:2016ezb}. Forthcoming data will be able to resolve the proper lineshapes and give a definitive answer to this question. For the time being, we decide to trust the results currently provided by the experiments, and take $\delta>0$.

If on the other hand the $X^+(\sim 3872)$ will  eventually be discovered, suppose with a larger width than $X^0$ for example, or $X_b$ in the beauty, or any other among  the states predicted by the diquarkonium model, that would challenge or simply invalidate the given description in favor of the  diquarkonium tetraquark model, or, depending on the results, in terms of  other approaches. 

Secondly, it is extremely plausible that compact tetraquarks are indeed part of the QCD spectrum, as understood also on the basis of large-$N$ QCD considerations. Therefore, one might expect to observe them even in absence of close by meson-meson thresholds. Our assumption is that, unless enhanced by the mechanism described, the probability of producing a diquarkonium tetraquark on his own is very small. Exception made for large-$N$ QCD, we do not have a more rigorous explanation, from first principles, of this point, but we rather consider it mostly motivated, \emph{a posteriori}, by the persisting experimental lack of several diquarkonium states~\footnote{On the other hand, the absence of meson molecules from the ensemble of observed states is clearly explained by the production issues reported in \sectionsname{\ref{xdeut} and \ref{appmol}}.}.

It is not clear yet how to treat those resonances that could be formed in correspondence of radial excitations of diquarkonia because  we are not sure about how the mechanism works, and if it does, in the case in which transitions are possible within the closed channel (in the potential $V_2$ of \figurename{~\ref{pot2x}}).  Also neutral  $J^{PC}=1^{--}$ $Y$  resonances, identified as orbital excitations of diquarkonia (\sectionname{\ref{yresos}}),   have not been included in this discussion for the same reason. %

\section{Other models and approaches} \label{others}
In this Section we illustrate other options for the description of \XYZ phenomenology. These were proposed since the early days after the discovery of the $X(3872)$ and the $Y(4260)$ and might have a somewhat more restricted application since they do not apply `systematically' to all of the observed states. The fact that \XYZ observed resonances should all be of the same nature contains admittedly some prejudice, which motivated many of the arguments made in this review. Until a widely accepted picture is not reached, we have to consider the possibility that not all of them can be understood with the same scheme.         

\subsection{Hadroquarkonium}
\label{sec:hadrocharmonium}
The peculiar property of the vector $Y$ states and of the $Z(4430)$ motivated their interpretation in terms of another possible hadron configuration, the so-called hadroquarkonium~\cite{Dubynskiy:2008mq,Dubynskiy:2008di,Li:2013ssa,Wang:2013kra}. These states decay predominantly into a particular charmonium + light mesons, $\jpsi$ for the $Y(4260)$ and the $Z_c(3900)$ and $\psiprime$ for the others. In particular, they are not observed decaying into open charm final states, despite these modes are expected to dominate the width in all ordinary charmonium models. Therefore, one of their possible interpretations is that of a `core' composed by a compact $Q\bar Q$ state, surrounded by a ``cloud" of light hadronic matter. While meson molecules (\sectionname{\ref{mol2}}) and diquarkonia (\sectionname{\ref{diquark}}) are discriminated by the overlapping of their constituents --- largely separated mesons for the former and compact diquarkonia for the latter --- hadroquarkonia would be characterized by the relative sizes of the core and the cloud. The light matter cloud is thought to have a typical radius much larger than that of the central quarkonium.

The hadroquarkonium model shares several qualitative features with ordinary atomic physics. The potential between its two components would be a QCD analogous of the van der Waals force. The key assumption is that such interaction is strong enough to allow for a bound state but not enough to substantially modify the nature of the compact $Q\bar Q$ core. If this is verified, then the decay pattern of the above mentioned resonances would be simply explained. However, as we will see shortly, it is difficult to reach a full assessment of these hypotheses.

An analysis to look for hadroquarkonia on the Lattice has been presented in~\cite{gunnar}.

\subsubsection{The Hamiltonian}
Since the $Q\bar Q$ core is a color neutral object, its interaction with the gluonic field generated by the light degrees of freedom can be treated using a QCD multipole expansion~\cite{Gottfried:1977gp}. The effective Hamiltonian for such an interaction can be written as~\cite{Gottfried:1977gp,Voloshin:1978hc,Peskin:1979va,Bhanot:1979vb}
\begin{align} \label{Heffhadroquark}
H_\text{eff}=-\frac{1}{2}\alpha^{(\psi_1\psi_2)}E_i^aE_i^a
\end{align}
We adopt the Wilsonian normalization of the gluonic fields~\footnote{This means that the lagrangian is normalized as $\mathcal{L} = - 1/4g^2 F^a_{\mu\nu} F^{a\mu\nu}$. }. The previous equation is simple to understand in analogy with the electromagnetic case. The leading interaction between an overall neutral composite object and an external electric field is proportional to $-\bbb d\cdot \bbb E$, with $\bbb d$ the dipole moment of the object. At the same time, the induced dipole must be proportional to the external field itself, $\bbb d\propto\alpha \bbb E$, with $\alpha$ being the polarizability. Therefore, the chromo-polarizability $\alpha^{(\psi_1\psi_2)}$ is essentially a measure of the deformability of the central $Q\bar Q$. A less heuristic derivation can be found in~\ref{app:Heff}.

Very little is know \emph{ab initio} about the chromo-polarizability~\footnote{It has been shown in~\cite{Voloshin:2004un} that the diagonal element $\alpha^{(\psi\psi)}$ for $\psi=\jpsi,\,\Upsilon$ can be measured directly through the $\psi\to\pi\pi\ell^+\ell^-$ process.}. The only two quantities that have been estimated are the off-diagonal elements $\alpha^{(\jpsi \,\psiprime)}$ and $\alpha^{(\Upsilon \, \Upsilon(2S))}$, which have been measured from the $\psiprime\to \jpsi\,\pi^+\pi^-$ and $\Upsilon(2S)\to \Upsilon \,\pi^+\pi^-$ decays obtaining~\cite{Voloshin:2004un}
\begin{align}
\left|\alpha^{(\jpsi\,\psiprime)}\right|\simeq2.00\gev^{-3} \quad \text{ and } \quad \left|\alpha^{(\Upsilon\,\Upsilon(2S))}\right|\simeq0.66\gev^{-3}
\end{align}
However, in light of the definition~\eqref{alpha}, the chromo-polarizability should also satisfy the Schwartz inequality
\begin{align}
\alpha^{(\psi\psi)}\alpha^{(\psi^\prime\psi^\prime)}\geq \left|\alpha^{(\psi\psi^\prime)}\right|^2
\end{align}
with $\psi=(\jpsi,\,\Upsilon)$ and $\psi^\prime=(\psiprime,\,\Upsilon(2S))$. It is also generally expected for each diagonal term to be larger than the off-diagonal one~\cite{Dubynskiy:2008mq}. It is natural to expect for the QCD Van der Waals force to  deform the core rather than change its nature.

A first challenge to this picture comes from the $Y(4260)\to Z_c(3900)\,\pi\to (J/\psi\,\pi)\,\pi$ decay observed both at \belle~\cite{Liu:2013dau} and \bes~\cite{Ablikim:2013mio}. According to the description above, the $Y(4260)$ would be a hadrocharmonium with a $J/\psi$ core surrounded by an isoscalar pion cloud. In this context, it is hard to understand why a decay into its constituents should happen via an intermediate isovector state as the $Z_c$~\cite{Wang:2013kra}. This issue could be overcome if the $Z_c(3900)$ itself has a hadrocharmonium structure, with the same heavy quark core. This interpretation might be motivated by the suppression of the $Z_c\to h_c\pi$ decay, which was unexpected both in the molecular and in the tetraquark models. Nevertheless, as we will see shortly, it is seriously put in jeopardy by the observation of the $Z_c\to DD^*$ decay.

Even though suppressed, the chromoelectric field generated by the hadronic cloud might also induce a transition from the core quarkonium to another, $\psi_1\to\psi_2$, with rate mostly determined by the off-diagonal chromo-polarizability $\alpha^{(\psi_1\psi_2)}$. Consequently, small but detectable decay rates into quarkonia different from the original core are expected.

Decays into open flavor mesons should be suppressed as well, since they require the splitting of the $Q\bar Q$ by means of the soft gluons generated by the surrounding cloud. Indeed, using the holographic soft wall model for QCD~\cite{Karch:2006pv}, it has been argued that the decay of the hadroquarkonium into open flavor mesons is exponentially suppressed by the mass of the heavy quark, $\exp\left(-b \sqrt{M_Q/\Lambda_{QCD}}\right)$, with $b$ being a constant. 

Although in good agreement with the $Y$ states and with the $Z(4430)$, this result challenges the interpretation of the $Z_c(3900)$ in terms of hadrocharmonium~\cite{Voloshin:2013dpa}, since a prominent decay into $DD^*$ has been observed~\cite{Ablikim:2013xfr}.

\subsubsection{To bind, or not to bind\ldots}
Let us now determine under which conditions the interaction~\eqref{Heffhadroquark} can allow bound states. It is easy to show that the potential is indeed attractive. Neglecting light quark masses, the trace of the Yang-Mills stress-energy tensor can be written as
\begin{align}
T_\mu^{\,\,\mu}=\frac{\beta(g)}{2g^3}F_{\mu\nu}^aF^{a\mu\nu}=-\frac{9}{32\pi^2}F_{\mu\nu}^aF^{a\mu\nu}=\frac{9}{16\pi^2}\left(E_i^aE_i^a-B_i^aB_i^a\right)
\end{align}
where we used the one-loop expression for the beta function $\beta = \left(-\tfrac{11}{3} N + \tfrac{2}{3} n_f\right)g^3/(4\pi)^2$ for 3 light flavors. and $\bbb B^a$ is the chromo-magnetic field. At $q=0$, its average over a state $|\mathcal{X}\rangle$ of \emph{light} hadronic matter is
\begin{align}
\langle \mathcal{X}|T_\mu^{\,\,\mu}(\bbb q=0)|\mathcal{X}\rangle=M_\mathcal{X}
\end{align}
assuming a normalization $\langle \mathcal{X}|\mathcal{X}\rangle=1$.
The potential acting on the light matter by the static pair is obtained considering the diagonal interaction $H_\text{eff}^{\psi\psi}$: 
\begin{align}
\mathcal{V} \langle \mathcal{X}|H_\text{eff}|\mathcal{X}\rangle=-\frac{\mathcal{V}}{2}\alpha^{(\psi\psi)}\langle \mathcal{X}|E_i^aE_i^a|\mathcal{X}\rangle=-\frac{8\pi^2}{9}\alpha^{(\psi\psi)}M_\mathcal{X}-\frac{\mathcal{V}}{2}\alpha^{(\psi\psi)}\langle \mathcal{X}|B_i^aB_i^a|\mathcal{X}\rangle\leq -\frac{8\pi^2}{9}\alpha^{(\psi\psi)}M_\mathcal{X}
\end{align}
where $\mathcal{V}$ is a volume factor.
Now, just like we did in \sectionname{\ref{sec:1pi}}, the matrix element of $H_\text{eff}$  provides the potential in momentum space, here also in the static limit. Therefore
\begin{align}
\mathcal{V} \langle \mathcal{X}|H_\text{eff}|\mathcal{X}\rangle=V(\bbb q=0)=\int d^3x\,V(\bbb x)\leq-\frac{8\pi^2}{9}\alpha^{(\psi\psi)}M_\mathcal{X}
\end{align}
This shows that the interaction is of attractive nature. It is now a matter of understanding under which conditions it also allows for bound states. This will turn out to be a much more subtle issue. 

The kinetic energy of the system can be written in a non-relativistic fashion as $T\sim p^2/2\mu\sim1/(R_\mathcal{X}^2\mu)$, where $R_\mathcal{X}$ is the characteristic size of the light cloud and $\mu=M_\mathcal{X}M_\psi/(M_\mathcal{X}+M_\psi)$ is the reduced mass of the quarkonium + light matter system. Its integral will therefore be $\sim R_\mathcal{X}/\mu$. We also used the fact that the momentum will typically be $p\sim 1/R_\mathcal{X}$. It then follows that the total energy will be
\begin{align}
E=T+V\sim \frac{R_\mathcal{X}}{\mu}+\int d^3x \,V(\bbb x)\leq \frac{R_\mathcal{X}}{\mu}-\frac{8\pi^2}{9}\alpha^{(\psi\psi)}M_\mathcal{X}
\end{align}
Requiring this to be negative gives the following constraint for the presence of bound states
\begin{align}
\alpha^{(\psi\psi)}\frac{M_\mathcal{X} \, \mu}{R_\mathcal{X}}\geq C
\end{align}
where $C$ is some constant of order one, which will eventually depend on the details of the system and on the actual definition of the characteristic size. 

The previous inequality is more easily satisfied for larger values of the chromo-polarizability. Being excited quarkonia more extended than the ground state, it is extremely reasonable to expect their diagonal chromo-polarizability to be larger than that of the ground state, $\alpha^{(\psi(nS)\psi(nS))}>\alpha^{(\psi\psi)}$. This could explain why most of the resonances mentioned at the beginning of the section decay into $\psiprime$ rather than $\jpsi$.

One might also argue~\cite{Dubynskiy:2008mq} that if $R_\mathcal{X}$ grows slower than $M_\mathcal{X}$ then a bound state will necessarily appear for massive enough light excitations. However, the models which accurately reproduce the linear Regge trajectories --- see \eg~\cite{Karch:2006pv} --- usually find $R_\mathcal{X}\propto M_\mathcal{X}$. The conclusions about the possible existence of bound states are therefore substantially weakened since they crucially depend on the precise knowledge of $C$ and $\alpha^{(\psi\psi)}$, which are hardly accessible quantities.

Nevertheless, if one assumes the existence of such state, then a full zoology of them is expected. Not only they could contain different central $Q\bar Q$ cores, \eg $\chi_{cJ}$, $\eta_c$, $\eta_c(2S)$ and $h_c$ in the charm sector\footnote{HQSS also implies that hadroquarkonia having cores belonging to the same spin multiplet should be almost degenerate~\cite{Cleven:2015era}.}, but the light excitation $|X\rangle$ could even be baryonic, giving rise to a so-called \emph{baryo-quarkonium}. The $P_c$ pentaquarks might be good candidates for these~\cite{Kubarovsky:2015aaa}.

\subsubsection{A case for heavy HQSS violation} 
So far there is no evidence for possible states containing $\eta_c$ and $\eta_c(2S)$~\footnote{But the $Z_1(4050)$ and $Z_2(4250)$ seen by \belle in $\chi_{c1} \,\pi$ might be good candidate for hadrocharmonia.}. However,  \bes has observed a large cross section for $e^+ e^- \to h_c \,\pi\pi$, comparable with the $\to Y(4260)\to \jpsi\,\pi\pi$ one~\cite{Ablikim:2013wzq,Yuan:2014rta,Chang-Zheng:2014haa}. Although the resonant content is not clear (see \sectionname{\ref{sec:vectors}}), a relevant component for $Y(4260)$ and $Y(4260)\to h_c\,\pi\pi$ is allowed. 
If these decays were confirmed, they would imply a strong violation of the HQSS (the $h_c$ is a heavy spin singlet) and challenge the interpretation of the $Y(4260)$ and $Y(4360)$ as hadrocharmonia made of a $\jpsi$ and $\psiprime$ core, respectively. 

A possible solution has been proposed in~\cite{Li:2013ssa}. The two states are considered as a mixture of both $S_{c\bar c}=0$ and $S_{c\bar c}=1$ hadrocharmonia
\begin{align} \label{Ymix}
|Y(4260)\rangle=\cos\theta|\Psi_1\rangle-\sin\theta|\Psi_0\rangle \quad \text{ and } \quad |Y(4360)\rangle=\sin\theta|\Psi_1\rangle+\cos\theta|\Psi_0\rangle
\end{align}
where $|\Psi_{S_{c\bar c}}\rangle$ is the hadrocharmonium state\footnote{The $J^{PC}=1^{++}$ quantum numbers can arise from an $S_{c\bar c}=0$ hadrocharmonium only thanks to the light degrees of freedom.} and $\theta$ is the mixing angle. 

This hypothesis allows for interesting predictions for the pattern of $e^+e^-$ annihilation into $\jpsi\,\pi\pi$, $\psiprime\,\pi\pi$ and $h_c\,\pi\pi$ final states~\cite{Li:2013ssa}. The electromagnetic current can only create the $|\Psi_1\rangle$ state and hence while the  $\jpsi\,\pi\pi$ and $\psiprime\,\pi\pi$ channels can be produced directly, the $h_c\,\pi\pi$ one needs to happen through the $\Psi_1-\Psi_0$ mixing. The authors also estimate the mixing angle to be large, $\theta\simeq40^\circ$.

Note, however, that such a large mixing implies that the two states $|\Psi_1\rangle$ and $|\Psi_0\rangle$ are almost degenerate. Their masses are in fact given by
\begin{subequations}
\begin{align}
M_1&=\langle\Psi_1|H|\Psi_1\rangle=\cos^2\theta \,M(Y(4260))+\sin^2\theta \,M(Y(4360))\simeq 4293 \text{\mev} \\
M_2&=\langle\Psi_0|H|\Psi_0\rangle=\cos^2\theta \,M(Y(4360))+\sin^2\theta \,M(Y(4260))\simeq 4311 \text{\mev}
\end{align}
\end{subequations}
where $H$ is the unknown Hamiltonian determining the energy of the system. The closeness of these values prevents from interpreting $\Psi_1$ and $\Psi_0$ as hadrocharmonia having a $\jpsi$ and $h_c$ core respectively, since their masses are more than $400$\mev apart. Their $Q\bar Q$ centers might instead correspond to $\psiprime$ and $h_c$ respectively. The decay into $\jpsi$ would happen via the $\psiprime\to \jpsi$ transition~\cite{Cleven:2015era}. This is in contrast with the first picture proposed in~\cite{Dubynskiy:2008mq}, where the $Y(4260)$ was thought of as having a $\jpsi$ core. As explained at the beginning of the section, the QCD van der Waals force should induce the decay into a charmonium different from the core with a suppressed rate. Indeed, for the $Y(4360)$ we have $\mathcal{B}(Y(4360)\to \jpsi\pi^+\pi^-)/\mathcal{B}(Y(4360)\to \psiprime\pi^+\pi^-)<3.4\times10^{-3}$ --- see Sec.~\ref{sec:vectors}. Since in Eq.~\eqref{Ymix} the weights for the $\Psi_1$ component are comparable in the two states, one would expect a similar ratio for the $Y(4260)$ as well. This is not supported by data since its decay into $\psi(2S)$ has never been observed. One might argue that this is because the process has a fairly low $Q$-value. Nevertheless, this should imply the $Y(4260)$ to be narrower than the $Y(4360)$.

\subsection{Hybrids}
\label{sec:hybrids}
The constituent quark model allows to consider static gluons as  additional degrees of freedom belonging to the adjoint representation $\bm 8_c$ of the color group. The first consequence is the presence in the QCD spectrum of the so-called \emph{glueballs}, \ie bound states of gluons only. Indeed a product of an arbitrary number of adjoint fields always contains a color singlet, $\bm 8_c\otimes\dots\otimes \bm 8_c=\bm 1_c\oplus\dots\;$~\footnote{One expects massive glueballs also in pure Yang-Mills theory without matter fields. This makes them particularly interesting from a more fundamental point of view. For a review, see~\cite{Mathieu:2008me}.}. The presence of glueballs has been confirmed by several Lattice QCD analyses --- see \eg~\cite{Bali:1993fb,Morningstar:1999rf,Meyer:2002mk,Gregory:2012hu}.

Another intriguing possibility is that of joining together a $q\bar q$ in the octet representation and a constituent gluon
\begin{align}
\bm 3_c\otimes\bar{\bm 3}_c\otimes\bm 8_c=(\bm 8_c\oplus \bm 1_c)\otimes \bm 8_c=\bm{27}_c\oplus \overline{\bm{10}}_c\oplus\bm{10}_c\oplus \bm 8_c\oplus \bm8_c\oplus \bm8_c\oplus \bm 1_c
\end{align}
These particles are generally referred  to as \emph{hybrid mesons}. 
They were studied in the context of a constituent quark model~\cite{Horn:1977rq,Tanimoto:1982wy,Tanimoto:1982eh,Iddir:1988jd,Ishida:1989xh,Ishida:1991mx},  of the MIT bag model, both for the light~\cite{Jaffe:1975fd} and heavy sector~\cite{Hasenfratz:1980jv}, or within an effective field theory approach~\cite{Berwein:2015vca}.
In the constituent models, if the angular momentum of the gluon around the $q\bar q$ pair is $L_g$, the orbital angular momentum of the $q\bar q$ pair $L_{q\bar q}$ and its spin is $S_{q\bar q}$, then the parity and charge conjugation of the hybrid are
\begin{align}
P={(-1)}^{L_g+L_{q\bar q}}\quad \text{ and }\quad C={(-1)}^{L_{q\bar q}+S_{q\bar q}+1}
\end{align}

The lightest of these particles has $L_g=0$. For example, we can achieve exotic quantum numbers such as $J^{PC}=1^{-+}$ through a $q\bar q$ pair with $L_{q\bar q}=1$ and $S_{q\bar q}=0$.

\subsubsection{Modeling the hybrids: flux tube} \label{sec:hybridmod}
The constituent quark model is just one of several ways of describing hybrid mesons. The interaction between the quark, the antiquark and the gluon can be most easily understood via the phenomenological \emph{flux tube} picture. It is well-known that at large separations, the $q\bar q$ pair is confined in an approximately thin cylindrical region of color fields. One can attempt to describe this configuration of the QCD gauge fields using an effective string theory approach. In particular, assuming a Nambu-Goto action,\footnote{The Nambu-Goto (NG) action for a relativistic string is proportional to the area of the string itself. If $\tau$ and $\lambda$ are the proper time and the intrinsic coordinate along the string, and $X_\mu(\tau,\lambda)$ the spacetime position of its line element, then the NG action is given by
\begin{align}
S_{NG}\propto \int d\lambda\,d\tau \sqrt{-\text{det}\left(\partial_\alpha X_\mu\partial_\beta X^\mu\right)}
\end{align}
} the potential at large relative distances, $r$, is found to be~\cite{Arvis:1983fp,Juge:2002br} (see also~\cite{polchinski1998string} for a textbook treatment). %
\begin{align} \label{fluxtube}
V_n(r)=\sqrt{\sigma^2r^2+\frac{\pi\sigma}{6}(12n-1)}
\end{align}
Here $\sigma$ is the string tension and $n$ is the radial quantum number of the string modes. 

For the $n=0$ string ground state and for large $r$ one obtains the well known $V_0(r)= \sigma r$~\footnote{Still for $n=0$, the first $1/r$ correction is $\pi/12r$, the so-called L\"uscher term~\cite{Luscher:1980ac}. This behavior has been confirmed by lattice calculations with remarkable precision~\cite{Necco:2001xg}. }, typical of a confining theory, and used in different phenomenological parametrizations for describing the quarkonia spectrum (see \eg the Cornell potential~\cite{Eichten:1978tg}). Higher values of $n$ correspond to quantized excitations of the string between the quarks and hence to hybrid mesons.

It should be stressed that, despite being easily visualized, the description of the color flux tube in terms of effective string theory is problematic. Not only its quantization in four dimensions presents issues, but the energies of the gluonic excitations are not in agreement with the non-perturbative results obtained from Lattice QCD. On the other hand, the potential calculated within the AdS/QCD framework~\cite{Andreev:2012mc} is in quite good agreement with lattice data.

An approximation often used to determine the $Q\bar Q$ potential for the case of heavy quarks is that of the time-honored Born-Oppenheimer (BO) approximation. It is allowed since, in the large $m_Q$ limit, the typical time scale of the dynamics of the light degrees of freedom is much smaller than that of the heavy ones. The $Q\bar Q$ pair is treated as static at some fixed distance $r$ (now considered as a parameter) and the Schr\"odinger equation is solved for the gluonic degrees of freedom with energy $E_{\alpha}(r)$, $\alpha$ being the set of quantum numbers describing the gluon state. Alternatively, the nonperturbative static potential can be directly measured on the lattice. $E_\alpha(r)=V_\alpha(r)$ is then used as the effective potential for the heavy quark pair. Once the potential has been determined it is possible to look for bound states. If the static quark pair is in a color singlet configuration, one gets the usual quarkonium spectrum, while if the pair is in a color octet, one gets the hybrid spectrum.

This approach has been applied together with the previous flux tube model for the gluonic degrees of freedom in several papers, see \eg~\cite{Isgur:1984bm,Kokoski:1985is,Close:1994hc}. Some preliminary studies performed in order to relax the BO approximation were done in~\cite{Isgur:1999kx}, while the analysis of energy levels fully beyond it has been done in~\cite{Barnes:1995hc}. The authors found that the spectrum of the lowest hybrid mesons is robust. Some searches in this direction have been performed in Lattice QCD --- see \sectionname{\ref{sec:lqcd}}. Note that in this model the first excitation of the flux tube has quantum numbers $1^{+-}$ or $1^{-+}$, in contrast with the usual $1^{--}$ of the gluons. Therefore, a meson with exotic signature $1^{-+}$ is obtained when the quark pair has $S_{q \bar q}= 1$. This suggests that photoproduction can be the best production modes for these states. Finally, the production of charmonium hybrids in $B$ decays has been studied in~\cite{Chiladze:1998ti}.

\subsubsection{Modeling the hybrids: constituent gluons in Coulomb gauge} \label{sec:coulomb}
Another possibility is to describe the hybrid mesons using the Hamiltonian QCD approach in Coulomb gauge, and treating the gluonic modes as quasi-particles moving in a non-perturbative QCD vacuum~\cite{Swanson:1998kx,Guo:2008yz,Szczepaniak:2006nx,Guo:2014zva}. The main idea is to work in Coulomb gauge, $\bbb\nabla\cdot\bbb A^a=0$, so that the dynamical gluons can be separated from the instantaneous Coulomb-type forces that act between color charges. A rigorous derivation of the Coulomb gauge QCD Hamiltonian can be found in~\cite{Christ:1980ku}. The QCD spectrum is formally found by solving the Schr\"odinger equation
\begin{align} \label{HA}
H_\text{\tiny QCD}[\bbb A^a,\bbb \Pi^a] \, \psi_n[\bbb A^a]=E_n \, \psi_n[\bbb A^a]
\end{align}
with $\bbb \Pi^a(x)=-i\partial/\partial \bbb A^a(x)$. 

The peculiar feature of this model is that the constituent gluon mass is generated dynamically. To solve Eq.~\eqref{HA} one typically makes an ansatz for the BCS vacuum wave functional as~\cite{Szczepaniak:2006nx}
\begin{align} \label{psi0}
\psi_0[\bbb A^a]=\langle \bbb A^a|\Omega\rangle=\exp\left[ -\int\frac{d^3q}{{(2\pi)}^3} \bbb A^a(\bbb q) \, \omega(q) \, \bbb A^a(-\bbb q) \right]
\end{align}
This represents the \emph{non-trivial} gluon field distribution in the vacuum. The parameter $\omega(q)$ can be determined variationally by minimizing the vacuum expectation value (VEV) of the Hamiltonian, $\partial \langle \psi_0 |H_\text{\tiny QCD} | \psi_0 \rangle/\partial \omega(q)=0$. The solution is typically well approximated by $\omega(q)=m_g$ for $q\leq m_g$ and by $\omega(q)=q$ for $q>m_g$,
with the constituent gluon mass given by $m_g\simeq600\mev$~\cite{Szczepaniak:2001rg}.

As usual, the gauge field is expressed as a superposition of plane wave modes with creation and annihilation operators $\alpha^{a\dagger}_\lambda(\bbb q)$ and $\alpha^{a}_\lambda(\bbb q)$, with $\lambda=\pm1$ the polarization of the gluon. A hybrid $Q\bar Q g$ state is then obtained from the BCS vacuum as
\begin{align}
|r,\bbb q,\lambda\rangle\propto Q^\dagger_{\frac{r}{2}\bbb e_z} \alpha^{a\dagger}_\lambda(\bbb q)T^a \bar Q^\dagger_{-\frac{r}{2}\bbb e_z}|\Omega\rangle
\end{align}
where $r$ is the separation between the heavy quark and the heavy antiquark (lying along the $z$-axis), and we introduced the corresponding creation operators. The energies $E_n$ are then obtained from Eq.~\eqref{HA}.

This framework has been employed in~\cite{Szczepaniak:2006nx} to compute the spectrum for the lowest lying hybrids as a function of the $Q\bar Q$ separation, and compare it with Lattice QCD data.
 It nicely fits the data for the ground state but it consistently overestimates them for excited gluonic fields. However, it correctly reproduces the ordering of the levels with quantum numbers. It was shown in~\cite{Swanson:1998kx} that this is a non-trivial result since many phenomenological models based on the flux tube description or on the Nambu-Goto action fail at this task.

The analysis has been improved in~\cite{Guo:2008yz}. In particular, it is argued that the solution of the $\partial \langle \psi_0 |H_\text{\tiny QCD} | \psi_0 \rangle/\partial \omega(q)=0$ equation presents some ambiguities and it is therefore more sound to compute $\omega(q)$ phenomenologically from the spectrum of single gluon excitations, the so-called \emph{gluelumps}~\cite{Guo:2007sm}. Despite the improvement, the predictions are still systematically higher than the Lattice results. This might be due to intrinsic limitations of the gaussian, mean field, vacuum functional in Eq.~\eqref{psi0}. These techniques have also been used in~\cite{Guo:2014zva} to study the radiative decays of hybrids to charmonia.

\subsubsection{The hybrid meson description of the \texorpdfstring{\XYZ}{XYZ} states}
The hybrid meson model described so far has been mostly employed to explain the nature of the $Y(4260)$~\cite{Close:2005iz,Zhu:2005hp,Braaten:2013boa}, but some proposal has also been made for the $Y(4360)$~\cite{Kalashnikova:2008qr}, the $X(3872)$~\cite{Li:2004sta} and the $X(4140)$~\cite{Mahajan:2009pj}. 

In~\cite{Close:2005iz} arguments in favor of a hybrid interpretation of the $Y(4260)$ are made. In particular, it would be constituted  of a $J^{PC}=0^{-+}$ $c\bar c$ pair together with a $J^{PC}=1^{+-}$ gluonic excitation. Estimates obtained from the previously explained flux tube model~\cite{Barnes:1995hc} as well as lattice QCD simulations~\cite{Lacock:1996ny} give the mass of the lowest hybrid to be around $M\simeq4.0-4.2\gev$, in agreement with that of the $Y(4260)$. The fact that the decay width $\Gamma(Y(4260)\to e^+e^-)$ is a factor of 4 smaller than that for the $\psi(3770)$ is explained with a general suppression of the lepton-lepton coupling for hybrid mesons~\cite{Ono:1984dp}. This is mostly due to the smallness of the $c\bar c$ wave function at $r=0$. Also, the decay of hybrids into pairs of $1S$ ordinary mesons (\eg $D\bar D$, $D_s^+ D_s^-$, etc.) is expected to be largely suppressed~\cite{Page:1996rj,Kou:2005gt}, hence explaining the missing observation of such channels. Lastly, it is noted that a mixing of the $c\bar c g$ component of the $Y(4260)$ with the ordinary $c\bar c$ would not be allowed. In the BO approximation, in fact, the $Q\bar Q$ state and the gluon one decouple and states with different gluon occupation numbers are orthogonal. 

One issue with the $c\bar c g$ interpretation of the $Y(4260)$ is that the decay into charmonium + light hadrons was also expected to be fairly small. This is in contrast with the observation of the $\jpsi\,\pi\pi$ final state with large branching ratio. 

\subsubsection{An alternative kind of hybrid?} \label{sec:tetrahybrid}
A possible solution to this problem has been proposed in~\cite{Braaten:2013boa}. The $Z_c(3900)$ is tentatively interpreted as a different kind of hybrid, where the excited gluon is replaced by a light quark pair in the $\bm 8_c$ representation. We will refer to this kind of particle as a \emph{BO tetraquarks}.\footnote{Note that the ordinary $c\bar c g$ hybrid cannot accommodate charged states.} In this framework the large branching fraction for the decay of the $Y(4260)$ into $\jpsi\,\pi\pi$ would simply be explained by the $Y(4260)\to Z_c\,\pi$ process, followed by the copious $Z_c\to \jpsi\,\pi$ decay. The first reaction simply happens via a conversion of the gluon into a pion and an adjoint $[q\bar q]_{\bm 8_c}$ pair.

It is important to note that the BO tetraquarks presented in~\cite{Braaten:2013boa} are nothing but an alternative model for the description of the compact tetraquarks. For small objects of typical size around 1 fm, the distinction between different color arrangements of the internal quarks is artificial, since they all mix with each other. The separation between the diquarkonium presented in \sectionname{\ref{diquark}} and the present BO tetraquarks is made just for the sake of model building.

In any case, assuming the previously explained structures for the $Y(4260)$ and $Z_c(3900)$, the BO approximation has been used in~\cite{Braaten:2013boa,Braaten:2014qka} to compute the spectrum of hybrids and BO tetraquarks. 

The first thing to do is to find a set of quantum numbers of the gluonic excitation which are conserved in presence of the static $Q\bar Q$ pair. The first one is the eigenvalue, $\Lambda$, of $|\bbb e_r\cdot \bbb J_\ell|$, where $\bbb e_r$ is the unit vector pointing along the separation between the $Q$ and the $\bar Q$ and $\bbb J_\ell$ is the angular momentum of the light degree of freedom, \ie either $g$ or $[q\bar q]_{\bm 8_c}$. We then consider the product of the parity and charge conjugation of the light excitation, $\eta=(CP)_\text{light}$, and, when $\Lambda=0$, the eigenvalue, $\varepsilon=\pm$, of  the reflection of the light fields with respect to a plane containing the $Q\bar Q$ system. 

The notation adopted in~\cite{Braaten:2013boa} is that of molecular physics. In particular, the $\Lambda=0,1,2,\dots$ states are represented as $\Sigma,\,\Pi,\,\Delta,\dots$, and $\eta=+1$ and $-1$ are called respectively \emph{gerade} ($g$) and \emph{ungerade} ($u$). The quantum numbers of the light degrees of freedom are therefore indicated as $\alpha=\Sigma^+_\eta,\,\Sigma^-_\eta,\,\Pi_\eta,\,\Delta_\eta,\dots\;$.

The strategy to compute the spectrum is the one already anticipated. In particular, the hybrid potential at some fixed separation $r$ of the $Q\bar Q$ pair is given by $V_\alpha\equiv V_{n_\alpha}$ as in Eq.~\eqref{fluxtube} for large $r$, and by the following color-Coulomb potential for small $r$
\begin{align}
V_\alpha(r)\to\frac{\alpha_s(\mu)}{6r}+E_\alpha
\end{align}
Here the typical scale is $\mu=1/r$ and $E_\alpha$ is an additive term depending on the quantum numbers $\alpha$ and usually called \emph{gluelump} (see \eg~\cite{Marsh:2013xsa}). It can be thought as a massive quasi-particle excitation of the gauge field, a sort of constituent gluon. The different parameters of the potential can be obtained from Lattice QCD (see again \sectionname{\ref{sec:lqcd}}).

Given the potential, the energy levels of the ordinary hybrid mesons are found by solving the following Schr\"odinger equations for the heavy pair
\begin{align} \label{schrodinger}
\left[-\frac{1}{m_Q}\left(\frac{d}{dr}\right)^2+\frac{\left\langle \bbb L_{Q\bar Q}\right\rangle_{\alpha,r}}{m_Qr^2}+V_\alpha(r)\right]\psi_n(r)=E_n\psi_n(r)
\end{align}
where $\left\langle \bbb L_{Q\bar Q}\right\rangle_{\alpha,r}$ is the angular momentum of the heavy quark pair computed at a certain separation $r$ and for a given set of quantum numbers $\alpha$ of the gluonic degrees of freedom, and $m_Q$ is the reduced mass of the $Q\bar Q$ system.

\begin{figure}[t!]
\centering
\includegraphics[width=0.48\textwidth]{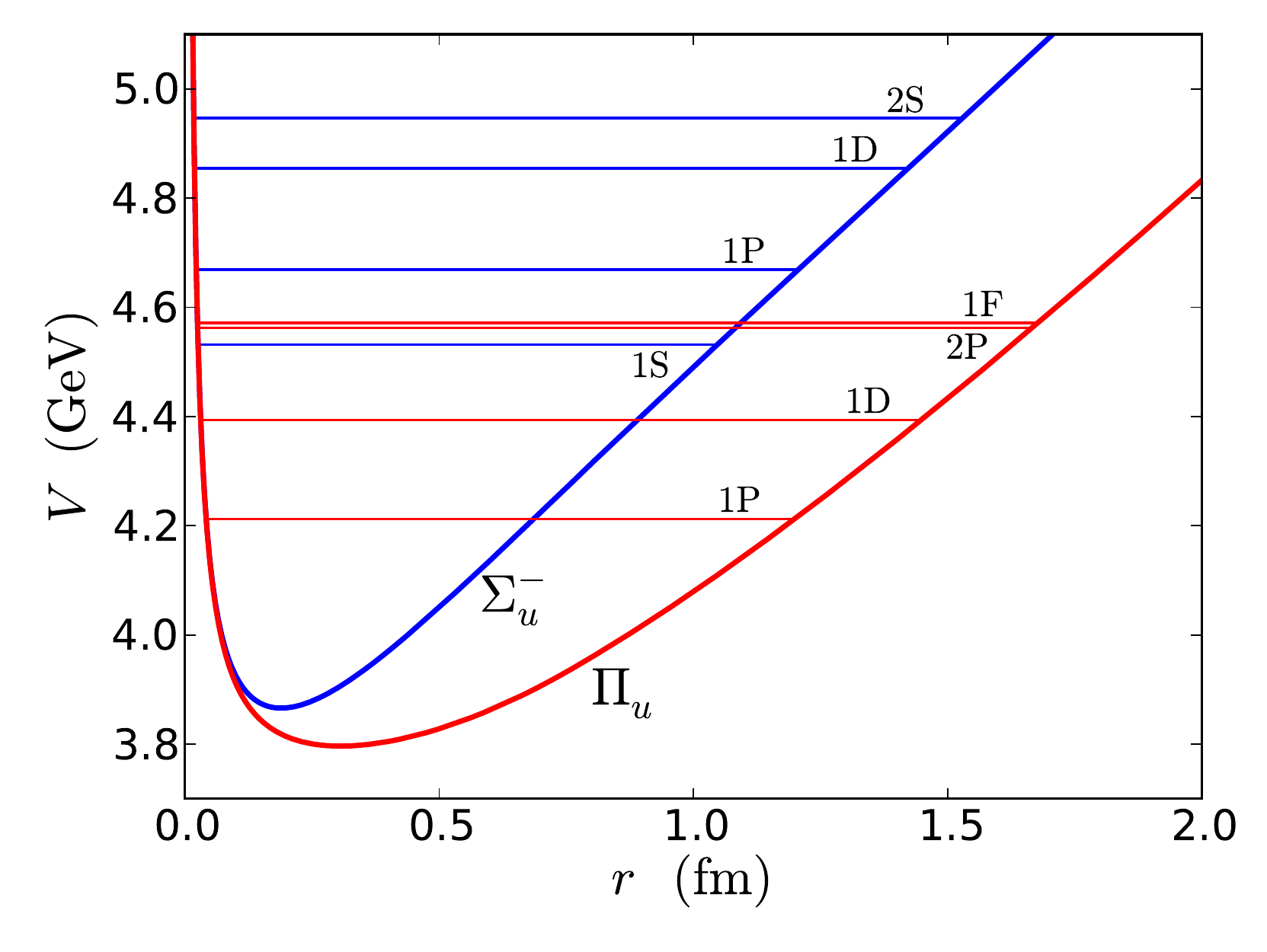}
\includegraphics[width=0.48\textwidth]{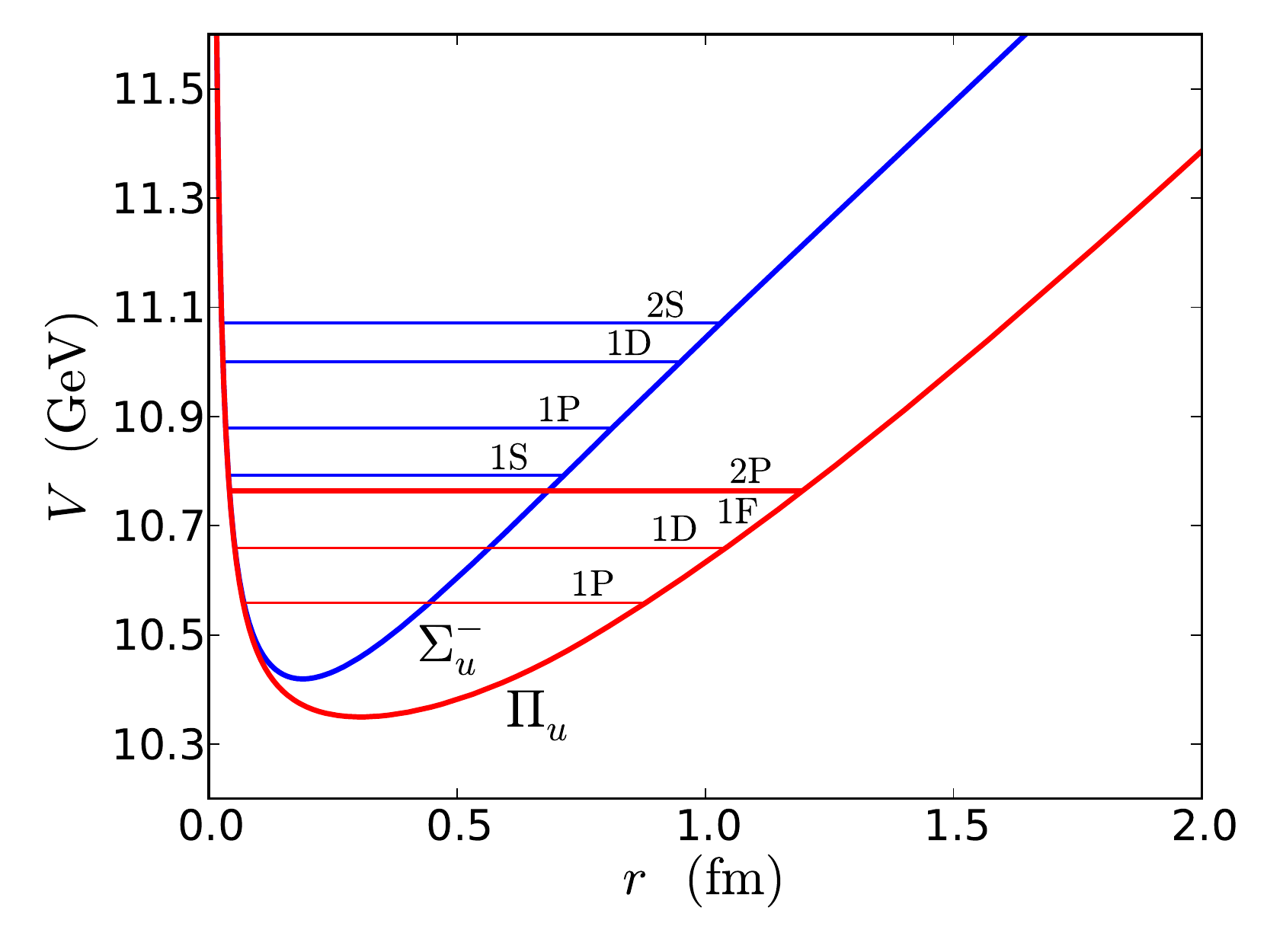}
\caption{Lowest energy levels obtained from the solutions to Eq.~\eqref{schrodinger} for the charm (left panel) and bottom (right panel) sectors. The notation is explained in the text. The analogous levels for the ground state of the gluon field, $\Sigma_g^+$, are just the ordinary quarkonia. Taken from~\cite{Braaten:2014qka}.} \label{fig:hybrid}
\end{figure}

The $Q\bar Q g$ energy levels resulting from this calculation are reported in \figurename{~\ref{fig:hybrid}}.
For the $c\bar c g$ case the lowest state is expected at $M\simeq4246$\mev, while for the $b\bar b g$ one at $M\simeq10559$\mev. While the first one is in agreement with the $Y(4260)$, no known resonance compatible with the hybrid interpretation has been found around the second mass.

To extract the spectrum of BO tetraquarks it is assumed in~\cite{Braaten:2013boa,Braaten:2014qka} that their potential has the same qualitative behavior as that of $Q\bar Q g$ hybrids. This is a strong assumption since, except for the asymptotic behavior at small and larger $r$, very little is know about such potential. This makes the predictions substantially model-dependent. Moreover, it is also assumed that the BO tetraquarks potential for the $[q\bar q]_{\bm 8_c}$ pair in the ground state, $\Sigma_g^+$, is completely unstable with respect to the transition into ordinary hybrid plus a light meson. It is therefore ignored. Again, this statement finds no rigorous ground but the practicality of the computation. 

In \tablename{\ref{tab:hybrids}} we reported the spectrum predicted in~\cite{Braaten:2013boa} for the BO tetraquarks in the charm sector. In particular, for each ordinary hybrid with quantum number $J^{PC}$ there will be a BO tetraquarks with the same $J^P$, with $I=1$ and hence $G=C(-1)^I=-C$. Because of HQSS --- see \sectionname{\ref{HQSS}} --- they will also organize themselves in quasi-degenerate spin multiples $T_n$. While the $Z_c$ is used as an input, the $Z_c^\prime$ would correspond to the $J^P=1^+$ member of either $T_3$ or $T_4$.
\begin{table} [ht]
\centering
\begin{tabular}{cc|cc|cc|cc}
\hline\hline
\multicolumn{2}{c|}{$T_1$} & \multicolumn{2}{c|}{$T_2$} & \multicolumn{2}{c|}{$T_3$} & \multicolumn{2}{c}{$T_4$} \\
\hline
$I^G(J^P)$ & Mass (\mevnospace) & $I^G(J^P)$ & Mass (\mevnospace) & $I^G(J^P)$ & Mass (\mevnospace) & $I^G(J^P)$ & Mass (\mevnospace)\\
\hline
$1^+(1^-)$ & 3839 & $1^-(1^+)$ & 3952 & $1^-(0^+)$ & 4025 & $1^-(2^+)$ & 4045 \\
$1^-(0^-)$ & 3748 & $1^+(0^+)$ & 3939 & $1^+(1^+)$ & 4030 & $1^+(1^+)$ & 4050 \\ 
$1^-(1^-)$ & 3770 & $1^+(1^+)$ & $\bbb{3897}$ & & & $1^+(2^+)$ & 4065 \\
$1^-(2^-)$ & 3887 & $1^+(2^+)$ & 3948 & & & $1^+(3^+)$ & 4101 \\
\hline\hline
\end{tabular}
\caption{Predicted masses for the BO tetraquarks in the charm sector as reported in~\cite{Braaten:2013boa}. The bold mass corresponds to the $Z_c(3900)$ and is used as input.} \label{tab:hybrids}
\end{table}

As one can see, the BO tetraquarks model also suffers from the same drawbacks as for the diquarkonium. The zoology of predicted states is copious and largely unobserved.

\subsubsection{\texorpdfstring{$X(3872)$ and $X(4140)$}{X(3872) and X(4140)} are not hybrids}

Lastly, it is worth mentioning that the interpretation of the $X(3872)$ in terms of a $c\bar cg$ proposed soon after its discovery~\cite{Li:2004sta} has been ruled out by the confirmation of the $X\to \jpsi\,\rho$ decay. The hybrid model cannot account for isospin violation since $Q\bar Q g$ is always an $I=0$ state. 

The interpretation of the $X(4140)$ as an hybrid~\cite{Mahajan:2009pj} relies, instead, on its quantum numbers being $J^{PC}=1^{-+}$. This possibility has been ruled out by the recent \lhcb analysis, which assigns $1^{++}$ to the state, see \sectionname{\ref{sec:ccss}}. Moreover, the decay of the gluon should be dominated by the creation of $u$ and $d$ quarks rather than $s$, as for the $X(4140) \to\jpsi\,\phi$ decay.
\begin{figure}[b]
\centering
\vcenteredhbox{\includegraphics[width=.25\textwidth]{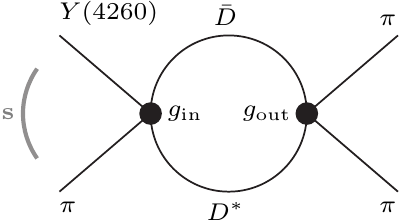}}\hspace{2cm}
\vcenteredhbox{\includegraphics[width=.60\textwidth]{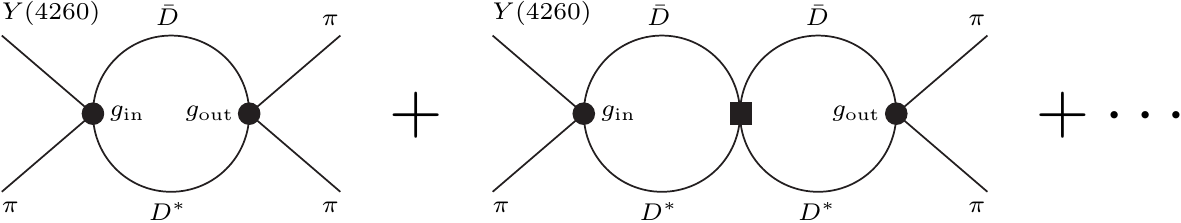}}
\caption{Left panel: One-loop diagram for the scattering $Y(4260)\,\pi\to\jpsi \pi$. Right panel: Unitarization of the one-loop diagram. } \label{fig:cuspdiagrams}
\end{figure}

\subsection{Nothing: kinematical effects}
\label{sec:cusps}
Another option that has been widely discussed in the literature arises from the  properties of the scattering amplitudes. Imposing unitarity and analyticity can indeed give rise to singularities in the complex plane, that are not related to resonances, but might mimic their behavior creating peaks and rapid motion of the phase shift. The idea and the formalism is known since long time~\cite{smatrix}. Work has been done in the context of \XYZ physics  in~\cite{Bugg:2008wu,Bugg:2004rk,Bugg:2011jr,Swanson:2014tra}. To simplify the discussion, in the following we will neglect all the spins, and we will take the $D$ and $D^*$ to be degenerate in mass. 

Consider, for example, the process $Y(4260) \to \jpsi\pi \pi$.  We discuss the kinematic effects that can give origin to a peak in the $\jpsi \pi$ invariant mass compatible with the $Z_c(3900)$ (see \sectionname{\ref{sec:Zc}}) but without the presence of any real resonance. 
More precisely, consider the scattering process $Y \pi \to \jpsi\,\pi$, which is described by the same amplitude as the decay process (as a consequence of crossing symmetry). One can assume for the process to be dominated by the diagram in the left panel of \figurename{~\ref{fig:cuspdiagrams}}.  The amplitude can thus be calculated with the usual Feynman rules, but the analytical structure emerges more transparently by using the Cutkosky rules, which give the imaginary part of the diagram, and the dispersion relations to calculate the real part.  The former is
\begin{equation}
 \Im \Pi(s) = \frac{k_\text{i}(s) }{4\pi\sqrt{s}} g_\text{in}(s) g_\text{out}(s) \,\theta\left(s - 4 m_D^2\right)
\end{equation}
where $k_\text{i} = \tfrac{1}{2}\sqrt{s - 4m_D^2}$ is the 3-momentum of the $\bar D$ and $D^*$ in the center-of-mass frame, and $g_\text{in,out}$ are form factors. The dispersive integral relating the full amplitude to its imaginary part is
\begin{equation} \label{dispersion}
 \Pi (s)= \frac{1}{\pi} \int_{4m_D^2}^\infty ds^\prime \frac{\Im \Pi\!\left(s^\prime\right)}{s^\prime - s - i \epsilon}
\end{equation}
The integral can be solved, even explicitly for the case of constant $g$'s, in which case the solution is given by $\sqrt{s - 4m_D^2}$ times a logarithm.
Because of the square root, this function has a cusp exactly at threshold, as shown in the left panel of \figurename{~\ref{fig:hanhart}}.  Form factors suggested by the quark model, like $g \propto \exp\left(- s / \beta^2\right)$ have often been  used, even though they are not consistent with the analytical behavior of the amplitude (which cannot grow faster than a polynomial at infinity in the complex plane). 

However, this simple model does not take into account the unitarization. The simplest way to achieve a unitary amplitude is to resum the 1-loop diagrams %
(\figurename{~\ref{fig:cuspdiagrams}}, right panel), the one-loop expression goes to the denominator of the amplitude, $A(s) = \sum_n \left[\Pi(s)\right]^n = 1/\left(1 - \Pi(s)\right)$, which develops a pole when $\Pi(s)=1$, \ie a proper QCD state. 
The right panel of \figurename{~\ref{fig:hanhart}} shows that, if one still wants to stick to the one-loop diagram only, the restriction to a coupling small enough to give meaning to the perturbative expansion does not allow to fit data correctly~\cite{Guo:2014iya}. If the coupling is instead nonpertubative, each order of the expansion gives a drastically different result (dotted and dashed curves in \figurename{~\ref{fig:hanhart}) and one cannot avoid unitarization.

\begin{figure}[t]
\centering
\includegraphics[width=.45\textwidth]{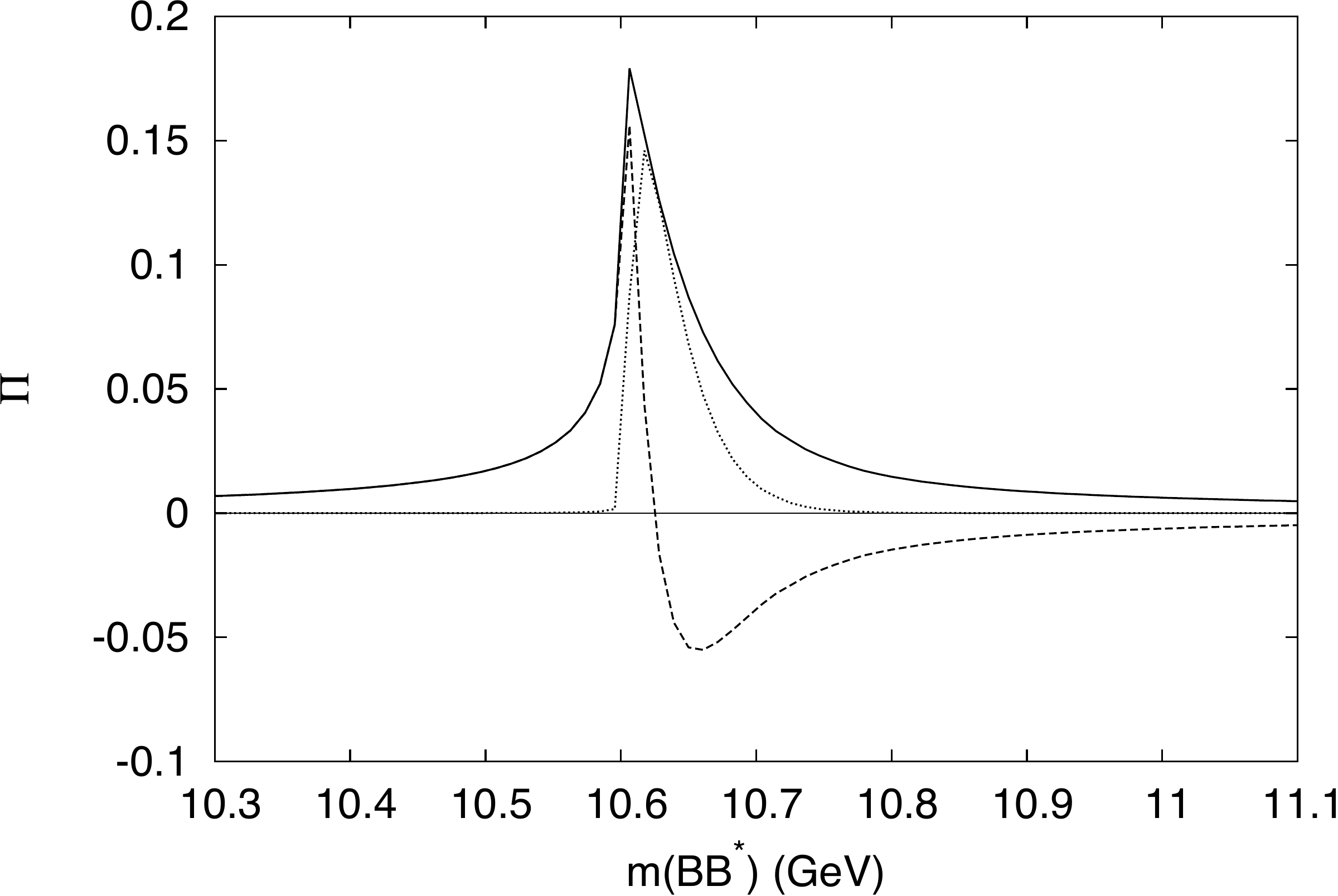} 
\includegraphics[width=.45\textwidth]{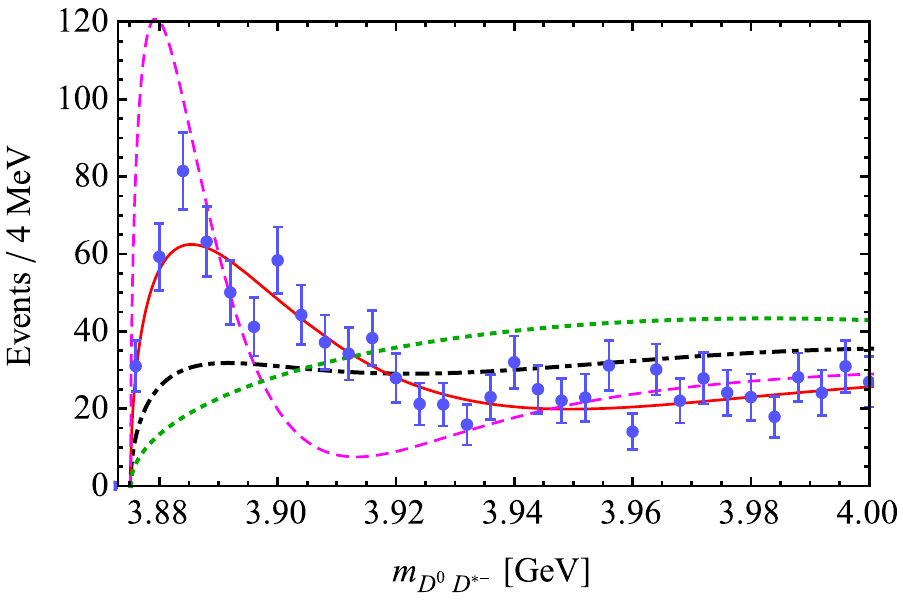}
\caption{Left panel: Plot of the $\Pi$ amplitude, showing the square root behavior of the imaginary part (dotted), the real part obtained via dispersion relation (dashed), and the modulus (solid), from~\cite{Swanson:2014tra}. Right Panel: The invariant $\bar D D^*$ mass distribution in $Y(4260)\to \pi \bar D D^*$ fitted in~\cite{Guo:2014iya}. The results from the tree level, one-loop and two-loop calculations are shown by the dotted, solid and dashed curves, respectively. The dot-dashed line shows the one-loop result
with the strength of the rescattering requested to be small to
justify a perturbative treatment.} \label{fig:hanhart}
\end{figure}

\begin{figure}[t]
\centering
\vcenteredhbox{\includegraphics[width=.40\textwidth]{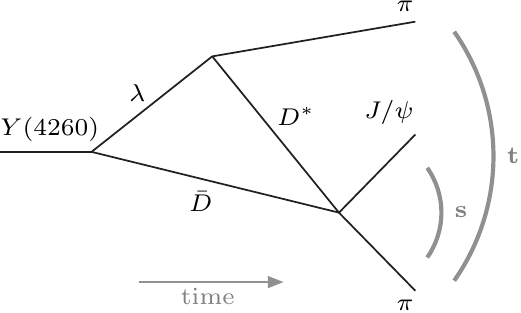}} \hspace{1cm}
\vcenteredhbox{\includegraphics[width=.50\textwidth]{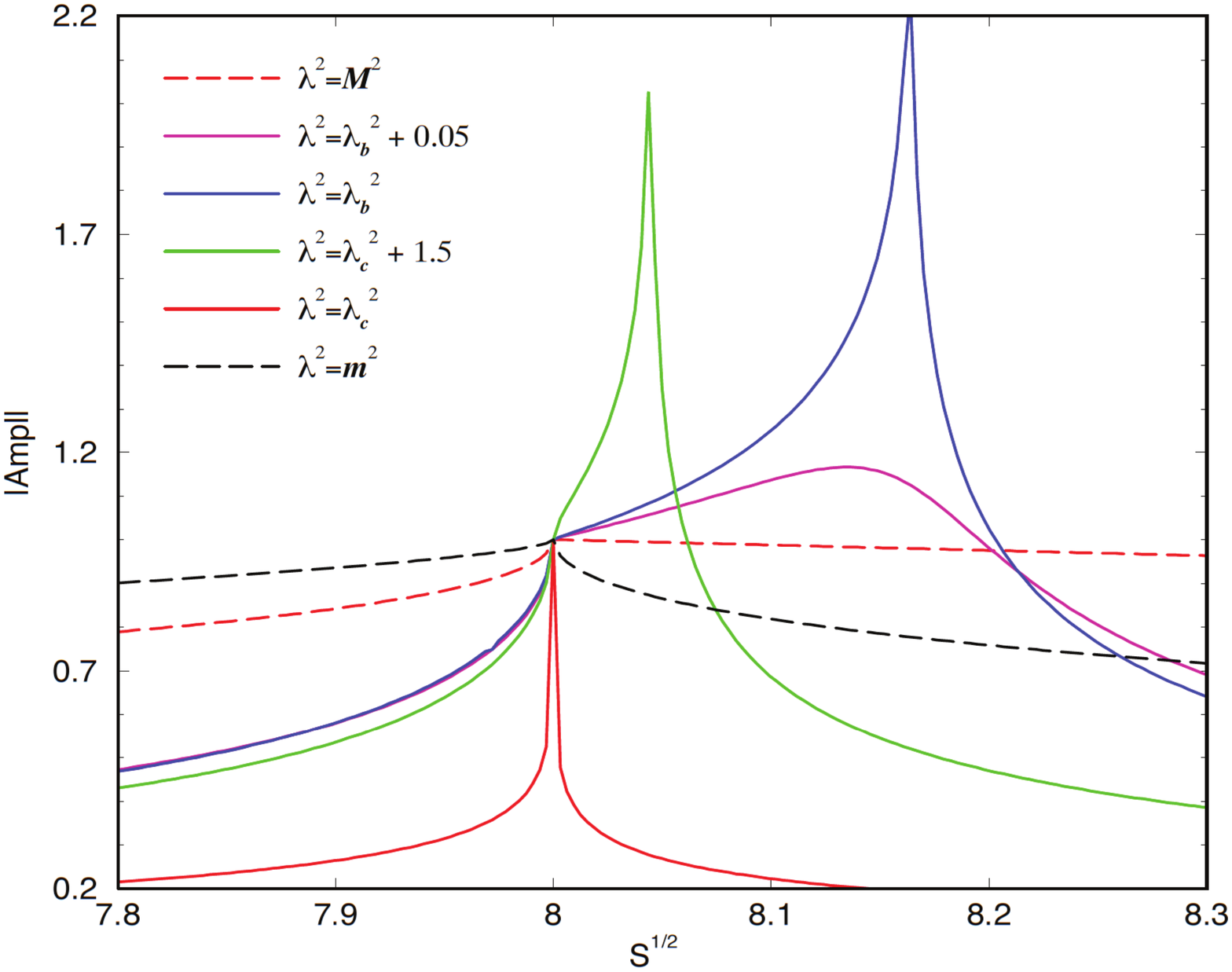}}
\caption{Left panel: Triangle diagram occurring in $Y(4260)\to \jpsi\,\pi\pi$ process. Right panel: values of the amplitude for different values of $\lambda$. If an actual state happens to fall in the small window which fulfills the Coleman-Norton conditions, a cusp appears. From~\cite{Szczepaniak:2015eza}.} \label{fig:triangle}
\end{figure}
In the previous discussion, the $Y \pi \to \bar D D^*$ vertex was parametrized with a smooth function $g_\text{in}(s)$. However, this is just an approximation for an actual exchange of particles, if their masses are heavy enough to contract their propagator. One can investigate instead if an appropriate model for the particle exchange produces any singularity close to the physical region. Triangle diagrams develop a logarithmic branch point~\footnote{Landau classified the leading singularities that a $n$-point amplitude can have in perturbation theory. The diagrams in \figurename{~\ref{fig:cuspdiagrams}} are basically two-points function (the four momenta enter only through the $s$ Mandelstam variable), and they exhibit a square root singularities. The triangle diagrams exhibit a logarithmic singularities, and so on.}. Consider again the $Y \to \jpsi \pi \pi$ process, but now described by the diagram in the left panel of \figurename{~\ref{fig:triangle}}. We consider the mesons in the loop to be the $D$, the $D^*$, and a generic charmed meson of mass $\lambda$. We want to study the behavior of the amplitude as a function of the mass $\lambda$, and see if any peak in the $\jpsi \pi$ invariant mass can occur. Again, we focus on the crossed process $Y\pi \to \jpsi \pi$. Since the centrifugal barriers will suppress any amplitude close to threshold, we consider only the  $S$-wave projection. The $\lambda$ meson is exchanged in the $t$ channel, so the projection of the propagator gives %
\begin{equation}
 C(s) = \frac{1}{2} \int_{-1}^1 d\cos\!\theta \frac{C}{t(s, \cos \theta) - \lambda^2 + i \epsilon}\,P_0(\cos \theta) 
 \end{equation}
where $C$ is a coupling constant, $P_0(\cos \theta) = 1$ is the $L=0$ Legendre polynomial. In the center of mass frame, $t = 2m_\pi^2 - 2 E_\pi^\text{in} E_\pi^\text{out} + 2 p^\text{in} p^\text{out} \cos\theta$, which is linear in $\cos \theta$. The integral thus gives a logarithm, 
\begin{equation}
 C(s) = -\frac{C}{2p^\text{in} p^\text{out}} \,Q_0\!\left(\frac{E_\pi^\text{in} E_\pi^\text{out}  - m_\pi^2 - \tfrac{1}{2}\lambda^2 - i \epsilon}{p^\text{in} p^\text{fin}}\right) 
 \end{equation}
where
$Q_0(x) = \tfrac{1}{2} \log \frac{x+1}{x-1}$ is the Legendre function of the second kind. The unitarity of the $S$-matrix can be rewritten as an equation for the discontinuity  of the analytic continuation of the amplitude across the physical axis %
\begin{equation}
\disc A(s) = \frac{1}{2i} \left[A\left(s + i\epsilon\right) - A\left(s - i\epsilon\right)\right] = C(s) \rho(s) B^*(s)\, \theta\left(s - 4m_D^2\right)
\end{equation}
where $\rho(s) = \sqrt{1 - 4 m_D^2 / s}$ is the 2-body phase space, and $B$ the scattering amplitude $\bar D D^* \to \jpsi \pi$, which we approximate as a constant. The dispersion relation on $A$ reads 
\begin{equation}
A(s) = \frac{1}{\pi} \int_{4M_D^2}^\infty ds^\prime \frac{C(s^\prime) \rho(s^\prime) B}{s^\prime - s}
\end{equation}
The analytical structure of $C(s^\prime)$ is rather complicated, and can contain up to four different branch points. In general they do not produce sizable effect in $A(s)$: even though one of the $s^\prime$ branch points is close to the real axis, one can deform the integration contour to distance it. However, Coleman and Norton show that when all the particles in the loop are simultaneously on-shell, {\em and} their velocities are such that the the rescattering can occur in real time, two branch points in $s^\prime$ pinch the $s$ contour of integration, preventing the deformation of the contour. This produces cusps in $A(s)$ close to the real axis~\cite{Coleman:1965xm}. This happens for a very small window of $\lambda$, see for example the effects in the right panel of \figurename{~\ref{fig:triangle}}. In particular, the $D_1(2420)$ mass happens to be in this interval, and might produce an enhancement in correspondence of the $Z_c$.

Now, let us consider the $Y \to \bar D D^* \pi$ process. If no actual $Z_c(3900)$ exists, the Dalitz plot will exhibit a band in the $D^* \pi$ channel corresponding to the $D_1(2420)$, and a band in the $\bar DD^*$ channel due to the triangle singularity. Longtime ago, Schmid proved that the interference between the two bands cancels the contribution of the $\bar D D^*$ band: the projection on the $\bar DD^*$ invariant mass does not produce any peaks~\cite{schmid}. The situation is more complicated when multiple channels are involved. In this case, the compensation happens among the various channel, and if one focuses only on a given channel, the peak can happen to be still be evident~\cite{Szczepaniak:2015hya}. If this is the mechanism behind the peak, the complete coupled channel analysis will show a rather peculiar behavior: one expects a depletion of events in the channel where the triangle singularity occurs, and an enhancement in the other channels.  Present data in the $Y\to \bar D D^* \pi$ channel do not seem to favor this hypothesis, but his can be verified in future high statistics experiments. The role of triangle singularities for \XYZP states has been discussed in~\cite{Pakhlov:2014qva,Szczepaniak:2015eza,Szczepaniak:2015hya,Aitchison:2015jxa,Guo:2015umn,Guo:2016bkl,zc3900}. 

\subsection{QCD sum rules}

A technique often used to compute the mass, width and coupling constants of the exotic states is the QCD Sum Rules (QCDSR) --- see~\cite{Nielsen:2009uh} for a review.
The method was first appeared in~\cite{Novikov:1976tn}, and was later developed in~\cite{Novikov:1977dq,Shifman:1978bx}. It was used to study the properties of mesons. The focus is on the evaluation of a two-point correlation function as
\begin{align} \label{QCDSRcorr}
\Pi(q)\equiv i\int d^4x \, e^{iq x}\langle0|T\left(j(x)j^\dagger(0)\right)|0\rangle
\end{align}
where $j(x)$ is a current with the quantum numbers of the hadron we want to study. The choice of the current $j(x)$ is only dictated by the $(I^G)J^{PC}$ quantum numbers of the hadron and by  assumptions on its nature. In general, it can be a linear superposition of currents corresponding to different structures (quarkonium, diquarkonium, meson molecule, \emph{etc.}). 

The important assumption is that this correlator can be evaluated both from the fundamental degrees of freedom of QCD (the so-called \emph{OPE side}) and from an effective meson theory (the so-called \emph{phenomenological side}). This means that, one starts from the asymptotically free theory of quarks and gluons  towards non-perturbative scales. The assumption is that, in a certain range of scales, the two theories will give the same result for the correlator~\eqref{QCDSRcorr}.

On the OPE side,  one expands the function as a series of local operators
\begin{align}
\Pi_\text{\tiny OPE}(q^2)=\sum_n C_n(Q^2) \, \mathcal{O}_n
\end{align}
with $Q^2=-q^2$ and where the set $\{\mathcal{O}_n\}$ includes all the local, gauge-invariant operators that can be written in terms of the gluon and quark fields. They are also ordered by mass dimension. The information about the short-range, perturbative part of the correlator is (by construction) contained in the Wilson coefficients $C_n(Q^2)$. The matrix elements for the operators $\mathcal{O}_n$ are nonpertubative and must be evaluated through Lattice QCD or using some phenomenological model. They are universal, \ie once they are fixed using a certain observable like the mass of \emph{a} particle they cannot be changed. The same thing is true for the quark masses.

On the phenomenological side, instead, one writes the two-point function in terms of a spectral density $\rho(s)$
\begin{align} \label{phenside}
\Pi_\text{phen}(q^2)=-\int ds\,\frac{\rho(s)}{(q^2-s+i\epsilon)}+\cdots
\end{align}
with the dots representing subtraction terms. In order for the previous equation to be of practical use, the spectral density has to be parametrized in terms of a small number of parameters. One usually assumes that it has a pole corresponding to the mass of the ground-state hadron, while higher mass states are contained in a smooth, continuous part
\begin{align} \label{rho}
\rho(s)=\lambda^2\delta(s-m^2)+\rho_\text{cont}(s)
\end{align}
$\lambda$ being the coupling of the current to the lowest mass hadron, $H$, $\langle0| \, j \, |H\rangle=\lambda$. Moreover, the continuum contribution is taken to be zero below a certain threshold $s_0$, and to it coincide with the result obtained from the OPE side above it, \ie one makes the ansatz
\begin{align} \label{rhocont}
\rho_\text{cont}(s)=\rho_\text{{\tiny OPE}}(s) \, \theta(s-s_0)
\end{align}
where $\pi\,\rho_\text{{\tiny OPE}}(s)=\Im\Pi_\text{{\tiny OPE}}(s)$. The threshold parameter $s_0$ is again universal. Combining Eqs.~\eqref{phenside}, \eqref{rho} and \eqref{rhocont} allows to extract the mass of the resonance.

The decay widths, \ie the coupling constants, can be estimated with similar procedures but starting from the three-point correlator.

The previous formalism has been applied to study the properties of the $X(3872)$~\cite{Matheus:2009vq}, of the $Z_c^{(\prime)}$~\cite{Navarra:2014aba,Chen:2015fsa,Agaev:2016dev}, of the $Z_b^{(\prime)}$~\cite{Chen:2015ata,Wang:2013zra} and of the $Y$ states~\cite{Wang:2016mmg}. The literature on the topic is impressively vast. We will report here just a limited number of results. The interested reader should refer to other reviews~\cite{Nielsen:2009uh}. 

It should be stressed that QCD sum rules are supposed to be the result of first principle, field theoretic QCD, and as such they should provide reliable, univocal nonpertubative results. Unfortunately, there is some freedom in choosing the rules  to compute the different observables. Moreover, some quantities can often be reproduced by suitably tuning the different compositions of the state under consideration. In this respects, the technique is often able to reproduce existing data, but its predictive power is somehow limited. In general, QCD sum rules are applicable to the lowest states in a channel with fixed quantum numbers. This is not the case for the exotic states discussed in this work. The truncated OPE expansion should lack of the theoretical precision necessary to extract informations relative to excited states. The following results should therefore always be taken with a grain of salt, as it will soon be clear from some of the following conclusions.

In~\cite{Matheus:2009vq} the mass of the $X(3872)$ and its $\Gamma(X\to J/\psi\pi^+\pi^-)$ decay width are analyzed. They assumed  the state to be a mixture of ordinary charmonium and  meson molecule. The current in Eq.~\eqref{QCDSRcorr} is taken to be
\begin{align}
J^u_\mu(x)=\sin\theta j_\mu^{\,(4u)}(x)+\cos\theta j_\mu^{\,(2u)}(x)
\end{align}
where $\theta$ is a mixing angle and $j_\mu^{(2u)}$ and $j_\mu^{(4u)}$ are current with the quantum numbers of charmonium and meson molecules respectively. They are given by
\begin{subequations}
\begin{align}
j^{\,(2u)}_\mu(x)&=\frac{1}{6\sqrt{2}}\langle \bar u u \rangle \, \bar c_\alpha(x)\gamma_\mu\gamma_5 c^\alpha(x) \\
j^{\,(4u)}=\mu(x)&=\frac{1}{\sqrt{2}}\left[ \bar u_\alpha(x)\gamma_5 c^\alpha(x)\,\bar c_\beta(x)\gamma_\mu u^\beta(x) - \bar u_\alpha(x)\gamma_\mu c^\alpha(x)\,\bar c_\beta(x)\gamma_5 u^\beta(x) \right]
\end{align}
\end{subequations}
Using the previous currents and the techniques explained above they estimated the mass and decay width of the $X$ to be $M_X=(3.77\pm0.18)\gev$ and $\Gamma(X\to J/\psi\pi^+\pi^-)=(9.3\pm6.9)\mev$, with a small mixing angle, $\theta\in[5^\circ,13^\circ]$. They also managed to estimate the  weights of the different components of the $X$ wave function. They found that it is composed for the $\sim97\%$ by ordinary charmonium and for the remaining $\sim3\%$ by a molecule. Of the latter one $\sim88\%$ is $D^0\bar D^{0*}$ and $\sim12\%$ is $D^+D^{*-}$. Of course, today we know that these results are not compatible with experimental data. The $X(3872)$ is not a charmonium. The same authors studied the $X$ under the assumption of a diquarkonium internal structure~\cite{Matheus:2006xi} and found a mass in agreement with the observation.

In~\cite{Navarra:2014aba} the decays of $Z_c(3900)$ were studied assuming a diquarkonium structure. The authors found a total width $\Gamma=(63.0\pm18.1)\mev$, and the following partial widths: $\Gamma(Z_c\to J/\psi\pi)=(29.1\pm8.2)\mev$, $\Gamma(Z_c\to \eta_c\rho)=(27.5\pm8.5)\mev$ and $\Gamma(Z_c\to DD^*)=(3.2\pm0.7)\mev$. The width is indeed compatible with the experimental one within $2\sigma$, but the ratio of the decay rates into $DD^*$ and into $J/\psi\,\pi$ is much smaller than the one reported in Eq.~\eqref{Zcratio}. A similar analysis was performed in~\cite{Agaev:2016dev} with analogous results.

The $Z_b$ is instead studied in~\cite{Chen:2015ata} using a variety of molecular interpolating currents. The extracted masses are nicely compatible with experiment. On the other hand, the properties of the $Z_b$ and $Z_b^\prime$ are analyzed in~\cite{Wang:2013zra} assuming a diquarkonium nature. The masses are very much in agreement with the observation and the predictions for the decay widths of the $Z_b$ give $\Gamma(Z_b\to\Upsilon\pi)=(4.77^{+3.27}_{-2.46})\mev$ and $\Gamma(Z_b\to\eta_c\rho^+)=(13.52^{+8.89}_{-6.93})\mev$. While the first one agrees with the the experiment, no data on the second decay channel are available yet.

Lastly, the application of QCD sum rules to the $Y$ states has been throughly done in~\cite{Wang:2016mmg}. The authors find support for the diquarkonium nature of the $Y(4660)$ and for the mixed charmonium-diquarkonium for the $Y(4260)$ and $Y(4360)$.

\subsection{Lattice QCD} \label{sec:lqcd}

Some efforts in trying to confirm the existence and understand the nature of the exotic states have been made using Lattice QCD techniques --- see~\cite{Gupta:1997nd,gattringer2009quantum} for an extensive treatment. Let us briefly recall the underlying strategy.

For a given set of quantum numbers one usually considers an ensemble of interpolating operators, $O_i(t)$, compatible with those quantum numbers.
Once they are identified, it is possible to study their two-point correlators, which can be written as
\begin{equation}
\langle 0 |O_i^\dagger(t)O_j(0)|0\rangle=\SumInt_n\langle 0|e^{iHt}O_i^\dagger(0)e^{-iHt} |n\rangle\langle n|O_j(0)|0\rangle = \SumInt_n Z^{n*}_i Z_j^n e^{-iE_nt}
\end{equation}
where we defined the \emph{overlaps}  $Z_i^n\equiv\langle n|O_i(0)|0\rangle$. The $\left|n\right\rangle$ are the multiparticle eigenstates with given quantum numbers. We remark that in the infinite volume their spectrum is obviously continuous.  

To make the previous oscillating sum suitable to numerical studies, one performs an analytic continuation to Euclidean time ($t\to-i\tau $). Spacetime is then discretized, with a lattice spacing $a$, and the analysis is restricted to a finite box with size $L$, in which case the correlator becomes
\begin{align} \label{Latticecorr}
C_{ij}(N_\tau)=\langle O_i^\dagger(\tau)O_j(0)\rangle=\sum_{n=1}^\infty Z^{n*}_iZ_j^n e^{-E_n a \tau},
\end{align}
with $\tau$ integer, $0 \le \tau \le L/a$. Since the allowed momenta are now discretized, the physical states are now a discrete set.
To extract the mass of the lowest lying level, the so-called \emph{effective mass} method can be used. One defines
\begin{align}
M_{(ij)}^\text{eff}(\tau)=\log\frac{C_{ij}(\tau-1)}{C_{ij}(\tau)}
\end{align}
For large $\tau$, the previous expression converges to the mass of the lowest energy state. For a realistic simulation with a finite range in the Euclidean time $\tau$, it will exhibit a plateau, from which the spectrum can be fitted. To improve the reliability of the results one usually combines  data obtained from different interpolating operators.

Other ways to determine the presence of a resonance and the value of its mass are to solve the \emph{generalized eigenvalue problem} (GEP)~\cite{Luscher:1990ck}, or to employ the so-called \emph{L\"uscher's method}~\cite{Luscher:1990ux,Luscher:1991cf}, which uses the dependence of the correlators from the size of the box to extract information about the scattering phase shift in the infinite volume limit.

It should be stressed that the impact that Lattice QCD analyses can have on the study of the \XYZ states is limited by some technical issues, which can hardly be overcome with the present resources. Some of these problems are
\begin{itemize}
\item Any operator $O_i$ able to resolve an exotic state is also able to resolve other states with the same quantum numbers. For example, if $O_i$ interpolates the charged $Z_c(3900)^+$ with $J^{PC}=1^{+-}$ then it will also give access to the $b_1^+$ and all its excitations. In principle, one should extract the full spectrum of the $b_1$ up to the mass of the $Z_c$, which is not feasible. In practice, one neglects the charm quark annihilation diagrams, which are expected to give a small contribution because of the OZI rule. This also prevents the mixing with hadrons made of light quarks only (like the $b_1$). However, the relevance of these diagrams for such fine-tuned systems like the \XYZ states is presently unknown.  

\item It should be clear by now that the problem at hand is made even more non-trivial by the presence of several close by meson-meson threshold. To be able to distinguish between one of the \XYZ resonances and a simple threshold requires a resolution that is right now far from achievable. For example, to have a precision of, say, $\Delta E\sim10$~MeV one would need a lattice spacing of at least $a\sim1/\Delta E\sim20$~fm. Realistic lattices have typically $a$ of the order of a few tens of a fm, with a total box size of a few fm itself.

\item If these exotic resonances are indeed molecular states, as proposed in part of the literature, the smallness of their binding energy would also imply a large size in real space --- see Sections~\ref{ascattering} and \ref{xdeut}. In this case, the total size of the box would be even smaller than the total extension of the state.
\end{itemize}
Despite the previous limitations, some attempts to look for four-quark resonances have been made. 

The simplest lattice calculations involve infinitely heavy quarks (static limit). Although in this limit it is not possible to have reliable predictions on the spectrum and decay properties of the states, it is a clean environment to study the interquark potential (which in this limit is a well defined quantity), and to provide a ground basis for the different models and phenomenology. We already cited the work by Lucini \etal~\cite{reticolo}, which observed some evidence for the formation of a scalar diquark, and the work by Cardoso \etal~\cite{Cardoso:2011fq}, where there is evidence that the $qq\bar q\bar q$ system indeed arranges itself in the expected H shaped configuration.  A new method to study this setup has been proposed in~\cite{Giusti:2015rra}, and might give new information about the diquark formations. Also systems with two static quarks and two light dynamical antiquarks have been explored~\cite{Bicudo:2012qt,Brown:2012tm,Bicudo:2015vta,Francis:2016hui}.

The only positive result for the existence of tetraquark resonances has been obtained in~\cite{Prelovsek:2013cra} for the $X(3872)$. The authors performed an analysis with valence and dynamical $u$ and $d$ quarks, with $m_u=m_d$ and a pion mass $m_\pi=266$\mev. Charm annihilation diagrams were also neglected. The authors did not consider the mixing of different partial waves due to the nonzero spin, or the proper identification of the spin in terms of representations of the cubic group. They studied both the isospin $I=0$ and $I=1$ cases. Among the obtained levels they observe one very close to the $DD^*$ threshold, and which they interpret as the $X(3872)$. It should be stressed that, for the reasons explained above, there is also the possibility for such level to correspond to the ordinary $\chi_{c1}(2P)$ charmonium, which carries that same quantum numbers as the $X$.

All the other searches for \XYZ states on the lattice (including manifestly exotic states with non-zero flavor charge~\cite{Esposito:2013fma,Guerrieri:2014nxa,Ikeda:2013vwa}) have returned inconclusive results.

Some analyses have also been done to study hybrid mesons. In this case, the results are more reliable since one can employ the Born-Oppenheimer approximation to describe a system with a heavy $Q\bar Q$ pair and a light degree of freedom (see Section~\ref{sec:hybrids}). In particular, although the phenomenological application of these states to the study of \XYZ resonances is problematic, their existence in the QCD spectrum has now been well assessed. In the bottom sector, the hybrid potential and the corresponding energy levels have been found in~\cite{Juge:1997nc,Juge:1999ie}.
The state of the art for the $c\bar c$ spectroscopy, including hybrid states, has instead been set by~\cite{Liu:2012ze}, albeit in the one-particle (zero width) approximation.

\begin{figure}[p]
\centering
\includegraphics[width=.75\textwidth]{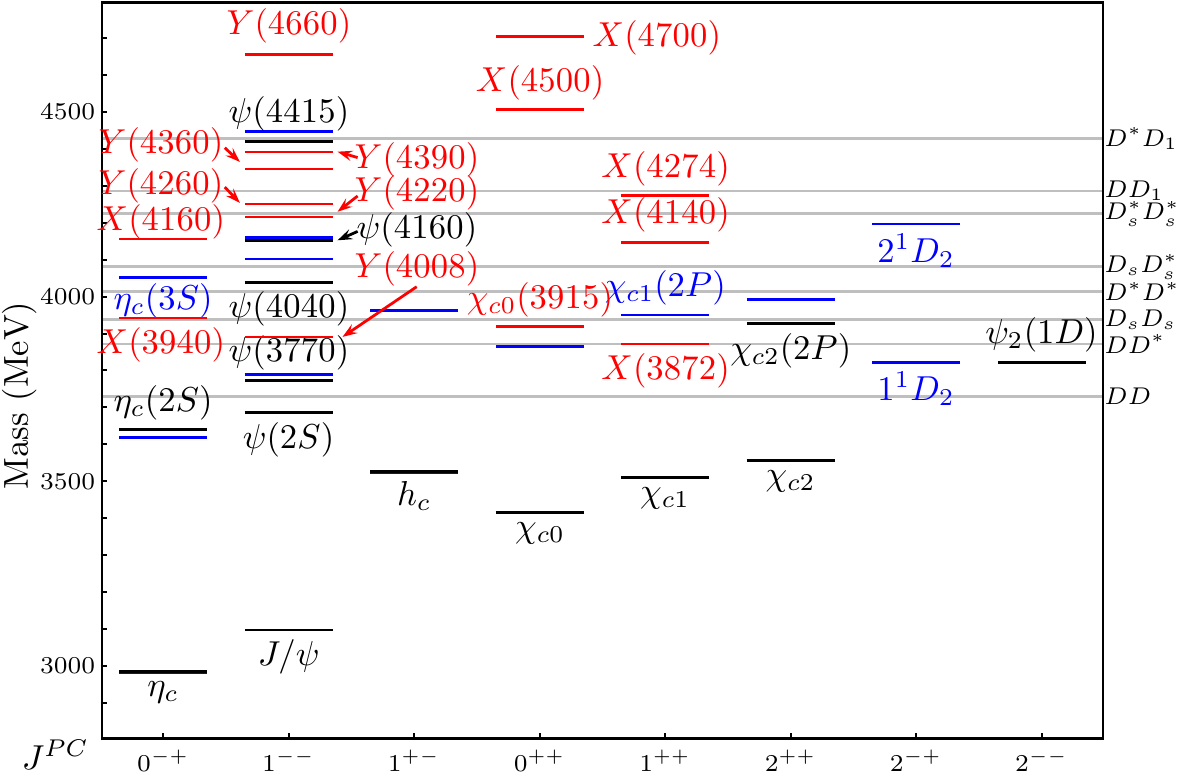} \\ \vspace{1cm}
\includegraphics[width=.75\textwidth]{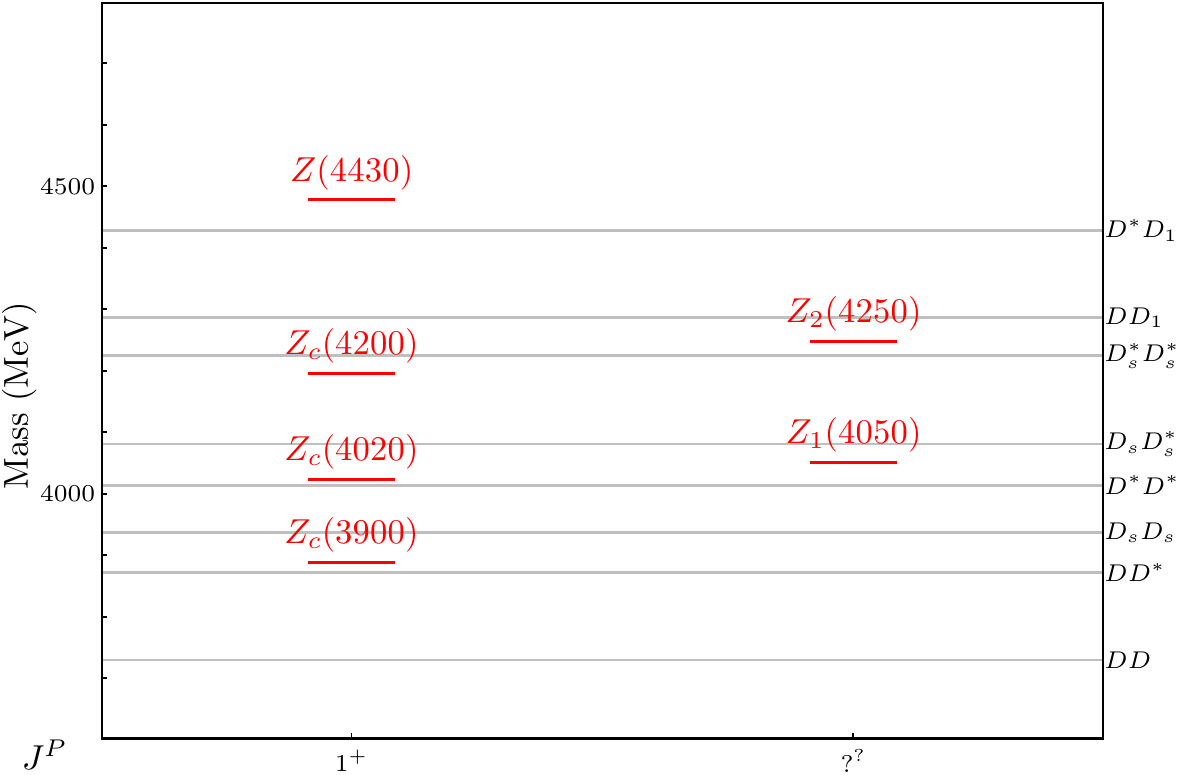}
\caption{Charmonium sector. In the upper panel, we show ordinary charmonia 
and neutral exotic states, in the lower panel charged exotic states. 
Black lines represent observed charmonium levels, blue lines represent 
predicted levels according to Radford and Repko~\cite{Radford:2007vd}, 
red lines are exotic states. The open charm thresholds are reported on the right.}
 \label{fig:charmonium_levels}
\end{figure}
 \begin{table}[p]\small\centering
\begin{tabular}{lrcllc}\hline\hline
State & $M$ (\mev) & $\Gamma$ (\mev) & $J^{PC}$ & Process (mode) & Experiment (\#$\sigma$) \\
\hline
      $X(3872)$ & $3871.69\pm0.17$ & $<1.2$ & $1^{++}$ & See \tablename{\ref{tab:xdecays}, \ref{tab:xdecays2}} & See \tablename{\ref{tab:xdecays}, \ref{tab:xdecays2}}\\
      $Z_c(3900)^+$ & $3888.4\pm1.6$ & $27.9\pm2.7$ & $1^{+-}$ &
      $Y(4260)\to\pi^-(D\bar{D}^*)^+$ & \bes~\cite{Ablikim:2013xfr,Ablikim:2015swa} ($>10$) \\
      & & & &
      $Y(4260)\to\pi^-(\pi^+\jpsi)$ & \bes~\cite{Ablikim:2013mio} (8), \belle~\cite{Liu:2013dau} (5.2) \\
      & & & & & NU group~\cite{Xiao:2013iha} ($3.5$) \\
      $Z_c(3900)^0$ & $3893.6\pm 3.7$ & $31\pm 10$ & $1^{+-}$ &
      $Y(4260)\to\pi^0(D\bar{D}^*)^0$ & \bes~\cite{Ablikim:2015gda} ($10$) \\
      & & & &
      $Y(4260)\to\pi^0(\pi^0\jpsi)$ & \bes~\cite{Ablikim:2013mio} ($10.4$)\\
      & & & & & NU group~\cite{Xiao:2013iha} ($5.7$) \\
      $Z_c^\prime(4020)^+$ & $4023.9\pm2.4$ & $10\pm6$ & $1^{+-}$ &
      $e^+e^-\to\pi^-(\pi^+h_c)$ & \bes~\cite{Ablikim:2013wzq} ($8.9$) \\
      & & & &
      $e^+e^-\to\pi^-(D^*\bar{D}^*)^+$ & \bes~\cite{Ablikim:2013emm} ($10$) \\
      $Z_c^\prime(4020)^0$ & $4024.5\pm3.1$ & $23\pm6 \pm 1$ & $1^{+-}$ &
      $e^+e^-\to\pi^0(\pi^0h_c)$ & \bes~\cite{Ablikim:2014dxl} ($5$) \\
      & & & &
      $e^+e^-\to\pi^0(D^*\bar{D}^*)^0$ & \bes~\cite{Ablikim:2015vvn} ($5.9$) \\

      $\chi_{c0}(3915)$ & $3918.4\pm1.9$ & $20\pm5$ & $0^{++}$ &
      $B\to K(\omega \jpsi)$ & \belle~\cite{Abe:2004zs} ($8$), \babar~\cite{Aubert:2007vj,delAmoSanchez:2010jr} ($19$) \\
      & & & &
      $e^+e^-\to e^+e^-(\omega \jpsi)$ & \belle~\cite{Uehara:2009tx} ($7.7$), \babar~\cite{Lees:2012xs} ($7.6$) \\
      $X(3940)$ & $3942^{+9}_{-8}$ & $37^{+27}_{-17}$ & $(0^{-+})$ &
      $e^+e^-\to \jpsi\,(D\bar{D}^*)$ & \belle~\cite{Abe:2007jn,Abe:2007sya} ($6$) \\
      \multirow{2}{*}{($Y(4008)$?)} & $3891\pm42$ & $255\pm42$ & \multirow{2}{*}{$1^{--}$}  &
      \multirow{2}{*}{$e^+e^-\to (\pi^+\pi^-\jpsi)$} & \belle~\cite{Yuan:2007sj,Liu:2013dau} ($7.4$) \\
        & $3813^{+62}_{-97}$ & $477^{+78}_{-65}$ & &
       & \bes~\cite{Ablikim:2016qzw} (np) \\
      $Z(4050)^+$ & $4051^{+24}_{-43}$ & $82^{+51}_{-55}$ & $?^{?+}$ &
      $\bar{B}^0\to K^-(\pi^+\chi_{c1})$ & \belle~\cite{Mizuk:2008me} ($5.0$), \babar~\cite{Lees:2011ik} ($1.1$) \\
      ($Z(4050)^+$?) & $4054 \pm 3$ & $45 \pm 13$ & $?^{?-}$ & $\epem \to (\psiprime\,\pi^+)\,\pi^-$ & \belle~\cite{Wang:2014hta} ($3.5$) \\
      $X(4140)$ & $4146.5^{+6.4}_{-5.3}$ & $83^{+30}_{-25}$ & $1^{++}$ &
      $B^+ \to (\jpsi \phi ) K^+$ & \lhcb~\cite{Aaij:2016nsc,Aaij:2016iza} ($8.4$), see \tablename{\ref{tab:ccss}}\\
      $X(4160)$ & $4156^{+29}_{-25}$ & $139^{+113}_{-65}$ & $(0^{-+})$ &
      $e^+e^-\to \jpsi\,(D^*\bar{D}^*)$ & \belle~\cite{Abe:2007sya} ($5.5$) \\
     $Z(4200)^+$ & $4196^{+35}_{-30}$ & $370^{+99}_{-110}$ & $1^{+-}$ &
     $\bar{B}^0\to K^-(\pi^+\jpsi)$ & \belle~\cite{Chilikin:2014bkk} ($7.2$) \\
     $Y(4220)$ & $4218.4 \pm 4.1$ & $66.0 \pm 9.0$ & $1^{--}$ &
     $e^+e^-\to (\pi^+\pi^- h_c)$ & \bes~\cite{Ablikim:2016qzw} (np) \\
     $Y(4230)$ & $4230 \pm 8$ & $38\pm12$ & $1^{--}$ &
     $e^+e^-\to (\chi_{c0}\omega)$ & \bes~\cite{Ablikim:2014qwy} ($>9$) \\
      $Z(4250)^+$ & $4248^{+185}_{-45}$ & $177^{+321}_{-72}$ & $?^{?+}$ &
      $\bar{B}^0\to K^-(\pi^+\chi_{c1})$ & \belle~\cite{Mizuk:2008me} ($5.0$), \babar~\cite{Lees:2011ik} ($2.0$) \\
      $Y(4260)$ & $4251\pm9$ & $120\pm12$ & $1^{--}$ &
      $e^+e^-\to (\pi\pi \jpsi)$ & \babar~\cite{Aubert:2005rm,Lees:2012cn} ($8$), \cleo~\cite{Coan:2006rv,He:2006kg} (11) \\
      & & & & & \belle~\cite{Yuan:2007sj,Liu:2013dau} ($15$), \bes~\cite{Ablikim:2013mio} (np) \\
      & & & &
      $e^+e^-\to(f_0(980)\jpsi)$ & \babar~\cite{Lees:2012cn} (np), \belle~\cite{Liu:2013dau} (np) \\
      & & & &
      $e^+e^-\to(\pi^-Z_c(3900)^+)$ & \bes~\cite{Ablikim:2013mio} ($8$), \belle~\cite{Liu:2013dau} ($5.2$) \\
      & & & &
      $e^+e^-\to(\gamma\,X(3872))$ & \bes~\cite{Ablikim:2013dyn} ($5.3$) \\
      $X(4274)$ & $4273^{+19}_{-9}$ & $56^{+13}_{-16}$  & $1^{++}$ &
      $B^+ \to (\jpsi \phi ) K^+$ & \lhcb~\cite{Aaij:2016nsc,Aaij:2016iza} ($6.0$), see \tablename{\ref{tab:ccss}}\\
      ($X(4350)$?) & $4350.6^{+4.6}_{-5.1}$ & $13^{+18}_{-10}$ & $0/2^{?+}$ &
      $e^+e^-\to e^+e^-(\phi \jpsi)$ & \belle~\cite{Shen:2009vs} ($3.2$) \\
      $Y(4360)$ & $4346.2\pm6.3$ & $102.3\pm9.9$ & $1^{--}$ &
      $e^+e^-\to(\pi^+\pi^-\psiprime)$ & \belle~\cite{Wang:2007ea,Wang:2014hta} ($8$), \babar~\cite{Lees:2012pv} (np) \\
      $P_c(4380)^+$ & $4380\pm 30$ & $205\pm88$ & $3/2^{-}$ &
      $\Lzb\to K^-(\jpsi\,p)$ & \lhcb~\cite{Aaij:2015tga} ($9$) \\
     $Y(4390)$ & $4391.6\pm6.4$ & $139.5\pm16.1$ & $1^{--}$ &
     $e^+e^-\to (\pi^+\pi^- h_c)$ & \bes~\cite{Ablikim:2016qzw} ($10$) \\
      \hline\hline
\end{tabular}
\caption{Summary of quarkoniumlike mesons and pentaquarks. 
For charged states, the $C$-parity is given for the neutral members of the corresponding 
isotriplets. The states in parentheses are mere evidence of states, with low significance. The controversy on the $Y(4008)$ will be discussed in \sectionname{~\ref{sec:vectors}}. Also, new data about the $Y(4260)$ by \bes~\cite{Ablikim:2016qzw} are not included in the average.  The signature assignments in parentheses are just tentative, or deduced by the measurement of the charged partners.}
\label{tab:allexp} 
\end{table}

\begin{table}[t]\small\centering
\begin{tabular}{lrcllc}\hline\hline
State & $M$ (\mev) & $\Gamma$ (\mev) & $J^{PC}$ & Process (mode) & Experiment (\#$\sigma$) \\
\hline
     $Z(4430)^+$ & $4478\pm17$ & $180\pm 31$ & $1^{+-}$ &
      $\bar{B}^0\to K^-(\pi^+\psiprime)$ & \belle~\cite{Mizuk:2009da,Chilikin:2013tch} ($6.4$), \babar~\cite{Aubert:2008aa} ($2.4$) \\
      & & & & & \lhcb~\cite{Aaij:2014jqa} ($13.9$) \\
      & & & &
      $\bar{B}^0\to K^-(\pi^+\jpsi)$ & \belle~\cite{Chilikin:2014bkk} ($4.0$) \\
       $P_c(4450)^+$ & $4449.8\pm 3.0$ & $39\pm 20$ & $5/2^{+}$ &
      $\Lzb\to K^-(\jpsi\,p)$ & \lhcb~\cite{Aaij:2015tga} ($12$) \\
      $X(4500)$ & $4506^{+16}_{-19}$ & $92^{+30}_{-29}$  & $0^{++}$ &
      $B^+ \to (\jpsi \phi ) K^+$ & \lhcb~\cite{Aaij:2016nsc,Aaij:2016iza} ($6.1$), see \tablename{\ref{tab:ccss}}\\
      $Y(4630)$ & $4634^{+9}_{-11}$ & $92^{+41}_{-32}$ & $1^{--}$ &
      $e^+e^-\to(\Lambda_c^+\bar{\Lambda}_c^-)$ & \belle~\cite{Pakhlova:2008vn} ($8.2$) \\
      $Y(4660)$ & $4657\pm11$ & $70\pm11$ & $1^{--}$ &
      $e^+e^-\to(\pi^+\pi^-\psiprime)$ & \belle~\cite{Wang:2007ea,Wang:2014hta} ($5.8$), \babar~\cite{Lees:2012pv} ($5$) \\ 
      $X(4700)$ & $4704^{+17}_{-26}$ & $120^{+52}_{-45}$ & $0^{++}$ &
      $B^+ \to (\jpsi \phi ) K^+$ & \lhcb~\cite{Aaij:2016nsc,Aaij:2016iza} ($5.6$), see \tablename{\ref{tab:ccss}}\\
\hline
($X(5568)$?) & $5567.8^{+3.0}_{-3.4}$ & $21.9^{+8.1}_{-6.9}$ & $?^?$ & $p\bar p \to (B_s^0 \pi^\pm) + \text{All}$ & \Dzero~\cite{D0:2016mwd} ($3.9$), \lhcb~\cite{LHCb:2016ppf} (not seen)\\
      $Z_b(10610)^+$ & $10607.2\pm2.0$ & $18.4\pm2.4$ & $1^{+-}$ &
      $\Upsilon(5S)\to\pi(\pi\Upsilon(nS))$ & \belle~\cite{Belle:2011aa} ($>10$) \\
      & & & & 
      $\Upsilon(5S)\to\pi^-(\pi^+h_b(nP))$ & \belle~\cite{Belle:2011aa} ($16$) \\
      & & & & 
      $\Upsilon(5S)\to\pi^-(B\bar{B}^*)^+$ & \belle~\cite{Adachi:2012cx,Garmash:2015rfd} ($9.3$) \\
      $Z_b(10610)^0$ & $10609\pm 4 \pm 4$ & $18.4$ (fixed) & $(1^{+-})$ &
      $\Upsilon(5S)\to\pi(\pi\Upsilon(nS))$ & \belle~\cite{Krokovny:2013mgx} ($6.1$) \\
      $Z_b(10650)^+$ & $10652.2\pm1.5$ & $11.5\pm2.2$ & $1^{+-}$ &
      $\Upsilon(5S)\to\pi^-(\pi^+\Upsilon(nS))$ & \belle~\cite{Belle:2011aa} ($>10$) \\
      & & & & 
      $\Upsilon(5S)\to\pi^-(\pi^+h_b(nP))$ & \belle~\cite{Belle:2011aa} ($16$) \\
      & & & & 
      $\Upsilon(5S)\to\pi^-(B^*\bar{B}^*)^+$ & \belle~\cite{Adachi:2012cx,Garmash:2015rfd} ($8.1$) \\
      \hline\hline
\end{tabular}
\caption{({\em Continued}) Summary of quarkoniumlike mesons and pentaquarks. 
For charged states, the $C$-parity is given for the neutral members of the corresponding 
isotriplets. The states in parentheses are mere evidence of states, with low significance. The signature assignments in parentheses are just tentative, or deduced by the measurement of the charged partners.}
\label{tab:allexp3} 
\end{table}

\section{A travel guide to experimental results}
\label{sec:experiment}
We present here a short summary of the most important experimental results in the exotic \XYZP sector. This chapter does not aim to cover entirely  the large number of papers published on this topic, but rather to give a synthetic review of the aspects which appear more relevant from the theoretical point of view. In doing so, we will not follow the chronological order.
Whenever available, we will calculate averages using the new published measurement, to update the results shown in the PDG 2014. For the $X(3872)$, we show the result of a global fit to all available data, to provide an estimate of the absolute branching fractions.

The landscape of exotic resonances is summarized in \figurename{~\ref{fig:charmonium_levels}}, and in \tablesname{\ref{tab:allexp} and \ref{tab:allexp3}}.
Some states, 
like the $X(3872)$ or the $X(3915)$, have more or less the correct mass 
and quantum numbers of some missing ordinary 
charmonia; on the other hand, in the vector sector we have much more levels 
than expected. In any case, the decay pattern of all these states is not 
compatible with charmonia predictions, and so it calls for  exotic interpretations.
No ordinary assignment is possible for the charged charmoniumlike and bottomoniumlike states, or for the new hidden-charm pentaquarks. 

Depending on the available energies and colliding particles, 
the various experiments can exploit different production modes of the exotic states.
Some of these can constrain the quantum numbers assignments. A generic state $X$ 
can be produced
\begin{itemize}
 \item Directly with $e^+e^- \to X$, 
or in association with Initial State Radiation (ISR) which lowers the center 
of mass energy, $e^+e^- \to e^+e^- \gamma_\text{ISR} \to X \gamma_\text{ISR}$. The 
quantum numbers must be the same as the photon, $J^{PC} = 1^{--}$. The $Y$ states discovered with this technique are discussed in \sectionname{\ref{sec:vectors}}.
The $\tau$-$c$ and the $B$-factories can directly explore the charmonium and bottomonium vectors, respectively. This is what \bes did, collecting a large data sample at the $Y(4260)$ peak. 
The $B$-factories (\babar and \belle) can also perform ISR analyses on the charmonium sector.
  \item In the fusion of two quasi-real photons, 
$e^+e^- \to e^+ e^- \gamma\gamma \to e^+ e^- X$, where $e^+$ and $e^-$ are 
scattered at a small angle and are lost in the beampipe; the signal events have no 
tracks and neutral particles but the daughters of $X$. If the photons are 
quasi-real, Landau-Yang theorem holds~\cite{Yang:1950rg}, and $J \neq 1$; 
moreover $C= +$ is constrained. The $B$-factories discovered the $\chi_{c0}(3915)$ and the $X(4350)$ with this technique (\sectionname{\ref{sec:others}}). 
  \item In double charmonium production, usually $e^+e^- \to \jpsi X$, 
which constrains $X$ to have $C=+$. This allows for a partial reconstruction of the $X$ state. The $B$-factories discovered the $X(3940)$ and the $X(4160)$ with this technique (\sectionname{\ref{sec:others}}). 
\item In $B$ decays, which allow $X$ to have any $J^{PC}$, 
albeit low values of the spin are preferred. Both exclusive and inclusive analyses can be performed. The $B$-factories provide a clean environment, but the huge amount of data collected at \lhcb allow for very-high statistics, multidimensional analyses, in particular if no neutral particles are involved. The $X(3872)$ (\sectionname{\ref{sec:X3872}}), the $Z(4430)$ (\sectionname{\ref{sec:Z4430}}) and the $[cs][\bar c \bar s]$  (\sectionname{\ref{sec:ccss}}) states have been discovered in $B$ decays.
\item In $B_c$ and bottom baryon decays. \lhcb is the main performer in these channels. The general-purpose detectors \cms and \atlas can in principle look for these channels, but they suffer from the lack of particle identification, and a much larger background. These channels have led to the discovery of the pentaquarks, see \sectionname{\ref{sec:pentaexp}}.
\item In $\psi(nS)$, $Y(4260)$ and $\Upsilon(nS)$ decays. 
The $Z_c$s and $Z_b$s states have been discovered in this way by \bes and \belle, respectively (see \sectionname{\ref{sec:Zc} and \ref{sec:Zb}}).
\item In inclusive analyses at hadron colliders. Until now, only the $X(3872)$ and the $X(4140)$ have been observed inclusively. It is usually possible to separate the non-prompt (due to the decay of some long-lived meson or baryon) and prompt fractions. These analyses have been performed by \cdf, \Dzero, \lhcb, \cms and \atlas, which testifies for the growing interest the exotic charmonium physics has raised in the high energy physics community.
\end{itemize}
Although the \cleo collaboration officially no longer exists, a group based in Northwestern University still publishes analyses based on \cleoc data. We will quote their results as the NU group.

The readers more interested in experimental details are invited to look at~\cite{Chen:2016qju,Esposito:2014rxa,pbf,Faccini:2012pj,Drenska:2010kg}. Prospects for future searches have been recently exposed in~\cite{Briceno:2015rlt,Lutz:2015ejy}.

\subsection{The \texorpdfstring{$X(3872)$}{X(3872)}}
\label{sec:X3872}
\begin{figure}[t]
\begin{center}
\includegraphics[width=.7\textwidth]{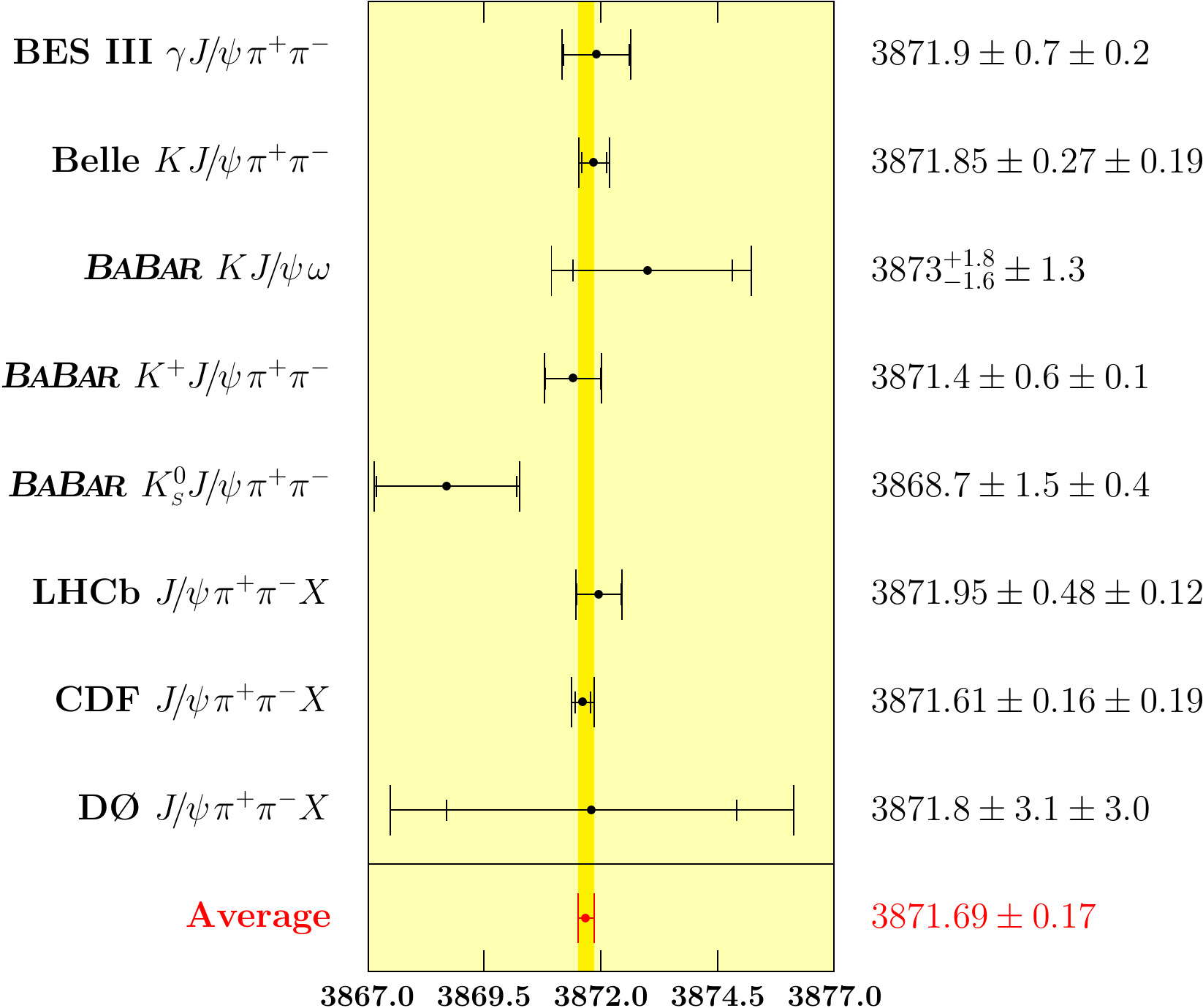} 
 \caption{Averaged mass of the $X(3872)$, according to the measurements taken into account in the PDG~\cite{pdg}.}
    \label{FIG:XYZ:X-MASS}
    \end{center}
\end{figure}
The discovery of an unexpected charmonium state in 2003 gave birth to the long saga of exotic quarkonia. The \belle collaboration announced the observation of a narrow resonance in the $B \to K (\jpsi\,\pi^+\pi^-)$ channel, dubbed $X(3872)$ to stress its mysterious nature.~\footnote{It is worth noticing that the \belle paper on the $X(3872)$ discovery~\cite{Choi:2003ue} has $50\%$ more citations than the one on the CP violation in the $B^0$ system~\cite{Abe:2001xe}, for which the $B$-factories were designed and financed. This should give an idea of the broad interest this new field piqued in the high energy physics community.} 
In this decade, this state has been confirmed in many different production channels: in $B$ decays~\cite{Aubert:2004ns}, in inclusive  $p\bar{p}$~\cite{Acosta:2003zx,Abazov:2004kp} and $pp$ collisions~\cite{Aaij:2011sn,Chatrchyan:2013cld}, and more recently in the radiative decay of the $Y(4260)$~\cite{Ablikim:2013dyn}.
The measured mass is $M=(3871.69 \pm 0.17) \mev$~\cite{pdg}, while the best available upper bound on the width is given by $\Gamma < 1.2\mev$ at 90\% C.L.~\cite{Choi:2011fc}. The mass of the state is some 100\mev lighter than expected for a $\chi_{c1}(2P)$ states, although this discrepancy is reduced in some calculations with unquenched light quarks~\cite{Ferretti:2013faa}. The $\pi^+\pi^-$ invariant mass distribution~\cite{Choi:2003ue,Abulencia:2005zc} and angular analyses~\cite{Abulencia:2006ma} showed that the $\pi^+\pi^-$ amplitude is dominated by the $\rho$ meson. If this state were an ordinary charmonium, the transition via an isovector light meson should be highly suppressed.
The size of the isospin breaking is indeed signaled by the ratio between the radiative decay width~\cite{pdg,Abe:2005ix,Aubert:2006aj,Aubert:2008rn,Bhardwaj:2011dj}, with respect to the isospin violating mode, 
\begin{equation}
 \frac{\Gamma\left(X(3872)\to\jpsi\gamma\right)}{\Gamma\left(X(3872)\to\jpsi\rho(\to \pi^+\pi^-)\right)} = 0.24 \pm 0.05,
\end{equation}
to compare, for example, to an expected ratio of $O(10)$ if the $X(3872)$ were the $\chi_{c1}(2P)$~\cite{Olsen:2004fp}. Finally, the ratio~\cite{delAmoSanchez:2010jr}
\begin{equation}
 \frac{\Gamma\left(X(3872)\to\jpsi\omega(\to \pi^+\pi^-\pi^0)\right)}{\Gamma\left(X(3872)\to\jpsi\rho(\to \pi^+\pi^-)\right)} = 0.8 \pm 0.3,
\end{equation}
manifestly shows an isospin violation of $O(1)$.

The observation of the radiative decay, and the non-observation of $X(3872)\to \chi_{c1}\gamma$~\cite{Choi:2003ue}, established the $C=+$ assignment. 
For $J^P$, the first angular analyses of the $\jpsi\,\pi^+\pi^-$ decay by \belle~\cite{Abe:2005iya} and \cdf~\cite{Abulencia:2006ma} were able to rule out all but the $1^{++}$ and $2^{-+}$ assignments. The latter could not be excluded because of the two independent helicity amplitudes for $X(2^{-+})\to\jpsi\rho$, which turn out in an additional complex parameter, hard to constrain in inclusive $X(3872)$ production. On the other hand, the axial assignment was preferred by theoretical models. The analysis of the $\jpsi \omega$ invariant mass distribution by \babar~\cite{delAmoSanchez:2010jr} favored the $2^{-+}$ hypothesis, and stimulated a discussion on its theoretical feasibility~\cite{Burns:2010qq,Hanhart:2011tn,Faccini:2012zv,Faccini:2012pj,Braaten:2013poa}. The negative search of $X(3872)$ in $\gamma\gamma$ fusion at \cleo made the pseudotensor hypothesis unlikely.
Later, \belle updated the angular analysis with the final dataset~\cite{Choi:2011fc}, still not being able to rule out $2^{-+}$. 
Finally, \lhcb published two analyses of a large $B^+ \to K^+ X(3872)$ sample~\cite{Aaij:2013zoa,Aaij:2015eva}, respectively restricting or not to the lowest available partial wave. These studies are based on an event-by-event likelihood ratio test of the two
hypotheses on the full 5D angular distribution, and favor the $1^{++}$ over $2^{-+}$ at $8\sigma$ level (\figurename{~\ref{fig:xlikel}}, left panel).

In~\figurename{~\ref{FIG:XYZ:X-MASS}} we report the list of the mass measurements used for the PDG average. The current world average takes into account the $X(3872)\to \jpsi \rho,\omega$ decays. 
The mass coincides with the $D^0 D^{*0}$ threshold within errors, which gave birth to all the speculations about the molecular nature of this state --- see \sectionname{\ref{mol2}}. The most accurate estimate of the `binding energy' is~\cite{Tomaradze:2015cza}
\begin{equation}
 B = M_D + M_{D^*} - M_X = (3 \pm 192)\kev.
\end{equation}
A precise determination of the mass, together with the precise determination of the width, will be able to constrain the loosely bound molecule hypothesis~\cite{Polosa:2015tra}. The mass observed in the $X(3872)\to\Dstarz\Dzb$ decay~\cite{Gokhroo:2006bt,Aubert:2007rva,Adachi:2008su} is significantly higher, $M = (3873.8 \pm 0.5)\mev$.
The possibility 
that $X(3875)\to\Dstarz\Dzb$ and $X(3872)\to\jpsi\,\pi^+\pi^-$ are distinct particles, for example the two almost-degenerate $[cu][\bar c \bar u]$ and $[cd][\bar c \bar d]$ tetraquarks, was discussed in~\cite{Maiani:2007vr}. Some papers~\cite{Artoisenet:2010va,Hanhart:2010wh,Hanhart:2011jz} argued instead that, since the $\Dstarz$ is in general off-shell, only the detailed study of the $\pi^0\Dz\Dzb$ and $\gamma\Dz\Dzb$ lineshapes can distinguish between a below- and above-threshold $X(3872)$. Moreover, in order to improve the 
resolution, the experimental analyses constrain the $\Dstar$ mass, 
and this yields to a reconstructed $X(3872)$ mass which is above threshold by construction. Because of these biases, this channel has been dropped from the mass averages in the PDG. 

\begin{figure}[t]
\begin{center}
\vcenteredhbox{\includegraphics[width=.40\textwidth]{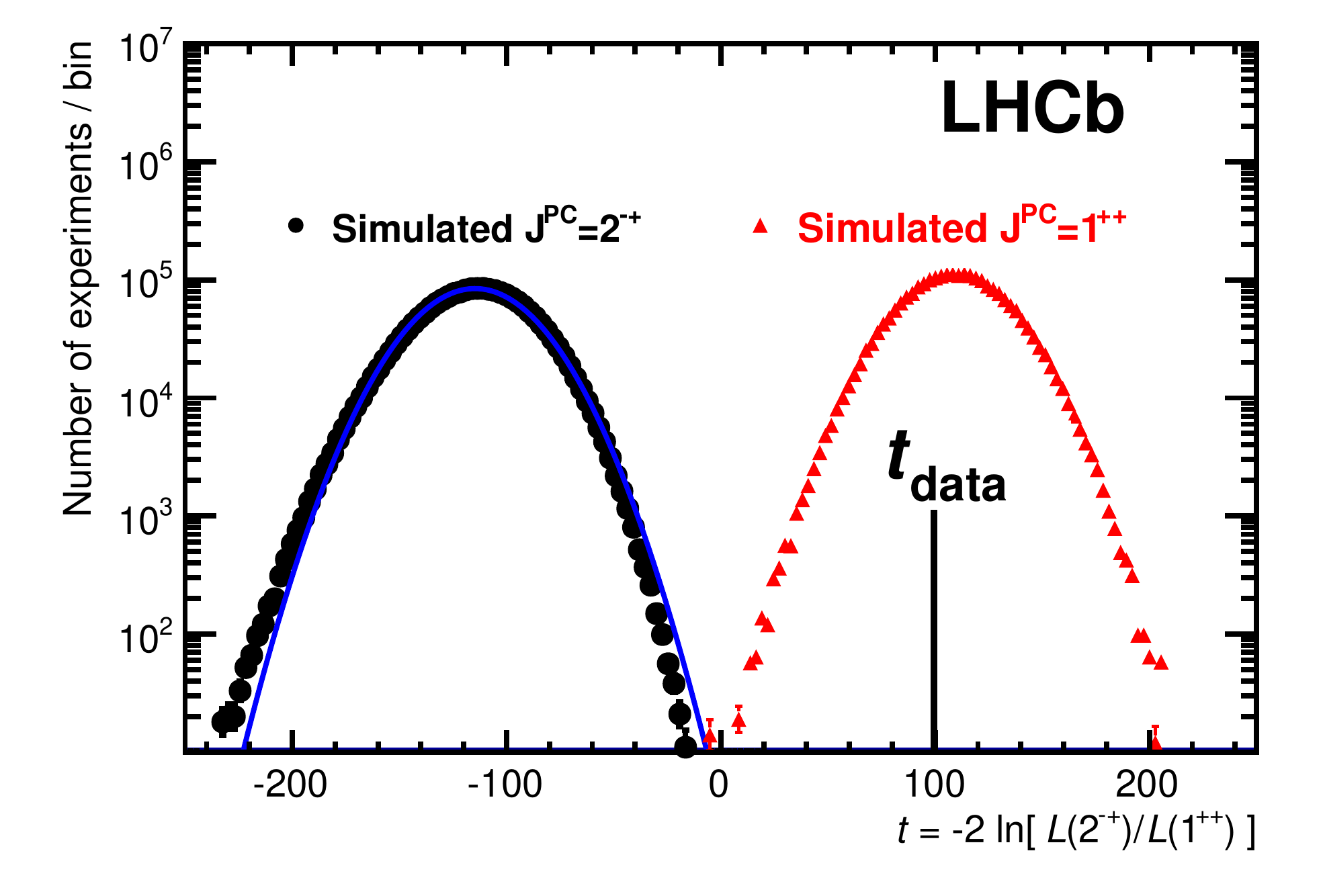}}
\vcenteredhbox{\includegraphics[width=.59\textwidth]{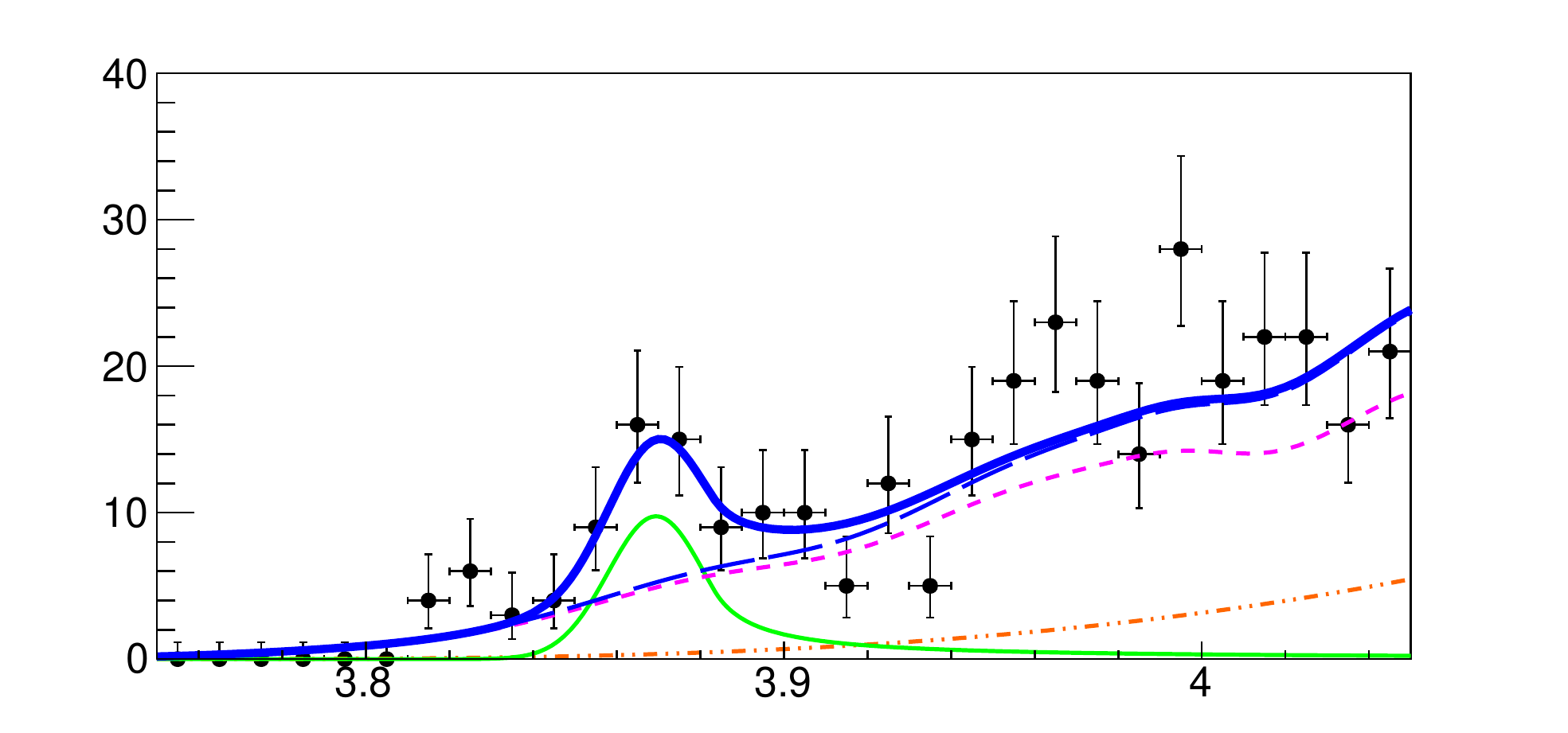}}
 \caption{Left panel: Likelihood ratio test for the spin of the $X(3872)$, favoring the $1^{++}$ assignment, by \lhcb~\cite{Aaij:2013zoa}. Right panel: mass projection of the $\psiprime\,\gamma$ invariant mass, where the $X(3872)$ is evident, by \lhcb~\cite{Aaij:2014ala}.}
    \label{fig:xlikel}
    \end{center}
\end{figure}

The measurement of radiative decays can be useful to constrain some of the available models.
\babar~\cite{Aubert:2008rn} and \lhcb~\cite{Aaij:2014ala} observed a $X(3872)$ signal in the $\psiprime \gamma$ final state (\figurename{~\ref{fig:xlikel}}, right panel), giving a relative branching fraction of 
\begin{equation}
\frac{\Gamma\left(X(3872)\to\psiprime\gamma\right)}{\Gamma\left(X(3872)\to\jpsi\gamma\right)} = 2.6 \pm 0.6.
\end{equation}
On the other hand, \belle saw no significant signal in this channel, and put a $< 2.1$ 90\% \CL upper limit on the ratio.
In the original molecular model proposed by Swanson, the radiative transition to $\psiprime$ was predicted to be suppressed~\cite{Swanson:2004pp,Swanson:2006st}. In an analysis based on NREFT (see \sectionname{\ref{sec:NREFT}}), Hanhart~\etal claim that the radiative decays depend on the short-range structure of the molecule, and give up on predicting the ratio, thus `reconciling' the molecular hypothesis with experimental data. A calculation by Braaten~\cite{Braaten:2007sh} based on universal low energy scattering theory (see \sectionname{\ref{ascattering}}), predicted
\begin{equation} 
\Gamma\left(X(3872) \to p\bar p\right) = \left(\frac{\Lambda}{m_\pi}\right)^2 \sqrt{\frac{B}{600\kev}}\times 28\ev
\end{equation}
where $\Lambda$ is the scale up to which one can neglect the effective range of the interaction (say $\sim 3m_\pi$), and $B$ the molecule binding energy. Using $B = 190\kev$  (the available upper limit) and $\Gamma \gtrsim \Gamma(\Dstarz) \sim 70\kev$, one gets $\BR\left(X\to p \bar p\right) \lesssim 2\times 10^{-3}$, which coincides with the experimental bound by \lhcb~\cite{Aaij:2013rha,Aaij:2016kxn}. Further improvement of the precision on this branching ratio, together with a better determination of the binding energy and width of the $X(3872)$ is needed to constrain the molecular hypothesis.
The decay $Y(4260) \to \gamma X(3872)$ has been observed by \bes~\cite{Ablikim:2013dyn}. We will discuss this channel in the \sectionname{\ref{sec:vectors}}.

Other production mechanisms like $B^0\to K^+ \pi^- X(3872)$
have also been studied. Such decays are seen~\cite{Adachi:2008te,Bala:2015wep}, but the non-resonant $K\pi$ dominates.
This is in contrast to ordinary charmonium states,
where the $B\to\Kstar\ccbar$ and $B\to K\ccbar$ branching fractions are comparable,
and \Kstar\ dominates over non-resonant $K\pi$. 
Also, a preliminary analysis by COMPASS shows the production of $\psiprime$ and $X(3872)$ in the $\mu^+ N \to \mu^+ (\jpsi\,\pi^+ \pi^-)\,\pi^\pm N^\prime$ process~\cite{compass}. The ratio of events for the two states is $N(X(3872)) / N(\psiprime) = 0.9 \pm 0.4$. If the production mechanism is assumed to be the same (a vector meson dominated virtual photon), this ratio equals $\Gamma(X(3872) \to \jpsi\,\pi^+\pi^-) \times \BR(X(3872) \to \jpsi\,\pi^+\pi^-) / \Gamma(\psiprime \to \jpsi\,\pi^+\pi^-) \times \BR(\psiprime \to \jpsi\,\pi^+\pi^-)$, and one might use this to constrain the total width of the $X(3872)$.

In a decade of experimental activity, the $X(3872)$ has been searched in a plethora of different final states, some of which allowed us to establish the correct quantum numbers. 
In most of these analyses, the $X(3872)$ is produced in $B$ decays, so that it is not possible to disentangle the production branching fraction $\BR(B \to K X)$ from the final $\BR(X \to f)$. These  factors cancel in the ratio, but some model have predictions for the absolute branching fraction. An inclusive \babar analysis of $B^+ \to K^+ (c\bar c)$ states provides an upper limit for the production branching fraction, $\BR(B \to K X(3872)) < 3.2\times 10^{-4}$ at 90\% \CL~\cite{Aubert:2005vi}. 
Combining the likelihood from the
measurements of the product branching fractions in the observed channels, the $B \to X(3872) K$
upper limit, the $X(3872)$ width distribution~\cite{Liu:2013dau}, we extract
the likelihood for the absolute $X(3872)$ branching fractions and the widths in each of the decay modes with a bayesian procedure. 
We then use these distributions to set limits on the
not observed channels. The full shape of the experimental likelihoods was used whenever
available, while gaussian errors and poissonian counting distributions have been assumed
elsewhere. We summarize the results in \tablename{\ref{tab:xdecays} and \ref{tab:xdecays2}}. The plot of the likelihoods and the code are available at~\cite{miofit}.

\begin{figure}[t]
\centering
\includegraphics[width=.50\textwidth]{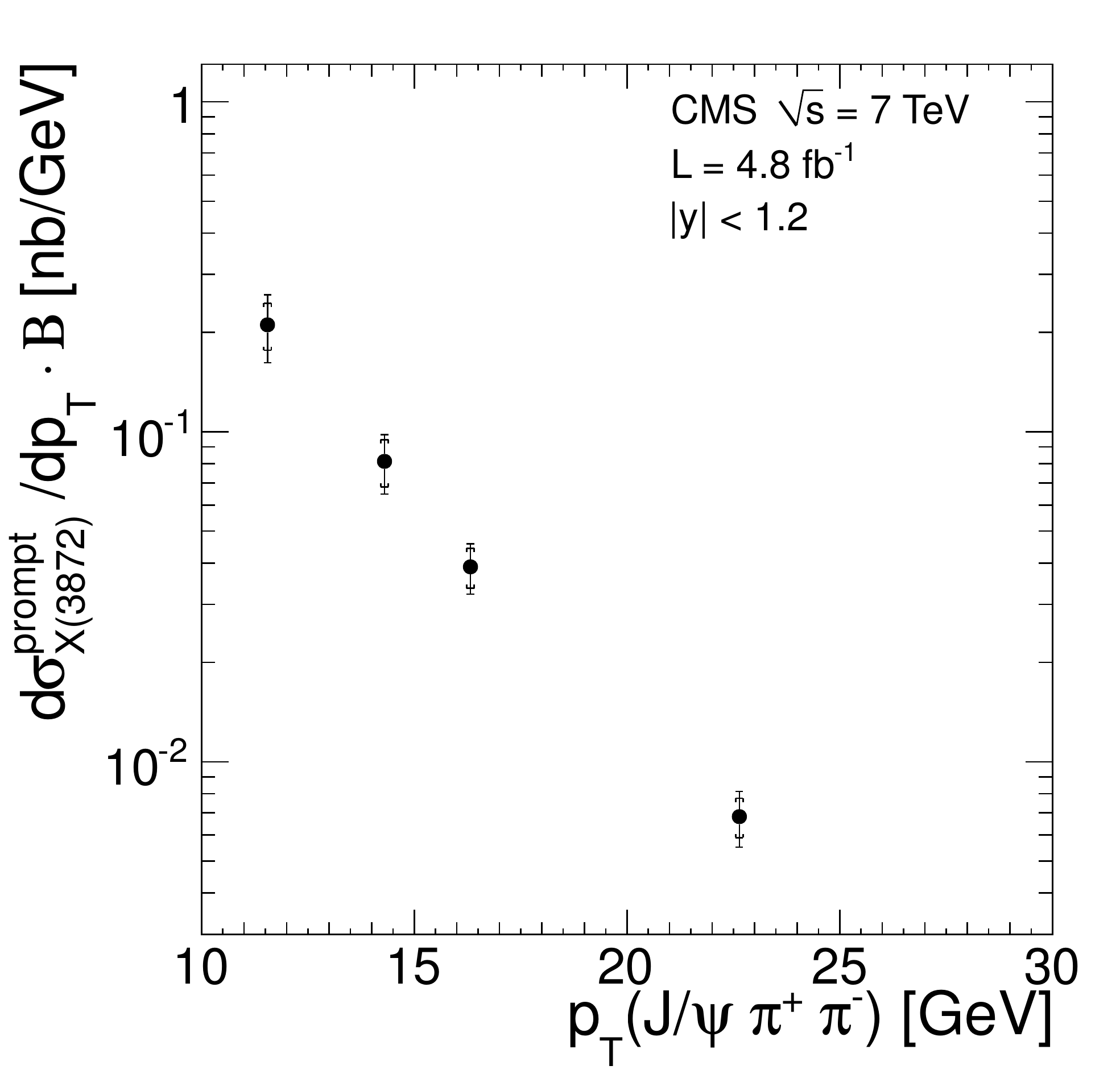}\hspace{1cm}
\includegraphics[width=.43\textwidth]{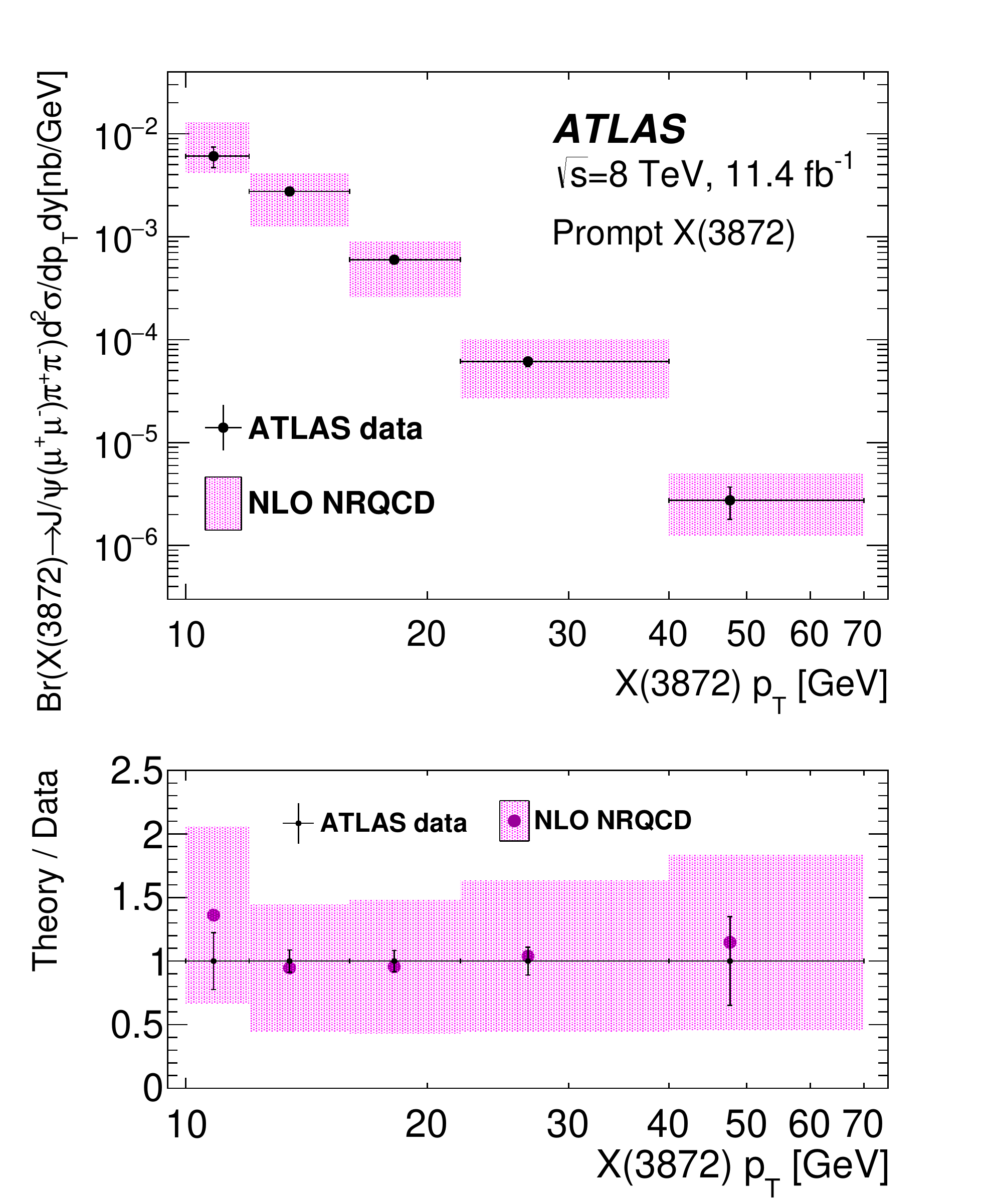}
\caption{Measured differential cross section for prompt $X(3872)$ production times branching
fraction $\BR( X(3872) \to \jpsi\,\pi^+\pi^-)$ as a function of $p_T$, from \cms~\cite{Chatrchyan:2013cld} (left panel), and \atlas~\cite{ATLAS:2016jtu} (right panel).}
 \label{fig:xcmscrosssection}
\end{figure}

The searches for neutral and charged partner states of the $X(3872)$ have been motivated by the predictions of the tetraquark model. %
For example, Ref.~\cite{Maiani:2004vq} proposed that the $X$ states produced in $B^+$ and $B^0$ decays were different. If so, the masses of the two $X$ should differ by a few{\mev}s. 
The analyses by \babar~\cite{Aubert:2005zh,Aubert:2008gu}
and \belle~\cite{Adachi:2008te,Choi:2011fc} distinguish the two samples, and give for the mass difference $M(X\,|\,\Bp\to\Kp X) - M(X\,|\,\Bz\to\Kz X) = (0.2 \pm 0.8)\mev$, and for the ratio of product branching fractions
\begin{equation}
  \frac	{\BR(\Bz\to\Kz X) \times \BR(X\to\pi^+\pi^-\jpsi)}{\BR(\Bp\to\Kp X) \times \BR(X\to\pi^+\pi^-\jpsi)} =0.47 \pm 0.13.
\end{equation}
Similarly, the inclusive analysis by \cdf~\cite{Aaltonen:2009vj},
of the $\jpsi\,\pi^+\pi^-$ spectrum, gave no evidence for
a second neutral state, setting an upper limit on the mass difference of
$3.6\mev$ at the 95\% C.L. 
The charged partners  have been searched at the $B$-factories~\cite{Aubert:2004zr,Choi:2011fc}, but no evidence is seen. The strictest limits on the product branching
fractions are
\begin{subequations}
\begin{align}
 \BR(\Bzb\to\Km X^+) \times \BR(X^+ \to \rho^+\jpsi)%
 & < 4.2 \times 10^{-6},\\
 \BR(\Bp\to\Kz X^+) \times \BR(X^+ \to \rho^+\jpsi)%
  & < 6.1 \times 10^{-6},%
\end{align}
\end{subequations}
but they are not so severe with respect to the neutral channel~\cite{Aubert:2008gu,Choi:2011fc},
  $\BR(\Bp\to\Kp X) \times \BR(X \to \rho^0\jpsi) = (8.6 \pm 0.81) \times 10^{-6}$. Recently, Voloshin proposed that these charged states could mix with the $Z_c(3900)$ because of isospin breaking, suggesting to resolve two close peaks in the $Y(4260)\to \pi^+ ( DD^*)^-$ process~\cite{Voloshin:2016xth}.

We conclude this section on the $X(3872)$ with the inclusive production at hadron colliders, which has been extensively studied in the literature (see \ref{appmol}). The prompt production has been studied at \cdf~\cite{cdfnote} and \cms~\cite{Chatrchyan:2013cld}, giving
\begin{subequations}
\begin{align}
 \frac{\sigma^\text{prompt}\left(p\bar p \to X(3872) + \text{all}\right)}{\sigma\left(p\bar p \to X(3872) + \text{all}\right)} &= (83.9 \pm 4.9 \pm 2.0)\%\; \text{ at }\; \sqrt{s}=1.96\tev,\\
 \frac{\sigma^\text{prompt}\left(p p \to X(3872) + \text{all}\right)}{\sigma\left(p p \to X(3872) + \text{all}\right)} &= (73.7 \pm 2.3 \pm 1.6)\%\; \text{ at }\; \sqrt{s}=7\tev.
\end{align}
\end{subequations}

\begin{table}[p]
\small
  \centering 
  \begin{tabular}{llllcc} \hline\hline 
$B$ decay mode & $X$ decay mode & \multicolumn{2}{c}{product branching fraction ($\times 10^5$)} & $B_\text{fit}$ ($\%$) & $R_\text{fit}$ \\ \hline
$\Kp X$ & $X\to \pi\pi\jpsi$ & $\mathbf{0.86\pm0.08}$ & average & $ 7.3_{-2.3}^{+2.8}$ & 1\\
 &  & $0.84\pm 0.15\pm 0.07$ & \babar~\cite{Aubert:2008gu} &  & \\
 &  & $0.86\pm 0.08\pm 0.05$ & \belle~\cite{Choi:2011fc} &  & \\
$K^0 X$ & $X\to \pi\pi\jpsi$ & $\mathbf{0.41\pm0.11}$ & average &  & \\
 &  & $0.35\pm 0.19\pm 0.04$ & \babar~\cite{Aubert:2008gu} &  & \\
 &  & $0.43\pm 0.12\pm 0.04$ & \belle~\cite{Choi:2011fc} &  & \\
$(K^+\pi^-)_{NR}X$ & $X\to \pi\pi\jpsi$ & $0.79\pm 0.13\pm0.04$ & \belle~\cite{Bala:2015wep} &  & \\
$\Kstarz X$ & $X\to \pi\pi\jpsi$ & $R^{\prime}=(34\pm9\pm2)\%$ & \belle~\cite{Bala:2015wep} &  & \\ 
$(K^0\pi^+)_{NR}X$ & $X\to \pi\pi\jpsi$ & $1.06\pm 0.30\pm0.09$ & \belle~\cite{Bala:2015wep} &  & \\ \hline
$K X$ & $X\to\omega\jpsi$ & $R=0.8\pm0.3$ & \babar~\cite{delAmoSanchez:2010jr} & $5.3\pm 2.8$ & $0.65_{-0.20}^{+0.41}$\\
$\Kp X$ &  & $0.6\pm 0.2\pm 0.1$ & \babar~\cite{delAmoSanchez:2010jr} &  & \\
$\Kz X$ &  & $0.6\pm 0.3\pm 0.1$ & \babar~\cite{delAmoSanchez:2010jr} &  & \\ 
$K X$ & $X\to\pi\pi\pi^0\jpsi$ & $R= 1.0 \pm 0.4 \pm 0.3$ & \belle~\cite{Abe:2005ix} &  & \\ 
\hline
$\Kp X$ & $X\to \Dstarz\Dzb$ & $\mathbf{8.5\pm2.6}$ & average & $66_{-12}^{+13}$ & $8.2_{-1.8}^{+1.9}$ \\
 &  & $16.7\pm 3.6\pm 4.7$ & \babar~\cite{Aubert:2007rva} &  & \\
 &  & $7.7\pm 1.6 \pm 1.0$ & \belle~\cite{Adachi:2008su} &  & \\
$\Kz X$ & $X\to \Dstarz\Dzb$ & $\mathbf{12\pm4}$ & average &  & \\
 &  & $22\pm 10 \pm 4$ & \babar~\cite{Aubert:2007rva} &  & \\
 &  & $9.7\pm 4.6\pm1.3$ & \belle~\cite{Adachi:2008su} &  & \\ \hline
$\Kp X$ & $X\to \gamma\jpsi$ & $\mathbf{0.202\pm0.038}$ & average & $1.73 \pm 0.68$ & $0.24 \pm 0.06$\\
$\Kp X$ &  & $0.28\pm 0.08\pm 0.01$ & \babar~\cite{Aubert:2008rn} &  & \\
 &  & $0.178^{+0.048}_{-0.044}\pm 0.012$ & \belle~\cite{Bhardwaj:2011dj} &  & \\
$\Kz X$ &  & $0.26\pm 0.18\pm 0.02$ & \babar~\cite{Aubert:2008rn} &  & \\
 &  & $0.124^{+0.076}_{-0.061}\pm 0.011$ & \belle~\cite{Bhardwaj:2011dj} &  & \\ \hline
$\Kp X$ & $X\to \gamma\psiprime$ & $\mathbf{0.44\pm0.12}$ & \babar~\cite{Aubert:2008rn} & $3.8\pm 1.8$ & $0.51 \pm 0.15$\\
$\Kp X$ &  & $0.95\pm 0.27\pm 0.06$ & \babar~\cite{Aubert:2008rn} &  \multicolumn{2}{c}{$R^{\prime\prime}_\text{fit} = 2.25_{-0.65}^{+0.55}$}  \\
 &  & $0.083^{+0.198}_{-0.183}\pm 0.044$ & \belle~\cite{Bhardwaj:2011dj} &  & \\
 &  & $R^{\prime\prime}=2.46 \pm 0.64 \pm 0.29$ & \lhcb~\cite{Aaij:2014ala} &  & \\
 $\Kz X$ &  & $1.14\pm 0.55\pm 0.10$ & \babar~\cite{Aubert:2008rn} &  & \\
 &  & $0.112^{+0.357}_{-0.290}\pm 0.057$ & \belle~\cite{Bhardwaj:2011dj} &  & \\ 
\hline\hline

					\end{tabular} 
	\caption{Measured $X(3872)$ product branching fractions,
	separated by production and decay channel.
	The last two columns report the results in terms of
	absolute $X(3872)$ branching fraction ($B_\text{fit}$) and in terms of 
	the branching fraction normalized to $\jpsi\,\pi\pi$ ($R_\text{fit}$)
	as obtained from the global likelihood fit described in the text.
	For non-zero measurements we report the mean value, and the 68\% \CL range in form of asymmetric errors. The limits are provided at 90\% \CL. The details of the fit are discussed in the text, the code is available on~\cite{miofit}.
	The $X(3872)\to\pi\pi\pi^0\jpsi$ is dominated by $\omega\jpsi$, but no limits on the non-resonant $\pi\pi\pi^0\jpsi$ component have been set.
	The ratio $R^{\prime}$ given by \belle~\cite{Bala:2015wep} is $\BR\left(\Bz \to X \Kstarz(892))\times\BR(\Kstarz(892) \to K^+ \pi^-)\right)/\BR\left(\Bz \to X  K^+ \pi^-)\right)$. 
	The ratio $R^{\prime\prime}$ is $\BR\left(X(3872)\to \psiprime\gamma\right)/ \BR\left(X(3872)\to \jpsi\gamma\right)$.
  }\label{tab:xdecays}
\end{table}
\begin{table}[t]
\small
  \centering 
  \begin{tabular}{llllcc} \hline\hline 
$B$ decay mode & $X$ decay mode & \multicolumn{2}{c}{product branching fraction ($\times 10^5$)} & $B_{fit}$ & $R_{fit}$ \\ \hline
$\Kp X$ & $X\to p\bar p$ & $<9.6 \times 10^{-4}$ & \lhcb~\cite{Aaij:2013rha,Aaij:2016kxn} & $< 1.7\times10^{-4}$ & $<2.4 \times 10^{-3}$\\ \hline
 $\Kp X$ & $X\to\gamma\chi_{c1}{}^{\,\S}$ & $< 9.6 \times 10^{-3}$ & \belle~\cite{Bhardwaj:2013rmw} & $< 1.0 \times 10^{-3}$ & $<0.014$\\ 
$\Kp X$ & $X\to\gamma\chi_{c2}{}^{\,\S}$ & $< 0.016$ & \belle~\cite{Bhardwaj:2013rmw} & $< 1.7 \times 10^{-3}$ & $<0.024$\\ 
$\kaon X$ & $X\to\gamma\gamma^{\,\S}$ & $<4.5\times 10^{-3}$ & \belle~\cite{Abe:2006gn} & $< 4.5 \times 10^{-4}$ & $< 6.6 \times 10^{-3}$\\ 
$\kaon X$ & $X\to \eta\jpsi^{\,\S}$ & $<1.05$ & \babar~\cite{Aubert:2004fc} & $< 0.11$ & $<1.6$\\ 
$\Kp X$ & $X\to \Dz\Dzb{}^{\,\S}$ & $<6 \times 10^{-5}$ & \belle~\cite{Abe:2003zv} & $< 0.29 $ & $< 4$\\ 
$\Kp X$ & $X\to \Dp\Dm{}^{\,\S}$ & $<4 \times 10^{-5}$ & \belle~\cite{Abe:2003zv} & $< 0.25$ & $<3.6$\\ 
\hline\hline
\multicolumn{2}{l}{Other modes}   & \multicolumn{2}{l}{Fraction/Width} & &  \\ \hline
$\gamma\gamma \to X^{\,\S}$ & $X\to \jpsi\,\pi\pi$ & $\Gamma_{\gamma\gamma} \times \BR_{\psi\pi\pi} <12.9\ev$ & \cleo~\cite{Dobbs:2004di} & \multicolumn{2}{c}{$-$}\\ 
$\epem \to X$ & $X \to \jpsi\,\pi\pi$ & $\Gamma_{ee} \times \BR_{\psi\pi\pi} < 0.13\ev$ & \bes~\cite{Ablikim:2015ain} & \multicolumn{2}{c}{$\Gamma_{ee} < 2.34\ev$}\\
$\mu N\to \mu N^\prime \pi X$ & $X\to \jpsi\,\pi\pi$ & $R^{\prime\prime\prime} = 0.9\pm 0.4$ & COMPASS~\cite{compass} &\multicolumn{2}{c}{$-$}\\ 
\hline\hline

					\end{tabular} 
	\caption{Continuation of \tablename{\ref{tab:xdecays}}. Measured $X(3872)$ product branching fractions,
	separated by production and decay channel.
	The last two columns report the results in terms of
	absolute $X(3872)$ branching fraction ($B_{fit}$) and in terms of 
	the branching fraction normalized to $\jpsi\,\pi\pi$ ($R_{fit}$)
	as obtained from the global likelihood fit described in the text.
	For non-zero measurements we report the mean value, and the 68\% \CL range in form of asymmetric errors. The limits are provided at 90\% \CL. The details of the fit are discussed in the text, the code is available on~\cite{miofit}.
	Given the preliminary nature of the result by COMPASS~\cite{compass}, the ratio $R^{\prime\prime\prime}=\displaystyle{\frac{\BR(X\to \psi \pi\pi)^2\,\Gamma_{X}}{\BR(\psiprime \to \psi \pi\pi)^2 \,\Gamma_\psiprime}}$ is not used to constrain the fit. The channels with  $\S$ are forbidden by the signature $J^{PC} = 1^{++}$.
  }\label{tab:xdecays2}
\end{table}
The same measurement is not explicitly presented in the \cdf note, 
but Bignamini \etal estimated~\cite{Bignamini:2009sk}: 
\begin{equation}
\sigma^\text{prompt}\left( pp \to X(3872) + \text{all}\right) \times \BR\left( X(3872) \to \jpsi\,\pi^+\pi^- \right) = (3.1 \pm 0.7)\nb\; \text{ at }\; \sqrt{s}=1.96\tev.
\end{equation}

\cms published the value for the integrated prompt production cross section, $\sigma^\text{prompt}\left( pp \to X(3872) + \text{all}\right) \times \BR\left( X(3872) \to \jpsi\,\pi^+\pi^- \right) = (1.06 \pm 0.11\pm 0.15)\nb$ in the region $10 < \pt < 30\gev$ and $\left|y\right|< 1.2$, at $\sqrt{s}=7\tev$. The differential cross section as a function of $\pt$ is shown in \figurename{~\ref{fig:xcmscrosssection}}, left panel. 
\lhcb measured the inclusive cross section (without separating the prompt contribution) to be $\sigma^\text{incl}\left( pp \to B_c + \text{all}\right) \times \BR\left( B_c \to X(3872) + \text{all} \right) = (5.4 \pm 1.3\pm 0.8)\nb$ in the region $5 < \pt < 20\gev$ and $2.5 < y < 4.5$, at $\sqrt{s}=7\tev$~\cite{Aaij:2011sn}.

Very recently, \atlas published a preliminary study on the prompt production of the $X(3872)$. 
In the $|y| < 0.75$, $10 < \pt < 70\gev$  region, they measured the prompt and non-prompt fractions with respect to the $\psiprime$, and distinguished between long-lived (\ie from $B$, $B_s$ and $\Lambda_b$ decays), and short-lived (\ie from $B_c$ decay) non-prompt production
\begin{subequations}
 \begin{align}
  \frac{\BR\left(B,B_s,\Lambda_b \to X(3872) + \text{All}\right) \times \BR\left( X(3872) \to \jpsi\,\pi^+\pi^-\right)}{\BR\left(B,B_s,\Lambda_b \to \psiprime + \text{All}\right) \times \BR\left( \psiprime \to \jpsi\,\pi^+\pi^-\right)} &= (3.57\pm0.33\pm0.11)\% \\
  \frac{\sigma\left( pp \to B_c + \text{all}\right) \times \BR\left( B_c \to X(3872) + \text{all} \right)}{\sigma\left( pp \to X(3872) \text{ non-prompt} + \text{all} \right)} &= (25\pm13\pm2\pm5)\%
 \end{align}

\end{subequations}

We report the differential cross section as a function of $\pt$, in \figurename{~\ref{fig:xcmscrosssection}}, right panel.

\subsection{The \texorpdfstring{$Z_c(3900)$ and $Z_c^\prime(4020)$}{Zc(3900) and Zc'(4020)}}
\label{sec:Zc}
\begin{figure}[t]
  \begin{center}
     \includegraphics[width=.45\textwidth]{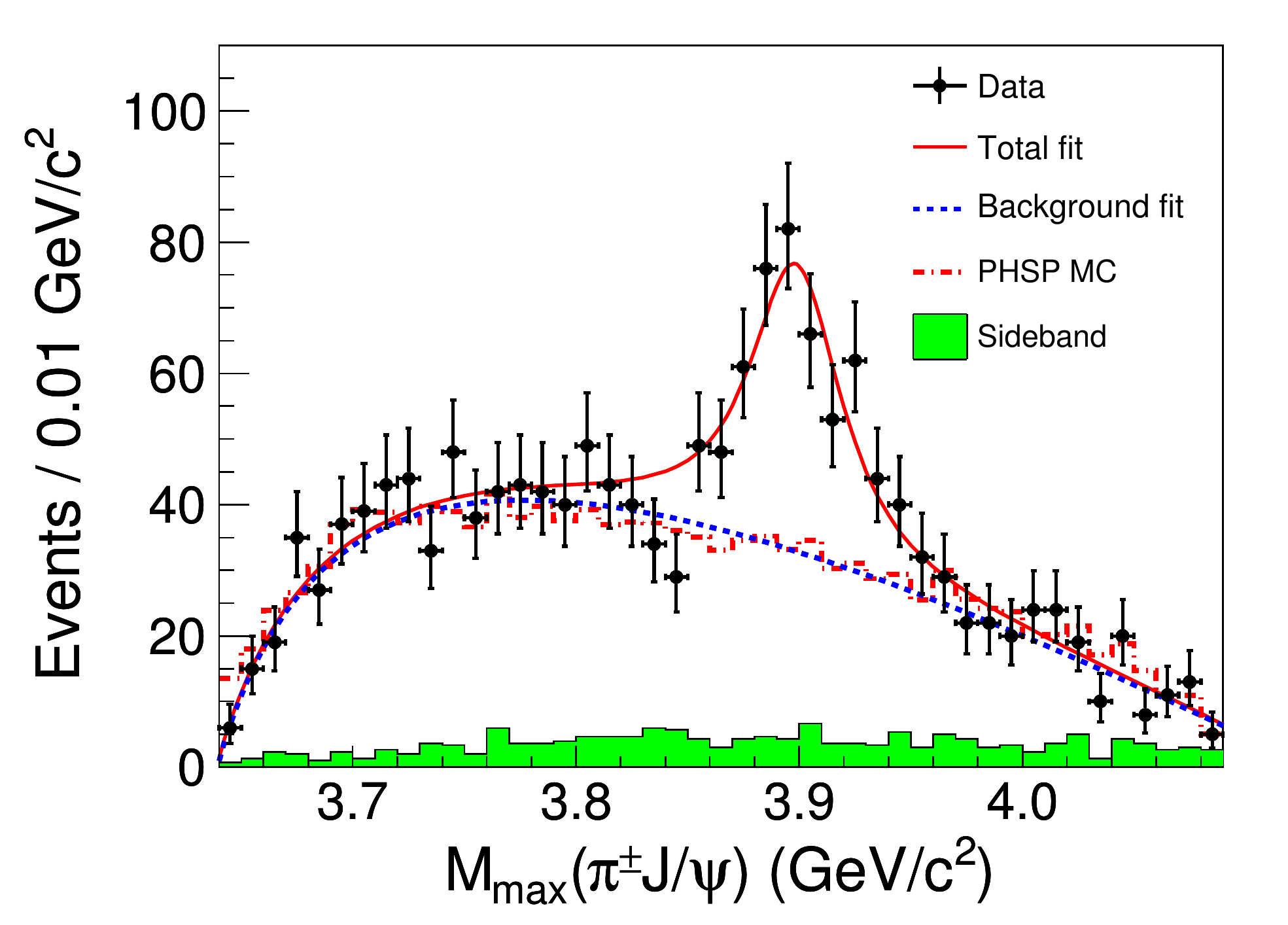} 
     \includegraphics[width=.45\textwidth]{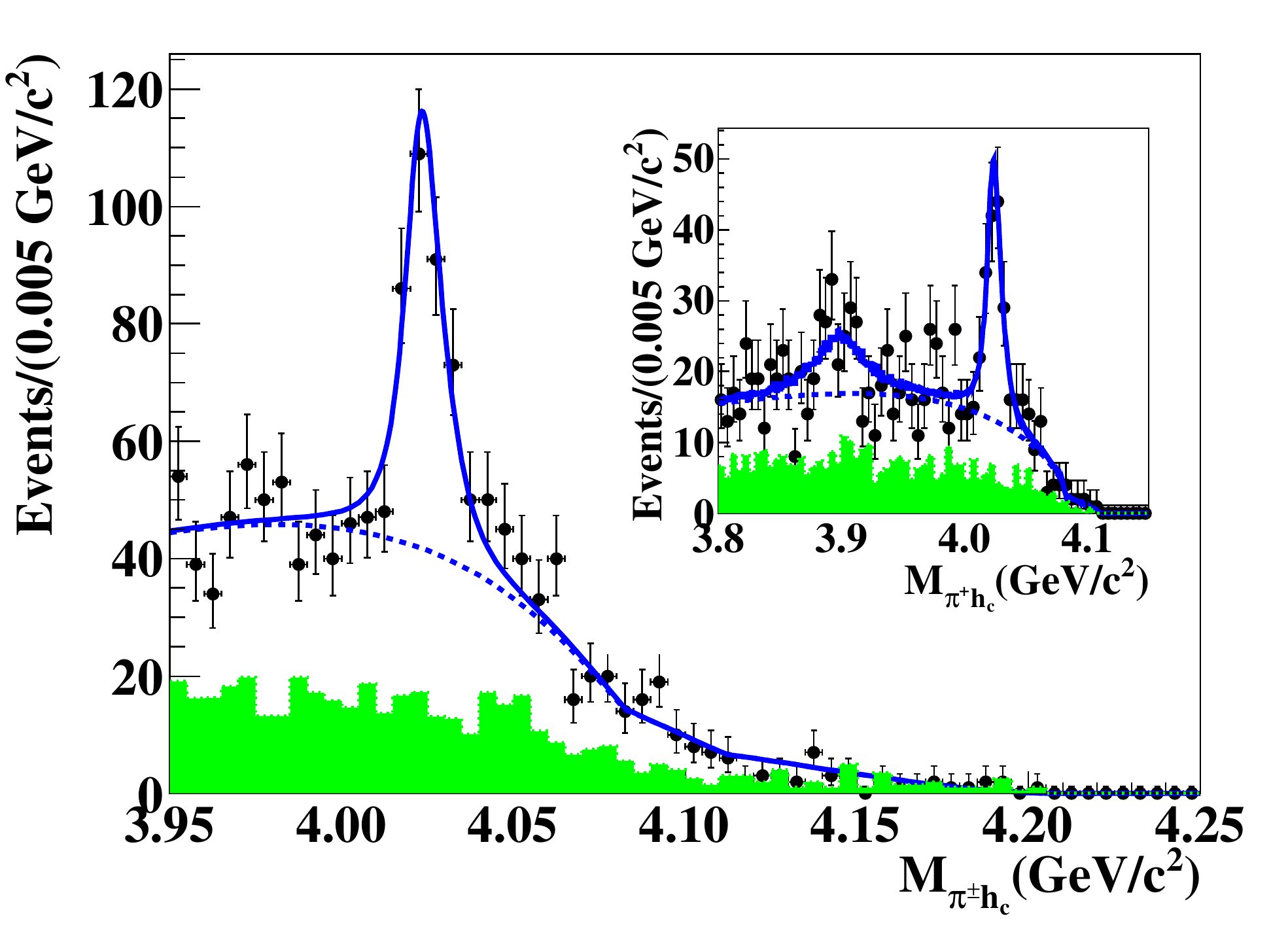} 
  \end{center}
   \caption{Left panel: Distributions of $M_\text{max}(\jpsi\,\pi^\pm)$, \ie the larger one of the two $M(\jpsi\,\pi^\pm)$ in each event, according to \bes~\cite{Ablikim:2013mio} (left) in the $Y(4260)\to\jpsi\,\pi^+\pi^-$ decay. The red solid curve is the result of the fit, the blue dotted curve is the background component, the green histogram shows the normalized \jpsi sideband events. Right panel: fits to the $M(h_c\,\pi)$ distributions by \bes~\cite{Ablikim:2013wzq}; the inset shows the sum fits if allowing for an additional $Z_c(3900)$ resonance.}
  \label{fig:z3900jpsipi}
\end{figure}
\begin{figure}[t]
  \begin{center}
     \includegraphics[width=6cm]{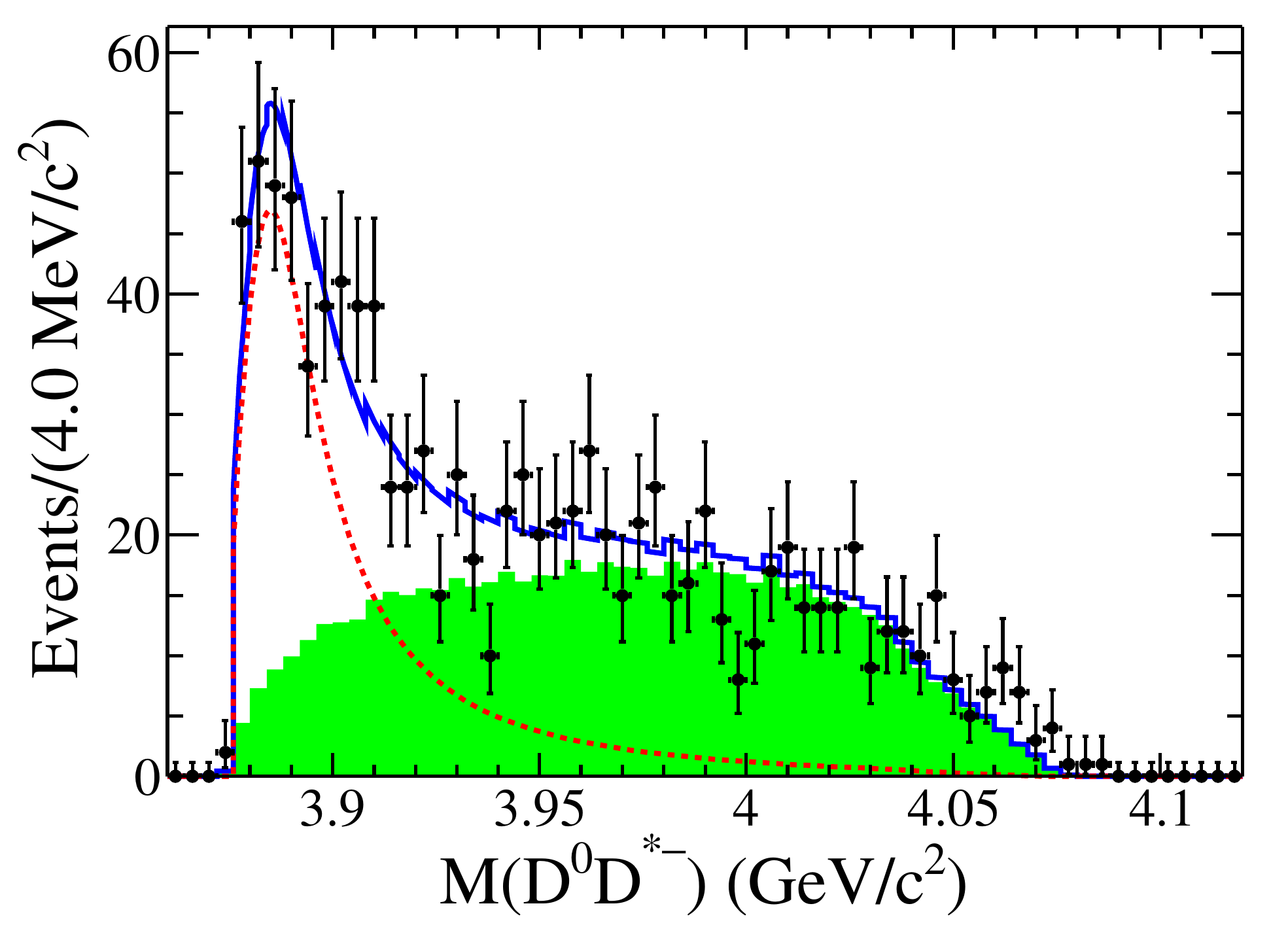} 
     \includegraphics[width=6cm]{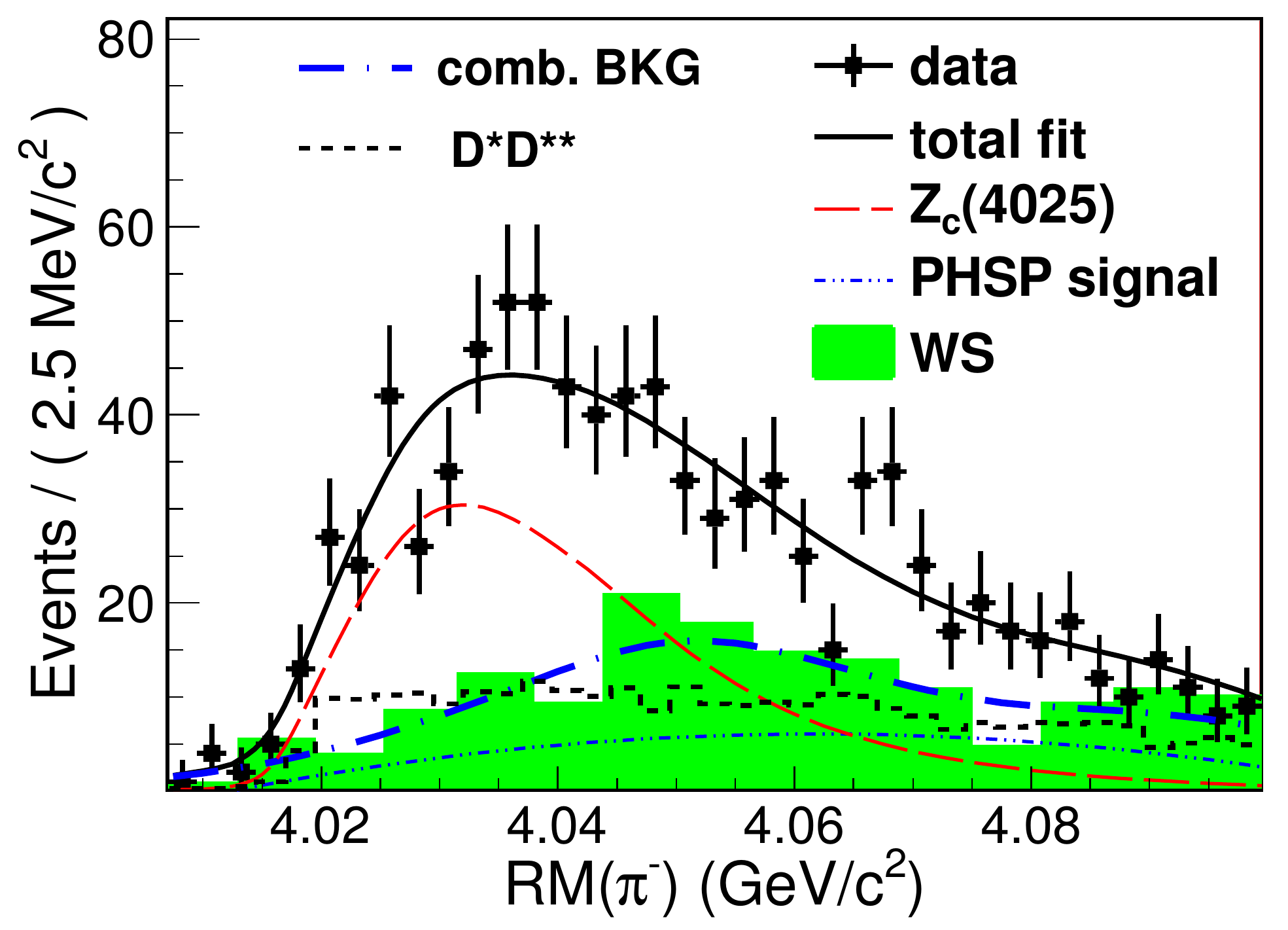} 
  \end{center}
   \caption{Left panel: Invariant mass distributions of $D^0 D^{*-}$, in the \bes double tag analysis~\cite{Ablikim:2015swa}. The solid curve is the result of the fit, the blue dotted curve is the background component. Right panel: Unbinned maximum likelihood fit to the $\pi$ recoil mass spectrum, in the $e^+e^- \to (\Dstar\Dstarb)^+\pi^-$ analysis by \bes~\cite{Ablikim:2013emm}. }
  \label{fig:z3900DDstar}
\end{figure}

The discovery of two charged axial states in the $X(3872)$ region renewed the interest into the exotic hadron spectroscopy. We have discussed that the tetraquark model predicts two $J^{PC} = 1^{+-}$ states. Similarly, molecules with the same signature are expected to form at the $DD^*$ and $D^*D^*$ threshold.
\bes~\cite{Ablikim:2013mio} and \belle~\cite{Liu:2013dau} observed an intense peak in the $\jpsi\,\pi^+$ invariant mass, while studying the $Y(4260) \to \jpsi\,\pi^+\pi^-$ channel~\footnote{Some contamination from continuum events $e^+e^- \to \gamma^* \to Z_c(3900)^+ \pi^-$ can be present, and an energy dependent analysis is needed to enlighten on this.} \figref{fig:z3900jpsipi}; \bes by analyzing data at the $Y(4260)$ peak, and \belle studying events associated with an undetected ISR photon (for more details on the ISR physics, see \sectionname{\ref{sec:vectors}}). 

The minimal quark content for such a state is $c\bar c u \bar d$, thus it is manifestly exotic.
The averaged measured mass is $M = (3884.4 \pm 1.6) \mev$, roughly $10~\mev$ above the $DD^*$ threshold, and the width is $\Gamma = (27.9 \pm 2.7) \mev$. This state was dubbed $Z_c(3900)$. The average production branching fraction is 
\begin{equation}
 \frac{\BR\left(Y(4260)\to Z_c(3900)^+ \pi^-\right)\times\BR\left(Z_c(3900)^+ \to\jpsi\,\pi^+\right)}{\BR\left(Y(4260) \to\jpsi\,\pi^+\pi^-\right)} = (22.4\pm 3.1)\%.
\end{equation}
Soon after, \bes observed a similar signal in $Y(4260)\to (D\Dstarb)^+ \pi^-$, both with single~\cite{Ablikim:2013xfr} and double tag~\cite{Ablikim:2015swa} techniques~\figref{fig:z3900DDstar}. The signature $J^P=1^+$ is favored by angular distributions, and the relative branching ratio is
\begin{equation} \label{Zcratio}
 \frac{\BR\left(Z_c(3900) \to D\Dstarb\right)}{\BR\left(Z_c(3900) \to\jpsi\,\pi\right)} = 7.7 \pm 1.3 \pm 2.8.
\end{equation}

An analysis of \cleoc data by the NU group~\cite{Xiao:2013iha} confirmed the $Z_c(3900)$ in the decay of the $\psi(4160)$, and found a hint of a neutral $Z_c(3900)^0$ decaying to $\jpsi\,\pi^0$.
\bes confirmed this neutral state, in both $\jpsi\,\pi^0$ and $(\Dbar \Dstar)^0$ channels~\cite{Ablikim:2015tbp,Ablikim:2015gda}, thus establishing the complete isospin triplet. This channel fixes the $C = -$, $G = +$ assignment.

The $Z_c(3900)$ has been searched with no luck in the OZI-suppressed $e^+e^- \to (\omega \pi^+ \pi^-)$ channel by \bes~\cite{Ablikim:2015cag}, in the $B \to (\jpsi\,\pi^+) K$ channel by \belle~\cite{Chilikin:2014bkk}, and in photoproduction $\mu^+ N \to \mu^+ (\jpsi\,\pi^+) N^\prime$ at COMPASS~\cite{Adolph:2014hba}. 
 
A close sibling, dubbed $Z_c^\prime(4020)$, has been found by \bes in $e^+e^- \to (\Dstarp\Dstarzb)\,\pi^-$~\cite{Ablikim:2013emm} and $e^+e^- \to (h_c\,\pi^+)\,\pi^-$ \cite{Ablikim:2013wzq}, with mass $M = (4023.9 \pm 2.4) \mev$, slightly above the $D^*D^*$ threshold, and width $\Gamma= (10\pm 6)\mev$~\footnote{Note the 2$\sigma$ disagreement between the width measured in $h_c\,\pi^+$, $\Gamma = (7.9 \pm 2.7 \pm 2.6)\mev$~\cite{Ablikim:2013wzq}, with the one measured in $D^* \bar D^*$, $\Gamma = (24.8 \pm 5.6 \pm 7.7)\mev$~\cite{Ablikim:2013emm}.}. The quantum numbers are $J^{PC}=1^{+-}$ as well. There is also evidence for the neutral partner $Z_c^\prime(4020)^0$, in both closed and open charm channels~\cite{Ablikim:2014dxl,Ablikim:2015vvn}.

The $Z_c(3900)$ is also searched in the $h_c\,\pi$ 
final state~\cite{Ablikim:2013wzq}, and a bump occurs at $2.1\sigma$ level, thus not statistically significant. Similarly, no $Z_c^\prime(4020)$ peak is visible in the $\jpsi\,\pi$ spectrum~\cite{Faccini:2013lda}.
This has to be compared with the $Z_b^{(\prime)}$ system, where both states appear in the same final states (see \sectionname{\ref{sec:Zb}}). This fact is of particular interest to constrain the NREFT molecular models (\sectionname{\ref{sec:NREFT}}), which in general favor the decay into $h_c$ with respect to $\jpsi$~\cite{Cleven:2013sq,Wang:2013cya,Esposito:2014hsa}: more statistics is needed to enlighten on this.

\subsection{The bottomonium system: \texorpdfstring{$Z_b(10610)$ and $Z_b^\prime(10650)$, and search for $X_b$ and $Y_b$}{Zb(10610) and Zb'(10650), and search for Xb and Yb}}
\label{sec:Zb}
Heavy quark flavor symmetry suggests that a heavy replica of the $Z_c(3900)$ and $Z_c^\prime(4020)$ system has to occur in the bottomonium sector.
Indeed, \belle observed two narrow resonances in the $\Upsilon(5S) \to (\Upsilon(nS)\,\pi^+)\,\pi^-$ and  $\Upsilon(5S)\to (h_b(nP)\,\pi^+)\,\pi^-$ channels~\cite{Belle:2011aa}. A resonant intermediate structure explains both the anomalously high rate for $\Upsilon(5S) \to \Upsilon(nS)\,\pi^+ \pi^-$ (the partial widths are about two orders of magnitude larger than typical width for dipion
transitions among the lower $(nS)$ states)~\cite{Abe:2007tk}, and the heavy quark spin symmetry violation in the decay into a $h_b$~\cite{Adachi:2011ji}.

\begin{figure}[t]
  \centering
  \includegraphics[width=0.30\textwidth]{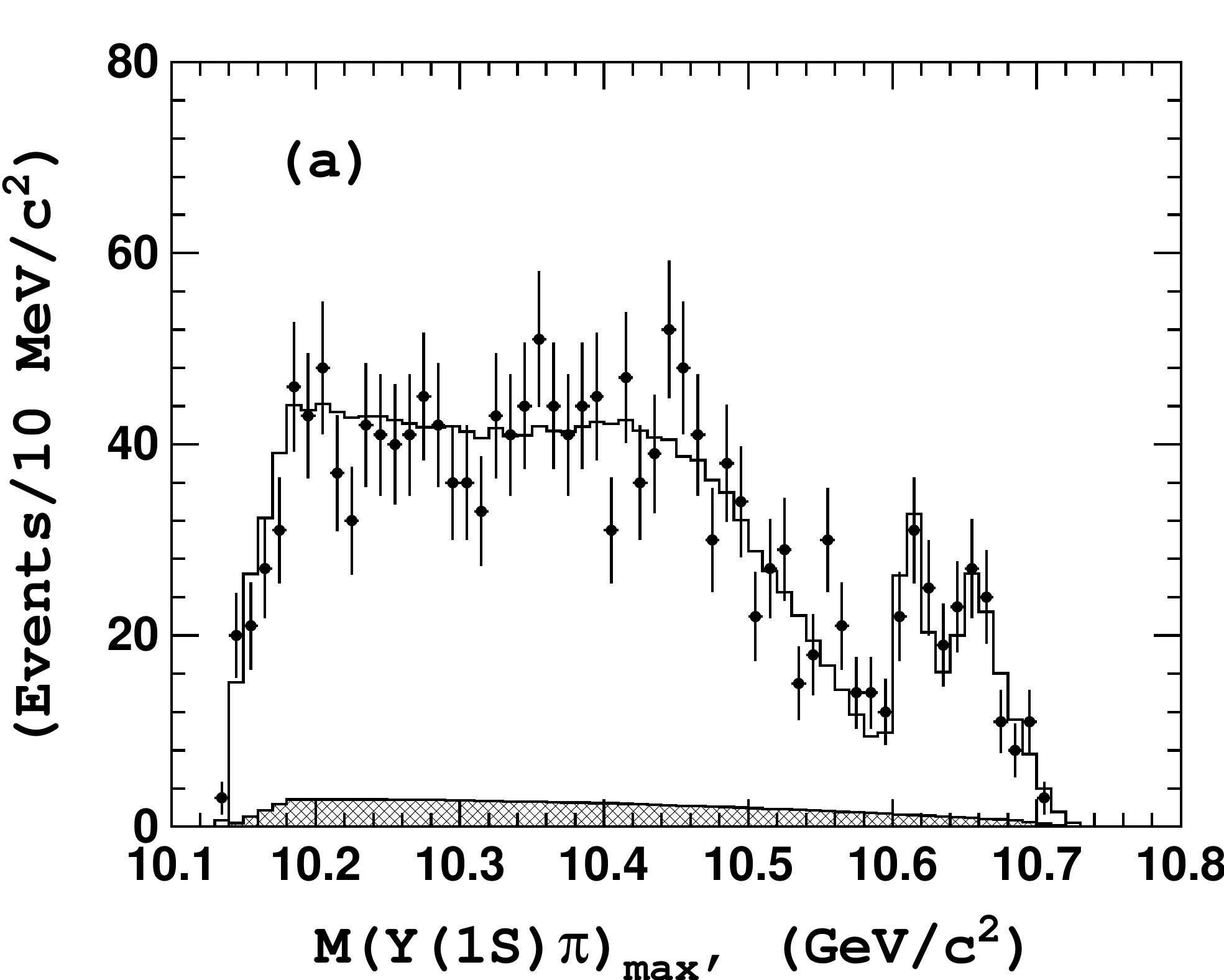} 
  \includegraphics[width=0.30\textwidth]{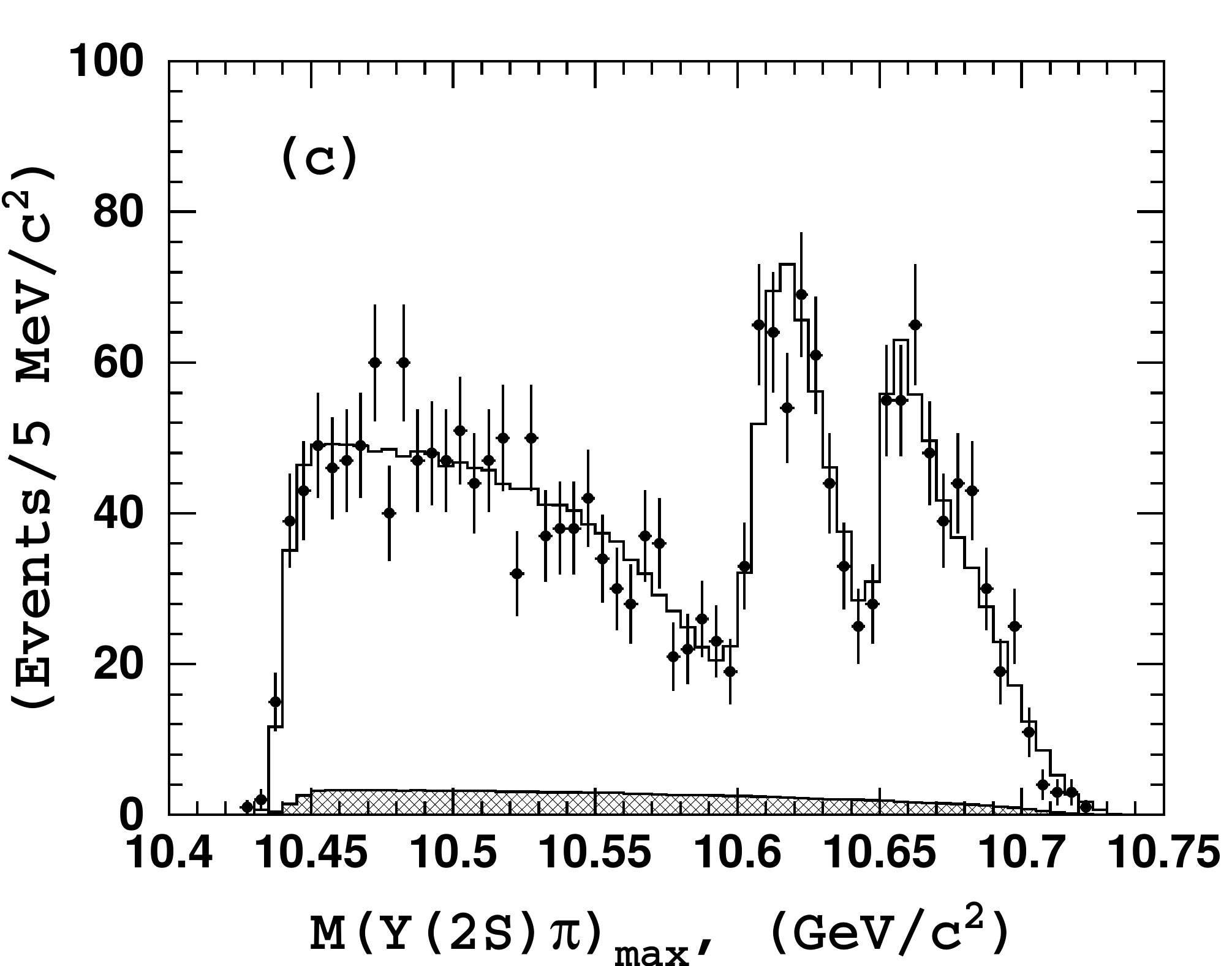} 
  \includegraphics[width=0.30\textwidth]{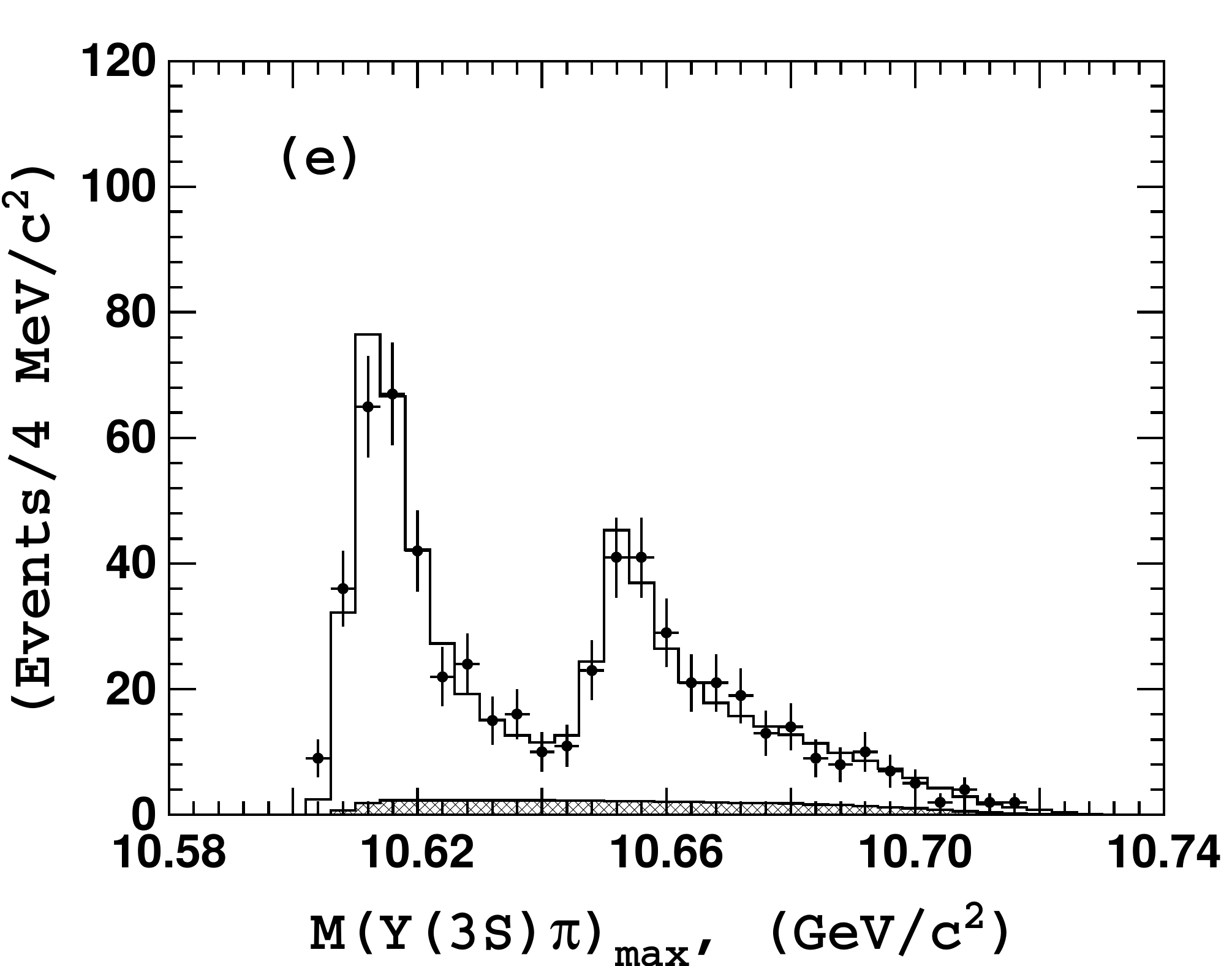} %
\includegraphics[width=0.30\textwidth]{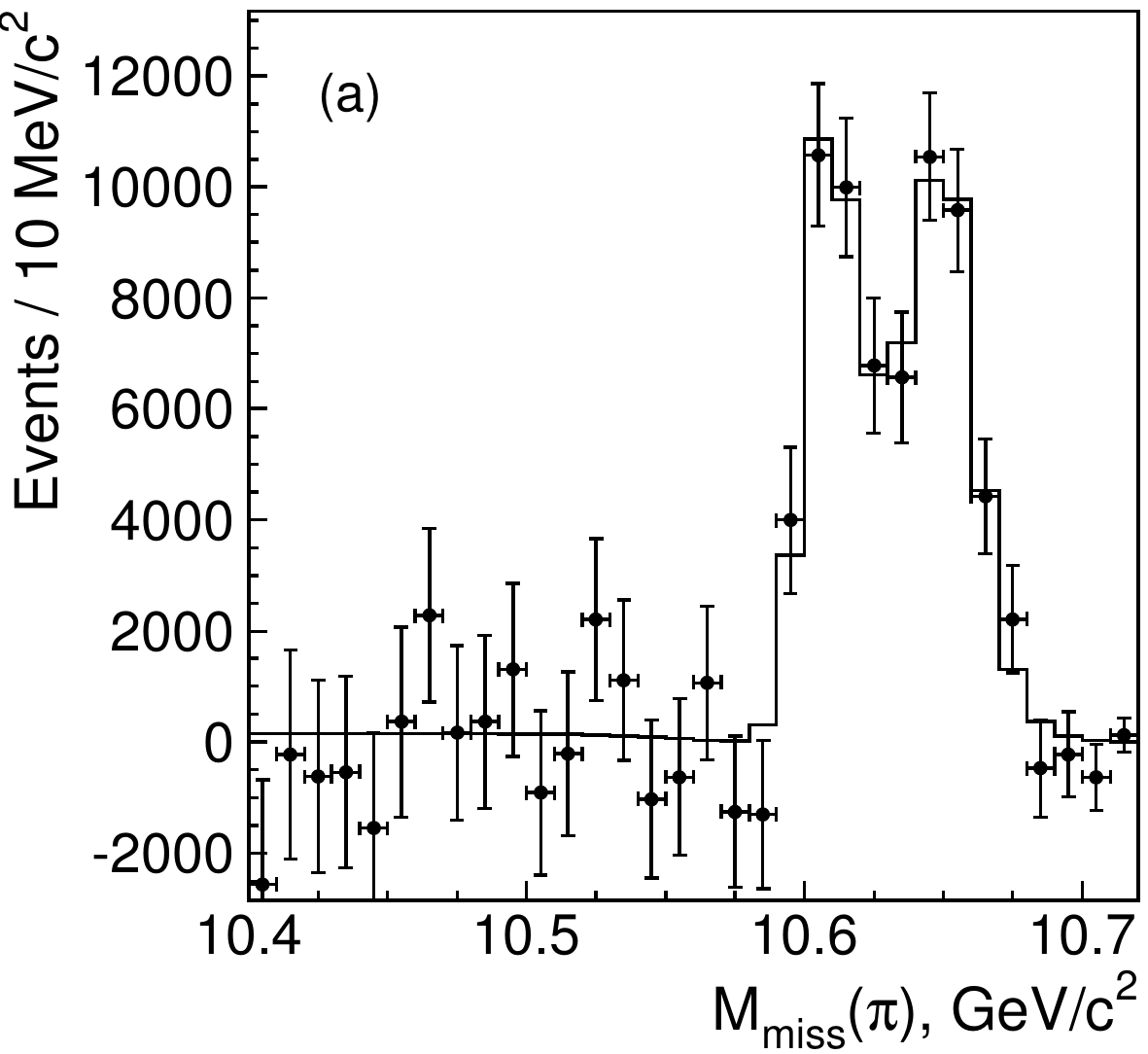} 
\includegraphics[width=0.30\textwidth]{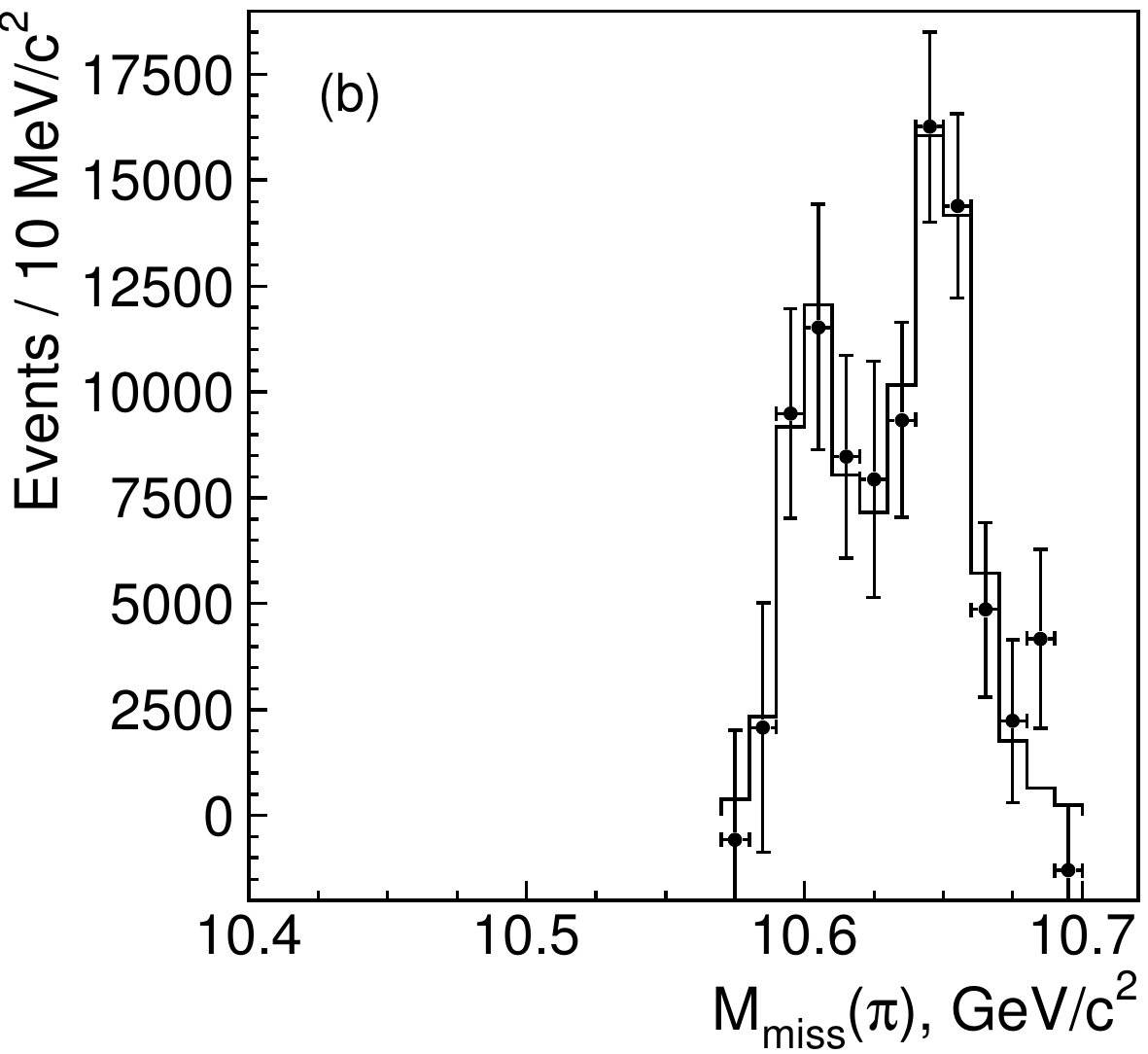}
  \caption{Comparison of fit results (open histogram) with
    experimental data (points with error bars) for events in the $\Upsilon(nS)$ (first 3 plots) and $h_b(nP)$ (last 2 plots) regions. From \belle~\cite{Belle:2011aa}.
}
\label{fig:mhbpi}
\end{figure}

\begin{table}[b]\centering
\begin{tabular}{lccccc} \hline \hline
Final state & $\Upsilon(1S)\,\pi^+\pi^-$      &
              $\Upsilon(2S)\,\pi^+\pi^-$      &
              $\Upsilon(3S)\,\pi^+\pi^-$      &
              $h_b(1P)\,\pi^+\pi^-$           &
              $h_b(2P)\,\pi^+\pi^-$
\\ \hline
           $M[Z_b(10610)]$, \!\mev        &
           $10611\pm4\pm3$                  &
           $10609\pm2\pm3$                  &
           $10608\pm2\pm3$                  &
           $10605\pm2^{+3}_{-1}$            &
           $10599{^{+6+5}_{-3-4}}$
 \\
           $\Gamma[Z_b(10610)]$, \!\mev  &
           $22.3\pm7.7^{+3.0}_{-4.0}$       &
           $24.2\pm3.1^{+2.0}_{-3.0}$       &
           $17.6\pm3.0\pm3.0$               &
           $11.4\,^{+4.5+2.1}_{-3.9-1.2}$   &
           $13\,^{+10+9}_{-8-7}$
 \\
           $M[Z_b(10650)]$, \!\mev        &
           $10657\pm6\pm3$                  &
           $10651\pm2\pm3$                  &
           $10652\pm1\pm2$                  &
           $10654\pm3\,{^{+1}_{-2}}$        &
           $10651{^{+2+3}_{-3-2}}$
 \\
           $\Gamma[Z_b(10650)]$, \!\mev     &
           $16.3\pm9.8^{+6.0}_{-2.0}$~      &
           $13.3\pm3.3^{+4.0}_{-3.0}$       &
           $8.4\pm2.0\pm2.0$                &  
           $20.9\,^{+5.4+2.1}_{-4.7-5.7}$   & 
           $19\pm7\,^{+11}_{-7}$ 
 \\
           Rel. normalization               &
           $0.57\pm0.21^{+0.19}_{-0.04}$    &
           $0.86\pm0.11^{+0.04}_{-0.10}$    &
           $0.96\pm0.14^{+0.08}_{-0.05}$    &
           $1.39\pm0.37^{+0.05}_{-0.15}$    &
           $1.6^{+0.6+0.4}_{-0.4-0.6}$
 \\
           Rel. phase, degrees              &
           $58\pm43^{+4}_{-9}$              &
           $-13\pm13^{+17}_{-8}$            &
           $-9\pm19^{+11}_{-26}$            &
           $187^{+44+3}_{-57-12}$           &
           $181^{+65+74}_{-105-109}$   
\\
\hline \hline
\end{tabular}
  \caption{Comparison of results on $Z_b(10610)$ and $Z_b^\prime(10650)$ parameters
           obtained from $\Upsilon(5S)\to\Upsilon(nS)\,\pi^+\pi^-$ and $\Upsilon(5S) \to h_b(nP)\,\pi^+\pi^-$ 
analyses~\cite{Belle:2011aa}. }
\label{tab:zb}
\end{table}

\begin{figure}[t]
  \centering
 \includegraphics[width=0.40\textwidth]{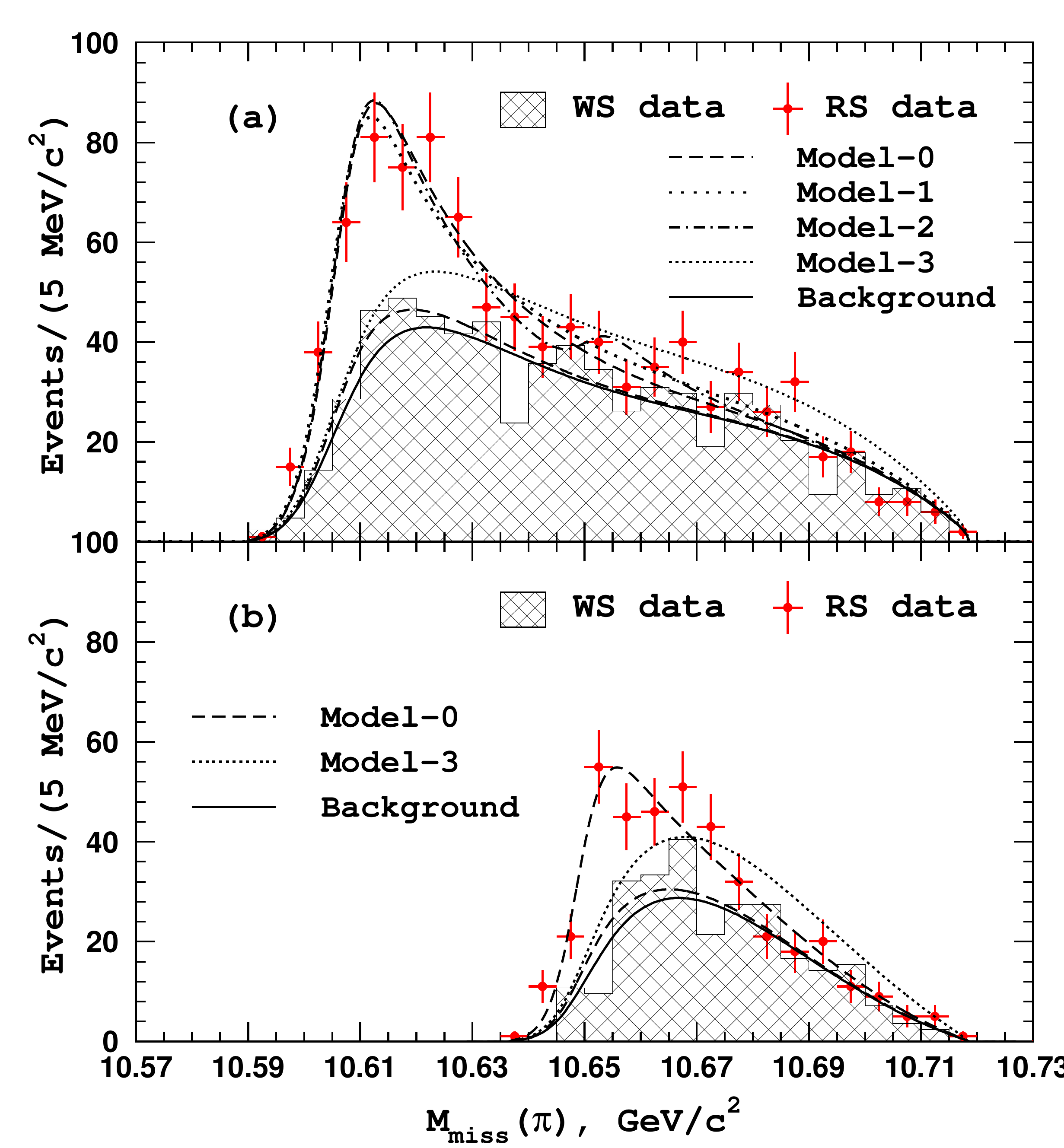} 
  \caption{The $\Upsilon(5S) \to \bar B^{(*)} B^* \pi$ channel measured by \belle~\cite{Garmash:2015rfd}. The $Z_b(10610)$ and $Z_b^\prime(10650)$ dominate the $\bar B B^*$ and the $\bar B^* B^*$, respectively.
}
\label{fig:zbopen}
\end{figure}
We report the result of the Dalitz plot fits in \tablename{\ref{tab:zb}}. All the studied channels show the presence of two charged resonances, dubbed $Z_b(10610)$ and $Z_b^\prime(10650)$, with compatible masses and widths. The one-dimensional invariant mass projections for events in each $\Upsilon(nS)$ and $h_b(nP)$ signal region are shown in \figurename{~\ref{fig:mhbpi}}. 
The averaged masses and widths are $M=(10607.2 \pm 2.0) \mev$,  $\Gamma = (18.4 \pm 2.4) \mev$, 
and $M^\prime=(10652.2 \pm 1.5) \mev$, $\Gamma^\prime = (11.5 \pm 2.2) \mev$, respectively.
The $Z_b(10610)$
production rate is similar to that of the $Z_b^\prime(10650)$ for each of the five decay
channels. The fit shows that the relative phase between the two resonances is zero for the final states
with the $\Upsilon(nS)$, and $180^\circ$ for the final states with
$h_b(nP)$, as expected according to both the tetraquark~\cite{Ali:2014dva}, and the molecular model~\cite{Bondar:2011ev}. The production of the $Z_b$s saturates the transitions to $h_b(nP)\,\pi^+\pi^-$, which indicates for the $Z_b$s to be superpositions of different heavy quark spin eigenstates. The analyses of the  dipion angular 
distributions favor the $J^P=1^+$ spin-parity assignment
for both the states~\cite{Belle:2011aa,Garmash:2014dhx}. The decay modes constrain $C = -$, $G=+$.

Recently, \belle has been able to find the neutral isospin partner $Z_b(10610)^0$~\cite{Krokovny:2013mgx} in $\Upsilon(5S)\to\Upsilon(nS)\,\pi^0\pi^0$ decays. No significant signal of $Z_b^{\prime}(10650)^0$ is seen.

The proximity of these two peaks to the $BB^*$ and $B^* B^*$ thresholds stimulated the search of the $Z_b$s in pairs of open bottom mesons~\cite{Adachi:2012cx, Garmash:2015rfd}. The Dalitz plots of $\Upsilon(5S)\to (\B\Bstar)^- \pi^+$ and  $\Upsilon(5S)\to (\Bstar\Bstar)^- \pi^+$ report a highly significant signal of $Z_b^-(10610) \to (\B \Bstar)^-$ and  $Z_b^{\prime-}(10650) \to (\Bstar \Bstar)^-$, 
respectively, with negligible continuum contribution. The $Z_b^{\prime-}(10650) \to (\B \Bstar)^-$ component, favored by phase space, is instead compatible with zero\footnote{$Z_b^{-}(10610) \to (\Bstar \Bstar)^-$ is phase-space forbidden.}. The plot of the invariant masses recoiling off the pion is shown in \figurename{~\ref{fig:zbopen}}. The best estimate for the branching ratios is reported in \tablename{\ref{tab:zbratios}}, under the assumption the observed channels to saturate the resonance widths. This decay pattern is natural in the molecular picture, whereas Ali~\etal had predicted the decay $Z_b^\prime(10650) \to \bar B^* B^*$ to be forbidden in the tetraquark model~\cite{Ali:2011ug}. However, this conclusion was later overturned, considering that light spin flip can naturally occur in the decay~\cite{Ali:2014dva}. A more detailed discussion on this subject is in \sectionname{\ref{diquark}}.

\begin{table}[t]
\centering
  \begin{tabular}{lcc}  \hline \hline
  ~Channel~\hspace*{5mm}  & \multicolumn{2}{c}{Fraction, \%}   \\
              & ~~~~~~~~$Z_b(10610)$~~~~~~~~  & ~~$Z_b^\prime(10650)$~~      \\
\hline 
 $\Upsilon(1{\rm S})\,\pi^+$ & $0.60\pm0.17\pm0.07$ & $0.17\pm0.06\pm0.02$ \\
 $\Upsilon(2{\rm S})\,\pi^+$ & $4.05\pm0.81\pm0.58$ & $1.38\pm0.45\pm0.21$ \\
 $\Upsilon(3{\rm S})\,\pi^+$ & $2.40\pm0.58\pm0.36$ & $1.62\pm0.50\pm0.24$ \\
 $h_b(1{\rm P})\,\pi^+$     & $4.26\pm1.28\pm1.10$ & $9.23\pm2.88\pm2.28$ \\
 $h_b(2{\rm P})\,\pi^+$     & $6.08\pm2.15\pm1.63$ & $17.0\pm3.74\pm4.1$  \\
 $B^+\bar{B}^{*0}+\bar{B}^0B^{*+}$
                          &  $82.6\pm2.9\pm2.3$  &         $-$         \\
 $B^{*+}\bar{B}^{*0}$      &         $-$          & $70.6\pm4.9\pm4.4$  \\
\hline \hline
  \end{tabular}
\caption{List of branching fractions for the $Z^+_b(10610)$ and 
         $Z^+_b(10650)$ decays. From \belle~\cite{Garmash:2015rfd}.}
  \label{tab:zbratios}
\end{table}

Before the discovery of the $Z_b$s, the anomalously high rate of $\Upsilon(5S)$ dipion transition leads to the speculation that two vector states were actually contributing: the actual $\Upsilon(5S)$, and an exotic vector tetraquark state, the so-called $Y_b(10890)$~\cite{Ali:2009pi,Ali:2009es}. A refit to the \babar $R_b = \sigma(e^+ e^- \to b \bar b) / \sigma(e^+ e^- \to \mu^+ \mu^-)$ data~\cite{Aubert:2008ab} allowed for the presence of a narrow peak slightly heavier than the $\Upsilon(5S)$. However the most recent analysis by \belle on the full dataset excluded the existence of such a narrow peak~\cite{Santel:2015qga}. The same analysis shows a $2\sigma$ discrepancy between the resonance parameters of the $\Upsilon(5S)$ measured in $R_b$, and in $R_{\Upsilon \pi\pi}$. This also has suggested for the states seen in the two channels to be different~\cite{Ali:2016gli}. Still, the \belle analysis comments that, while the $\Upsilon \pi\pi$ channel has basically no continuum contribution, the presence of many different open bottom threshold is expected to affect the lineshapes in $R_b$, leading to slightly different resonance parameters (see for example the model in~\cite{Tornqvist:1984fx}). Also, if no peaking background is included, the decay of the $\Upsilon(5S)$ is saturated by the existing $Z_b^{(\prime)} \pi$ and $\Upsilon \pi\pi$ final states, and no room for $B_s^{(*)} \bar B_s^{(*)}$ is left, despite  they represent the $20\%$ of on-peak events. This inconsistency led \belle not to use $R_b$ data for the precise extraction of resonance parameters, thus resolving the discrepancy with $R_{\Upsilon\pi\pi}$.  
We conclude that, in the present condition, data do not support the existence of two different states close to the $\Upsilon(5S)$. Instead, the possibility for the $\Upsilon(5S)$ to contain a sizable tetraquark component is still viable.

The existence of the $Z_b$s states immediately calls for a $X_b$, a hidden-bottom partner of the $X(3872)$, with $J^{PC}=1^{++}$. This is predicted by most of the models which fulfill heavy quark spin and flavor symmetry~\cite{Guo:2013sya,Ali:2014dva}, and it is generally expected to lay close to the $BB^*$ threshold. This state has been searched at \belle in the isospin-conserving $e^+e^- \to \gamma X_b (\to \Upsilon(1S) \omega)$ channel at the $\Upsilon(5S)$ peak~\cite{He:2014sqj}, and at \atlas in the isospin-violating inclusive $X_b \to \Upsilon(1S)\,\pi^+ \pi^-$  channel~\cite{Aad:2014ama}~\footnote{This very channel has been searched in analogy to the $X(3872) \to \jpsi\,\pi^+\pi^-$ decay. It is worth noticing that, if the isospin violation happens at the hadron level, it will be sizable in the charm sector (since the difference in mass being $M(D^+ D^{*-}) - M(D^0 \bar D^{*0}) \sim 8\mev$), but negligible in the bottom sector ($M(B^+ B^{*-}) \simeq M(B^0 \bar B^{*0})$). However, if the isospin violation is driven by the heavy scale at a more fundamental level~\cite{Rossi:1977dp}, it can occur in the $X_b$ as well.}.

\subsection{Vector \texorpdfstring{$Y$}{Y} states}
\label{sec:vectors}
\begin{figure}[t]
\centering
\includegraphics[width=.6\textwidth]{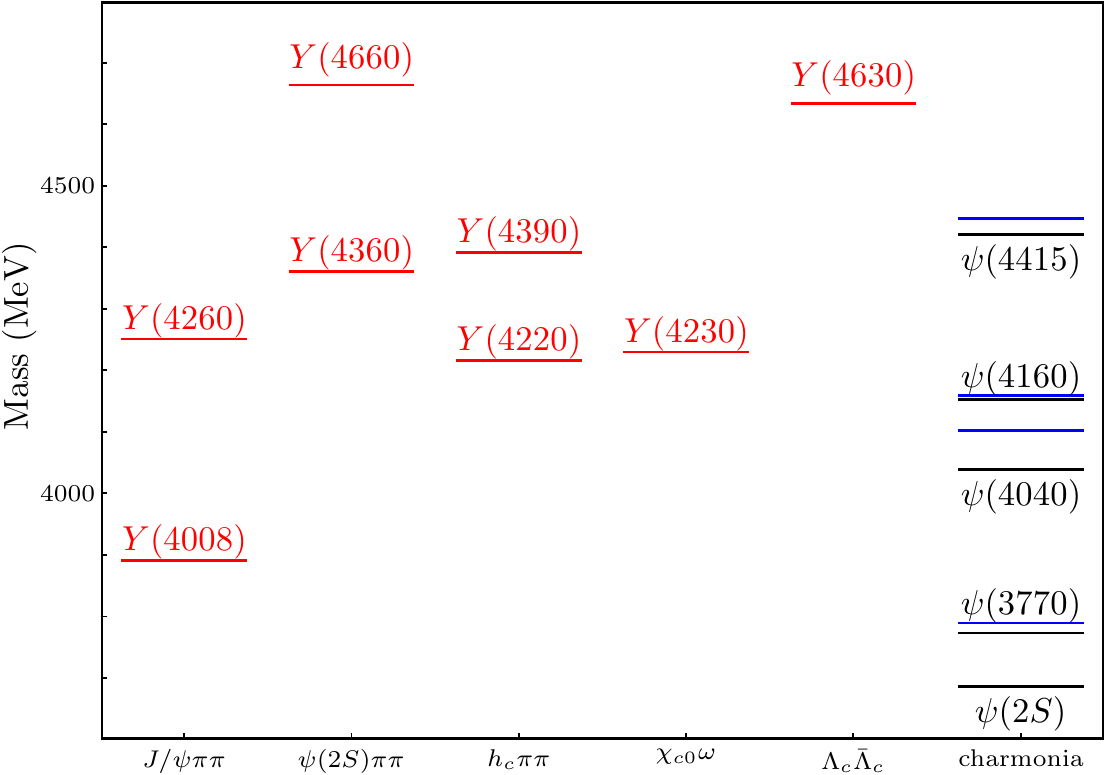}
\caption{Exotic vector states divided by decay channel. In the right column, we report observed (black) and predicted (blue) charmonium levels. 
Red lines are exotic states.}
 \label{fig:ipsilon}
\end{figure}

The $e^+ e^-$ collider can directly produce states with unambiguous signature $J^{PC}=1^{--}$. This happens if the center-of-mass energy coincides with the position of a resonance, which requires the knowledge of the mass of the resonance itself~\footnote{But fortune favors the bold~\cite{Augustin:1974xw}\ldots}. Alternatively, a fraction of the events contains an additional energetic photon $\gamma_\text{ISR}$ emitted by the initial state, that effectively decreases the center-of-mass energy.
One can thus scan a large range of $\sqrt{s}$ (which is function of the energy of the $\gamma_\text{ISR}$) without varying the energy of the beams. The $\gamma_\text{ISR}$ can be either detected (large angle ISR, good for exploring low $\sqrt{s}$), or lost in the beampipe (small angle ISR, good for exploring higher $\sqrt{s}$). In the latter case, the event is identified by requiring that the missing invariant mass is compatible with zero.
This technique allowed the $B$-factories to search and discover many unexpected vectors, usually called $Y$ states. Their exotic assignment is mainly due to two reasons: $i)$~the ordinary $\psi$ charmonia in that mass region are already well established, and there are no free slots for  these new $Y$s , and $ii)$ they do not decay into open charm final states, which instead dominate the decay width of the ordinary $\psi$ above-threshold.

In \figurename{~\ref{fig:ipsilon}} we show the summary of the detected vectors, together with their main decay channel. The tetraquark picture is able to frame most of these states in a comprehensive picture --- see \sectionname{\ref{sec:tetray}} --- identifying all the $Y$ states either as the lightest $L=1$ states, or as their radial excitations~\cite{Maiani:2014aja}. Other models provide different descriptions, on a case-by-case basis. For example, the $Y(4260)$ has been described in terms of a $\bar D D_1(2420)$ molecule, which would appear in the slightly displaced $Y(4220)$ and $Y(4230)$ peaks. However, it is important to stress that the resulting large binding energy of $\sim 70\mev$ pushes the state out of the range of applicability of the shallow states theory (\sectionname~\ref{mol2}). The $Y(4360)$ would be the $Y(4260)$ heavy quark spin partner, $\bar D^* D_1(2420)$~\cite{Wang:2013cya,Cleven:2013mka}.
The $Y(4660)/Y(4630)$ would be instead a $\psiprime \,f_0(980)$ molecule~\cite{Guo:2008zg}. %
Voloshin~\cite{Li:2013ssa} proposed the hadrocharmonium model to describe the decay pattern of the $Y(4260)$ and $Y(4360)$ (see \sectionname{\ref{sec:hadrocharmonium}}). Lattice QCD and constituent models usually identify a $c\bar c g$ hybrid vector state in the region of the $Y(4260)$ --- see \sectionname{\ref{sec:hybrids}}.
\begin{figure}[t]
\centering
\includegraphics[width=.45\textwidth]{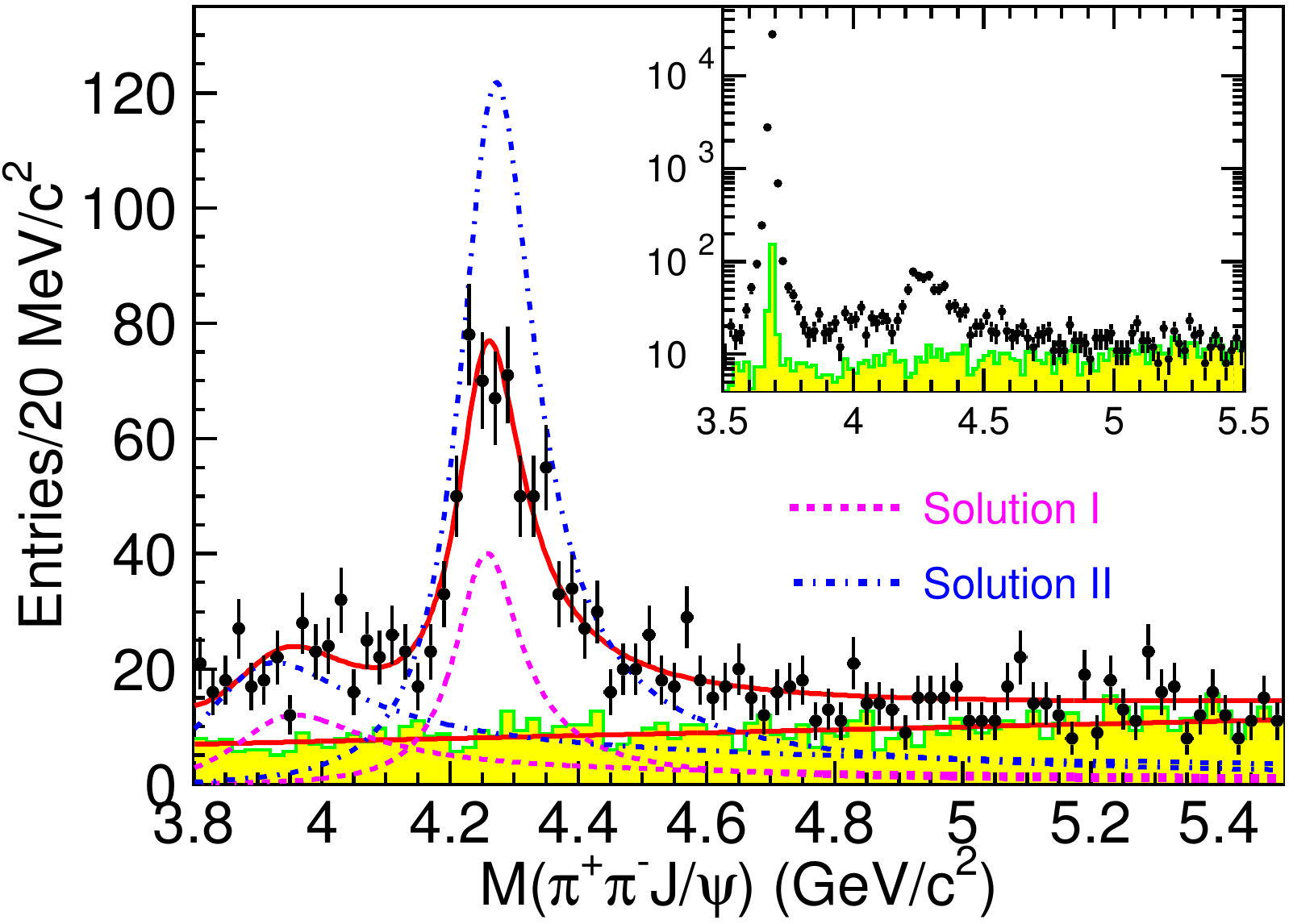} \hspace{1cm}
\includegraphics[width=.45\textwidth]{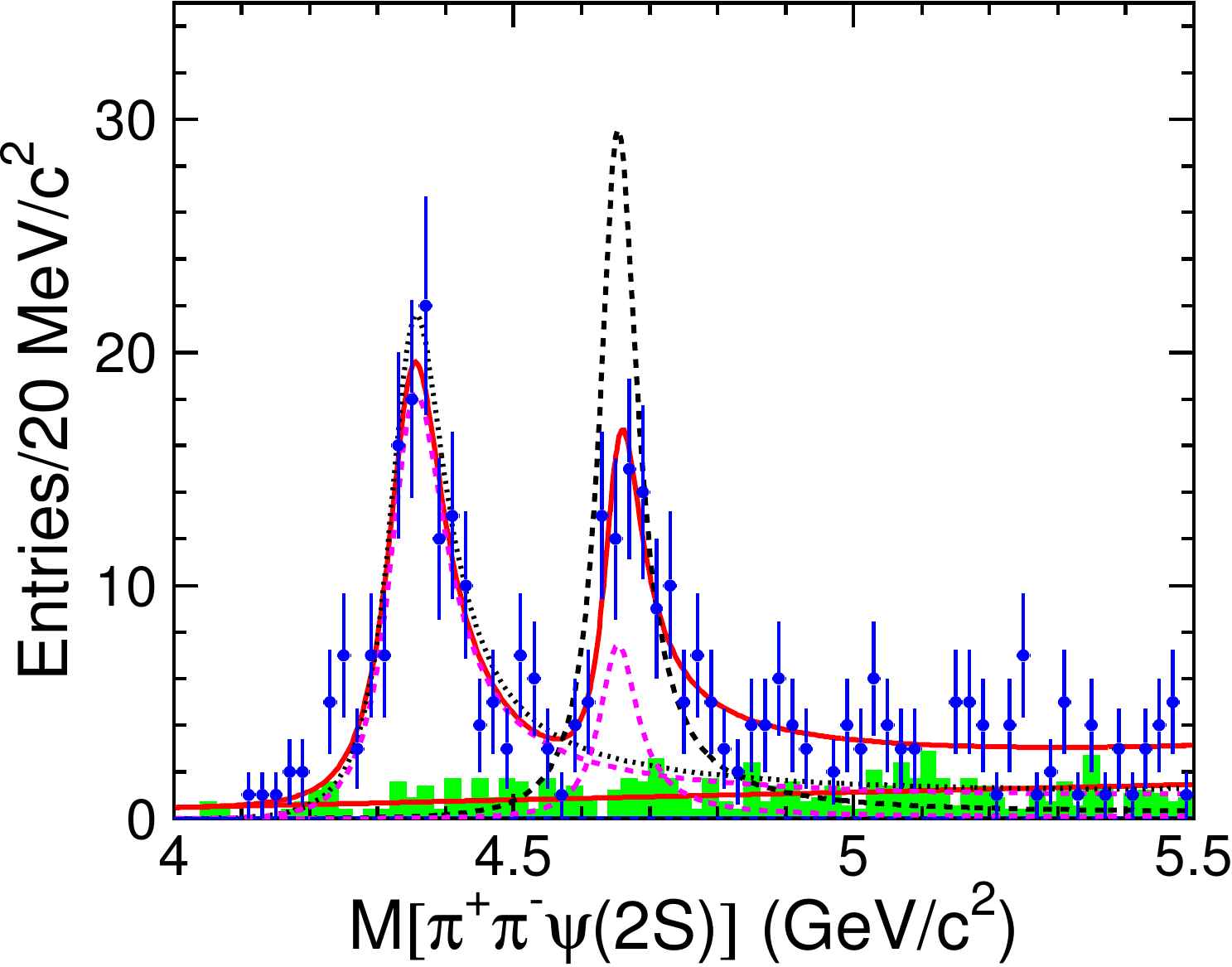}
\caption{\belle analyses of $e^+e^- \to \jpsi\,\pi^+\pi^-$ (left)~\cite{Liu:2013dau} and $\to \psiprime\,\pi^+\pi^-$ (right)~\cite{Wang:2014hta}.}
 \label{fig:ipsilonspectra}
\end{figure}

The first in the list is the $Y(4260)$, found at \babar in 2005 decaying into $\jpsi\,\pi^+\pi^-$, and later confirmed by \cleo and \belle~\cite{Aubert:2005rm,He:2006kg,Yuan:2007sj,Lees:2012pv,Liu:2013dau}. Evidence for $Y(4260)\to \jpsi\,\pi^0\pi^0$ was also reported by \cleo~\cite{Coan:2006rv}, whereas no clear sign of $Y(4260) \to\jpsi  K^+ K^-$ or $\to \KS \KS$ has been seen so far~\cite{Coan:2006rv,Yuan:2007bt,Shen:2014gdm}. The averaged values for mass and width are $M= (4251 \pm9)\mev$ and $\Gamma = (120 \pm 12)\mev$~\footnote{This average does not include the new results by \bes~\cite{Ablikim:2016qzw}, which assume a different lineshape for the cross section, due to the interference of two different states. This is commented later on.}. 
A thorough search of $Y(4260)$ in open charm final states has been performed by \babar, \belle and \cleoc, with no success. We report the list of most of the channels where the $Y(4260)$ has been searched, and the  90\% upper limits, in \tablename{\ref{tab:y4260}}. 
Instead, for the discovery mode, we have
 $\Gamma\left(Y(4260)\to e^+ e^-\right) \times \BR\left(Y(4260)\to \jpsi\,\pi^+\pi^-\right) = (9.2 \pm 1.0)~\ev$.
The electronic width can be constrained by looking at the inclusive $\sigma\left(\epem\to\text{hadrons}\right)$, giving $\Gamma\left(Y(4260)\to e^+ e^-\right) < 580\ev$ at 90\% C.L.~\cite{Mo:2006ss}. This estimate will be improved with the forthcoming data on the inclusive hadron cross section by \bes. This turns out to a dipion transition of $\Gamma\left(Y(4260)\to \jpsi\,\pi^+\pi^-\right)\sim 1\mev$, one order of magnitude larger than similar transitions of ordinary charmonia (and somehow similar to the dipion abundance in the $\Upsilon(5S)$ transitions, see \sectionname{\ref{sec:Zb}}).
\begin{table}[t]
\centering
\begin{tabular}{lll}\hline\hline
  Final state & Upper limit (90\% C.L.)& Experiment \\ \hline
 \multicolumn{3}{l}{ $\Gamma_{ee} \times \BR\left(Y(4260)\to f\right)$ ($\mathrm{e\!V}$) } \\ \hline
  $\jpsi K^+K^-$ & $1.2$ & \belle~\cite{Yuan:2007bt} \\
  $\jpsi\eta$ & $14.2$ & \belle~\cite{Wang:2012bgc} \\
  $\psiprime\,\pi^+\pi^-$ & $\sim 12$ & Reanalysis~\cite{Liu:2008hja} \\
  $\phi\pi^+\pi^-$ & $0.4$ & \babar~\cite{Aubert:2007ur} \\
  $\KS K^+\pi^-$ & $0.5$ & \babar~\cite{Aubert:2007ym} \\
  $K^+ K^-\pi^0$ & $0.6$ & \babar~\cite{Aubert:2007ym} \\ \hline
  \multicolumn{3}{l}{  $\BR\left(Y(4260)\to f\right) / \BR\left(Y(4260)\to\jpsi\,\pi^+\pi^-\right)$ } \\ \hline
  $\D\Dbar$ & $1.0$ & \babar~\cite{Aubert:2006mi}, \\
$\Dstar\Dbar$ & $34$ & \babar~\cite{Aubert:2009aq}  \\
$\Dstar\Dstarb$	& $11$ & \cleoc~\cite{CroninHennessy:2008yi}  \\
$\Dsp\Dsm$	& $0.7$ & \babar~\cite{delAmoSanchez:2010aa} \\
$\Dssp\Dsm$	& $0.8$ & \cleoc~\cite{CroninHennessy:2008yi}  \\
$\Dssp\Dssm$	& $9.5$ & \cleoc~\cite{CroninHennessy:2008yi} \\
$\Dz\Dstarm\pi^+$ & $9$ & \belle~\cite{Pakhlova:2009jv} \\
$\Dstar\Db\,\pi$	& $15$ & \cleoc~\cite{CroninHennessy:2008yi}  \\
$\Dstar\Dstarb\pi$	& $8.2$ & \cleoc~\cite{CroninHennessy:2008yi} \\
  $h_c\,\pi^+\pi^-$ & $1.0^{\,\dagger}$& \cleo~\cite{CLEO:2011aa} \\
$p\bar p$ & $0.13$ & \babar~\cite{Aubert:2005cb} \\ \hline
  \multicolumn{3}{l}{  $\sigma\left(e^+e^-\to f\right)$ ($\mathrm{pb}$) } \\ \hline
  $\chi_{c1}\omega$ & $18$ $(\sqrt{s}=4.31\gev)$ & \bes~\cite{Ablikim:2014qwy} \\
$\chi_{c2}\omega$ & $11$ $(\sqrt{s}=4.36\gev)$ & \bes~\cite{Ablikim:2014qwy} \\ \hline\hline
 \end{tabular}
\caption{Upper limits for $Y(4260)$ into different final states. $^\dagger$ This has to be updated with new \bes results~\cite{Ablikim:2016qzw,BESIII:2016adj}.}
\label{tab:y4260}

\end{table}

To investigate the nature of this resonance, it is important to understand whether the 
pion pair is dominated by any resonance. %
The updated \babar analysis finds some evidence of $Y(4260)\to\jpsi f_0(980)$, although a relevant nonresonant component is still needed~\cite{Lees:2012cn}. This might explain why the state does not decay into $\psiprime$~\footnote{The significance of a $Y(4260)\to \psiprime\,\pi^+\pi^-$ is $2.4\sigma$ only~\cite{Wang:2014hta}.}, since the decay $Y(4260)\to \psiprime f_0(980)$ is phase-space forbidden (see also \sectionname{\ref{HQSS}}). In this case, a sizable $Y(4260) \to \jpsi K \bar K$ is expected, which calls for a new analysis in this channel.
For the transition to the exotic $Z_c$s states, see \sectionname{\ref{sec:Zc}}, but a proper energy scan is still due to understand whether the production of $Z_c$s is dominated by the $Y(4260)$ or by continuum events.

The identification of the $Y(4260)$ as $\bar D D_1(2420)$ molecule calls for the search of the decay into these constituents. The analysis of the Dalitz plot $Y(4260) \to \bar D D^* \pi$ allows to search the intermediate narrow $D_1(2420)$ as a $D$-wave $D^*\pi$ resonance. \bes performed the analysis on both the single and double tagged samples, finding no $D_1$ component, but limited statistics prevented to draw strong conclusions~\cite{Ablikim:2013xfr,Ablikim:2015swa}.

The radiative decays of the $Y(4260)$ can shed some light on the nature of this state. The production cross section $\sigma\left(e^+ e^-\to \gamma X(3872)\right) \times \BR\left(X(3872) \to \jpsi\,\pi^+ \pi^-)\right)$ measured by \bes scales as a function of the center-of-mass energy consistently with a Breit-Wigner with $Y(4260)$ mass and width as parameters, consequently the observed events come from the intermediate resonant state and not from the continuum~\cite{Ablikim:2013dyn}. The estimate for the relative branching ratio is $\BR\left(Y(4260) \to \gamma X(3872) \right) $ $/  \BR\left(Y(4260) \to \jpsi\,\pi^+\pi^- \right) \sim 0.1$~\footnote{Assuming for the branching ratio $\BR\left(X(3872)\to \jpsi\,\pi^+\pi^-\right) = 5\%$.}, which is relatively large. This radiative decay is naturally expected in the tetraquark model as an E1 transition of the orbitally excited tetraquark to its ground state, with no spin change~\cite{Maiani:2014aja}. These predictions, as well as the ones under the molecular~\cite{Guo:2013nza} and hybrid hypotheses~\cite{Guo:2014zva}, can be tested by more precise measurements of these branching ratios.

\begin{figure}[t]
\centering
\includegraphics[width=.9\textwidth]{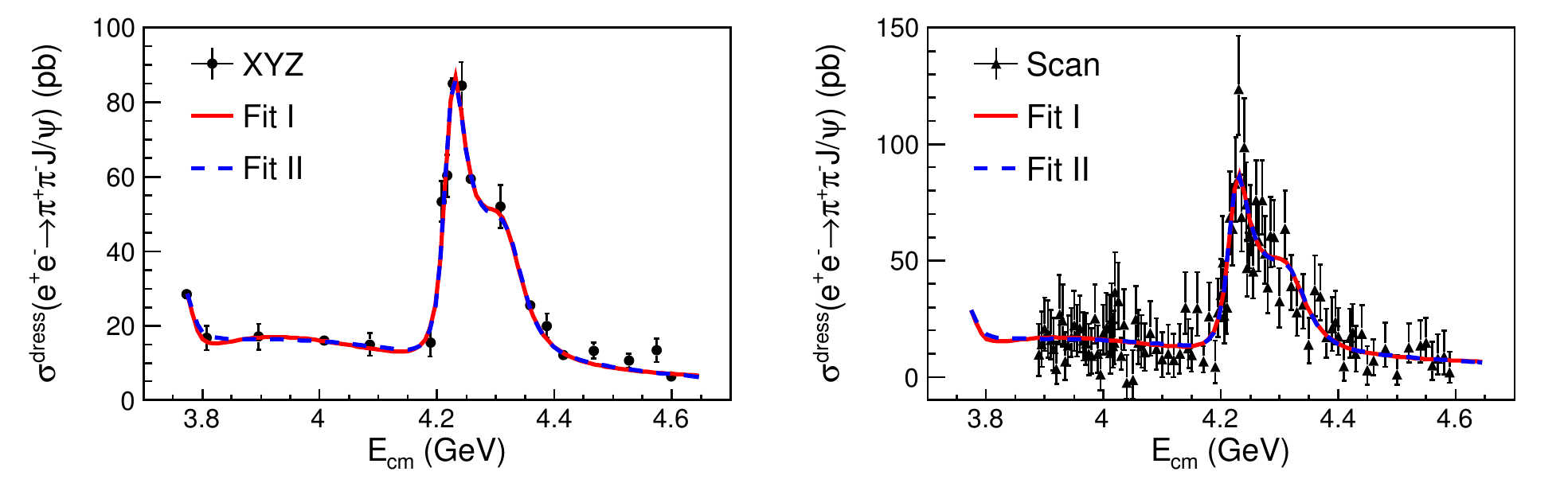}
\caption{Measured dressed cross section $\sigma^\text{dress}(e^+e^-\to\pi^+\pi^- \jpsi$ at \bes~\cite{Ablikim:2016qzw},  and simultaneous fit to the high-statistics datasets (left) and to the energy scan data (right) with
the coherent sum of the Breit-Wigner functions for the $Y(4008)$, the $Y(4260)$, and the $Y(4360)$ (red solid curves), or substituting the $Y(4008)$ with  a coherent  exponential continuum. }
 \label{fig:besscan}
\end{figure}
In the same $\jpsi\,\pi^+\pi^-$ final state, \belle reported also a broad structure named $Y(4008)$~\cite{Yuan:2007sj,Liu:2013dau}, with $M=(3890.8 \pm 40.5 \pm 11.5)\mev$ and $\Gamma = (254.5 \pm 39.5 \pm 13.6)\mev$. This is at odds with \babar~\cite{Lees:2012cn}, which does not observe any structure in that region. Very recently, \bes published the results of a precise scan analysis~\cite{Ablikim:2016qzw} (see \figurename{~\ref{fig:besscan}}), which finds a much broader state, $M = (3812.6^{+61.9}_{-96.6})\mev$ and $\Gamma = (476.9^{+78.4}_{-64.8})\mev$. Alternatively, it is substituted by an exponential background, obtaining an equally good fit. In the same analysis, the asymmetric shape of the $Y(4260)$ turns out to be due to the interference of a narrower $Y(4260)$, with mass and width $M=(4222.0 \pm 3.1 \pm 1.4)\mev$ and $\Gamma=(44.1\pm4.3\pm2.0)\mev$, with a slightly heavier one, with $M=(4320.0 \pm 10.4 \pm 7.0)\mev$ and $\Gamma=(101.4^{+25.3}_{-19.7}\pm10.2)\mev$. The latter is compatible with the state observed in $e^+e^- \to \psiprime \pi\pi$ discussed below. 
More statistics will help in solving the controversy about the $Y(4008)$. 

The analysis of the $\psiprime\,\pi^+ \pi^-$ channel gave some surprises: instead of finding the $Y(4260)$, \babar and \belle observed other two resonances: the $Y(4360)$ and the $Y(4660)$~\cite{Aubert:2006ge,Lees:2012pv,Wang:2007ea,Wang:2014hta}, with masses and widths $M = (4346.2 \pm 6.3)\mev$, $\Gamma = (102 \pm 10)\mev$, and $M=(4657 \pm 11)\mev$, $\Gamma = ( 70 \pm 11 \pm 3)\mev$, respectively. The latter has not been seen in $\jpsi\,\pi\pi$, despite the larger phase space. The decay $Y(4360) \to \jpsi \,\pi^+\pi^-$ has been recently observed by \bes~\cite{Ablikim:2016qzw}, which is at odds with the combined analysis of \babar and \belle data~\cite{Cotugno:2009ys}.
In \figurename{~\ref{fig:ipsilonspectra}} we report some distributions of $\jpsi\,\pi^+\pi^-$ and $\psiprime\,\pi^+\pi^-$ by \belle.

The preference of tetraquark states to decay into baryons was understood since the origins of the model~\cite{Rossi:1977cy,Montanet:1980te,Maiani:2004vq,Rossi:2016szw}. Motivated by this, \belle searched for vector resonances decaying into $\Lambda_c^+ \Lambda_c^-$~\cite{Pakhlova:2008vn}.
A structure (the $Y(4630)$) has actually been found near the baryon threshold, with Breit-Wigner parameters $M=(4634^{+8}_{-7}{}_{-8}^{+5})\mev$ and $\Gamma = (92^{+40}_{-24}{}^{+10}_{-21})\mev$.
A combined fit of the $\psiprime\,\pi^+\pi^-$ and $\Lambda_c^+\Lambda_c^-$ spectra concluded that the two structures $Y(4630)$ and $Y(4660)$
can indeed be the same state, with a strong preference for the baryonic decay mode:
$\BR(Y(4660)\to\Lambda_c^+\Lambda_c^-)/ \BR(Y(4660)\to\psiprime\,\pi^+\pi^-)= 25\pm7$~\cite{Cotugno:2009ys}. The possibility these two structures to be actually distinct has been explored in~\cite{Maiani:2014aja}.

\bes also measured the $e^+e^- \to h_c\,\pi^+\pi^-$ cross sections at center-of-mass energies varying from $3.896$ to $4.600\gev$ \cite{Ablikim:2013wzq,BESIII:2016adj}. The values of the cross sections are similar to the $e^+ e^- \to \jpsi\,\pi^+\pi^-$ one, but the lineshape looks  different (see \figurename{~\ref{fig:yuan}}). 
The $h_c\,\pi^+\pi^-$ spectrum has been fitted first by Yuan~\cite{Chang-Zheng:2014haa,Yuan:2014rta}, which found a significant signal for a $Y(4220)$, while the higher mass region could be described either by a $Y(4290)$ state, or by a phase-space background. The very recent energy scan by \bes~\cite{BESIII:2016adj} has confirmed the former hypothesis, giving for the two resonances masses and widths $M = (4218.4 \pm 4.0 \pm 0.9)\mev$ and $\Gamma =  (66.0 \pm 9.0 \pm 0.4)\mev$,  $M=(4391.6 \pm 6.3 \pm 1.0)\mev$ and  $\Gamma = (139.5 \pm 16.1 \pm 0.6)\mev$.
A somewhat similar signal has been seen by \bes in $e^+ e^-\to \chi_{c0}\omega$~\cite{Ablikim:2014qwy,Ablikim:2015uix} at a mass of $M = (4230 \pm 8)\mev$ and a width of $\Gamma = (38\pm12)\mev$, again not compatible  with $Y(4260)$ parameters. The hypothesis that the $Y(4220)$ and $Y(4230)$ are indeed the same tetraquark state has been studied in~\cite{Faccini:2014pma}. The possibility that, instead, the state in $\chi_{c0} \omega$ is the ordinary $\psi(4160)$ has been proposed in~\cite{Li:2014jja}.
New \bes data suggest for this $Y(4220)$ to coincide with the narrow one observed in $\jpsi \, \pi^+\pi^-$. It is clear that a combined refit of all available data in this sector is needed to establish the existence of many of the unsettled $Y$ states.

\begin{figure}[t]
\centering
\includegraphics[width=.45\textwidth]{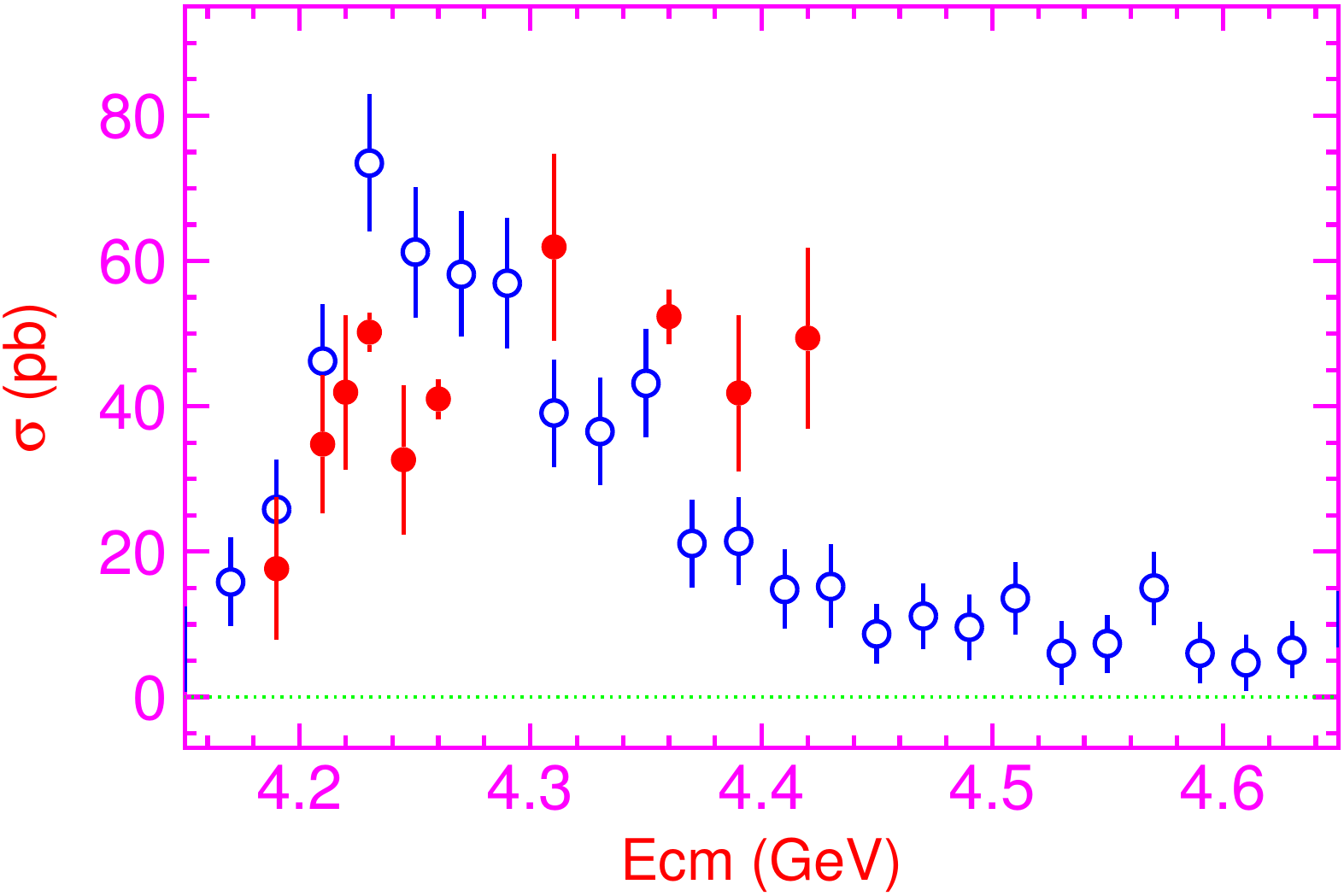}\hspace{1cm}
\includegraphics[width=.45\textwidth]{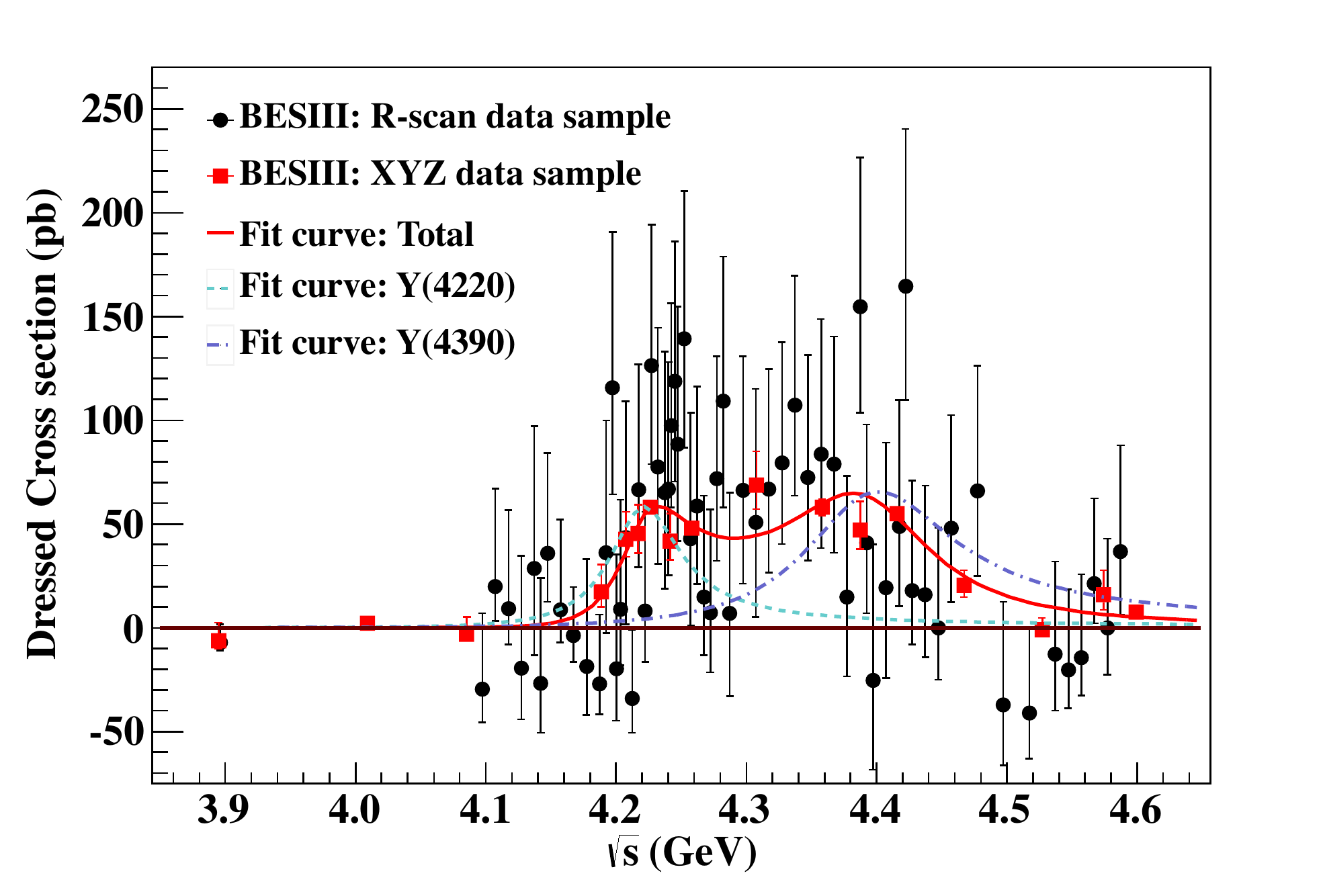}
\caption{Left panel: \bes data of $e^+e^-\to h_c\,\pi^+\pi^-$ (red dots)~\cite{Ablikim:2013wzq} compared to \belle data of $e^+e^-\to \jpsi\,\pi^+\pi^-$ (blue circles)~\cite{Liu:2013dau}. From Yuan~\cite{Chang-Zheng:2014haa,Yuan:2014rta}. Right panel: new \bes data of $e^+e^-\to h_c\,\pi^+\pi^-$~\cite{BESIII:2016adj}. The dash
(dash-dot) curve shows the contribution from the two structures $Y(4220)$ ($Y(4390)$). }
 \label{fig:yuan}
\end{figure}

\subsection{The \texorpdfstring{$Z(4430)$}{Z(4430)}}
\label{sec:Z4430}
The childbirth of the $Z(4430)$ was long and troubled. The state was claimed in 2007 by \belle,
as a peak in the $B \to K (\psiprime\,\pi^+)$ channel~\cite{Choi:2007wga}, and it  was the first observation of a charged charmoniumlike state.
Such analyses can suffer from the rich structure of $K\pi$ resonances, which reflect into the $\psiprime\,\pi$ channel and might create fake peaks. However, \belle considered that the events with $M(\psiprime\,\pi^-)\sim 4430\mev$ correspond to events with $\cos\theta_{K\pi}\simeq 0.25$, \ie an angular region where interfering  $L=0,1,2$ partial waves cannot produce a single peak without creating other larger structures elsewhere. 
The subsequent \babar analysis parametrized the $K\pi$ spectrum with a model-independent technique, \ie just projecting the $K\pi$ spectrum over partial waves without assuming any specific $K^*$ content. The reflection of this system could fit the $\psiprime\,\pi$ spectrum without needing any additional $Z(4430)$ resonance. The final dataset \belle's Dalitz analysis confirmed the presence of a resonance~\cite{Mizuk:2009da,Chilikin:2013tch}. Finally, \lhcb puts beyond doubt the existence of this state, first with a high-statistics 4D fit to the usual isobar model, and later applying the \babar's model-independent analysis~\cite{Aaij:2014jqa,Aaij:2015zxa}.
This state is extremely interesting, because is far from any reasonable open-charm threshold with the correct quantum numbers~\footnote{Unless one resorts to the radial excitations $D(2S)$ and $D^*(2S)$, but their broad width of $\sim 100\mev$ challenges any proper definition of threshold.}.
The averaged mass and width are 
$M=(4478 \pm 17)\mev$ and $\Gamma =(180\pm31) \mev$, whereas the favored signature is $J^{PC} = 1^{+-}$. The \lhcb analysis also stresses the resonant nature of the $Z(4430)$, by fitting six independent complex numbers in bins of $M^2(\psiprime\,\pi)$, instead of forcing the Breit-Wigner lineshape. The result is consistent with the Breit-Wigner expectation (see \figurename{~\ref{fig:z4430lhcb}}), although other mechanisms can mimic a similar resonant behavior --- see \sectionname{\ref{sec:cusps}}.
\begin{figure}[h]
  \begin{center}
    \vcenteredhbox{\includegraphics[width=.52\textwidth]{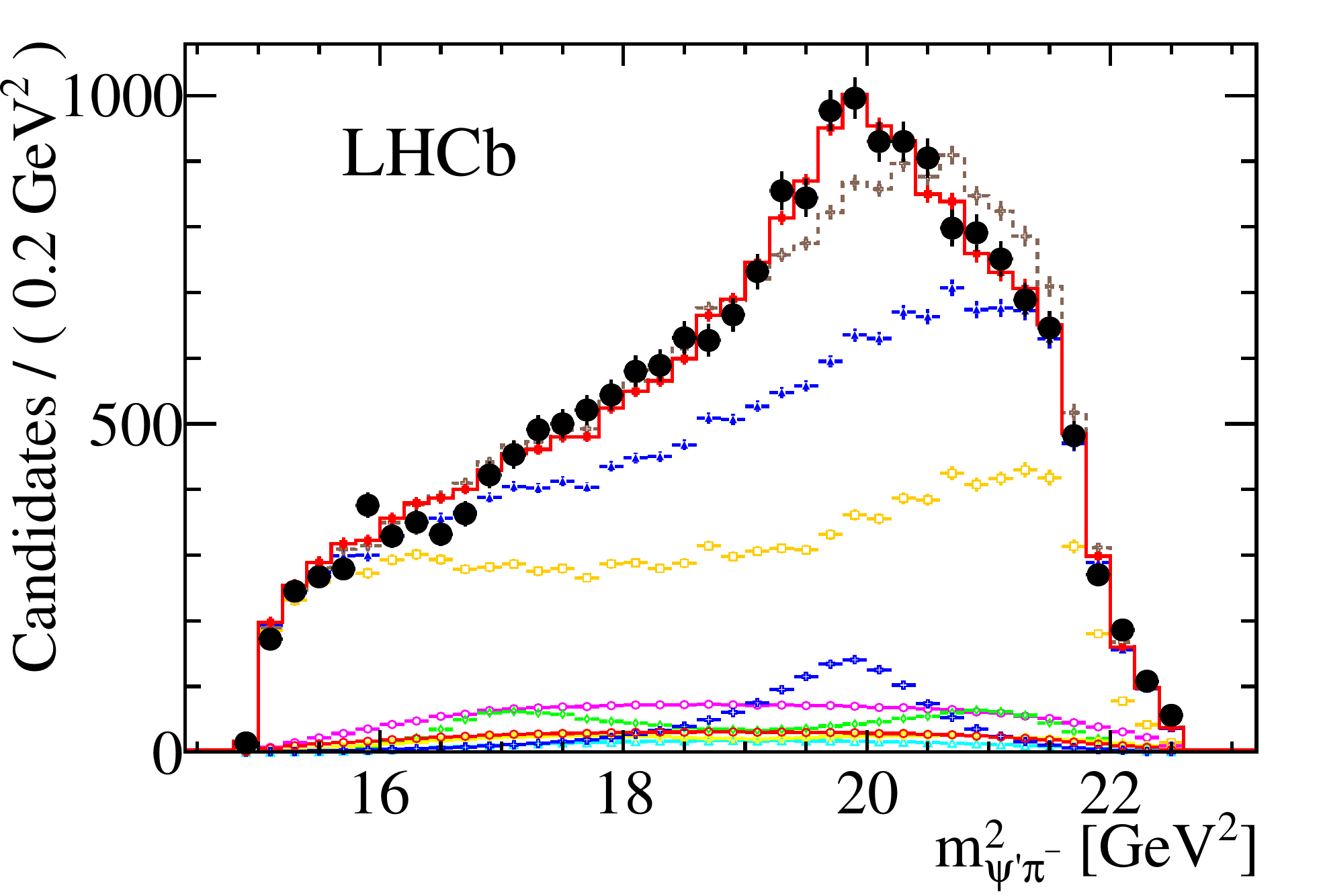} \hspace{1cm}}
    \vcenteredhbox{\includegraphics[width=.40\textwidth]{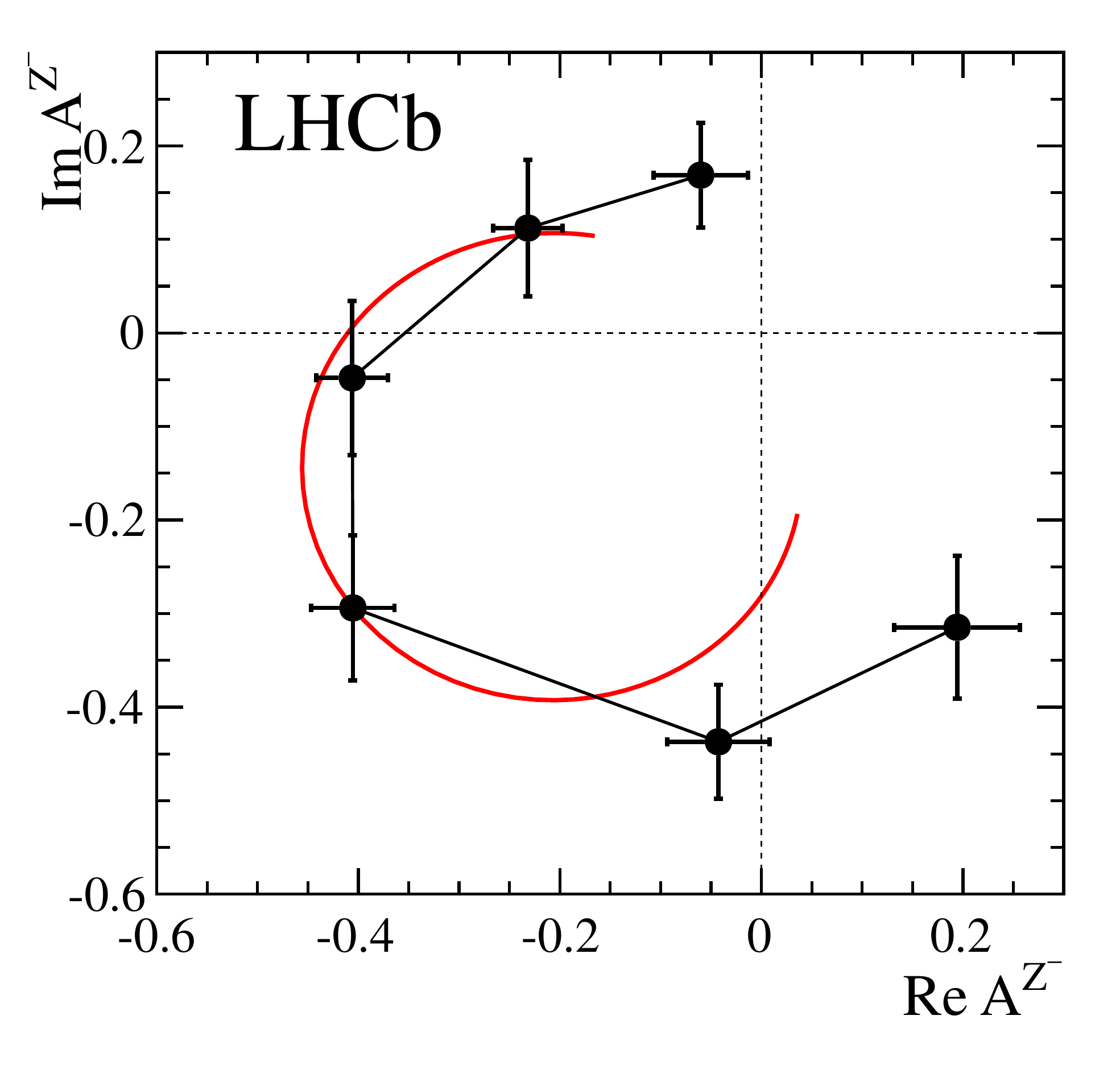} }
  \end{center}
   \caption{Invariant mass distributions in $\psiprime\,\pi^-$ channel (left) and resonant behavior (right) according to \lhcb~\cite{Aaij:2014jqa}. In the left panel, the red solid (brown dashed) curve shows the fit with (without) the additional $Z(4430)$ resonance. In the right panel, the complex value of the $Z(4430)$ fitted amplitude for six bins of $M(\psiprime\,\pi)$ is shown. The red curve is the
prediction from the Breit-Wigner formula with a resonance mass (width) of $4475$ ($172$)\mev.}
  \label{fig:z4430lhcb}
\end{figure}

\belle also searched for the same state in the $B \to K (\jpsi\,\pi^+)$ channel~\cite{Chilikin:2014bkk}. A $4\sigma$ signal consistent with the $Z(4430)$ is found. We can thus compare
\begin{subequations}
 \begin{align}
\BR\left(\Bzb\to K^- Z(4430)^+\right)\times\BR\left(Z(4430)^+ \to\psiprime\,\pi^+\right) &= \left(6.0^{+1.7}_{-2.0}{}^{+2.5}_{-1.4}\right)\times10^{-5}.\\
\intertext{with}
\BR\left(\Bzb\to K^- Z(4430)^+\right)\times\BR\left(Z(4430)^+ \to\jpsi\,\pi^+\right) &= \left(5.4^{+4.0}_{-1.0}{}^{+1.1}_{-0.6}\right)\times10^{-6},
\end{align}
\end{subequations}
\ie the decay into \psiprime is larger by one order of magnitude, despite the smaller phase space.

\subsection{The \texorpdfstring{$[cs][ \bar c  \bar s]$}{[cs][cbar sbar]} sector}
\label{sec:ccss}
The \cdf experiment announced a resonance close to threshold in $\jpsi \phi$ invariant mass, in the channel $B^+\to \jpsi \phi K^+$~\cite{Aaltonen:2009tz,Aaltonen:2011at}. Since the creation of a $s\bar s$ pair is OZI suppressed, the very existence of such states likely requires exotic interpretations. This state was called $X(4140)$, or sometimes $Y(4140)$.

\begin{figure}[t]
\centering
  	\vcenteredhbox{\includegraphics[width=.44\textwidth]{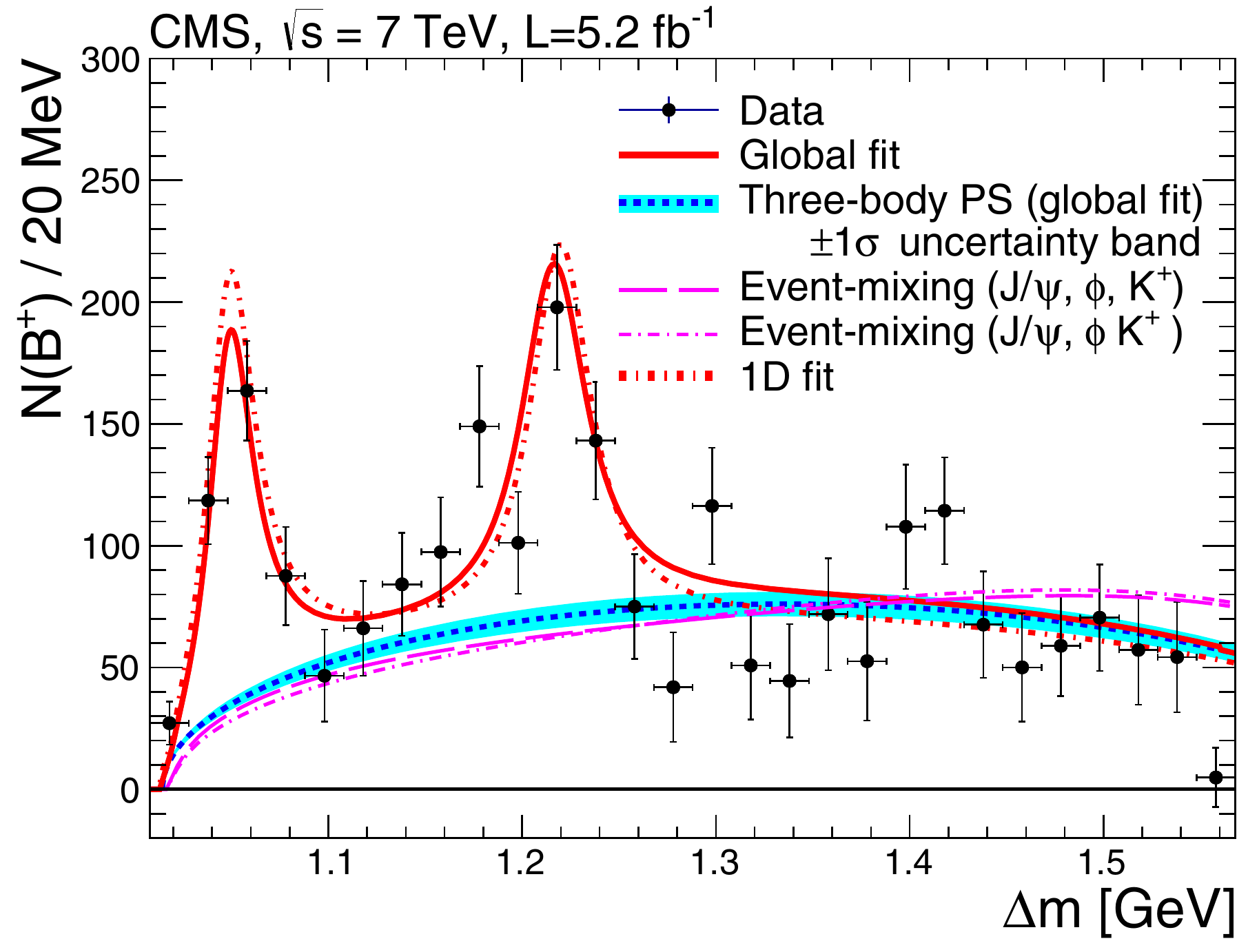}}
  	\vcenteredhbox{\includegraphics[width=.54\textwidth]{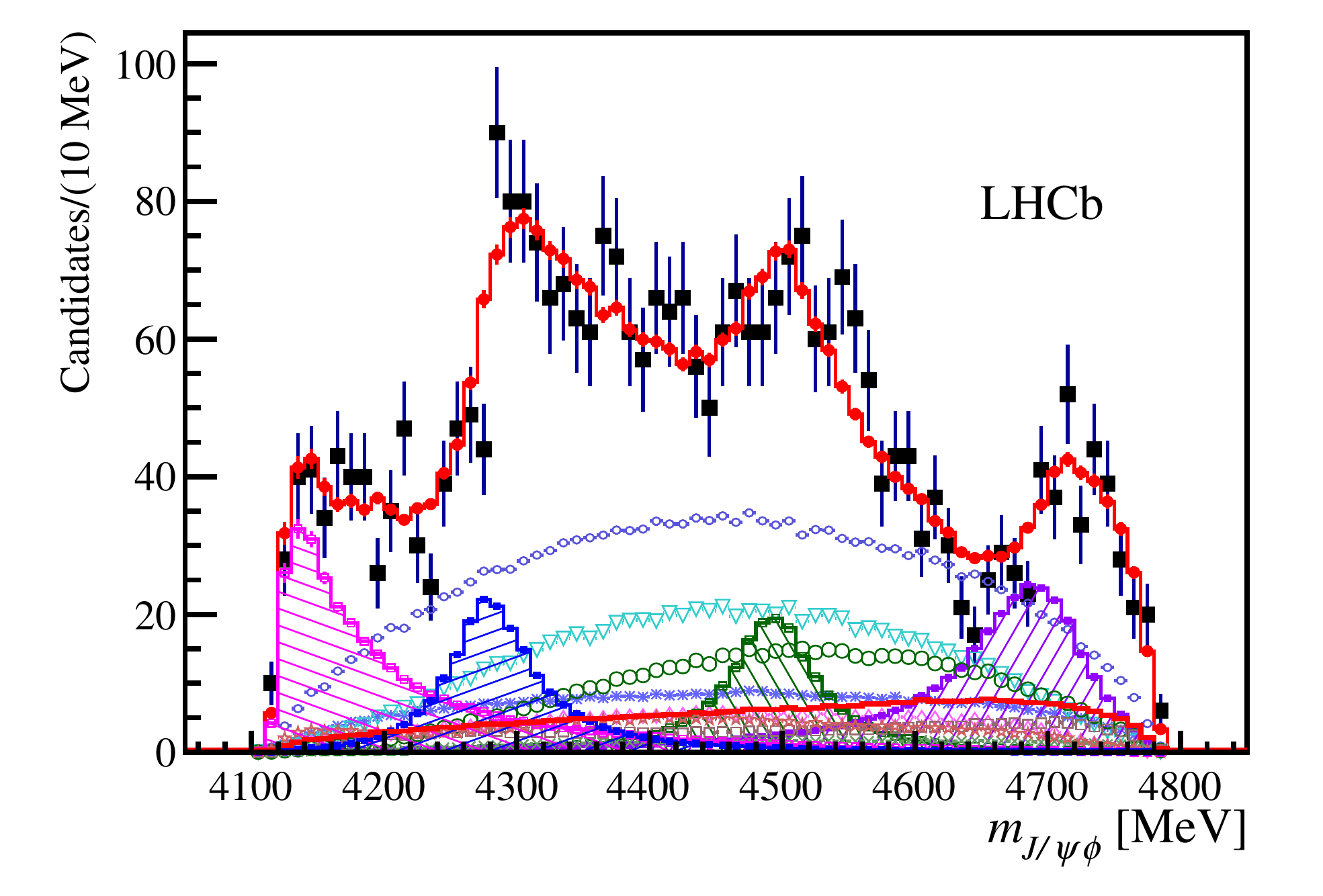}}
  \caption{Invariant mass distributions of $\jpsi\phi$, from the $B^+ \to K^+ \jpsi \phi$ data by \cms (\cite{Chatrchyan:2013dma}, left panel) and \lhcb (\cite{Aaij:2016nsc,Aaij:2016iza}, right panel). }
  \label{fig:jpsiphilhcb}
\end{figure}
\begin{table}[p]\small
\centering
 \begin{tabular}{lcllcl}
\hline\hline
Experiment  & Events & Mass (Me\!V) & Width (Me\!V) & 
Sign. & Fraction \% \\
\hline\hline
\multicolumn{6}{c}{$X(4140)$ peak}\\ \hline
\cdf (2008), 2.7 fb$^{-1}$  \cite{Aaltonen:2009tz} & 
$58\pm10$  &  
$4143.0\pm2.9\pm1.2$  &
$11.7^{+8.3}_{-5.0}\pm3.7$ &
$3.8\sigma$ & \\
\belle (2009) \cite{Brodzicka:2010zz} &
$325\pm21$  &
$4143.0$ {fixed} & 
$11$ {fixed} &  
$1.9\sigma$ & \\
\cdf (2011) 6.0 fb$^{-1}$  \cite{Aaltonen:2011at} & 
$115\pm12$  &
$4143.4^{+2.9}_{-3.0}\pm0.6$ &
$15.3^{+10.4}_{-6.1}\pm2.5$ & 
$5.0\sigma$ &
$14.9\pm3.9\pm2.4$ \\
\lhcb (2011) 0.37 fb$^{-1}$  \cite{Aaij:2012pz} &
$346\pm20$  &
$4143.4$ fixed & 
$15.3$ fixed   &
$1.4\sigma$    &
$<7$ @~90\% \CL \\
\cms (2013) 5.2 fb$^{-1}$  \cite{Chatrchyan:2013dma} &
$2480\pm160$ &
$4148.0\pm2.4\pm6.3$ &
$28^{+15}_{-11}\pm19$ & 
$5.0\sigma$ &
$10\pm3$ (stat.)  \\
\Dzero (2013) 10.4 fb$^{-1}$  \cite{Abazov:2013xda} &
$215\pm37$  &
$4159.0\pm4.3\pm6.6$ &
$19.9\pm12.6^{+1.0}_{-8.0}$ &
$3.1\sigma$ &
$21\pm8\pm4$ \\
\babar  (2014) \cite{Lees:2014lra} &
$189\pm14$  &
$4143.4$ fixed & 
$15.3$ fixed  &
$1.6\sigma$ &
$<13.3$ @~90\% \CL \\
\hline
\multicolumn{2}{l}{\Dzero (2015) 10.4 fb$^{-1}$  \cite{Abazov:2015sxa}}
&
&
&
&
\\

\multicolumn{2}{l}{\quad$p\bar p \to\jpsi\phi + \text{All}$ (prompt)} &
\multirow{2}{*}{$4152.5\pm1.7^{+6.2}_{-5.4}$} &
\multirow{2}{*}{$16.3\pm5.6\pm11.4$} & $4.7\sigma$  & \\
\multicolumn{2}{l}{\quad$p\bar p\to\jpsi\phi + \text{All}$  (non-prompt)} & & &
$5.7\sigma$ &
 \\
\hline
Average  &  & $4145.5 \pm 2.3$ & $14.9 \pm 4.4$ & & \\ \hline\hline
\multicolumn{6}{c}{Higher peaks}\\ \hline
\cdf 6.0 fb$^{-1}$  \cite{Aaltonen:2011at} & 
$115\pm12$ &
$4274.4^{+8.4}_{-6.7}\pm 1.9$ &
$32.3^{+21.9}_{-15.3}\pm7.6$ &
$3.1\sigma$ & \\
\lhcb 0.37 fb$^{-1}$  \cite{Aaij:2012pz} &
$346\pm20$  &
$4274.4$ fixed &
$32.3$ fixed  &
  &
$<8$ @~90\%C.L. \\
\cms 5.2 fb$^{-1}$  \cite{Chatrchyan:2013dma} &
$2480\pm160$ &
$4313.8\pm5.3\pm7.3$ &
$38^{+30}_{-15}\pm16$ &
  & \\
\Dzero 10.4 fb$^{-1}$  \cite{Abazov:2013xda} &
$215\pm37$  &
$4328.5\pm12.0$ &
$30$ fixed &
  & \\
\babar \cite{Lees:2014lra} &
$189\pm14$ &
$4274.4$ fixed &
$32.3$ fixed  &
$1.2\sigma$  &
$<18.1$ @~90\% \CL \\
\hline
\belle \cite{Shen:2009vs} &
$\gamma\gamma\to\jpsi\phi$ &
$4350.6^{+4.6}_{-5.1}\pm0.7$ &
$13^{+18}_{-9}\pm4$ & 
$3.2\sigma$ & \\
\hline\hline
\multicolumn{6}{c}{New \lhcb results~\cite{Aaij:2016nsc,Aaij:2016iza}}\\ \hline
$X(4140)$, $J^{PC} = 1^{++}$ & $4289\pm 151$
& $4146.5\pm4.5^{+4.6}_{-2.8}$ & $83\pm21^{+21}_{-14}$ & $8.4\sigma$ & $13\pm3.2^{+4.8}_{-2.0}$ \\
$X(4274)$, $J^{PC} = 1^{++}$ & $4289\pm 151$ & $4273.3\pm8.3^{+17.2}_{-3.6}$ & $56\pm11^{+ 8}_{-11}$  
   & $6.0\sigma$  &  $7.1\pm2.5^{+3.5}_{-2.4}$ \\
$X(4500)$, $J^{PC} = 0^{++}$ & $4289\pm 151$ & $4506\pm11^{+12}_{-15}$ & $92\pm21^{+21}_{-20}$  & $6.1\sigma$ & $6.6\pm2.4^{+3.5}_{-2.3}$  \\
$X(4700)$, $J^{PC} = 0^{++}$ & $4289\pm 151$ & $4704\pm10^{+14}_{-24}$ & $120\pm31^{+42}_{-33}$  & $5.6\sigma$ & $12\pm 5^{+ 9}_{- 5}$ \\ \hline\hline
\end{tabular}
\caption{Results related to the $\jpsi\phi$ resonances. 
We show the results of the previous experiments, with measured masses, widths and fit fractions, and compare with latest \lhcb results. In the second column, we report the total number of $B^+ \to \jpsi \phi K^+$ events. 
}
\label{tab:ccss}
\end{table}

The first low-statistics analysis by \lhcb~\cite{Aaij:2012pz}, and the analyses by \belle and \babar~\cite{Brodzicka:2010zz,Lees:2014lra} were not able to see any significant signal in this channel.
\belle searched this state in $\gamma\gamma$ fusion, driven by a molecular prediction~\cite{Branz:2009yt}, but found no $X(4140)$~\cite{Shen:2009vs}. Similarly, \bes saw no $X(4140)$ signal in the  $e^+e^-\to \gamma (\jpsi \phi)$ process~\cite{Ablikim:2014atq}. Instead, \Dzero~\cite{Abazov:2013xda} and \cms~\cite{Chatrchyan:2013dma} have  confirmed the observation, with significances of $\sim 3\sigma$ and $>5\sigma$, respectively, although with slightly inconsistent values for the mass and width. \Dzero found the state also in prompt production~\cite{Abazov:2015sxa}. \cdf, \cms and \Dzero also reported evidence for a second peak at a mass of $M \sim 4300 \mev$ and $\Gamma \sim 35\mev$.

Very recently, \lhcb published the preliminary amplitude analysis of the full 6D Dalitz distributions~\cite{Aaij:2016nsc,Aaij:2016iza}. The results are impressive: they not only confirm both the $X(4140)$ and the second peak, although with much larger widths, but they also find other two heavier states. The situation is resumed in \tablename{\ref{tab:ccss}}. 
The two lighter states, $X(4140)$ and $X(4274)$, have $J^{PC} = 1^{++}$, and masses and widths 
$M = (4146.5\pm4.5^{+4.6}_{-2.8})\mev$, $\Gamma = (83\pm21^{+21}_{-14})\mev$, and 
$M = (4273.3\pm8.3^{+17.2}_{-3.6})\mev$, $\Gamma = (56\pm11^{+ 8}_{-11})\mev$, respectively.
The two heavier ones, $X(4500)$ and $X(4700)$, are scalars $J^{PC} = 0^{++}$, and have resonance parameters 
$M=(4506\pm11^{+12}_{-15})\mev$, $\Gamma=(92\pm21^{+21}_{-20})\mev$, and $M=(4704\pm10^{+14}_{-24})\mev$, $\Gamma = (120\pm31^{+42}_{-33})\mev$, respectively.

In \figurename{~\ref{fig:jpsiphilhcb}} we show the $M(\jpsi\phi)$ distribution measured at \cms and \lhcb. The discrepancy between the new \lhcb results and all the previous ones deserves some comments. The excellent particle ID, and the large statistics allowed \lhcb to have a very precise sample with low background, and to perform the full 6D fit. The fit included up to 5 $\phi K^+$ Breit-Wigners, consistent with known or predicted $K^*$ resonances. Still, a sizable nonresonant component is needed for a good fit. All these are summed coherently to the four $J/\psi \phi$ Breit-Wigners. On the other hand, the other analyses just fitted the 1D $\jpsi \phi$ projection because of limited statistics, adding incoherently one or two Breit-Wigners to a phase-space background. The incoherent background in general tends to produce more narrow widths than the coherent one (the same happened for the $Z(4430)$, \sectionname{\ref{sec:Z4430}}), and could justify the mismatch. However, the $X(4140)$ is observed by \Dzero in inclusive production~\cite{Abazov:2015sxa}. In this case, the coherent sum of signal and background is meaningless, and the incoherent sum gives a width of $\Gamma = (16.3 \pm 5.6 \pm 11.4)\mev$, much narrower than the one observed at \lhcb, and compatible with the previous observation. An independent observation of this state in inclusive production might help resolving this controversy.

\subsection{Pentaquark states}
\label{sec:pentaexp}
In July 2015, \lhcb reported the observation of two peaks, denoted $P_c(4380)^+$ and $P_c(4450)^+$, in the invariant mass of $\jpsi\,p$ in the $\Lambda_b^0 \to \jpsi\,p\,K^-$ decay~\cite{Aaij:2015tga}. In the Dalitz plot in \figurename{~\ref{fig:pentadalitz}} a band in the $\jpsi\,p$ is indeed visible, and clearly extends beyond the $\Lambda^*$ region ($m^2_{Kp} < 4 \gev^2$). The full amplitude fit cannot satisfactorily describe the data without including these two Breit-Wigner shaped resonances~(\figurename{~\ref{fig:projectionspenta}). The lighter state has a mass $M=(4380\pm 8\pm 29)\mev$ and a width $\Gamma=(205\pm 18\pm 86)$\mev, while the heavier one has a mass  $M=(4449.8\pm 1.7\pm 2.5)\mev$ and a width  $\Gamma=(39\pm 5\pm 19)\mev$. The interference pattern of the two resonances in different bins of the helicity angle require them to have opposite parities, and the preferred $J^P$ are $\left(\tfrac{3}{2}^-, \tfrac{5}{2}^+\right)$, although $\left(\tfrac{3}{2}^+, \tfrac{5}{2}^-\right)$ and $\left(\tfrac{5}{2}^+, \tfrac{3}{2}^-\right)$ are not excluded. 
The higher mass state has a fit fraction of  ($4.1\pm0.5\pm 1.1)$\%, and the lower mass state of  ($8.4\pm0.7\pm4.2$)\%, of the total sample.
To study the resonant behavior of the two states, the amplitudes are represented as the combination of independent
complex numbers at six equidistant points in the range $\pm \Gamma_0=39\mev$ around the  as determined in the default fit.
Real and imaginary parts of the amplitude are interpolated in mass between the fitted points.
The resulting Argand diagram, shown in \figurename{~\ref{fig:argandpenta}},
is consistent with a rapid motion of the $P_c(4450)^+$ phase when its magnitude
reaches the maximum, whereas no strong conclusion can be drawn for the wider $P_c(4380)^+$.

A new analysis by \lhcb has recently confirmed the existence of pentaquark-like structure with the same model-independent used in the analysis of the $Z(4430)$ (see \sectionname{\ref{sec:Z4430}}), \ie without assuming any Breit-Wigner lineshape for the $\Lambda^*$, but just projecting the $M(Kp)$ invariant mass in partial waves, and studying the reflection in $M(\jpsi p)$. If no pentaquark is included, the $M(\jpsi p)$ distribution cannot reproduce data~\cite{Aaij:2016phn}.
Finally, \lhcb studied also the Cabibbo-suppressed $\Lambda_b^0 \to \jpsi\,p\,\pi^+$ channel. Also in this case, the $N^*$ reflections are not able to reproduce the $\jpsi\,\pi^+$ spectrum, but lack of statistics prevents the resolving of different structures (possibly $P_c(4380)$, $P_c(4450)$, or $Z(4200)$)~\cite{Aaij:2016ymb}.

A diquark-diquark-antiquark description has been proposed in~\cite{Maiani:2015vwa,Anisovich:2015cia,Lebed:2015tna} (see also \sectionname{\ref{sec:pentatheor}}). The molecular models can describe the narrow state, either as a $\bar D^{(*)} \Sigma_c$ molecule~\cite{Roca:2015dva}, or as a $\chi_{c1} p$ molecule~\cite{Guo:2015umn}, but cannot at the same time describe the broader state, or with opposite parity to the narrow one.
To confirm the \lhcb discovery, a search in another channel is needed. For example, Refs.~\cite{Wang:2015jsa,Kubarovsky:2015aaa,Astrid:2016} have proposed to search the pentaquark in photoproduction, estimating sizable cross sections.
\begin{figure}[t]
\begin{center}
\includegraphics[width=.5\textwidth]{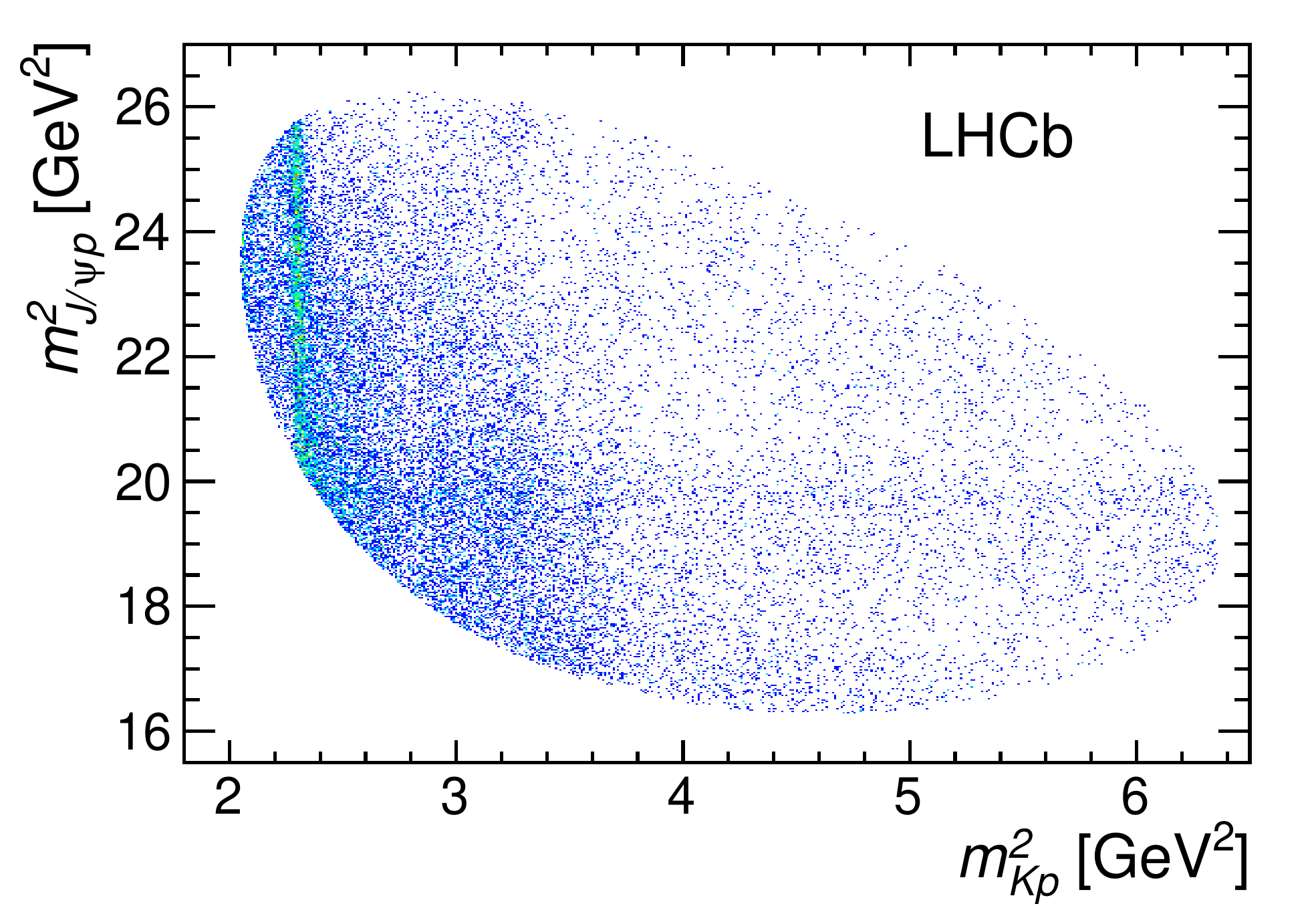} 
 \caption{Invariant mass squared of $Kp$ versus $\jpsi\,p$ for candidates within $\pm 15\mev$ of the $\Lzb$ mass~\cite{Aaij:2015tga}.}
    \label{fig:pentadalitz}
    \includegraphics[width=.44\textwidth]{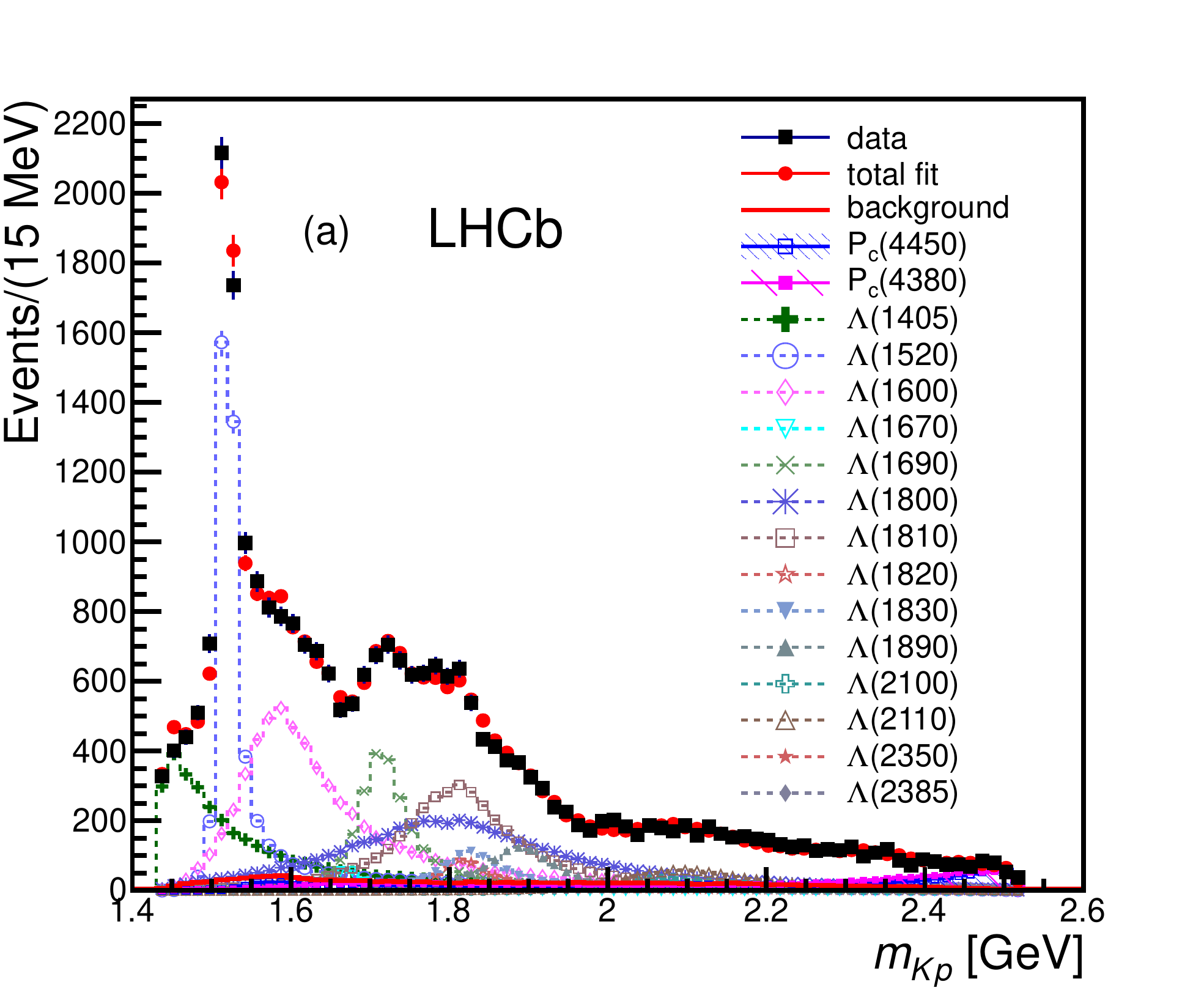} \hspace{1cm}
\includegraphics[width=.44\textwidth]{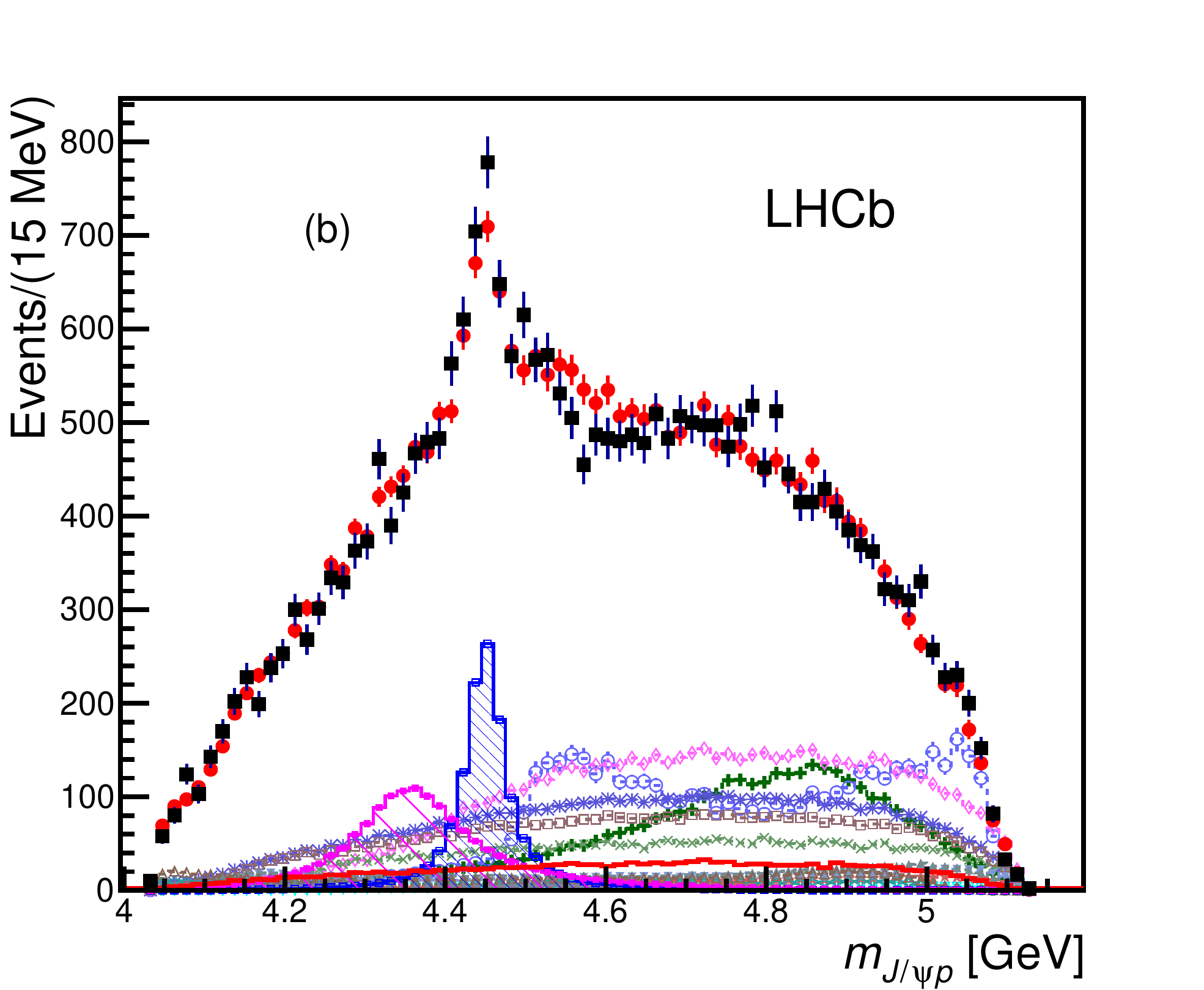} 
\caption{Results for the full amplitude fit in the $m_{Kp}$ (left) and $m_{\jpsi\,p}$ variables. 
The data are shown as (black) squares with error bars, while the (red) circles show the results of
the fit. The error bars on the points showing the fit results are due to simulation statistics~\cite{Aaij:2015tga}.}
     \label{fig:projectionspenta}
    \end{center}
\end{figure}

\clearpage
\begin{figure}[t]
\begin{center}
\includegraphics[width=.6\textwidth]{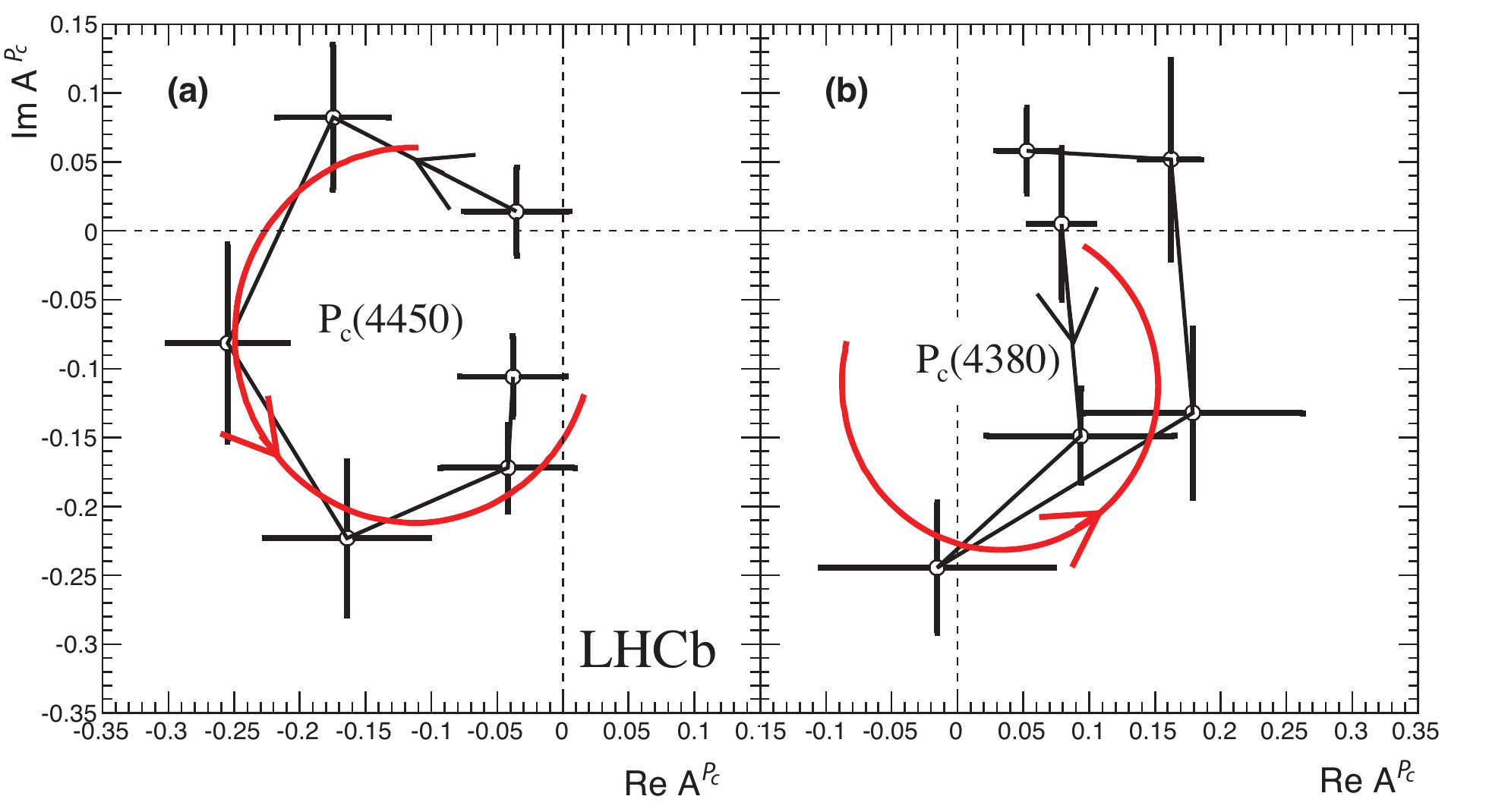} 
 \caption{Fitted values of the real and imaginary parts of the amplitudes for the favored spin fit for the $P_c(4450)^+$ state (left) and for the $P_c(4380)^+$ state (right), each divided into six $m_{\jpsi\,p}$ bins
of equal width shown in the Argand diagrams as connected points with
error bars ($m_{\jpsi\,p}$ increases counterclockwise). The solid (red) curves are the predictions from
the Breit-Wigner formula~\cite{Aaij:2015tga}.
}
    \label{fig:argandpenta}
    \end{center}
\end{figure}
\subsection{Other states}
\label{sec:others}
\smalltitle{$\bbb{\chi_{c0}(3915)}$}
A resonance decaying into $\jpsi \omega$ was observed by \belle and \babar in the $B \to K (\jpsi \omega)$ channel~\cite{Abe:2004zs,Aubert:2007vj,delAmoSanchez:2010jr}. Both experiments found the same state in $\gamma\gamma$ fusion~\cite{Uehara:2009tx,Lees:2012xs}. The latter production mechanism constrains $C = +$, $J \neq 1$, and the study of angular correlations favors a $J^{PC} = 0^{++}$ assignment. 
At the beginning \belle named this state $Y(3940)$, then renamed to $Y(3915)$ after a more precise determination of the mass.
The averaged mass and width are $M=(3918.4\pm 1.9)\mev$ and $\Gamma = (20 \pm 5)\mev$.
The PDG 2014 rebaptized to $\chi_{c0}(2P)$, choosing for an ordinary charmonium assignment for the state. However, the $\chi_{c0}(2P)$ is expected to have $\Gamma(\chi_{c0}(2P) \to \D\Dbar)\sim 30\mev$, \ie wider than the total width measured of the $Y(3915)$. Even if no upper bound on $\BR(Y(3915)\to \D\Dbar)$ has been reported, 
no signs of a signal for such a decay appear in the
measured $\D\Dbar$ invariant mass distributions for $B\to \D\Dbar K$ decays~\cite{Aubert:2007rva,Brodzicka:2007aa}. Moreover,  the hyperfine splitting $\chi_{c2}(2P) - \chi_{c0}(2P)$ would be only $6\%$ with respect to the $\chi_{c2}(1P) - \chi_{c0}(1P)$ splitting,  much smaller than the same ratio in the bottomonium system ($r \sim 0.7$), and than the potential model 
predictions~\cite{Barnes:2005pb} ($0.6 < r <0.9$). These facts challenge the ordinary charmonium interpretation~\cite{Guo:2012tv,Olsen:2014maa}. In the PDG 2015 online edition a compromise was reached, and it was called $\chi_{c0}(3915)$ eventually. This state has been identified as the lightest $[cs] [\bar c \bar s]$ tetraquark~\cite{Lebed:2016yvr}, although this interpretation looks at odds with the other states seen in $\jpsi \phi$ (see \sectionname{\ref{sec:ccss}}). Alternatively, Voloshin and Li have proposed a $D_s^+ D_s^-$ molecular assignment~\cite{Li:2015iga}.

\smalltitle{A $\bbb{X(5568)}$?} Recently \Dzero claimed the discovery of a narrow state in the $B_s^0 \pi^\pm$ invariant mass,
where the $B_s^0$ is reconstructed in the $\jpsi\,\phi$ final state. The mass and width are $M = (5567.8 \pm 2.9^{+0.9}_{-1.9})\mev$ and $\Gamma = (21.9 \pm 6.4^{+5.0}_{-2.5})\mev$, and the significance of the excess is $3.9\sigma$, increased to $5.1\sigma$ if an additional cut ($\Delta R < 0.3$) is considered~\cite{D0:2016mwd}. The ratio of the yield of resonant over background events in the detector acceptance region and $\pt >10\gev$ is $\rho =  (8.6 \pm 1.9 \pm 1.4)\%$. 
The use of this additional cut  has been criticized, even though the analysis was blind, and the cut was not optimized for this purpose. 
Right after, \lhcb explored the very same channel, finding no hints of any resonant structure, neither at the mass observed by \Dzero, nor all the way to the $BK$ threshold, and putting the 90\% C.L. upper limit  $\rho < 1.6\%$ in the detector acceptance region and $\pt >10\gev$~\cite{LHCb:2016ppf}. Even considering the differences between the two experiments in terms of acceptance, center-of-mass energy, and initial state, this gap makes the observation by \Dzero less plausible. In \figurename{~\ref{fig:BsD0LHCb}} we show the $B_s^0 \pi^\pm$ distributions measured by the two experiments. \Dzero has confirmed the observation when the $B_s^0$ is reconstructed in the $D_s^- \mu^+ \nu$ final state~\cite{Zieminska}. On the other hand, neither the recent preliminary analysis by \cms is able to see the state~\cite{CMS5568}. 
It is worth noticing that such a state was at odds with many exotic models~\cite{Esposito:2016itg,Ali:2016gdg,Burns:2016gvy,Guo:2016nhb}.

\begin{figure}[t]
\begin{center}
\vcenteredhbox{\includegraphics[width=.40\textwidth]{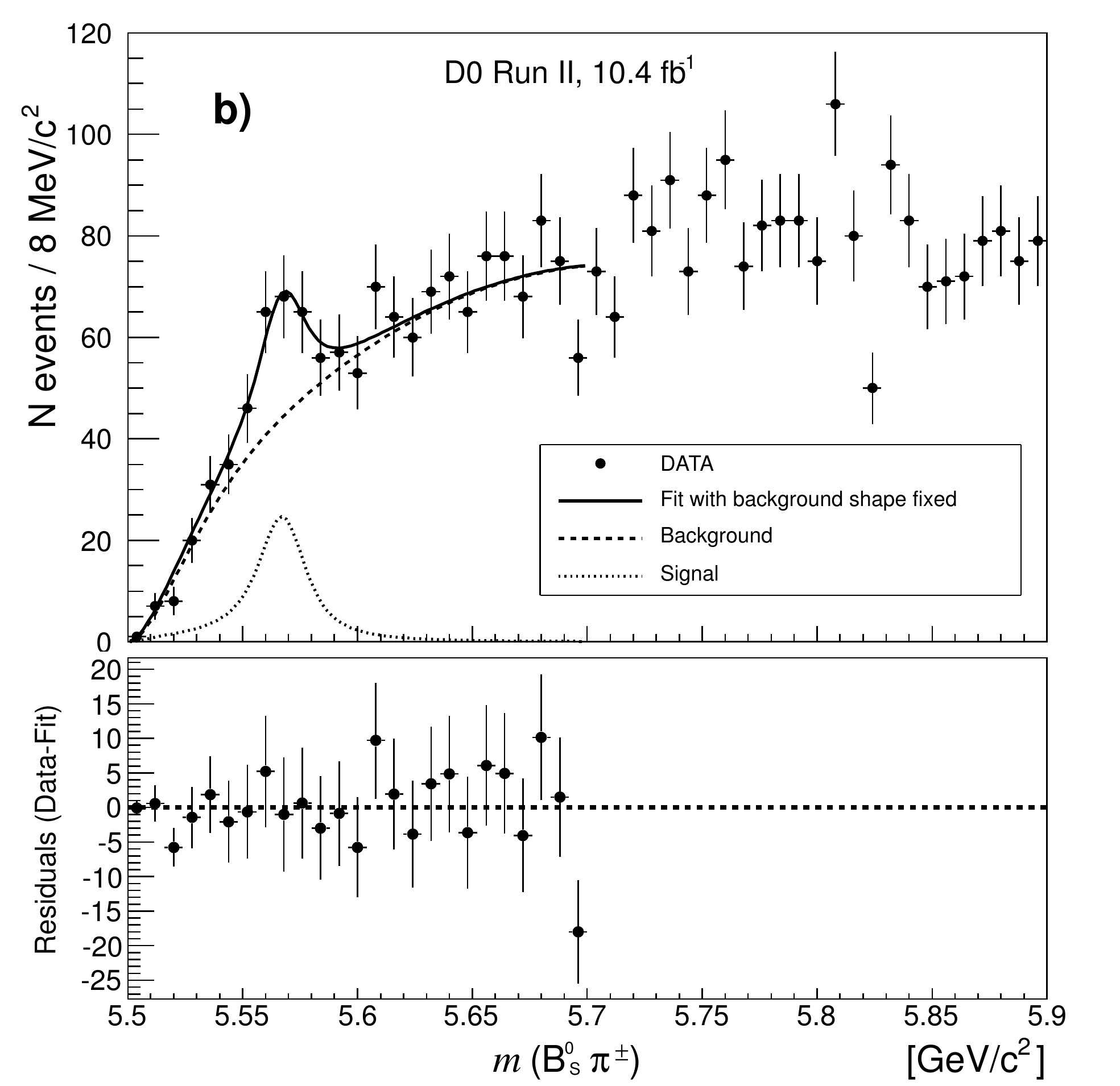} }
\vcenteredhbox{\includegraphics[width=.56\textwidth]{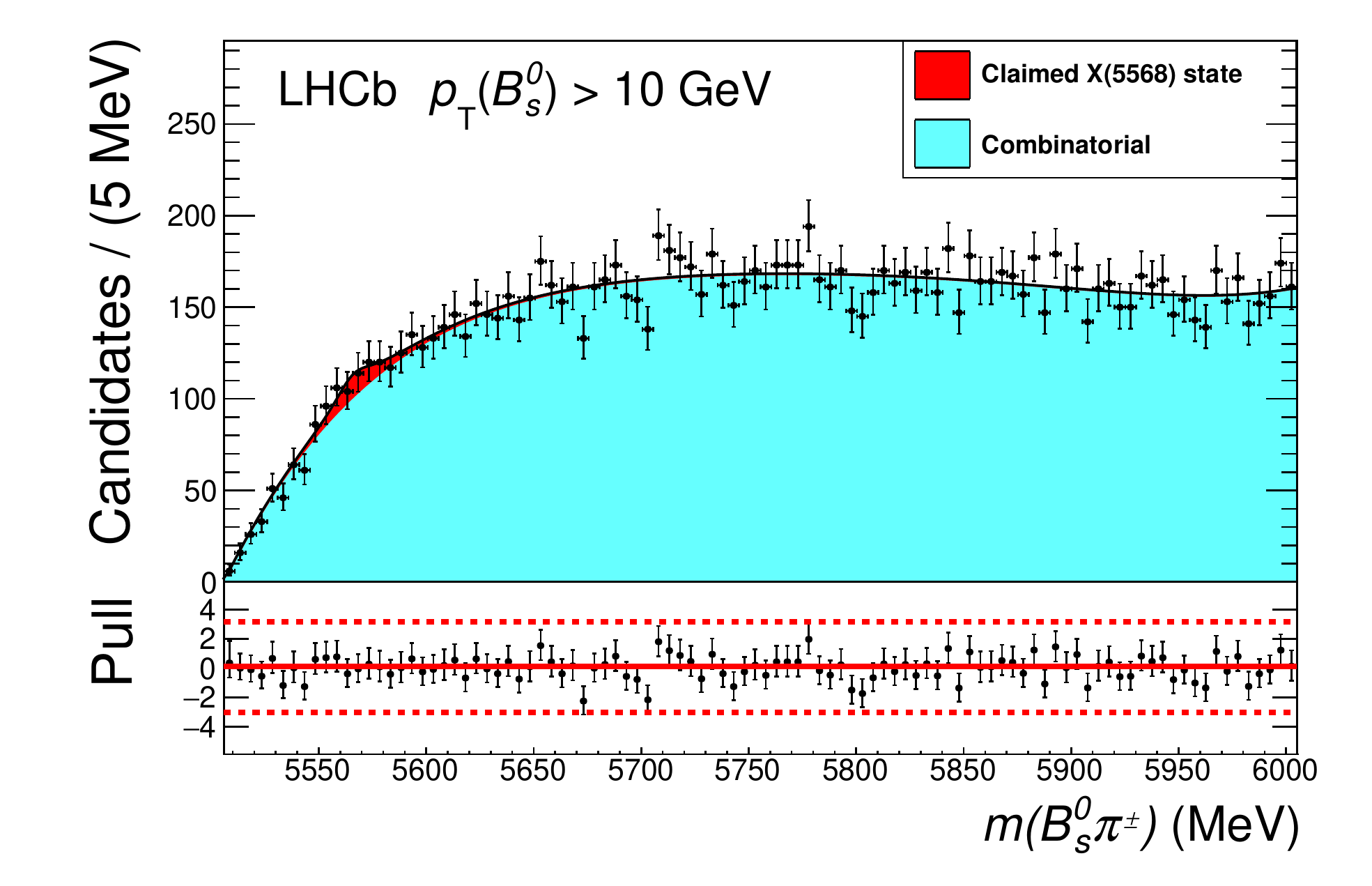} }
\caption{Left panel: The $m(B_s^0 \pi^{\pm})$ distribution
together with the background distribution and the fit results, with $\pt > 10\gev$ and no cone cut, at \Dzero~\cite{D0:2016mwd}. Right panel: the same  at \lhcb~\cite{LHCb:2016ppf} }
    \label{fig:BsD0LHCb}
    \end{center}
\end{figure}

\smalltitle{A $\bbb{X(4350)}$?} Triggered by the observation of a $X(4140)$ resonance by \cdf in $B \to K (\jpsi \phi))$, \belle explored the $\gamma\gamma \to \jpsi\phi$ channel. A peak with a $3.2\sigma$ significance was seen at $M=(4350.6^{+4.6}_{-5.1}\pm 0.7)\mev$ and
$\Gamma=(13^{+18}_{-9}\pm 4)\mev$%
, and dubbed $X(4350)$. The narrow width and the signature $J\neq 1$ constrained by the production mode prevent the identification of this state with the other ones observed by \lhcb in $B \to K (\jpsi \phi))$, see \sectionname{\ref{sec:ccss}}.

\smalltitle{$\bbb{X(3940)}$ and $\bbb{X(4160)}$}
The $X(3940)$ was observed by \belle in associated production with a $\jpsi$ with $M=(3942^{+7}_{-6} \pm 6)\mev$ and $\Gamma = (37^{+26}_{-15} \pm 8)\mev$~\cite{Abe:2007jn,Abe:2007sya}. A partial reconstruction technique 
showed that $X(3940)\to \Dstar\Dbar$ is the prominent decay mode
, whereas $X(3940)\to \D\Dbar$ and $\to \jpsi\omega$ show no signal.
The production mechanism $e^+e^- \to \gamma^* \to \jpsi X(3940)$ constrains the state to have $C=+$, and favors low values of the spin. The absence of the $\D\Dbar$ decay suggests unnatural parity, so the tentative $J^{PC}$ assignment for this state is $0^{-+}$.

The same analysis also observed a state called $X(4160)$ in the $\Dstar\Dstarb$ invariant mass. %
The fitted mass and width are $M=(4156^{+25}_{-20}\pm15)\mev$ and $\Gamma=(139^{+111}_{-61}\pm21)\mev$. 
Also in this case, the favored signature is $J^{PC}=0^{-+}$. These two states are good candidate for the $\eta_c(3S)$ and $\eta_c(4S)$ respectively, although their masses deviate from the ordinary predictions~\cite{Rosner:2005gf,Chao:2007it}.

\smalltitle{$\bbb{Z_1(4050)}$ and $\bbb{Z_2(4250)}$}
In a Dalitz-plot analysis of $B\to \chi_{c1}\pi^+ K$ decays, \belle could get an acceptable fit only by adding two resonances
in the  $\chi_{c1}\pi^+$ channel, called $Z_1(4050)$ and $Z_2(4250)$~\cite{Mizuk:2008me}. We report the Dalitz projection in \figurename{~\ref{fig:z12belle}}. The fitted masses and widths are
$M_1  =(4051 \pm 14^{+20}_{-41})\mev$, $\Gamma_1 =(82^{+21+47}_{-17-22})\mev$, and 
$M_2 =(4248^{+44+180}_{-29-35})\mev$, $\Gamma_2 =(177^{+54+316}_{-39-61})\mev$, respectively.
The same decay was investigated by \babar, which carefully
studied the effects of interference between resonances in the $K\pi$
system~\cite{Lees:2011ik}. Considering interfering resonances in the $K\pi$ channel only, \babar obtained good fits to data without adding any $\chi_c \pi$ resonance, but because of limited statistics, no strong conclusion could be made against the existence of these resonances. Their interpretation as $D$-wave tetraquarks has been suggested in a flux-tube model~\cite{Deng:2015lca} Also, if one assumes a $0/2^{++}$ assignment for the $Z_1(4050)$, the resonance can be identified as a member of the ground-state tetraquark multiplet --- see \sectionname{~\ref{scaltens}}. Similarly, the $Z_2(4250)$ might be the radial excitation of the lightest $0^{++}$ tetraquark.

\begin{figure}[t]
  \begin{center}
    \includegraphics[width=.5\textwidth]{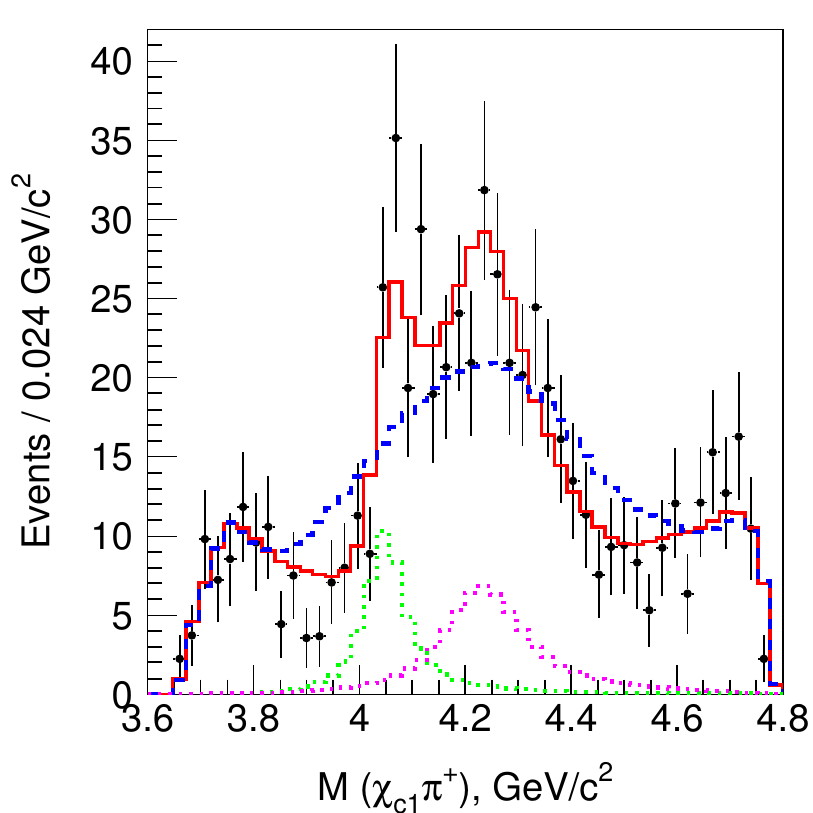} 
  \end{center}
   \caption{Invariant mass distributions of 
	$\chi_{c1}\pi^\pm$,
	with fit results showing the charged resonances in the \belle~\cite{Mizuk:2008me} analysis.
	The region of the
	$K^{\ast}(890)$ and $K^{\ast}(1410)$ peaks are removed.
	In left panel, the solid red histogram shows the results
	of the fit that includes coherent $Z_1$ and $Z_2$ amplitudes;
	the dashed blue curve is the result of the fit using $K\pi$
	amplitudes only.}
  \label{fig:z12belle}
\end{figure}

\smalltitle{A $\bbb{Z(4050)}$?} \belle report a $3.5\sigma$ excess in the $\psiprime\,\pi^+$ invariant mass, in the $e^+e^- \to \psiprime\,\pi^+\pi^-$~\cite{Wang:2014hta}, with best fit parameters $M=(4054 \pm 3 \pm 1)\mev$ and $\Gamma =(45 \pm 11\pm 6)\mev$.

\smalltitle{$\bbb{Z(4200)}$} 
In the same analysis of the $B\to K \psiprime\,\pi$ which confirmed the $Z(4430)$ (see \sectionname{\ref{sec:Z4430}}), \lhcb observed another significant broad structure, with mass and width $M= (4239 \pm 18 ^{+45}_{-10})\mev$, $\Gamma = (220 \pm 47^{+108}_{-74})\mev$, and likely $J^{PC} = 0^{-+}$. Since the resonant nature in the Argand plot was unclear, \lhcb prudently did not claim the discovery. Soon after, \belle claimed the discovery of a broad $Z(4200)$ state with likely $J^{PC}=1^{+-}$, mass and width $M= (4196^{+31}_{-29}{}^{+17}_{-13}) \mev$, $\Gamma=(370^{+70}_{-70}{}^{+70}_{-132}) \mev$, with a significance of $6.2\sigma$. The two observations might be related to the same state, although such large widths cast some doubts on the actual resonant nature of these states. The interpretation as a tetraquark state has been suggested in the context of a flux-tube model~\cite{Deng:2015lca}, or with QCD sum rules~\cite{Chen:2015fsa}.

\section{Conclusions and outlook}
A considerable part of the material presented in this review is selected from the vast literature on the \XYZP resonances, with the  scope of preparing the discussion 
appearing in the core sections in which we have much further elaborated on some new ideas  published recently by our group.  The main drive  of this paper is to present a clear identification of the most striking conflicts of models with data and the indication of one (or more) possible  ways to the solution of them.   In some few cases we have reported, and set in the context, even some textbook results and derivations  which we have used in diverse occasions. This was essentially done for the purpose of being self-contained. 

The connection between the diquarkonium discrete spectrum and open meson thresholds might provide a set of `selection rules' that fit the experimental data so far available. We presented  some predictions and the rules for formulating them. Elements of a unitary picture are provided, although in the form of work in progress. The discussion about \XYZP states is certainly not concluded here, and more  efforts, and hopefully new forthcoming data, are  needed. Therefore, our hope is that the arguments formulated in this paper might  attract and stimulate new research  and fresh ideas also from  outside the community which has traditionally worked in the field.  

Even though we believe that no `new physics' in the usual sense will be needed to definitely understand multiquark resonances, we cannot think that such a richness of experimental data and theoretical problems should pertain only to the practitioners of hadronic physics with a more or less high nuclear theory culture but could attract more attention in the high energy physics experimental/theoretical  communities. %

Coping and solving this problem could shed new light on 
non-perturbative QCD and in general on strong interactions, as documented in several valuable papers. We actually show here that, to some extent, this has already happened, and one of our aims was that of reporting on the diversity of ideas which have been advanced to solve the problems that data pose, and even to propose new directions of investigation which could arise exploiting methods proper of other branches of physics.%

Model builders and lattice gauge theorists could find a very stimulating `playground' in seeking solutions to the multiquark resonances problem, and this, in our opinion, is a concrete opportunity of progress in a field which is firmly connected to data and experimental physics practice.  

As shown in the sections more closely related to experiments, the complexity and variety of performed analyses is remarkable. Again this results from the work of a somewhat restricted community which we hope will benefit of more support and collaboration in the coming years. 

In a few concluding words, we observe that diquark-antidiquark states, hybrids and molecules, in various forms, all catch important aspects of the problem, being understood that the observed multiquark resonances are not the mere manifestation of kinematical effects, like cusps, as some authors claim. 

There might be  several ways to learn from what was understood with all these methods, not losing the contact to experimental evidence and preferring all those alternatives which appear to be formulated in a clearly falsifiable way.   In a sense, this is a privilege of the field: theoretical models can be checked against data on reasonable time scales.   

\section*{Acknowledgements}
Most of the work presented here derives from invaluable collaboration with Luciano Maiani, Fulvio Piccinini and Veronica Riquer. They are all implicitly coauthoring this paper, exception made for what is imprecise or even wrong, which is our full responsibility. Along the years we benefited of the collaboration of Riccardo Faccini, who helped us to  find the way in some experimental data analysis intricacies, and of a number of collaborators each of them contributing with their insight and work to solve the problems offered by the changing experimental picture --- we wish to thank F.~Brazzi, T.~Burns, G.~Cotugno, N.~Drenska, G.~Filaci, A.~L. Guerrieri, M.~Papinutto, V.~Prosperi, C.~Sabelli, and N.~Tantalo. ADP benefited from sporadic collaborations and discussions with A.~Ali, I.~Bigi, B.~Grinstein, and R.~Lebed and recent exchanges with G.~C. Rossi, G. Veneziano, and S. Peris. AP wishes to thank M.~Bochicchio and A.~Szczepaniak for many fruitful discussions. The support and dialog with experimentalists has been constant. We wish to thank S.~Stone for several decisive exchanges, together with A.~A.~Alves, M.~Battaglieri, G.~Cavoto, M.~Destefanis, R.~Mussa, A.~Palano, M.~Pappagallo, A.~Pompili, F.~Renga, M.~Rescigno, and U.~Tamponi. The confrontation with the ideas of E.~Braaten and M.~Voloshin has been very instructive to us. This material is based upon work supported in part by the U.S. Department of Energy, Office of Science, Office of Nuclear Physics under contract DE-AC05-06OR23177, and DE-FG02-11ER41743. %

\appendix
\clearpage

\section{\texorpdfstring{$X(3872)$}{X(3872)} production at hadron colliders\label{appmol}}
It is possible to estimate an upper bound for the production cross section of the $X(3872)$ as follows~\cite{Bignamini:2009sk}
\begin{equation}
\label{eq:upperbound}
\sigma(p\bar p\to X(3872))\lesssim \sigma^\text{max} (p\bar p\to \Dz\Dstarzb)\sim \int_{|\bm k|\in \mathcal{ R}}
\left|\langle \Dz\Dstarzb(\bm k)|p\bar p\rangle\right|^2,
\end{equation}
where $\bm k$ is the relative momentum between the two $D$ mesons in their center of mass and $\mathcal{R}$ is the domain where the two-body wave function for the molecular $X(3872)$ is significantly different from zero. 

Such an upper bound can  be estimated by  counting the number of $D^0\bar D^{*0}$ produced with a relative momentum lower than a certain  $k_0$ value. This has been done~\cite{Bignamini:2009sk} using HERWIG~\cite{Corcella:2000bw} and PYTHIA~\cite{Sjostrand:2006za}, taking $\mathcal{R}$ to be a ball or radius $[0,35]$\mev, on the basis of a na\"{i}ve gaussian shape for the two-body wave function of the $X$. 

Next we assume that all these pairs will convert into $X(3872)$.   

The result of the MC simulation (with a MC luminosity of $\sim100\nb^{-1}$) was a maximum production cross section of $0.071$ nb for HERWIG and $0.11$ nb for PYTHIA, which are both smaller than the experimental value ($\sim 30$~nb) by more than two orders of magnitude.  This seemed to be the definitive proof of the inconsistency of the molecular interpretation with the experimental data.

However, the previous approach was later criticized in~\cite{Artoisenet:2009wk} and it was shown that the theoretical and experimental cross sections might be matched resorting to Final State Interactions (FSI)~\cite{book:17198}. The possible presence of FSI, in fact, casts doubts on the applicability of the simple coalescence picture to the case of the $X(3872)$, since the two components of the molecule could be bound by final state rescattering even if their relative center of mass  momentum is large. In particular, the Migdal-Watson theory would change the previous results in two different ways
\begin{enumerate}
\item The cross section for the production of the $X$ should be modified to
\begin{align} \label{eq:FSI}
\sigma(p\bar p\to X(3872))\simeq\left[\sigma(p\bar p\to \Dz\Dstarzb)\right]_{k_0<k_0^\text{max}}\times\frac{6\pi\sqrt{-2\mu E_X}}{\Lambda}
\end{align}
where $\left[\sigma(p\bar p\to \Dz\Dstarzb)\right]_{k_0<k_0^\text{max}}$ is the upper bound evaluated in~\eqref{eq:upperbound} and $\Lambda~\sim~m_\pi$ is the typical range of the interaction between the components;
\item Instead of being taken as the inverse of the spread of the spatial wave function, the maximum value for the relative momentum should be given by the inverse of the range of the interaction, $k_0^\text{max}\simeq c\Lambda$, with $c=\mathcal{O}(1)$.
\end{enumerate}
By setting $k_0=2.7\,\Lambda\simeq 360$\mev one can increase the theoretical cross section up to $32$ nb, which is in agreement  with the experimental value.

\begin{figure}[t]
\centering
\includegraphics[width=\textwidth]{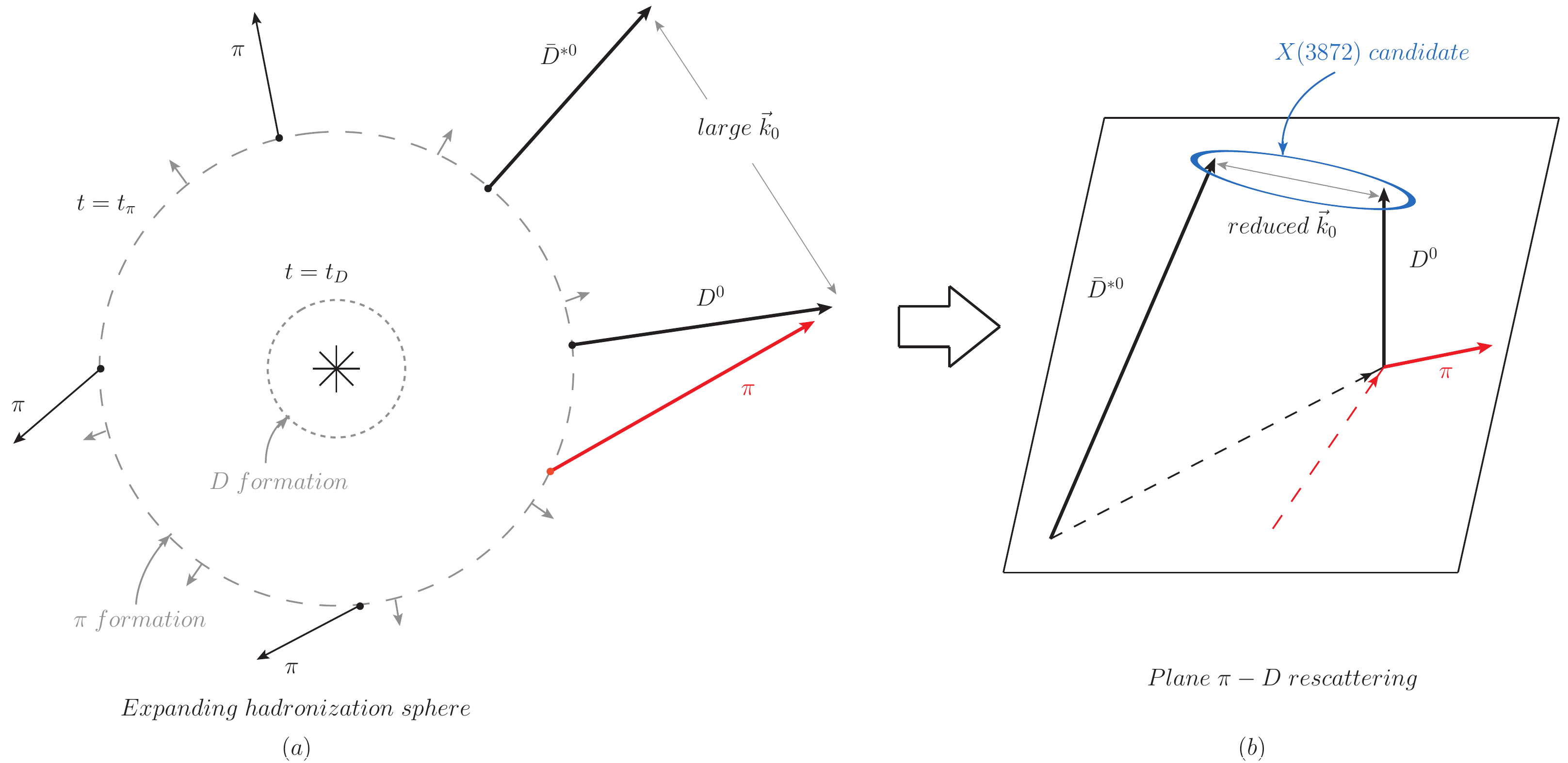}
\caption{Pictorial representation of the rescattering mechanism. After the main high-energy interaction has taken place, the final state particles can be thought of  as belonging to an expanding sphere. The hadronization time of a certain particle goes as $t_\text{hadr}\propto1/m$. Therefore the $D$ mesons hadronize at an earlier time $t_D$ whereas pions hadronize at a later time $t_\pi$ (dotted and dashed spheres respectively). In figure (a) the $D^0\bar D^{*0}$ pair starts with a large relative momentum $\bm k_0$. However, the $D^0$ might interact with one of the comoving pions (red arrow). The $\pi-D$ rescattering (figure (b)) can deviate the $D^0$ and reduce the relative momentum $\bm k_0$ thus producing a possible $X(3872)$ candidate.} \label{fig:rescattering}
\end{figure}
This approach, left alone, has some flaws~\cite{Bignamini:2009fn}: it can be shown that the use of Eq.~\eqref{eq:FSI} should enhance the occurrence of a new hypothetical molecule, the $D_s^+ D_s^{*-}$, which otherwise would be suppressed, as one could infer by looking at data on $D_s$ production at Tevatron~\cite{Acosta:2003ax}. In fact, the theoretical production cross section for this $X_s$ would be $\sigma\simeq 1\div 3$ nb and  should be detected by the \cdf experiment. No hint for such a particle has been found. Furthermore, the applicability of the Migdal-Watson theorem requires that, \emph{i)} the two final particles should be in an $S$-wave state and \emph{ii)} they should be free to interact with each other up to relative distances comparable to the interaction range.

The inclusion of relative momenta up to $k_0^\text{max}\simeq 360$\mev means to include relative orbital angular momenta up to $\ell\sim k_0^\text{max}/m_\pi\simeq 2\div 3$, thus violating the hypothesis \emph{i)}. Moreover, using again the MC softwares HERWIG and PYTHIA, one can show~\cite{Bignamini:2009fn} that in high energy collisions, such as those occurring at Tevatron and LHC, there are on average $2\div 3$ more hadrons having a relative momentum with respect to one of the two components smaller that $100$\mev; this extra hadron `pollution' challenges  the hypothesis \emph{ii)}. 

Even though the presence of other hadrons (mainly pions) surrounding the system might not allow the use of FSI, it might have  played an important role at explaining the unnaturally high prompt production of the $X(3872)$. 

It has been proposed~\cite{Esposito:2013ada} that the possible elastic scattering of  ``comoving'' pions with one of the components of the molecule might decrease their relative momentum, hence increasing the number of would-be molecules. The idea is that the interaction might push the pair both to higher and to lower values of $k_0$.  However, since the majority of would-be molecules are produced with high relative momenta, even if a small fraction of them would be pushed to smaller momenta, that could cause a feed-down of pairs towards the lower bins of the distribution, where the $X(3872)$ candidates should be found. For a pictorial representation of the considered rescattering mechanism see \figurename{~\ref{fig:rescattering}}.

It is worth noting that, if we assume  the initial total energy $E$ of the pair to be positive, the decrease in $k_0$ due to  elastic scattering may even bring $E$ to negative values, hence assuring the binding of the molecule  in a deuteron-like state (see \sectionname{\ref{xdeut}}). Therefore, in this model the $X(3872)$ would be a genuine, negative energy $D^0\bar D^{*0}$ bound state, whose lifetime would be entirely regulated by the lifetime of its shorter lived component, the $\bar D^{*0}$. Hence, this mechanism also predicts a narrow width, $\Gamma_X\sim\Gamma_{D^*}\simeq 65\kev$. Thus, when considering the interactions with comoving pions there are two possibilities
\begin{enumerate}
\item The energy $E$ in the center of mass is decreased so that to meet the condition $E\sim B$, where $B$ is the close-to-threshold discrete level discussed in \sectionname{\ref{shallowbs}};
\item Because of the interaction with a third body (the pion) of one of the constituents of the would-be-molecule, $E$ becomes small and negative: a deuteron-like state is formed.
\end{enumerate}

The interaction in the final state of the molecular constituents with pions  has been studied in~\cite{Guerrieri:2014gfa,Esposito:2013ada} with MC methods. The recipe used is as follows: the 10 most coplanar pions to the $D^0\bar D^{*0}$ plane are selected, then the pion which will interact with (say the $D^0$) is randomly chosen and lastly the most parallel pion to the non-interacting meson (say the $\bar D^{*0}$) is selected. One expects this configuration to be the most effective in physical events. Moreover, in order to prevent that $D$ mesons belonging to different jets (separated in coordinate space) would get closer by the scattering with a hard pion, one also requires $\Delta R_{DD^*}\equiv \sqrt{{(\Delta y_{DD^*})}^2+{(\Delta \phi_{DD^*})}^2}<0.7$. 

It has been checked~\cite{Guerrieri:2014gfa} that this mechanism does not spoil the high energy behavior of the relevant $D$ meson distribution. It was actually showed that the inclusion of one elastic scattering improves the agreement of the simulation with  experimental data. This is a strong hint of the fact that this mechanism actually takes place in real physical events and should hence be considered when studying final hadronic distributions.

As one can see from \figurename{~\ref{fig:markov}}, the proposed mechanism is actually effective in feeding down the lower $k_0<50$\mev bin. It is also possible to estimate how many of these interactions may take place. In particular, considering a model where all the produced hadrons are flying away from each other on the surface of a sphere  and taking into account the range of the interaction, one finds~\cite{Esposito:2013ada} that the simulations suggest an average of 3 scatterings per event. These consecutive interactions can be reproduced by implementing a Markov chain~\cite{Esposito:2013ada}.  
\begin{figure}
\centering
\includegraphics[width=.6\textwidth]{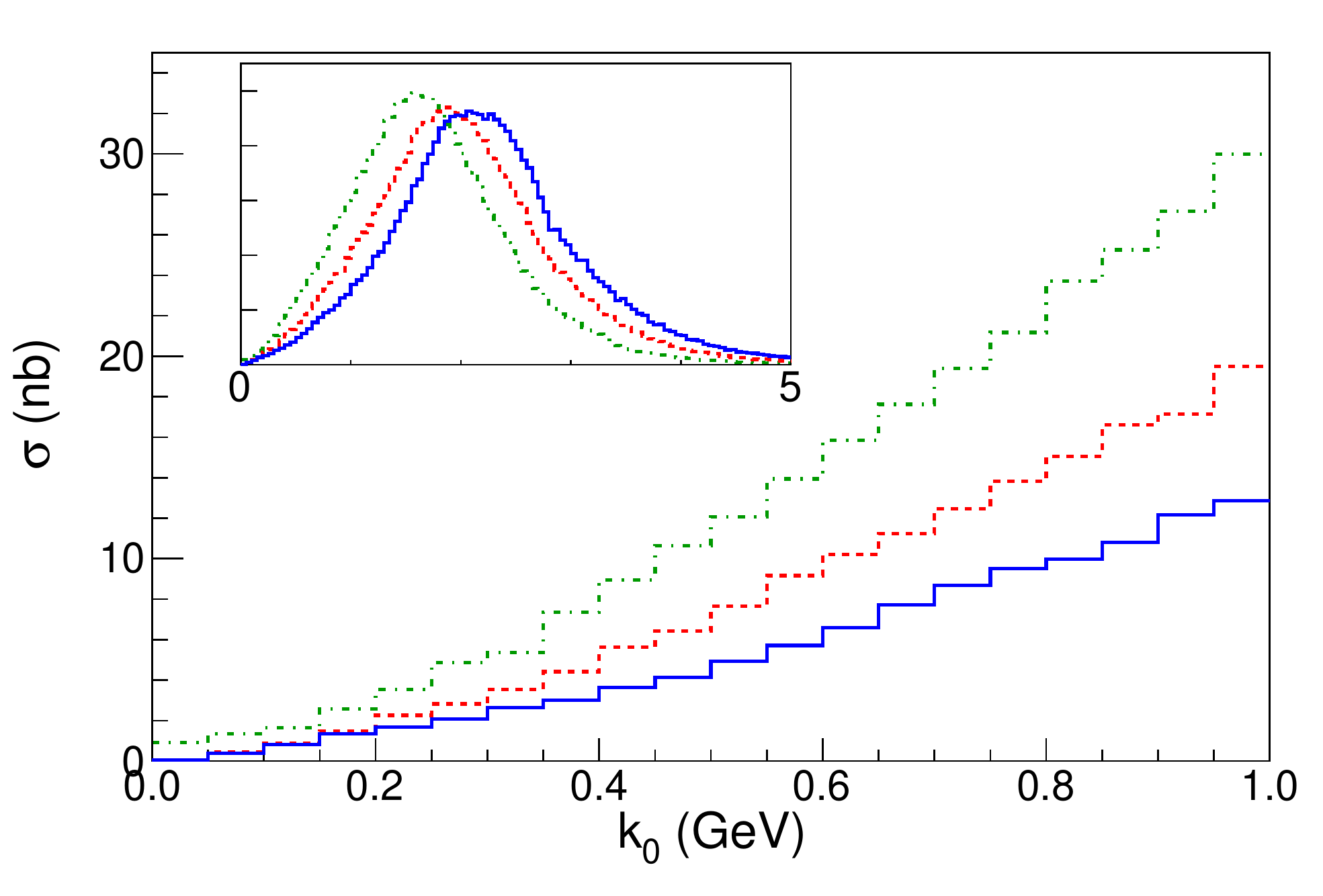}
\caption{Integrated cross section of $D^0\bar D^{*0}+h.c.$ pairs at \cdf obtained with HERWIG, without (blue, solid), with one (red, dashed) and with three (green, dot-dashed) interactions with pions, from~\cite{Guerrieri:2014gfa}. In the inset the same plot on a wider range of $k_0$ values.} \label{fig:markov}
\end{figure}

If one trusts the coalescence model for the $X(3872)$ and hence consider $k_0^\text{max}\simeq 50$\mev, not even the elastic scattering with three consecutive pions is able to enhance the production cross section up to the experimental one ($\sigma\simeq 30$ nb). Moreover, if one considers the use of FSI~\cite{Artoisenet:2010uu,Artoisenet:2009wk} as explained previously, then it should be $k_0^\text{max}\simeq 360$\mev. With this integration region, the simulations produce a cross section after the interaction with one pion --- and after a rescaling needed to take into account the different normalization factors between the two works~\cite{Artoisenet:2010uu,Guerrieri:2014gfa}-- that is equal to $\sigma(1\pi)\simeq 52$ nb, even larger than the experimental one!

In \tablename{\ref{tab:markov}} we report the values of the integrated cross section for the production of the $X(3872)$ varying both the number of interacting pions and the maximum $k_0$ allowed for the pair.
\begin{table}[t]
\centering
{\begin{tabular}{c|ccc}
\hline\hline
$k_0^\text{max}$ & 50\mev & 300\mev & 450\mev \\
\hline
$\sigma(0\pi)$ & 0.06 nb & 6 nb & 16 nb 	\\
$\sigma(1\pi)$ & 0.06 nb & 8 nb & 22 nb \\
$\sigma(3\pi)$ & 0.9 nb & 15 nb & 37 nb \\
\hline\hline
\end{tabular} 
\caption{Effect of multiple scatterings in $X(3872)$ production cross section. $k_0^\text{max}$ indicates the integration region $0<k_0<k_0^\text{max}]$.}
\label{tab:markov}}
\end{table}

To summarize, the experimental value of the prompt production cross section of the $X(3872)$ casted serious doubts on a simplified interpretation in terms of a $D^0\bar D^{*0}$ molecule. According to the expectations following from the phenomenological coalescence model, the production of such a weakly bound state should be strongly suppressed in high energy collisions. Even though many ideas and models have been proposed during the years none of them has successfully reconciled the theoretical expectations with the experimental results.

It should also be emphasized that the inclusion of possible interactions between comoving pions and final state mesons~\cite{Esposito:2013ada,Guerrieri:2014gfa} turned out to improve the accordance between the simulated MC distributions and the experimental ones.

\section{Diquarks in SU(N) \label{appN}}
Consider two quarks interacting through the exchange of one virtual gluon in $N=3$ QCD as in \figurename{~\ref{dq}} 
\begin{figure}[htb!]
 \begin{center}
   \includegraphics[width=6truecm]{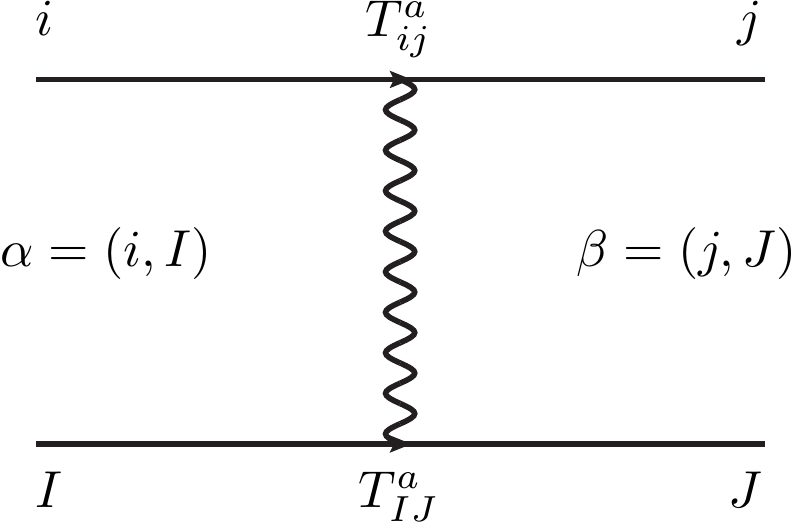}
 \end{center}
\caption{One-gluon exchange interaction.}   
\label{dq}
\end{figure}

The $T^a_{ij}T^a_{IJ}$ tensor product can be mapped into a $9\times 9$ matrix whose entries $A_{\alpha\beta}$ correspond to the 81 possible combinations of initial and final colors as in \figurename{~\ref{dq}}. The $v$ eigenvectors of $A$  identify  3  antisymmetric color configurations and 6 symmetric ones.  
For each $v$ the  $v^T A v$ product is a  superposition of the color diagrams in  \figurename{~\ref{dq}} 
defining amplitudes which are (anti-)symmetric under the simultaneous exchange of the colors 
$i\to I, j\to J$.

Each of these 9 color configurations is  weighted by a coefficient $h$, the  eigenvalue related to $v$. The $h$'s are found to be negative in the antisymmetric cases and positive in the symmetric ones: $h=-2/3$ and $h=1/3$ respectively for SU(3). 
The value of $h$ corresponds to the product of charges in a abelian theory --- thus we get repulsion in the symmetric eigenchannels~\footnote{To mean $i=I,j=J$.} and attraction in the antisymmetric ones. 

The eigenvalues $h$ are more conveniently computed through the quadratic Casimirs of the irreducible representations $\bm S_i$ obtained from the Kronecker decomposition of the product $\bm R_1\otimes \bm R_2=\bm S_1\oplus \bm S_2\oplus \ldots$. In the case of quark-quark interaction in SU(3), $\bm R_1= \bm R_2=\bm 3$ and $\bm S_1=\bar{\bm 3}$, $\bm S_2=\bm 6$. The formula for the eigenvalues $h_i$ in the various eigenchannels is in general
\begin{equation}
h_{i}=\frac{1}{2}(C_{\bm S_i}-C_{\bm R_1}-C_{\bm R_2})
\end{equation}
where $C_{\bm S_i}, C_{\bm R_1}, C_{\bm R_2}$ are the quadratic Casimirs in the 
$\bm S_i, \bm R_1, \bm R_2$ representations respectively.

In the generic case of SU(N) we have that 
\begin{equation}
\bm N\otimes \bm N=\bm{\frac{N(N-1)}{2}}\oplus \bm{\frac{N(N+1)}{2}}
\end{equation}
where $\bm{N(N-1)/2}$ is antisymmetric and $\bm{N(N+1)/2}$ is symmetric. 
 
 The Casimirs associated to these representations are given in the following table

 \begin{table}[htb!]
\centering
    \begin{tabular}{c|cc} \hline\hline
{\footnotesize Representation $\bm R$}  & {\footnotesize $C_{\bm R}$} &  {\footnotesize $h$ }\\ \hline
 {\footnotesize $\bm N$} & {\footnotesize $(N^2-1)/(2N)$} & $-$ \\ 
{\footnotesize $\bm{N(N+1)/2}$} & {\footnotesize $(N-1)(N+2)/N$} &{\footnotesize  $(N-1)/2N>0$}\\
{\footnotesize $\bm{N(N-1)/2}$} & {\footnotesize $(N+1)(N-2)/N$} & {\footnotesize $-(N+1)/2N<0$}\\
\hline\hline
\end{tabular}
 \caption{Quadratic Casimir operators for the fundamental, the two index symmetric and antisymmetric representations, in color SU(N), $N\geq 2 $. In the third column, the coefficient of the potential energy for color symmetric and antisymmetric diquarks in the one-gluon exchange approximation. Attraction in the antisymmetric channel persists at large $N$.}
\label{casimir}
\end{table}

In the singlet channel of $\bm N\otimes \overline{\bm N}$, the attraction is weighted by $h=-(N^2-1)/2N$. Therefore the singlet channel is $(N-1)$ more attractive than the antisymmetric 
$\bm{N(N-1)/2}$ channel reported in \tablename{\ref{casimir}}, in SU(3). In the one-gluon exchange approximation, the singlet channel in $q\bar q$ is (just) twice more attractive than the color antitriplet channel in $qq$.

Values in \tablesname{\ref{casimir}} can be obtained starting from formula for the diagonalization of the tensor product in SU(N)
\begin{equation}
T^a_{R_1}\otimes T^a_{R_2} = \bigoplus_{i } \frac{1}{2}(C_{S_i}-C_{R_1}-C_{R_2})\mathds{1}_{S_i}
\label{tensprodu}
\end{equation}
Take $S_1=\bm{N(N-1)/2}$, $S_2=\bm{N(N+1)/2}$ and $C_1=C_{S_1}$, $C_2=C_{S_2}$.  The trace of~\eqref{tensprodu} gives 
\begin{equation}
\frac{1}{2}x\frac{N(N-1)}{2}+\frac{1}{2}y\frac{N(N+1)}{2}=0
\label{yeqcn}
\end{equation}
where
\begin{subequations}
\begin{align}
x&=(C_1-2 C_{\bm N})\\
y&=(C_2-2 C_{\bm N})
\end{align}
\end{subequations}
and
\begin{equation}
C_{\bm N}=\frac{N^2-1}{2N}
\end{equation}

For the $\bm N$ of SU(N) we have the trace result
\begin{equation}
T^a_{ij}T^b_{jk} T^a_{kr}T^b_{ri}=\left(C_{\bm N}-\frac{1}{2} N\right)\frac{N^2-1}{2}=-\frac{1}{4}\left(\frac{N^2-1}{N}\right)
\label{da3vert}
\end{equation}
which can be written as a product of tensor products 
\begin{equation}
(T^a_{ij}T^a_{kr})(T^b_{jk} T^b_{ri})
\label{dafierzare}
\end{equation}
This can be mapped in the form 
\begin{equation}
A_{ik,jr}A_{jr,ki}
\end{equation}
which in turn becomes a trace by inverting $ki$ with $ik$.

Recalling the Fierz identity
\begin{equation}
T_{ir}^aT_{kj}^a-\frac{N-1}{2N}\delta_{ir}\delta_{kj}=\frac{N-1}{2N}\delta_{ij}\delta_{kr}-T_{ij}^a T_{kr}^a
\label{tfierzn}
\end{equation}
The term in~\eqref{dafierzare} can be rewritten as
\begin{equation}
(T^a_{ij}T^a_{kr})\left(-T^b_{ji} T^b_{rk}+\frac{N-1}{2N}\delta_{jk}\delta_{ri}\right)
\end{equation}
where the term neglected in parentheses gives zero. The latter, containing a term in  the form  $A_{ik,jr}A_{jr,ik}$ is a trace and can be re-written as 
\begin{equation}
-\frac{1}{4}x^2 \frac{N(N-1)}{2}-\frac{1}{4}y^2 \frac{N(N+1)}{2}+ \frac{N-1}{2N}\frac{N^2-1}{2}
\end{equation}
Plugging this in the {\it lhs} of~\eqref{da3vert} we get
\begin{equation}
-\frac{1}{4}x^2 \frac{N(N-1)}{2}-\frac{1}{4}y^2 \frac{N(N+1)}{2}=-\frac{1}{4}(N^2-1)
\end{equation}
From 
\begin{equation}
y=-x\,\frac{N-1}{N+1}
\label{rappyx}
\end{equation}
as obtained by~\eqref{yeqcn}, we finally get 
\begin{equation}
x^2\left( \frac{N(N-1)}{2} + \frac{N(N-1)^2}{2(N+1)}\right)=(N^2-1)
\end{equation}
with solutions
\begin{equation}
x=\pm \frac{N+1}{N}
\end{equation}
The negative sign solution gives 
\begin{equation}
C_1=\frac{(N+1)(N-2)}{N}
\end{equation}
In the case of SU(3), $R_1=\bar{\bm 3}$ and
\begin{equation}
C_1=C_{\bar{\bm 3}}=\frac{4}{3} 
\end{equation}
Computing  $y$ from~\eqref{rappyx} we get
\begin{equation}
C_2=\frac{(N-1)(N+2)}{N}
\end{equation}
In the case of SU(3) $R_2=\bm 6$ and 
\begin{equation}
C_2=C_{\bm 6}=\frac{10}{3}
\end{equation}

\section{Interaction Hamiltonian between open and closed channels in strong interactions \label{feshphen}}
We will now describe the formalism used to derive the interaction Hamiltonian between the open and closed channels. It first appeared in cold atom physics in~\cite{Fano:1961zz,Feshbach:1962ut} (see~\cite{leggett,petick} for a textbook treatment).
Suppose that ${\cal Q}$ is is the projection operator on the space of diquarkonia $\Psi_d$ and ${\cal P}$ is the one on the space of open charm/beauty meson-meson free states $\Psi_m$. Let $\Psi_m$ be the molecular threshold closer, from below, to the mass of the $\Psi_d$ state.

Assume the orthogonality of $P$ and $Q$ spaces 
\begin{equation}
{\cal Q}{\cal P}={\cal P}{\cal Q}=0
\label{orton}
\end{equation}
Let $H$ be the strong interaction Hamiltonian which determines the dynamics of the hadronization state $\Psi$ in~\eqref{statohadr}, which we can more shortly write as $\Psi=\Psi_Q+\Psi_P=({\cal P}+{\cal Q})\Psi$, since $({\cal P}+{\cal Q})=\mathds{1}$. Then
\begin{equation}
H\,\Psi=E\,\Psi
\end{equation}
can be projected into~\cite{Feshbach:1962ut,petick}
\begin{subequations}
\begin{align}
\label{psip}
(E-H_{PP})\Psi_P&=H_{PQ}\Psi_Q\\
(E-H_{QQ})\Psi_Q&=H_{QP}\Psi_P
\label{psiq}
\end{align}
\end{subequations}
with the notation of compact hadron $\Psi_Q$ states (diquarkonia) and free meson-meson states $\Psi_P$.%
The $H_{PQ}$ and $H_{QP}$ Hamiltonians represent the couplings between states in the $P$ and $Q$ sub-spaces
\begin{equation}
H_{PP}={\cal P}\, H{\cal P}\quad\quad H_{QQ}={\cal Q}\, H{\cal Q}\quad\quad H_{PQ}={\cal P}\, H{\cal Q}\quad\quad H_{QP}=H_{PQ}^\dagger={\cal Q}\, H{\cal P}
\end{equation}
Eq.~\eqref{psiq} has the formal solution (compare to~(\ref{ls}))
\begin{equation}
\Psi_Q=\frac{1}{E-H_{QQ}+i\epsilon} H_{QP}\Psi_P
\label{solpsiq}
\end{equation}
Despite the $+i\epsilon$ prescription, we do not actually distinguish between in- and out- ($\Psi^{\pm}$) scattering states as we assume to be at very low wave numbers $k$ where there are no $e^{\pm ikr}$ factors in the wave-mechanics description. The solution~\eqref{solpsiq} can be plugged back into~\eqref{psip} giving
\begin{equation}
(E-H_{PP}-H^\prime_{PP})\Psi_P=0
\end{equation}
where 
\begin{equation}
H_{PP}=H_0+V_1
\label{v12}
\end{equation}
$H_0$ is  the kinetic energy of the relative motion and $V_1$ are interactions in the $P$ space which are present also in absence of $Q$ levels. $H^\prime_{PP}$ interactions are  due instead to a spectrum of compact hadron states (diquarkonia) in the $Q$ space. The formal expression of $V_I$ is
\begin{equation}
V_I\equiv H^\prime_{PP}=H_{PQ}\frac{1}{E-H_{QQ}+i\epsilon}H_{QP}
\end{equation}
and represents an effective interaction in the $P$ space which   exists only because of discrete levels in the $Q$ space (one is enough!) being immersed in the continuum spectrum of $P$. It is the existence of such a discrete level which allows $P\to Q\to P$ 
transitions  regulated by a non-local potential in the space of  meson-meson states~\footnote{At a higher order of perturbation theory it is found
\begin{equation}
H^\prime_{QQ}=H_{QP}\frac{1}{E-H_{PP}+i\epsilon}H_{PQ}
\end{equation} 
which means that the interaction between the two channels shifts the levels in the closed channel.
}.  $H_{PQ}$ and $H_{QP}$ might be very small, as can be inferred from the discussion of the large-$N$ QCD expansion of \sectionname{\ref{sec:nonplanar}}. It is only the tuning of $E$ with the diquarkonium level which might abruptly enhance $V_I$ and produce the observed resonance effect.

Because of this, scattering in  $P$ space has a term in the $T$-matrix  
\begin{equation}
T_{\alpha\alpha}\sim \frac{|\langle \Psi_n|H_{QP}\,\Psi_\alpha\rangle|^2}{E_\alpha-E_n}
\label{boundfesh}
\end{equation}
where
\begin{equation}
E_n\Psi_n=H_{QQ}\Psi_n
\end{equation}
are the levels describing diquarkonia in $Q$ space. Eq.~(\ref{boundfesh}) should be confronted to Eq.~(\ref{bound}) noting that in the case of Eq.~(\ref{bound}) the discrete level is found in the negative energy spectrum of some potential $V$ whereas in the case of~(\ref{boundfesh}) the discrete level is {\it on the same side} of the continuum spectrum, but {\it pertains to a different potential} with respect to that giving the onset of continuum levels. This strongly differentiates our metastable state from a true resonance. The latter is above threshold but belongs to the same potential as for the two-meson state. 

$H_{QQ}$ can be described in the constituent quark model picture   by
\begin{equation}
H_{QQ}=H_0+V_2
\end{equation}
where $H_0$ is the kinetic term of diquarks --- compare with~\eqref{v12}.

\section{The effective hadroquarkonium Hamiltonian} \label{app:Heff}

In this section we will give a more rigorous derivation of the effective Hamiltonian~\eqref{Heffhadroquark} for the interaction between a compact quarkonium state and the gluonic field generated by the light matter around it. We will mostly follow the procedure in~\cite{Gottfried:1977gp,Voloshin:1978hc}. The full Hamiltonian of a system with an heavy quark pair $Q\bar Q$ and a gluonic field can be written as
\begin{align}
H=H_Q+H_g+H_\text{int}\equiv H_0+H_\text{int}
\end{align}
where $H_Q$ and $H_g$ describe the $Q\bar Q$ and the gluonic system in absence of their mutual interaction. The QCD multipole expansion for $H_\text{int}$ produces the leading terms~\cite{Gottfried:1977gp,Voloshin:1978hc,Peskin:1979va,Bhanot:1979vb}
\begin{align}
H_1=T^aV^a\quad\text{ and }\quad H_2=-\frac{1}{2}\Delta T^a\bbb r\cdot \bbb E^a
\end{align}
with $T^a=T^a_Q+T^a_{\bar Q}$, $\Delta T^a=T^a_Q-T^a_{\bar Q}$ and $\bbb E^a$ is the chromoelectric field generated by the light matter. The exact form of $V^a$ does not concern us as it will not give any contribution. $H_1$ corresponds to the ``charge" operator while $H_2$ to the dipole. It should also be noted that both $V^a$ and $\bbb E^a$ only act on the light degrees of freedom.

Consider now two states, $|n_1\rangle$ and $|n_2\rangle$ composed by a quarkonium, $|\psi_i\rangle$, and a light state, $|\phi_i\rangle$. In absence of interaction the complete state is simply the tensor product $|n_i\rangle=|\psi_i\rangle|\phi_i\rangle$. The amplitude $n_1\to n_2$ is then given by
\begin{align}
\mathcal{A}=\sum_{k=0}^\infty\langle n_1|H_\text{int}\frac{1}{E_{n_1}-H_0}\left(H_\text{int}\frac{1}{E_{n_1}-H_0}\right)^kH_\text{int}|n_2\rangle\approx \langle n_1|H_\text{int}\frac{1}{E_{n_1}-H_0}H_\text{int}|n_2\rangle=\langle n_1|H_\text{2}\frac{1}{E_{n_1}-H_0}H_\text{2}|n_2\rangle
\end{align}
where the first order transition is forbidden by color and in the last step we used the fact that $H_1$ always vanishes on color singlet states. We can now introduce a complete sum over the eigenstates of $H_0$, $|m\rangle=|m_Q\rangle|m_g\rangle$, where $|m_Q\rangle$ is a quarkonium eigenstate of $H_Q$ and $|m_g\rangle$ is a gluonic excitation, eigenstate of $H_g$. We obtain
\begin{align}
\mathcal{A}&=\sum_m\langle n_1|\frac{1}{2}\Delta T^a r_iE_i^a\frac{|m\rangle\langle m|}{E_{n_1}-H_0}\frac{1}{2}\Delta T^b r_jE_j^b|n_2 \rangle \notag \\
& = \sum_{m_Q,m_g}\langle \phi_1|E_i^a| m_g\rangle\langle m_g|E_j^b| \phi_2\rangle \langle \psi_1|\frac{1}{2}\Delta T^ar_i\frac{| m_Q\rangle\langle m_Q|}{E_{n_1}-E_{m_Q}-E_{m_g}}\frac{1}{2}\Delta T^b r_j|\psi_2\rangle
\end{align}
Note that since $\Delta T^a$ transforms a color singlet into an octet, the only states $|m_Q\rangle$ surviving in the sum are color octet. We will now make the approximation $E_{m_g}\ll E_{n_1}-E_{m_Q}$, \ie very soft gluonic mode. This is not necessarily a good approximation as shown in~\cite{Voloshin:1978hc} but it considerably simplifies the computation. The most general result can be found in~\cite{Peskin:1979va,Bhanot:1979vb}. We then have
\begin{align}
\mathcal{A}&\approx\sum_{m_Q,m_g}\langle \phi_1|E_i^a| m_g\rangle\langle m_g|E_j^b| \phi_2\rangle \langle \psi_1|\frac{1}{2}\Delta T^ar_i\frac{| m_Q\rangle\langle m_Q|}{E_{n_1}-E_{m_Q}}\frac{1}{2}\Delta T^b r_j|\psi_2\rangle \notag \\
&=\langle \phi_1|E_i^aE_j^b|\phi_2\rangle\sum_{m_Q} \langle \psi_1|\frac{1}{2}\Delta T^a r_i \frac{|m_Q\rangle\langle m_Q|}{E_{n_1}-E_{m_Q}}\frac{1}{2}\Delta T^b r_j|\psi_2\rangle \notag \\
&=-\langle \phi_1|E_i^aE_j^b|\phi_2\rangle \langle \psi_1|\frac{1}{2}\Delta T^a r_i G_{(8)}\frac{1}{2}\Delta T^b r_j|\psi_2\rangle
\end{align} 
In the last step we used the fact that $G_{(8)}=\sum_{m_Q}\frac{|m_Q\rangle\langle m_Q|}{E_{m_Q}-E_{n_1}}$ is the Green's function of a heavy quark pair in a color octet configuration. Since the states $|\phi_i\rangle$ are both color singlets, color and Lorentz invariance imply
\begin{align}
\langle \phi_1|E_i^aE_j^b|\phi_2\rangle=\frac{1}{8}\delta_{ab}\frac{1}{3}\delta_{ij}\langle \phi_1|\bbb E^c\cdot \bbb E^c|\phi_2\rangle
\end{align}
and therefore
\begin{align}
\mathcal{A}&\approx-\frac{1}{96}\langle \phi_1|E_j^bE_j^b|\phi_2\rangle \langle \psi_1|\Delta T^a r_i G_{(8)}\Delta T^a r_i|\psi_2\rangle
\end{align} 
Therefore, if we integrate out the quarkonium states and only leave the effective Hamiltonian for the light degrees of freedom, we arrive again to Eq.~\eqref{Heffhadroquark}
\begin{align} 
H_\text{eff}=-\frac{1}{2}\alpha^{(\psi_1\psi_2)}E_i^aE_i^a
\end{align}
where we defined the chromo-polarizability as
\begin{align} \label{alpha}
\alpha^{(\psi_1\psi_2)}=\frac{1}{48}\langle \psi_1|\Delta T^a r_i G_{(8)}r_i\Delta T^a|\psi_2\rangle
\end{align}
Note that, if $\psi_1 = \psi_2 = \psi$, $\alpha^{(\psi\psi)}$ is positive.

\bibliographystyle{elsarticle-num}
\bibliography{quattro}

\end{document}